\documentclass{illcdiss}

\usepackage[pdfpagelabels, bookmarks, pdftitle={Cryptography in a Quantum World}, pdfauthor={Stephanie Wehner}, pdfsubject={PhD Thesis}, pdfkeywords={quantum, quantum computing, quantum computation, cryptography, noisy-storage model}]{hyperref}

\usepackage{latexsym}
\usepackage{amsmath}
\usepackage{amssymb}
\usepackage{amsfonts}
\usepackage{ifthen}
\usepackage{revsymb}
\usepackage{yfonts}
\usepackage{graphicx}
\usepackage{backref}
\usepackage{longtable}
\usepackage{mathrsfs}

\def\Real{\mathbb{R}}
\def\Complex{\mathbb{C}}
\def\Natural{\mathbb{N}}

\newenvironment{sdp}[2]{
\smallskip
\begin{center}
\begin{tabular}{ll}
#1 & #2\\
subject to 
}
{
\end{tabular}
\end{center}
\smallskip
}
\def\01{\{0,1\}}

\newcommand{\floor}[1]{\lfloor{#1}\rfloor}
\newcommand{\eps}{\varepsilon}
\newcommand{\ket}[1]{|#1\rangle}
\newcommand{\bra}[1]{\langle#1|}
\newcommand{\outp}[2]{|#1\rangle\langle#2|}
\newcommand{\inp}[2]{\langle{#1}|{#2}\rangle} 
\newcommand{\ev}[2]{\langle #1#2 \rangle} 
\newcommand{\ketbra}[2]{|#1\rangle\langle#2|}
\newcommand{\braket}[2]{\langle #1|#2\rangle}

\newcommand{\Tr}{\mbox{\rm Tr}}

\newcommand{\assign}{:=}
\newcommand{\ol}[1]{\overline{#1}}
\newcommand{\set}[1]{\{#1\}}

\newcommand{\aoq}{X_s^a} 
\newcommand{\boq}{Y_t^b}
\newcommand{\tildeaoq}{\tilde X_s^a} 
 
\newcommand{\tildeboq}{\tilde Y_t^b}

\newcommand{\norm}[1]{\mbox{$\parallel{#1}\parallel$}}
\newcommand{\diag}{\mbox{\rm diag}}
\newcommand{\id}{\mathbb{I}}
\newcommand{\1}{\id}

\newtheorem{claim}{Claim}

\newenvironment{proof}
{\noindent {\bf Proof. }}
{{\hfill $\Box$}\\
 \smallskip}

\newcommand{\Tau}{\mathcal{T}}
\newcommand{\mA}{\mathcal{A}}
\newcommand{\mB}{\mathcal{B}}
\newcommand{\mV}{\mathcal{V}}
\newcommand{\mU}{\mathcal{U}}
\newcommand{\mM}{\mathcal{M}}
\newcommand{\mP}{\mathcal{P}}
\newcommand{\mY}{\mathcal{Y}}
\newcommand{\mN}{\mathcal{N}}
\newcommand{\hil}{\mathcal{H}}
\newcommand{\hilbert}{\mathcal{H}}
\newcommand{\ens}{\mathcal{E}}

\newcommand{\mE}{\mathcal{E}}

\newcommand{\mI}{\mathcal{I}}
\newcommand{\mX}{\mathcal{X}}
\newcommand{\mS}{\mathcal{S}}

\newcommand{\Y}{|\mY|}

\newcommand{\algA}{\mathscr{A}}
\newcommand{\algB}{\mathscr{B}}
\newcommand{\algZ}{\mathscr{Z}}
\newcommand{\setA}{\mathcal{A}}
\newcommand{\setB}{\mathcal{B}}
\newcommand{\setZ}{\mathcal{Z}}
\newcommand{\setS}{\mathcal{S}}
\newcommand{\setT}{\mathcal{T}}
\newcommand{\setY}{\mathcal{Y}}
\newcommand{\setF}{\mathcal{F}}
\newcommand{\setI}{\mathcal{I}}
\newcommand{\setX}{\mathcal{X}}
\newcommand{\setU}{\mathcal{U}}
\newcommand{\comm}{\mbox{Comm}}

\newcommand{\bop}{\mathbb{B}}
\newcommand{\isomorph}{\cong}
\newcommand{\Sp}{\mbox{Sp}}
\newcommand{\rank}{\mbox{rank}}
\newcommand{\dfdas}{\stackrel{\textrm{\tiny def}}{=}}
\newcommand{\spann}{\mathop{\mathrm{span}}\nolimits}  
\newcommand{\pr}{\prime}
\newcommand{\Comm}{\mbox{Comm}}

\newcommand{\mub}{b}
\newcommand{\Cn}{\mathbb{C}}
\newcommand{\N}{\mbox{N}}
\newcommand{\MOLS}{\mbox{MOLS}}
\newcommand{\U}{\textrm{U}}

\newcommand{\I}{\mathcal{I}}

\newcommand{\proj}[1]{\ket{#1}\bra{#1}}

\newcounter{protoCount}
\newcounter{protoList}
\newsavebox{\tmpbox}
\newlength{\protobox}
\newenvironment{protocol}[2]{
\bigskip
\addtocounter{protoCount}{1} \noindent \begin{lrbox}{\tmpbox}
\setlength{\protobox}{0.999\textwidth} \addtolength{\protobox}{-0.5cm}
\begin{minipage}[c]{\protobox}
\begin{bfseries}Protocol \theprotoCount: #1\end{bfseries}
\ifthenelse{\equal{#2}{\empty}}{}{\\Prerequisite: #2}
\begin{list}{\begin{bfseries}\arabic{protoList}:\end{bfseries}}
{\usecounter{protoList}} }{
\end{list}
\end{minipage}\end{lrbox}
\fbox{\usebox{\tmpbox}}
}

\usepackage{makeidx}     
\usepackage{appendix}
\setlongtables 
\makeindex

\begin{document}
\pagestyle{plain}
\pagenumbering{roman}


{\pagestyle{empty}
\newcommand{\printtitle}{%
{\Huge\textsf{Cryptography in a Quantum World}}}

\begin{titlepage}
\vspace*{-2cm}
\begin{flushleft}
\hspace*{-0.5cm}\printtitle
\end{flushleft}
\par\vskip 1.5cm
\begin{figure}[h!]
\begin{center}
\hspace*{-1.5cm}
\end{center}
\end{figure}
\vfill
\begin{flushright}
{\Large\textsf{Stephanie Wehner}}                           
\end{flushright}
\end{titlepage}

%
%
\mbox{}\newpage
\setcounter{page}{1}

\par\vskip 2cm
\begin{center}
\printtitle
\end{center}

\clearpage
\par\vskip 2cm
\begin{center}
ILLC Dissertation Series DS-2008-01
\par\vspace {4cm}
\illclogo{10cm}
\par\vspace {4cm}
\noindent%
For further information about ILLC-publications, please contact\\[2ex]
Institute for Logic, Language and Computation\\
Universiteit van Amsterdam\\
Plantage Muidergracht 24\\
1018 TV Amsterdam\\
phone: +31-20-525 6051\\
fax: +31-20-525 5206\\
e-mail: {\tt illc@science.uva.nl}\\
homepage: {\tt http://www.illc.uva.nl/}
\end{center}

\clearpage
\par\vskip 2cm
\begin{center}
\printtitle
\par\vspace {6cm}
{\large \sc Academisch Proefschrift}
\par\vspace {1cm}
{\large ter verkrijging van de graad van doctor aan de\\
Universiteit van Amsterdam\\
op gezag van de Rector Magnificus\\
prof.mr. P.F. van der Heijden\\                                 
ten overstaan van een door het college voor\\
promoties ingestelde commissie, in het openbaar\\
te verdedigen in de Aula der Universiteit \\        
op woensdag 27 februari 2008, te 14.00 uur \\ }        
\par\vspace {1cm} {\large door}
\par \vspace {1cm} 
{\Large Stephanie Dorothea Christine Wehner}                        
\par\vspace {1cm} 
{\large geboren te W\"urzburg, Duitsland.}
\end{center}

\clearpage
\noindent%
\begin{tabbing}
\hspace{4cm}\=\hspace{5cm}\\
Promotor:\> prof.dr.\ H.M.~Buhrman\\                      
\\
Promotiecommissie: 
\>prof.dr.ir.\ F.A.~Bais\\
\>prof.dr.\ R.J.F.~Cramer\\
\>prof.dr.\ R.H.~Dijkgraaf\\                        
\>prof.dr.\ A.J.~Winter\\
\>dr.\ R.M.~de Wolf
\end{tabbing}

\noindent
Faculteit der Natuurwetenschappen, Wiskunde en Informatica\\
Universiteit van Amsterdam\\
Plantage Muidergracht 24\\
1018 TV \ Amsterdam

\vfill

%
\noindent%
The investigations were supported by
EU projects RESQ IST-2001-37559, QAP IST 015848 and the NWO vici project 2004-2009.

\par\vspace {2cm}

%

%
\noindent%
Copyright \copyright\ 2008 by Stephanie Wehner\\[2ex] 
Cover design by Frans Bartels.\\                       
Printed and bound by PrintPartners Ipskamp.\\[2ex]           
ISBN: 90-6196-544-6


\clearpage
} 


\thispagestyle{plain}

\noindent
Parts of this thesis are based on material contained in the following papers:
\begin{itemize}

\item 
\textbf{Cryptography from noisy storage}\\
S. Wehner, C. Schaffner, and B. Terhal\\
Submitted\\
(Chapter~\ref{chapter:storageNoise})

\item 
\textbf{Higher entropic uncertainty relations for anti-commuting observables}\\
S. Wehner and A. Winter\\
Submitted\\
(Chapter~\ref{chapter:uncertainty})

\item 
\textbf{Security of Quantum Bit String Commitment depends on the information measure}\\
H. Buhrman, M. Christandl, P. Hayden, H.K. Lo and S. Wehner\\
In Physical Review Letters, 97, 250501 (2006)\\
(long version submitted to Physical Review A)\\
(Chapter~\ref{chapter:limitations})

\item
\textbf{State Discrimination with Post-Measurement Information}\\
M. Ballester, S. Wehner and A. Winter\\
To appear in IEEE Transactions on Information Theory\\
(Chapter~\ref{chapter:pistar})

\item
\textbf{Entropic uncertainty relations and locking: tight bounds for mutually unbiased bases}\\
M. Ballester and S. Wehner\\
In Physical Review A, 75, 022319 (2007)\\
(Chapters~\ref{chapter:uncertainty} and~\ref{chapter:locking}) 

\item
\textbf{Tsirelson bounds for generalized CHSH inequalities}\\
S. Wehner\\
In Physical Review A, 73, 022110 (2006)\\
(Chapter~\ref{chapter:optimalStrategies})

\item
\textbf{Entanglement in Interactive Proof Systems with Binary Answers}\\
S. Wehner\\
In Proceedings of STACS 2006, LNCS 3884, pages 162-171 (2006)\\
(Chapter~\ref{chapter:interactiveProofs})

\end{itemize}

\vspace{2cm}
\noindent
Other papers to which the author contributed during her time as a PhD student:
\begin{itemize}

\item 
\textbf{The quantum moment problem}\\
A. Doherty, Y. Liang, B. Toner and S. Wehner\\
Submitted

\item
\textbf{Security in the Bounded Quantum Storage Model}\\
S. Wehner and J. Wullschleger\\
Submitted

\item
\textbf{A simple family of non-additive codes}\\
J.A. Smolin, G. Smith and S. Wehner\\
In Physical Review Letters, 99, 130505 (2007)

\item
\textbf{Analyzing Worms and Network Traffic using Compression}\\
S. Wehner\\
Journal of Computer Security, Vol 15, Number 3, 303-320 (2007)

\item
\textbf{Implications of Superstrong Nonlocality for Cryptography}\\
H. Buhrman, M. Christandl, F. Unger, S. Wehner and A. Winter\\
In Proceedings of the Royal Society A, vol. 462 (2071), pages 1919-1932 (2006)

\item
\textbf{Quantum Anonymous Transmissions}\\
M. Christandl and S. Wehner\\
In Proceedings of ASIACRYPT 2005, LNCS 3788, pages 217-235 (2005)
\end{itemize}

\thispagestyle{plain}
\mbox{}
\vspace{2in}
\begin{flushright}
{\Large C'est v{\'e}ritablement utile puisque c'est joli.}\\
\emph{Le Petit Prince, Antoine de Saint-Exup{\'e}ry}
\end{flushright}

\thispagestyle{plain}
\mbox{}
\vspace{2in}
\begin{center}
{\em To my brother.}
\end{center}
\tableofcontents

\acknowledgments

Research has been an extremely enjoyable experience for me, and I had the opportunity to learn many 
exciting new things. However, none of this would have been possible without the help and support
of many people.

First, I would like to thank my supervisor Harry Buhrman for our interesting discussions
and for giving me the opportunity to be at CWI which is a truly great place to work. For the freedom
to pursue my own interests, I am deeply grateful.
My time as a PhD student would have been very different without Andreas Winter, and I would
especially like to thank him for our many enjoyable discussions and conversations.
I have learned about many interesting things from him, ranging from the beautiful topic
of algebras, that I discovered way too late, to his way of taking notes which I have shamelessly
adopted. I would also like to thank him for much encouragement, without which I may not have
dared to pursue my ideas about uncertainty relations much further. Much of Chapter~\ref{goodUncertainty} is
owed to him.
I would also like thank him, as well as Sander Bais, Ronald Cramer, Robbert Dijkgraaf, and Ronald de Wolf 
for taking part in my PhD committee.

Thanks also to Ronald de Wolf for supervising my Master's thesis, which was of tremendous help to
me during my time as a PhD student. Furthermore, I would like to thank Matthias Christandl
for our fun collaborations, a great trip to Copenhagen, and the many enjoyable visits to Cambridge.
Thanks also to Artur Ekert for making these visits possible, and for the very nice visit to Singapore. I am very
grateful for his persistent encouragement, and his advice
on giving talks is still extremely helpful to me.
For many interestings dicussions and insights I would furthermore 
like to thank 
Serge Fehr, Julia Kempe, Iordanis Kerenidis, 
Oded Regev, Renato Renner and Pranab Sen, 
as well as my collaborators Manuel Ballester, Harry Buhrman, Matthias 
Christandl, Andrew Doherty,
Patrick Hayden, 
Hoi-Kwong Lo, 
Christian Schaffner, Graeme Smith, John Smolin, Barbara Terhal, Ben Toner, Falk Unger, Andreas Winter, Ronald de Wolf, and J{\"u}rg Wullschleger.
Thanks also to Nebo{\v s}ja Gvozdenovi{\'c}, Dennis Hofheinz, Monique Laurent, Serge Massar and 
Frank Vallentin for useful pointers, and to Boris Tsirelson for supplying me with copies
of~\cite{tsirel:original} and~\cite{tsirel:separated}. Many thanks also to Tim van Erven, Peter Gr{\"u}nwald,
Peter Harremoes, Steven de Rooij, and Nitin Saxena for the enjoyable time at CWI, and to Paul Vit{\'a}nyi
who let me keep his comfy armchair on which many problems were solved.

Fortunately, I was able to visit many other places during my time as a PhD student. I am grateful to 
Dorit Aharanov, Claude Cr{\'e}peau, 
Artur Ekert, Julia Kempe, Iordanis Kerenidis, Michele Mosca, Michael Nielsen, David Poulin, John Preskill, Barbara Terhal, 
Oded Regev, Andreas Winter and Andrew Yao for their generous invitations. For making my visits so enjoyable,
I would furthermore like to thank Almut Beige, Agata Branczyk, Matthias Christandl, Andrew Doherty, 
Marie Ericsson, Alistair Kay, Julia Kempe, Jiannis Pachos, Oded Regev, Peter Rohde and Andreas Winter. 

Thanks to Manuel Ballester, Cor Bosman, Serge Fehr, Sandor H{\'e}man, Oded Regev, Peter Rohde, and especially Christian Schaffner
for many helpful comments on this thesis; any remaining errors are of 
course my own responsibility.
Thanks also to Frans Bartels for drawing the thesis cover and the illustrations of Alice
and Bob. I am still grateful to 
Torsten Grust and
Peter Honeyman who encouraged me to go to university in the first place.

Finally, many thanks to my family and friends for being who they are.
\\

\noindent
Amsterdam\hfill Stephanie Wehner\\
February, 2008.


\cleardoublepage
\pagestyle{headings}
\pagenumbering{arabic}

\part{Introduction}

\chapter{Quantum cryptography}\label{crytographyIntroChapter}\index{cryptography}\label{chapter:cryptoIntro}

Cryptography is the art of secrecy. Nearly as old as the art of writing itself, it concerns
itself with one of the most fundamental problems faced by any society
whose success crucially depends on knowledge and information:
With whom do we want to share information, and when, and how much? 

\section{Introduction}\label{qkd}

Starting with the first known encrypted texts from 1900 BC in Egypt~\cite{ancientCrypto}, 
cryptography has a fascinating history~\cite{cryptoHistory}. Its goal is simple: 
to protect secrets as best as is physically possible. Following our increased understanding
of physical processes with the advent of quantum mechanics, 
Wiesner~\cite{wiesner:conjugate}
proposed using quantum techniques for cryptography in the early 1970's. 
Unfortunately,
his groundbreaking work, which contained the seed for quantum key distribution, oblivious transfer (as described below),
and a form of quantum money, was initially met with rejection~\cite{brassard:history}. 
In 1982, Bennett, Brassard, Breitbart and Wiesner joined forces to publish ``Quantum cryptography, 
or unforgeable subway tokens''\index{unforgeable subway tokens} which luckily found acceptance~\cite{bbOne}, leading
to the by now vast field of research in quantum key distribution (QKD).\index{QKD}\index{quantum key distribution} 
Quantum key distribution allows two remote
parties who are only connected via a quantum channel to generate an arbitrarily long secret
key that they can then use to perfectly shield their messages from prying eyes. The idea is beautiful 
in its simplicity: unlike with classical data, quantum mechanics prevents us from copying an unknown 
quantum state. What's more is that any attempt to extract information from such a state can be detected!
That is, we can now determine whether an eavesdropper has been trying to intercept our secrets.
Possibly the most famous QKD protocol known to date was proposed in 1983 by Bennett and Brassard~\cite{bb84:isit},
and is more commonly known as BB84\index{BB84} from its 1984 full publication~\cite{bb84}. Indeed, many quantum cryptographic
protocols to date are inspired in some fashion by BB84. It saw its first experimental implementation
in 1989, when Bennett, Bessette, Brassard, Salvail and Smolin built the first QKD setup covering
a staggering distance of 32.5 cm~\cite{bb84:experimental,bb84:experimental2}! 
In 1991, Ekert proposed a beautiful alternative view of QKD based on quantum entanglement and the violation
of Bell's theorem, leading to the protocol now known as E91~\cite{ekert:91}.\index{E91} His work paved the way
to establishing the security of QKD protocols, and led to many other interesting tasks such as 
entanglement distillation. Since then, many other protocols such as B92~\cite{b92} have been suggested.
Today, QKD and its related problems form a well-established part of quantum 
information, with countless proposals and experimental implementations. It especially saw increased
interest after the discovery of Shor's quantum factoring algorithm in 1994~\cite{shor:factoring} that
renders almost all known classical encryption systems insecure, once a quantum computer is built. 
Some of the first security proofs were
provided by Mayers~\cite{mayers:qkdproof}, Lo and Chau~\cite{lochau:qkd}, and Shor and Preskill~\cite{sp:qkdproof}, finally
culminating in the wonderful work of Renner~\cite{renato:diss} who supplied the most general framework
for proving the security of any known QKD protocol.
QKD systems
are already available commercially today~\cite{idQuantique,magicq}. The best known
experimental implementations now cover distances of up to 148.7 km in optical 
fiber~\cite{qkd:optical}, and 144 km in free space~\cite{qkd:freespace} in an experiment
conducted between two Canary islands.

\begin{figure}[h]
\begin{minipage}{0.45\textwidth}
\begin{center}
\includegraphics[width=5cm]{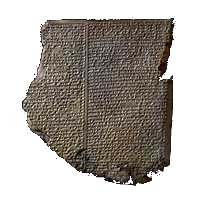}
\caption{Encrypted pottery glaze formula, Mesopotamia 1500 BC}
\end{center}
\end{minipage}
\begin{minipage}{0.45\textwidth}
\begin{center}
\includegraphics[width=5cm]{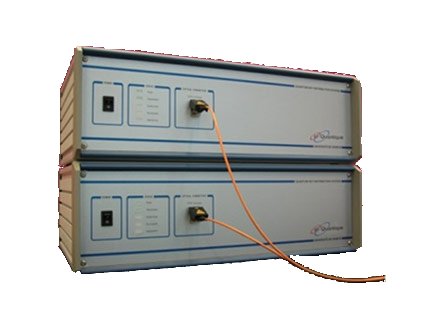}
\caption{QKD today}
\end{center}
\end{minipage}
\end{figure}

Traditional cryptography is concerned with the secure and reliable transmission of messages. 
With the advent of widespread electronic communication, however, new cryptographic tasks have become 
increasingly important. We would like to construct secure protocols for electronic voting, online auctions, contract signing and 
many other applications where the protocol participants themselves do not trust each other. Two primitives that can be used
to construct all such protocols are bit commitment and oblivious transfer. We will introduce both primitives
in detail below. Interestingly, it turns out that despite many initially suggested protocols~\cite{bbOne,crepeau:qot}, both primitives
are impossible to achieve when we ask for unconditional security. Luckily, as we will see in Chapter~\ref{chapter:noise}
we can still
implement both building blocks if we assume that our quantum operations are affected by noise. 
Here, the very problem that prevents us from implementing a full-scale quantum computer can be turned to 
our advantage. 

In this chapter, we give an informal introduction to cryptography in the quantum setting. We first introduce 
necessary terminology, before giving an overview over the most well-known cryptographic primitives. 
Since our goal is to give an overview, we will restrict ourselves to informal definitions. Surprisingly,
even definitions themselves turn out to be a tricky undertaking, especially when entering the quantum realm.
Finally, we discuss what makes the quantum setting so different from the classical one, and identify a range
of open problems. 

\section{Setting the state}

\subsection{Terminology}
In this text, we 
consider protocols among multiple participants $P_1,\ldots,P_n$, also called \emph{players}.\index{player} When considering
only two players, we generally identify them with the protagonists Alice and Bob. Each player may hold a 
\emph{private input},\index{private input} that is classical and quantum data unknown to the other players. In addition, the players
may have access to a shared resource such as classical shared randomness or quantum entanglement that has
been distributed before the start of the protocol. We will refer to any information that is available to
all players as \emph{public}.\index{public information} 
A subset of players may also have access to shared information that is known
only to them, but not to the remaining players. Such an input is called \emph{private shared input}.\index{private shared input} In the
case of shared randomness, this is also known as \emph{private shared randomness}.\index{private shared randomness}
The players can be connected by classical as well as quantum channels, and use them to exchange messages during 
the course of the protocol. A given protocol consists of a set of messages as well as a specification of
actions to be undertaken by the players. At the end of the protocol, each player may have a classical
as well as a quantum output.

A player is called \emph{honest},\index{honest} if he follows the protocol exactly as dictated. He is called \emph{honest-but-curious},\index{honest-but-curious}
if he follows the protocol, but nevertheless tries to gain additional information by processing the information supplied
by the protocol in a way which is not intended by the protocol. An honest player, for example, will simply ignore
parts of the information he is given, as he will do exactly as he is told. However, a player that is honest-but-curious
will take advantage of all information he is given, i.e., he may read and copy all messages as desired, and never forgets
any information he is given.\footnote{Note that since an honest-but-curious player never forgets any information, he effectively makes a copy of all messages. He will erase
his memory needed for the execution of the protocol if dictated by the protocol: his copy lies outside this memory.}
Yet, the execution of the protocol itself is unaffected as the player
does not change any information used in the protocol, he merely reads it. But what does this mean in a quantum setting? 
Indeed, this question appears to be a frequent point of debate.
We will see in Chapter~\ref{quantumIntro} that he cannot copy arbitrary quantum
information, and extracting non-classical information from a quantum state will necessarily lead to disturbance. 
Evidently, disturbance alters the quantum states during the protocol. Hence, the player actually took actions
to alter the execution of the protocol, and we can no longer regard him as honest. 
After examining quantum operations in Chapter~\ref{quantumIntro} we will return to the definition of an honest-but-curious player
in the quantum setting.
Finally, a player can also be \emph{dishonest}:\index{dishonest} 
he will do anything in his power to break the protocol. Evidently,
this is the most realistic setting, and we will always consider it here.

An \emph{adversary}\index{adversary} is someone who is trying to break the protocol. An adversary is generally modeled as an entity outside
of the protocol that can either be an eavesdropper, or take part in the protocol by taking control of specific players.
This makes it easier to model protocols among multiple players, where we assume that all dishonest players collaborate 
to form a single adversary.

\subsection{Assumptions}\label{sec:assumptions}

In an ideal world, we could implement any cryptographic protocol described below. Interestingly though, even in
the quantum world we encounter physical limits which prevent us from doing so with unconditional security.
\emph{Unconditional security}\index{unconditional security} most closely corresponds to the intuitive notion of ``secure''.
A protocol that is unconditionally secure fulfills its purpose and is secure even 
if an attacker is granted
unlimited resources. We happily provide him with the most powerful computer we could imagine and as much
memory space as he wants. The main question of unconditional security is thus whether the
attacker obtains enough information to defeat the security of the system. 
Unconditional security is also called \emph{perfect secrecy}\index{perfect secrecy} in the context
of encryption systems, and forms part of
\emph{information-theoretic security}\index{information theoretic security}.

Most often, however, unconditional security can never be achieved. We must therefore resign ourselves
to introducing additional limitations on the adversary: the protocol will only be secure if
certain assumptions hold. In practise, these assumptions can be divided into two big categories:
In the first, we assume that the players have access to a common resource with special properties.
This includes models such as a trusted initializer~\cite{trustedInit},\index{trusted initializer} or another source that provides the players
with shared randomness drawn from a fixed distribution. An example of this is also a noisy channel~\cite{crepeau:weakened}:\index{noisy channel}
Curiously, a noisy channel that neither player can influence too much turns out to be an incredibly powerful
resource. The second category consists of clear limitations on the ability of the adversary.
For example, the adversary may have limited storage space available~\cite{maurer:storage,serge:bounded},\index{bounded storage!classical}
or experience noise when trying to store qubits as we will see in Chapter~\ref{OTfromNoise}.
In multi-player protocols we can also demand that dishonest players cannot communicate during 
the course of the protocol, that messages between different players take a certain time to
be transmitted, or that only a minority of the players is dishonest. 
In the quantum case, other known assumptions include limiting the adversary to measure not more than
a certain number of qubits at a time~\cite{salvail:physical}, or introducing superselection rules~\cite{kitaev:super}, where the adversary
can only make a limited set of quantum measurements.\index{superselection rules}
When introducing such assumptions, we still speak of \emph{information-theoretic security}: 
Except for these limitations, the adversary remains all-powerful. In particular, he has
unlimited computational resources.

Classically, most forms of practical cryptography are shown to be \emph{computationally secure}.\index{computational security}
In this security model, we do not grant an adversary unlimited computational resources. Instead, 
we are concerned with the amount of computation required to break the security of a system.
We say that a system is \emph{computationally secure}, if the believed level of computation necessary
to defeat it exceeds the computational resources of any hypothetical adversary by a comfortable margin. 
The adversary is thereby allowed to use the best possible attacks against the system. 
Generally, the adversary
is modeled as having only polynomial computational power. This means that any attacks are restricted to time 
and space polynomial in the size of the underlying security parameters of the system. 
In this setting the difficulty of defeating the system's security is often proven to be as difficult
as solving a well-known problem which is believed to be hard.
The most popular problems are often number-theoretic
problems such as factoring. Note that for example in the case of factoring, it is not known whether
these problems are truly difficult to solve classically. Many such problems, such as factoring, fold
with the advent of a quantum computer~\cite{shor:factoring}. It is an interesting open problem
to find classical hardness assumptions, which are still secure given a quantum computer. 
Several proposals are known~\cite{oded:lattice}, but so far none of them have been proven secure.

In the realm of quantum cryptography, we are so far 
only interested in in\-forma\-tion-theo\-retic security:
we may introduce limitations on the adversary, but we do not resort to computational hardness assumptions. 

\subsection{Quantum properties}\label{section:properties}

Quantum mechanics introduces several exciting aspects to the realm of cryptography, which we can exploit
to our benefit, but which also introduce additional complications even in existing classical primitives
whose security does not depend on computational hardness assumptions. Here, we give a brief introduction to some
of the most striking aspects, which we will explain in detail later on. 

\begin{enumerate}

\item \textbf{Quantum states cannot be copied:}
In classical protocols, an adversary can always copy any messages and his classical data at will.
\index{no-cloning} Quantum states, however, differ: We will see in Chapter~\ref{chapter:quantumIntro}
that we cannot copy an arbitrary qubit. This property led to the construction of the unforgeable subway
tokens~\cite{bbOne} mentioned earlier. 

\item \textbf{Information gain can be detected:}
Classically there is no way for an honest player to determine 
whether messages have been read maliciously outside the scope of the protocol.
However, in a quantum setting we can detect whether
an adversary tried to extract information from a transmitted message. 
This property forms the heart
of quantum key distribution described below. It also allows us to construct \emph{cheat-sensitive} protocols, \index{cheat-sensitive}
a concept which is foreign to classical cryptography: even though we cannot prevent an adversary from
gaining information if he intends to do so, we will be able to detect such cheating and take
appropriate action. We will return to this aspect in Chapter~\ref{informationIntro}. 

\item \textbf{Uncertainty relations exist:}
Unlike in the classical world, quantum states allow us to encode multiple bits into a single state
in such a way that we cannot extract all of them simultaneously. This property is closely related
to cheat-sensitivity, and is a consequence of the existence of uncertainty relations\index{uncertainty relation} 
we will encounter in Chapter~\ref{chapter:uncertainty}.
It is also closely related to what is known as quantum random access codes, which will we employ in Chapter~\ref{chapter:bounding}.

\item \textbf{Information can be ``locked'':}
Another aspect we need to take into account when considering quantum protocols is an effect known as locking\index{locking}
classical information in quantum states. Surprisingly, the amount of correlation between two parties can
increase by much more than the data transmitted. We will examine this effect for a specific measure of correlation
in more detail in Chapter~\ref{chapter:locking}.

\item \textbf{Entanglement allows for stronger correlations:}
Entanglement\index{entanglement} is another concept absent from the classical realm. Whereas entanglement has many useful applications
such as quantum teleportation and can also be used to analyze the security of quantum key distribution, it also
requires us to be more cautious: In Chapter~\ref{chapter:interactiveProofs}, we will see that the parameters
of classical protocols can change dramatically if dishonest players share entanglement, even if
they do not have access to a full quantum computer. In Chapter~\ref{chapter:sc}, entanglement will 
enable an adversary to break any quantum string commitment protocol.

\item \textbf{Measurements can be delayed:}
Finally, we encounter an additional obstacle, which is also entirely missing from classical protocols: Players
may delay quantum measurements.\index{measurement!delay} In any classical protocol, we can be assured that any input and output is
fixed once the protocol ends. In the quantum case, however, players may alter their protocol input retroactively
by delaying quantum measurements that depend on their respective inputs. Essentially, in a classical protocol
the players will automatically be ``committed'' to the run of the protocol, whereas in the quantum setting
this property is entirely missing. This can make an important difference in reductions among several protocols
as we will see in Section~\ref{section:ot} below.
\end{enumerate}

\section{Primitives}

We now present an overview of the most common multi-party protocol primitives, and what is known about them
in the quantum setting. We already encountered quantum key distribution (QKD) in the 
introduction.\index{QKD}\index{quantum key distribution}
In this thesis, our focus lies on cryptographic protocols \emph{other} than QKD.

\subsection{Bit commitment}\label{sec:bcAssumptions}\index{bit commitment}

Possibly the most active area of quantum cryptography in the early stages next
to QKD was quantum bit commitment: 
Imagine two mutually distrustful parties Alice and Bob at distant
locations. They can only communicate over a channel, but want to
play the following game: Alice secretly chooses a bit $c$. Bob wants
to be sure that Alice indeed has made her choice. Yet, Alice wants
to keep $c$ hidden from Bob until she decides to reveal $c$. To
convince Bob that she made up her mind, Alice sends Bob a
commitment. From the commitment alone, Bob cannot deduce $c$. At a
later time, Alice reveals $c$ and enables Bob to open the
commitment. Bob can now check if Alice is telling the truth. This
scenario is known as \emph{bit commitment}. 
\begin{figure}[h]
\begin{center}
\includegraphics[scale=0.7]{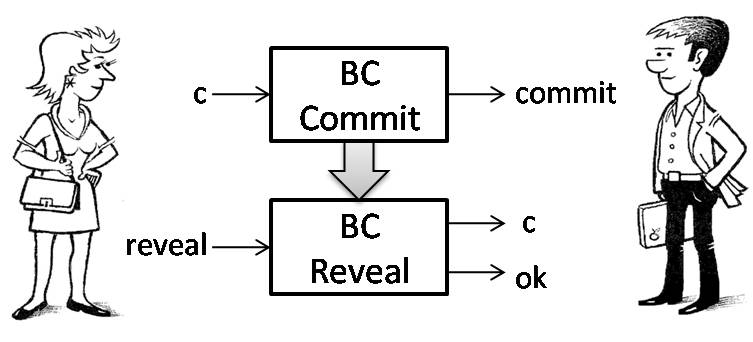}
\caption{Schematic run of a BC protocol when Alice and Bob
are honest.}
\end{center}
\end{figure}
Commitments play a
central role in modern-day cryptography. They form an important
building block in the construction of larger protocols in, for
example, gambling and electronic voting, and other instances of secure
two-party computation. In the realm of quantum mechanics, it has
been shown that oblivious transfer~\cite{crepeau:practicalOT} (defined in Section~\ref{section:ot}) can be
achieved provided there exists a secure bit commitment
scheme~\cite{yao:otFromBc,crepeau:qot}. In turn, classical oblivious
transfer can be used to perform any secure two-party
computation defined below~\cite{crepeau:committedOT}.
Commitments are also useful for constructing
zero-knowledge proofs~\cite{goldreich:book1} and lead to coin
tossing~\cite{blum:coin}. Informally, bit commitment can be defined as follows:
\begin{definition}\index{bit commitment}\index{BC}\index{commitment!bit}
\emph{Bit commitment} (BC) is a two-party protocol between Alice (the committer) and Bob (the verifier), which
consists of three stages, the committing and the revealing stage, and a final declaration stage in which
Bob declares ``accept'' or ``reject''. The following requirements should hold:
\begin{itemize}
\item (Correctness) If both Alice and Bob are honest, then before the committing stage Alice picks a bit $
c$.
Alice's protocol depends on $c$ and any randomness used. At the revealing
stage, Alice reveals to Bob the committed bit $c$. Bob accepts.
\item (Binding) 
If Alice wants to reveal a bit $c'$, then
$$
\Pr[\mbox{Bob accepts }|c'=0] + \Pr[\mbox{Bob accepts } |c'=1] \leq 1.
$$
\item (Concealing) If Alice is honest, Bob does not learn anything about $c$ before the revealing stage.
\end{itemize}
\end{definition}

Classically, unconditionally secure bit commitment
is known to be impossible. Indeed, this is very intuitive if we consider
the implications of the concealing condition: This condition implies
that exactly the same information exchange must have occurred
if Alice committed herself to $c=0$ or $c=1$, otherwise Bob would be
able to gain information about $c$. But this means that even if Alice
initially made a commitment to $c=0$, she can later reconstruct the run
of the protocol as if she had committed herself to $c=1$ and thus send
the right message to Bob to reveal $c=1$ instead.
Unfortunately, even quantum communication cannot help us to implement
unconditionally secure bit commitment without further assumptions:
After several quantum
schemes were suggested~\cite{bb84,brassard:bcAndCoin,brassard:bc}, 
quantum
bit commitment was shown to be impossible,
too~\cite{mayers:trouble,lo&chau:bitcom,mayers:bitcom,lo&chau:bitcom2,
brassard:brief,lo:promise,kretch:bc}, even in the presence of
superselection rules~\cite{kitaev:super}\index{superselection rules}, where the adversary can only
perform a certain restricted set of measurements.
In the face of the negative results, what can we
still hope to achieve?

Evidently, we need to assume that the adversary is limited in certain ways. In the classical case, 
bit commitment is possible if the adversary is \emph{computationally bounded}~\cite{goldreich:book1}, 
if \emph{one-way functions exist} \cite{Naor91,HaiRei07}, if Alice and Bob are connected via a
\emph{noisy channel}
that neither player can influence too much~\cite{crepeau:weakened,damgard:weakened,serge:unfairNoisy}, or if
the adversary is bounded in \emph{space} 
instead of time, i.e., he is only allowed to use a certain amount of storage space~\cite{maurer:storage}. 
Unfortunately, the security of the bounded classical storage model~\cite{maurer:storage,cachin:boundedOT}
is somewhat unsatisfactory:
First, a dishonest player needs only quadratically more memory than the honest one to break the security.
Second, as classical memory is very cheap, most of these protocols require huge amounts of
communication in order to achieve reasonable bounds on the adversaries memory.

Do we gain anything by using quantum communication? Interestingly, even without any further assumptions,
quantum cryptography at least allows us to implement imperfect forms of bit commitment, where Alice and
Bob both have a limited ability to cheat. That is, we allow Alice to change her mind, and Bob
to learn the committed bit with a small probability.
These protocols are based on the fact that quantum protocols
can exhibit a form of cheat sensitivity unavailable to classical communication~\cite{kent:cheatSensitive,dorit:bitEscrow}.
Exact tradeoffs on how well we can implement bit commitment in the quantum world can be found in~\cite{spekkens:tradeoffBc}.
Protocols that make use of this tradeoff are cheat-sensitive, as described in Section~\ref{sec:assumptions}.\index{cheat-sensitive}
Examples of such protocols have been used to implement coin tossing~\cite{ambainis:coin} 
as described in Section~\ref{sec:coin}.
In Chapter~\ref{chapter:sc}, we will consider commitments to an entire string of bits at once.\index{bit string commitment} Whereas
this task turns out to be impossible as well for a strong security definition, we will see that non-trivial quantum
protocols do exist for a very weak security definition.
Bit commitment can also be implemented under the assumption 
that faster than light communication is impossible, provided that Alice and Bob are located very 
far apart~\cite{kent:bc}, or if Alice and Bob are given access to non-local boxes~\cite{wehner05b} which provide
superstrong non-local correlations. 

But even a perfect commitment can be implemented, if we make quantum specific
assumptions.
For example, it is possible to securely implement BC provided that an adversary cannot 
measure more than a fixed number of qubits simultaneously~\cite{salvail:physical}. 
With current-day technology, 
it is \emph{very} difficult to store states even for a very short period of time. This
leads to the protocol presented in \cite{BBCS92,crepeau:qot}, which shows how to implement BC and OT (defined below) if the adversary is
not able to store \emph{any} qubits at all. 
In~\cite{serge:bounded,serge:new}, these ideas have been generalized in a very nice way
to the \emph{bounded-quantum-storage model},\index{bounded storage!quantum} where the adversary is computationally unbounded and is
allowed to have an unlimited amount of \emph{classical} memory. However, he is only allowed
a limited amount of \emph{quantum} memory.
The advantages over the classical bounded-storage model are two-fold:
First, given current day technology it is indeed very hard to store quantum states. Secondly, the honest
players do not require any quantum storage at all, making the protocol much more efficient.
It has been shown that such protocols remain secure when executed many times in a row~\cite{wehner07b}.

\subsection{Secure function evaluation}

An important aspect of modern day cryptography is the primitive known as secure function evaluation,
and its multi-player analogue, secure multi-party computation, first suggested by Yao~\cite{yao:sfe}.
Imagine that Alice and Bob are trying to decide whether to attend an unpopular administrative event. 
If Alice attends, Bob feels forced to attend as well and vice versa. However, neither of them wants to announce
publicly whether they are planning to attend or whether they would rather make up an excuse to remain at home, as this may have dire
consequences. How can Alice and
Bob solve their dilemma? Note that their problem can be phrased in the following form: Let $x$ be Alice's 
private input bit, where $x=1$ if Alice is planning to attend and $x=0$ if Alice skips the event.
Similarly, let $y$ be Bob's private input bit. Alice and Bob now want to compute $\mbox{OR}(x,y)$
in such a way that both of them learn the result, but neither of them learns anything more about the input
of the other player than can be inferred from the result. 
In our example, if $\mbox{OR}(x,y) = 1$, at least one of the players
is planning to attend the event. Both Alice and Bob now attend the event, and both of them can safely
claim that they really did plan to do so in the first place. If $\mbox{OR}(x,y)=0$, Alice and
Bob learn that they both agree, and do not need to fear any political consequences.

Secure function evaluation enables Alice and Bob to solve any such task. Protocols for secure
function evaluation enable us to construct protocols for electronic voting and secure auctions.
Informally, we define:
\begin{definition}\index{secure function evaluation}\index{SFE}
\emph{Secure function evaluation} (SFE) is a two-party protocol between Alice and Bob, 
where Alice holds a private input $x$ and Bob holds a private input $y$ such that
\begin{itemize}
\item (Correctness) If both Alice and Bob are honest, then they both output the same value $v = f(x,y)$. 
\item (Security) 
If Alice (Bob) is dishonest, then Alice (Bob) does not learn more about $x$ ($y$) then
can be inferred from $f(x,y)$.
\end{itemize}
\end{definition}

A common variant of SFE is so-called \emph{one-sided} SFE: Here, only one of the two players 
receives the result of the computation, $f(x,y)$.
Sadly, we cannot implement SFE for an arbitrary function $f$ classically without additional assumptions, 
akin to bit commitment.
Even in the quantum world, the situation is equally bleak: SFE remains impossible in the
quantum setting~\cite{lo:insecurity}!
Fortunately, the situations improves when we consider multi-party protocols as mentioned below.
 
\subsubsection{Oblivious transfer}\label{section:ot}

A special case of secure function evaluation is the problem of oblivious transfer, which was
first introduced by Rabin~\cite{rabin:ot}. The variant of 1-2 OT
appeared in a paper by Even, Goldreich and
Lempel~\cite{even:firstOT} and also, under a different name, in
the well-known paper by Wiesner~\cite{wiesner:conjugate}. 1-2 OT
allows Alice and Bob to solve a seemingly uninteresting problem:
The sender (Alice) secretly chooses two bits $s_0$ and $s_1$,
the receiver (Bob) secretly chooses a bit $c$. The primitive of oblivious transfer allows Bob to retrieve $s_c$
in such a way, that Alice cannot gain any information about $c$. At the same time, Alice is ensured
that Bob only retrieves $s_c$ and gets no information about the other input bit $s_{\bar{c}}$. Oblivious
transfer can be used to perform any secure two-party computation~\cite{kilian:foundingOnOT,crepeau:committedOT},
and is therefore a very important primitive.
\begin{figure}[h]
\begin{center}
\includegraphics[scale=0.7]{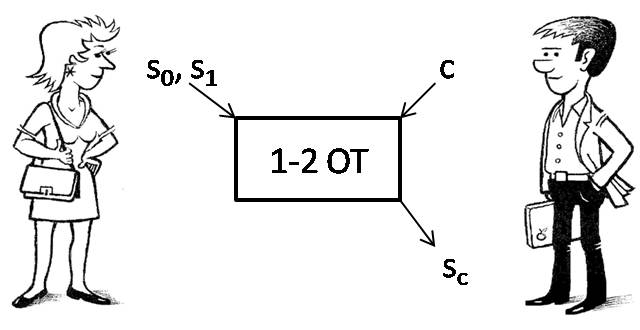}
\caption{Schematic run of a 1-2 OT protocol.}
\end{center}
\end{figure}

Unlike in the classical setting, oblivious transfer in the quantum world requires additional caution: We want that
after the protocol ends, both of Alice's inputs bits $s_0,s_1$ and Bob's choice bit $c$ have been determined. That is, they are fixed
and the players can no longer change their mind. In particular,
we do not want Bob to delay his choice of $c$ indefinitely, possibly by delaying a quantum measurement. Similarly, Alice
should not be able to change her mind about, for example, the parity of $s_0 \oplus s_1$ after the end of the protocol 
by delaying a measurement. Informally, we define

\begin{definition}\index{1-2 oblivious transfer}\index{1-2 OT}\index{oblivious transfer!1-2}\index{OT!1-2}
${2 \choose 1}$-\emph{oblivious transfer} (1-2 OT$(s_0,s_1)(c)$) is a two-party protocol between Alice (the sender)
and Bob (the receiver), such that
\begin{itemize}
\item (Correctness) If both Alice and Bob are honest, the protocol depends on Alice's two input bits $s_0,s_1 \in \01$
and Bob's input bit $c \in \01$. At the end of the protocol Bob knows $s_c$.
\item (Security against Alice) If Bob is honest, Alice does not learn $c$. 
\item (Security against Bob) If Alice is honest, Bob does not learn anything about $s_{\bar{c}}$.
\end{itemize}
After the protocol ends, $s_0,s_1$ and $c$ have been chosen.
\end{definition}
Classically, 1-2 OT can be obtained from the following simpler primitive, also known as Rabin-OT~\cite{rabin:ot} 
or erasure channel. Conversely, OT can be obtained from 1-2 OT.
\begin{definition}\index{OT!Rabin}\index{oblivious transfer!Rabin}
Rabin \emph{Oblivious transfer} (Rabin-OT) is a two-party protocol between Alice (the sender) and Bob (the receiver),
such that
\begin{itemize}
\item (Correctness) If both Alice and Bob are honest, the protocol depends on Alice's input bit $b \in \01$.
At the end of the protocol, Bob obtains $b$ with probability $1/2$ and knows whether he obtained $b$ or not.
\item (Security against Alice) If Bob is honest, Alice does not learn whether Bob obtained $b$.
\item (Security against Bob) If Alice is honest, Bob's probability of learning bit $b$ does not exceed $1/2$.
\end{itemize}
After the protocol ends, $b$ has been chosen.
\end{definition}

The fact that Alice and Bob may delay their measurements
makes an important difference, as the following simple example shows:
Consider the standard reduction of Rabin-OT to 1-2 OT:
Alice uses inputs $s_k = b$ and
$s_{\bar{k}} = 0$ with $k \in_R \01$. Bob 
uses input
$c \in_R \01$, for a randomly chosen $c$. 
The players now perform 1-2 OT$(s_0,s_1)(c)$ after
which the receiver holds $s_c$. Subsequently, Alice announces
$k$. If $k = c$, Bob succeeded in retrieving $b$ and
otherwise he learns nothing. This happens with probability
$p = 1/2$ and thus we have constructed Rabin-OT from one instance
of 1-2 OT. Clearly, this reduction fails if we use an 1-2 OT
protocol in which Bob can defer his choice of $c$, possibly by delaying a quantum measurement
that depends on $c$. He simply waits until Alice announces $k$, to retrieve $s_k$ with certainty.
This simple example makes it clear that implementing 1-2 OT is far from a trivial task in the quantum setting.
Even the classical definitions need to be revised carefully. In this brief overview, we restricted ourselves
to the informal definition given above, and refer to~\cite{Wullsc07} for an extensive treatment of the definition of
oblivious transfer.

Note that oblivious transfer forms an instance of secure function evaluation with $f: \01^2 \times \01 \rightarrow \01$
satisfying $f(s_0,s_1,c) = s_c$, where only one player (Bob) learns the output. Hence by Lo's impossibility result for SFE discussed earlier, 
oblivious transfer is not
possible in the quantum setting either without introducing additional assumptions. Indeed, note that there exists a classical
reduction of bit commitment to oblivious transfer (up to a vanishing probability), where we reverse the roles of Alice and
Bob for bit commitment: Alice simply chooses two $n$-bit strings $x_0 \in_R \01^n$, and $x_1 \in_R \01^n$.
Alice and Bob now use $n$ rounds of 1-2 OT, where Bob retrieves $x_c$ when he wants to commit to a bit $c$. To reveal,
he then sends $c$ and $x_c$ to Alice. Intuitively, one can thus hope to use the impossibility proof of bit commitment to
show that oblivious transfer is impossible as well, without resorting to~\cite{lo:insecurity}. 
However, note that we would first have to show the security of this reduction
with respect to a quantum adversary. 
Fortunately, oblivious transfer becomes possible if we make the same assumptions as for bit commitment described
in Section~\ref{sec:bcAssumptions}. We will consider
how to implement oblivious transfer if the adversary's quantum storage is subject to noise in Chapter~\ref{sec:cryptoFromNoise}.

\subsubsection{Coin tossing}\label{sec:coin}

Another example of SFE is the well-known primitive of coin tossing~\cite{blum:coin}, 
which can be viewed as an instance of
randomized secure function evaluation defined in~\cite{goldreich:book1}. Imagine that Alice and Bob want
to toss a coin, solely by communicating over a classical and a quantum channel. We thereby want to
ensure that neither party can influence the outcome of the coin toss by too much. Unfortunately,
we cannot implement this primitive classically without relying on additional assumptions. 

What assumptions do we need to implement coin tossing?
It is easy 
to see that we can implement one form of coin tossing, if we could perform bit commitment: Alice
chooses a random bit $b \in_R \01$ and commits herself to $b$. Subsequently, Bob chooses a random bit $b' \in_R \01$
and sends it to Alice. After receiving $b'$, Alice opens her commitment and reveals $b$. Both
parties now output $c = b \oplus b'$ as their outcome. Thus, any assumptions that enable us to implement
bit commitment also lead to coin tossing. Some assumptions even allow for very simple protocols: If we
assume that Alice and Bob are located far apart and faster-than-light communication is impossible, they can simply
both flip a coin themselves and send it over the channel. They then take the xor of the two bits
as the outcome of the coin flip. If Alice and Bob do not receive the other's bit within a certain
time frame they reject this execution of the protocol and restart. Since it takes the bit
a specific time to travel over the channel, both parties can be sure that it must have been
sent before a certain time, i.e., before receiving the other's bit.

Many definitions of coin tossing are known in the literature, which exhibit subtle differences
especially whether aborts are allowed during the protocol. In the quantum literature, \emph{strong
coin tossing}\footnote{Unfortunately, these names carry a slightly different
meaning in the classical literature.} has been informally defined as follows:
\begin{definition}\index{coin tossing!strong}
A quantum \emph{strong coin tossing} protocol with bias $\eps$ is a two-party protocol,
where Alice and Bob communicate and finally decide on a value $c \in \{0,1,\perp\}$ such that
\begin{itemize}
\item If both parties are honest, then $\Pr[c=0] = \Pr[c=1] = 1/2$.
\item If one party is honest, then for any strategy of the dishonest player
$\Pr[c=0] \leq 1/2 + \eps$ and $\Pr[c=1] \leq 1/2 + \eps$.
\end{itemize}
\end{definition}

Sadly, strong coin tossing cannot be implemented perfectly with bias $\eps=0$~\cite{lo:coin}. However,
one might hope that one could still achieve an arbitrarily small bias $\eps > 0$. 
Many protocols have been proposed for quantum strong coin tossing and subsequently been
broken~\cite{mayers:coin,zhang:coin}. Sadly, it was shown that strong coin tossing
cannot be implemented with an arbitrarily small bias, and $\eps=
1/\sqrt{2}-1/2\approx 0.207$ is the best we could hope to achieve~\cite{kitaev:coin}.
So far, quantum protocols for
strong coin tossing
with a bias of $\eps\approx 0.42$~\cite{dorit:bitEscrow} and finally
$\eps = 1/4$~\cite{ambainis:coin,spekkens:tradeoffBc,iordanis:coin,colbeck:coin} are known.
No formal definition of strong coin tossing in the quantum setting is known to date,
that specifies how to deal with an abort in the case when the protocol is executed
multiple times.

To circumvent this problem, a slightly weaker primitive has been proposed, which
carries the name \emph{weak coin tossing} in the quantum literature. Here, we explicitly
allow the dishonest party to bias the coin entirely in one direction, but limit his
ability to bias the coin the other way. This scenario corresponds to a setting
where, for example, Alice wins if the outcome is $c=0$ and Bob if $c=1$. However,
we do allow each player to give in and loose at will. Intuitively, this setting
makes more sense in all common practical examples when considering a standalone run of such a protocol, where each player has a preferred outcome.
Informally, we define
\begin{definition}\index{coin tossing!weak}
A quantum \emph{weak coin tossing} protocol with bias $\eps$ is a two-party protocol,
where Alice and Bob communicate and finally decide on a value $c \in 
\{0,1,\perp\}$ such that
\begin{itemize}
\item If both parties are honest, then $\Pr[c=0] = \Pr[c=1] = 1/2$.
\item If Alice is honest, then for any strategy of Bob 
$$\Pr[c=1] \leq 1/2 + \eps.$$
\item If Bob is honest, then for any strategy of Alice 
$$
\Pr[c=0] \leq 1/2 + \eps.
$$
\end{itemize}
\end{definition}

Weakening the definition in this way indeed helps us! It has been shown that we can construct
a quantum protocol for weak coin tossing that achieves a bias of 
$\eps\approx 0.239$~\cite{iordanis:coin},
$\eps\approx 0.207$~\cite{spekkens:coin},
$\eps\approx 0.192$~\cite{mochon:coin}, and
$\eps\approx 0.167$~\cite{mochon:coinFamily}. Very recently, however,
a protocol with an arbitrarily small bias has been 
suggested~\cite{mochon:newCoin}!
To date, there is also no 
formal definition of weak coin tossing in the quantum setting.

\subsubsection{Multiple players}\label{section:smp}\label{quantumSMP}

Secure multi-party computation (SMP)\index{SMP}\index{secure multi-party computation} 
concerns an analogous task to SFE, involving $n$ players
$P_1,\ldots,P_n$, where $P_j$ has a private input $x_j$. Their goal is to 
compute $f(x_1,\ldots,x_n)$, such that none of them can learn more about the input
of any other player than they can infer from $f(x_1,\ldots,x_n$). Fortunately,
the situation changes dramatically when extending the protocol to multiple players.
SMP can be implemented with unconditional security even classically, provided that 
$t < n/3$ of the players are dishonest~\cite{goldreich:book1}. If the adversary is not dishonest, but 
merely honest-but-curious, it is possible to increase $t$ up to $t < n/2$~\cite{goldreich:book1}. 
We refer to~\cite{cramer:smp} for an overview of classical secure multi-party computation.

Quantumly, one can generalize secure multi-party computation to the following setting.
Each player $P_j$ holds an input state $\rho_i \in \hil$ (see Chapter~\ref{chapter:informationIntro} for details). 
Let $\rho \in \hil_1 \otimes\ldots\otimes \hil_n$ 
denote the joint state of players $P_1,\ldots,P_n$. Then quantum secure multi-party computation (QSMP)
allows the players to compute any quantum transformation $U$ to obtain $U\rho U^\dagger$,
where player $P_j$ receives the quantum state on $\hil_j$ as his output. QSMP can be implemented
securely if $t < n/2$ of the players are dishonest~\cite{smith:qsmp,smith:qsmp2}.

Coin tossing has also been studied in the multi-party setting. Classically, 
multi-party coin tossing
forms part of secure multi-party computation~\cite{goldreich:book1}, and can thus be implemented
under the same assumptions. Quantumly, multi-party coin tossing has been studied in~\cite{hein:coin}.

\subsection{Secret sharing}\index{secret sharing}

Another interesting problem concerns the sharing of a classical or quantum secret.
Imagine Alice holding an important piece of information, for example the launch code
to her personal missile silo. Alice would like to enable members of her
community to gain access, but wants to prevent a single individual from launching a missile 
on his own. Secret sharing enables Alice to distribute some secret data $d$ among
a set of $n$ players, such that at least $t > 1$ players need to combine their
individual shares to reconstruct the original secret $d$. A trivial secret sharing
scheme for a bit $d \in \01$ involving just two players is as follows: Alice picks $r \in_R \01$
and hands $s_1 = d \oplus r$ to the first player, and $s_2 = r$ to the second player.
Clearly, if $r$ is chosen uniformly at random from $\01$, none of the individual players
can gain any information about $d$. Yet, when combining their individual shares
they can compute $s_1 \oplus s_2 = d$.

General secret-sharing schemes were introduced by Shamir~\cite{shamir:secret} and 
Blakey \cite{blakley:secret}.
They have found a wide range of applications, most notably to construct protocols for secure multi-party
computation as described in Section~\ref{section:smp}. Many classical secret sharing schemes are known 
today~\cite{handbook:crypto}. Quantum secret sharing was first introduced in~\cite{hillary:secret} and
shortly after in~\cite{gottesmann:secret}, which also formed a link between quantum secret sharing
schemes and error correcting codes. Quantumly, we can distinguish two types of secret sharing schemes:
The first allows to share a \emph{quantum} secret\index{secret!quantum}, i.e., Alice holds a quantum state $\rho$ 
and wants to construct $n$ quantum shares $\sigma_1,\ldots,\sigma_n$ such that when $t$ such shares
are combined $\rho$ can be reconstructed~\cite{hillary:secret,gottesmann:secret,gottesmann:secret2}.
The second allows us to share \emph{classical} secrets\index{secret!classical} using quantum states that have 
very nice \emph{data-hiding} 
properties~\cite{barbara:hiding1, barbara:hiding2, werner:hiding, graeme:hiding}:\index{data hiding}
it is not sufficient for $n$ parties to perform local measurements and communicate classically in order
to reconstruct the secret. To reconstruct the secret data they must communicate quantumly to perform
a coherent measurement on their states. It is an exciting open question whether such schemes can be
used to implement \emph{quantum} 
protocols for secure multi-party computations with \emph{classical} inputs that remain secure
as long as the dishonest players can only communicate classically, but not quantumly.

\subsection{Anonymous transmissions}\index{anonymous transmissions}
In all applications we considered so far, we were concerned with two aspects:
either, we wanted to protect protocol participants from being cheated by the
other players, or, we wanted to protect the secrecy of data from a third party 
as in the setting of key distribution
described in Section~\ref{qkd}.
In the problem of key distribution, sender and receiver know each other, but are
trying to protect their data exchange from prying eyes.
Anonymity, however, is the secrecy of identity. Primitives to hide
the sender and receiver of a transmission have received
considerable attention in classical computing. Such primitives
allow any member of a group to send and receive data
anonymously, even if all transmissions can be monitored.
They play
an important role in protocols for electronic
auctions~\cite{stajano99cocaine}, voting protocols and sending
anonymous email~\cite{chaum:mixnet}. 
An anonymous
channel which is completely immune to any active
attacks, would be a powerful primitive. It has been shown how two
parties can use such a channel to perform
key-exchange~\cite{alpern:keyless}.

A considerable number of classical schemes have been suggested for
anonymous transmissions. An unconditionally secure classical
protocol was introduced by Chaum~\cite{chaum:dc} in the context of
the Dining Cryptographers Problem. Such a protocol can also be considered
an instance of secure multi-party computation considered above.

Boykin~\cite{boykin:thesis}\index{Boykin, P. O.} considered a
quantum protocol to send classical information anonymously where
the players distribute and test pairwise shared EPR pairs, which
they then use to obtain key bits. His protocol is secure in the
presence of noise or attacks on the quantum channel. 
In~\cite{wehner04b}, we presented a protocol for anonymous
transmissions of classical data that achieves a novel property
that cannot be achieved classically: it is completely \emph{traceless}.\index{traceless}
This property is related, but stronger than the notion of
incoercibility in secure multi-party
protocols~\cite{canetti:coerce}.  
Informally, a protocol is
traceless, if a player cannot be forced to reveal his true input
at the end of the protocol. Even when forced to hand out his input,
output and randomness used during the course of the protocol, 
a player is able to generate fake input that
is consistent with all other data gathered from the run of the protocol.
The protocols suggested in \cite{boykin:thesis} are
not traceless, but can be modified to exhibit this property.
It would be interesting to see whether it is possible to make
general protocols for secure multi-party computation similarly traceless.

The first protocol for the anonymous trans\-mission of qubits was
constructed in~\cite{wehner04b}. 
Whereas the anonymous transmissions of classical bits can
be implemented via secure multi-party computation, the scenario 
is different when we wish to transmit qubits: as we will see in Chapter~\ref{quantumIntro},
qubits cannot be copied. Thus we cannot expect each player to obtain a copy of the output.
New protocols for creating anonymous entanglement and anonymously transmitting
qubits have since been suggested in~\cite{bouda:anon, anne:anon}.

\subsection{Other protocols}

Besides the protocols above, a variety of other primitives making use of particular quantum effects have been proposed.
One of the oldest suggested applications is the one of quantum money that is resistant to copying~\cite{wiesner:conjugate},
also proposed as unforgeable subway tokens~\cite{bbOne}.\index{unforgeable subway tokens}
Quantum seals~\cite{qseals1,qseals2,qseals3}\index{quantum seals} employ the notion of cheat sensitivity in order to provide data with a seal 
that is ``broken'' once the data is extracted. That is, we can detect whether the data has been read. Perfect quantum
seals that allow us to detect tampering with certainty have been shown to be impossible~\cite{qseals:nogo}. Nevertheless,
non-trivial constructions are can be implemented. 

Furthermore, quantum signature schemes~\cite{gottesman:signature}\index{quantum signatures} have been
proposed which exhibit unconditional security: here Bob can verify Alice's signature using a public key given to him ahead of time.
Sadly, such a scheme slowly consumes the necessary public key. Finally, protocols have been suggested for the encryption of
quantum data which allow $n$ qubits to be encoded using a $2n$ bit key achieving perfect secrecy~\cite{boykin:pqc,ronald:pqc}. 
Much smaller keys are possible, if we allow for small imperfections~\cite{nayak:smallKey,smith:smallKey}. Such encryption
schemes have also been used to allow for private circuit evaluation~\cite{childs:pce}: 
Here, Alice encrypts her quantum state before handing it to Bob who is capable of running a certain 
quantum operation that Alice would like to apply. This allows Alice to let her quantum operations be performed by Bob 
without revealing her quantum input.

\section{Challenges}

As we saw in Section~\ref{section:properties}, introducing quantum elements into cryptography leads to 
interesting new effects. Much progress has been made to exploit these quantum effects, although
many open questions remain. In particular, not much is known about how well quantum protocols compose. 
That is, when we use one protocol as a building block inside a larger application, does the protocol
still remain secure as expected? Recall from Section~\ref{section:properties} that especially our ability to delay quantum measurements
has a great influence on composition. 
Fortunately, quantum key distribution has been shown composable~\cite{benor:qkd,renato:diss,renato:compose}. 
However, composability remains a particularly tricky question in protocols where we are not faced 
with an external eavesdropper, but where the players themselves are dishonest.
Composability of quantum protocols was first considered in~\cite{vGraaf98}, followed 
by~\cite{smith:qsmp} who addressed the composability of QSMP, and the general composability frameworks
of~\cite{unruh:compose,benor:compose} applied to QKD~\cite{benor:qkd}. Great care must also be taken when
composing quantum protocols in the bounded quantum storage model~\cite{wehner07b}.\index{composability}\index{composability!bounded storage}
\index{bounded storage!composability}
Even though these composability frameworks exist, very few protocols have been proven secure
when composed.

Secondly, we need to consider what happens if an adversary is allowed to store even small amounts
of quantum information. There are many examples known where
quantum memory can prove much more useful to an adversary
than classical memory~\cite{ronald:qmem}, and
we will encounter such examples in Chapters~\ref{chapter:pistar} and~\ref{chapter:locking}.

Furthermore, it is often assumed that the downfall of computational assumptions such as 
factoring is the only consequence that quantum computing has on the security of classical protocols.
Sadly, this is by no means the only problem.
Classical protocols where the security depends on the fact that different players
cannot communicate during the course of the protocol may be broken when the players can share
quantum entanglement and perform even a very limited set of quantum operations, well within the
reach of current day technology. We will encounter such an example in 
Chapter~\ref{chapter:interactiveProofs}. 

Furthermore, we may conceive new primitives, unknown to the classical setting. One such primitive
is the distribution of shared quantum states in the presence of dishonest players. Here, our
goal is to create a protocol among $n$ players such that at the end of the
protocol $m \leq n$ players share a specified state $\rho$, where the dishonest players may 
apply any measurement to their share. It is conceivable to extend the QSMP protocol of~\cite{smith:qsmp}
to address this problem, yet, much more efficient protocols may be possible. Such a primitive
would also enable us to build up the resources needed by other protocols such as~\cite{wehner04b}.

Finally, it is an interesting question by itself, what cryptographic primitives are possible in a quantum mechanical
world. Conversely, it has even been shown that the axioms governing quantum mechanics can in part be obtained
from the premise that perfect bit commitment is impossible~\cite{qcVsbcnogo}. Perhaps such connections
may lead to novel insights.

\section{Conclusion}

Quantum cryptography beyond quantum key distribution is an exciting subject.
In this thesis, we will investigate several aspects that play an important role in nearly all cryptographic applications
in the quantum setting.
\vspace{0.5cm}

\textbf{In part I}, we will examine how to extract information from quantum states. We first consider
the problem of state discrimination. Here, our goal is to determine the identity of a state $\rho$ 
within a finite set of possible states $\{\rho_1,\ldots,\rho_n\}$. In Chapter~\ref{chapter:pistar}, we
will examine a special case of this problem that is of particular relevance to quantum cryptography in the
bounded quantum storage model: How well can we perform state discrimination if we are given additional
information after an initial quantum measurement, i.e., after a quantum memory bound is applied?
In Chapter~\ref{chapter:uncertainty}, we address uncertainty relations,
which play an important role in nearly all cryptographic applications. We will prove
tight bounds for uncertainty relations for certain mutually unbiased measurements. We will
also present optimal uncertainty relations for anti-commuting measurements.
Finally, in Chapter~\ref{chapter:locking}, we then examine a peculiar quantum effect known as locking 
classical information in quantum states. Such effects are important in the security of QKD,
and also play a role in quantum string commitments which we will encounter in part III. In particular,
we address the following question: Can we always obtain good locking effects for mutually unbiased
measurements?
\vspace{0.5cm}

\textbf{In part II}, we turn to investigate quantum entanglement. In Chapter~\ref{chapter:optimalStrategies},
we show how to find optimal quantum strategies for two parties who cannot communicate, but share quantum entanglement.
Understanding such strategies plays an important part in understanding the effect of entanglement in otherwise
classical protocols.
In Chapter~\ref{chapter:boundingEntanglement}, we then present some initial weak result on the amount of entanglement
such strategies require. Finally, in Chapter~\ref{chapter:interactiveProofs}, we show how the security of classical
protocols can be affected considerably in the presence of entanglement.
\vspace{0.5cm}

\textbf{In part III}, we investigate two cryptographic problems directly. In Chapter~\ref{chapter:qsc}, we
first consider commitments: Quantumly, one may hope that committing to an entire string of bits at once,
and allowing Alice and Bob a limited ability to cheat, may still be within the realm of possibilities. This does
not contradict that bit commitment itself is impossible. Unfortunately, we will see that for any reasonable security measure,
string commitments are also impossible. However, non-trivial protocols do become possible for very weak notions
of security. 

In Chapter~\ref{chapter:noise}, we then introduce 
the model of noisy-quantum storage that in spirit is very similar
to the setting of bounded-quantum storage: 
Here we assume that the adversary's quantum operations and storage
are subject to noise. We show that oblivious transfer can be implemented
securely in this model. We give an explicit tradeoff between
the amount of noise and the security of our protocol.

\part{Information in quantum states}\label{informationPart}

\chapter{Introduction}\label{informationIntro}\label{informationChapter}
\label{chapter:informationIntro}\label{chapter:quantumIntro}\label{classicalTheoriesChapter}\label{quantumIntro}\label{quantumVsclassical}
\label{introChapter}

To investigate the limitations and possibilities of cryptographic protocols in a physical world, we must familiarize
ourselves with its physical theory: quantum mechanics. What are quantum states and what sets them apart from the classical
scenario? Here, we briefly recount the most elementary facts that will be necessary for the remainder of this text.
We refer to~\cite{peres:book} for a more gentle introduction to quantum mechanics, to Appendix~\ref{appendix:semidef}
for linear algebra prerequisites, and to the symbol index on page~\pageref{symbolsPage} for unfamiliar notation. 
In later chapters, we examine some of the most striking aspects of quantum mechanics, such as 
uncertainty relations and entanglement in more detail.

\section{Quantum mechanics}

\subsection{Quantum states}
A $d$-dimensional quantum state is a positive semidefinite operator $\rho$ of norm 1 (i.e., $\rho$ has no
negative eigenvalues and $\Tr(\rho) = 1$)
living in a $d$-dimensional Hilbert space 
\footnote{A complete vector space with an inner product. Here, we
always consider a vector space over the complex numbers.} 
$\hil$.
We commonly refer to $\rho$ as a \emph{density operator}\index{density operator}
or \emph{density matrix}.\index{density matrix} A special case of a quantum state is a \emph{pure state}\index{pure state}\index{state!pure}, which has the property
that $\rank(\rho)=1$. That is, there exists some vector $\ket{\Psi} \in \hil$ such that we can write
$\rho = \outp{\Psi}{\Psi}$, where $\outp{\Psi}{\Psi}$ is a projector onto the vector $\ket{\Psi}$. 
If $\{\ket{0},\ldots,\ket{d-1}\}$ is a basis for $\hil$, we can thus
write $\ket{\Psi} = \sum_{j=0}^{d-1} \alpha_j \ket{j}$ for some coefficients $\alpha_j \in \Complex$.
Note that our normalization constraint implies that $\Tr(\rho) = \sum_j |\alpha_j|^2 = 1$. We 
also say that $\ket{\Psi}$ is in a \emph{superposition} of vectors $\ket{0},\ldots,\ket{d-1}$.
Clearly, for a pure state we have that $\rho^2 = \rho$ and thus $\Tr(\rho^2) = 1$.

Let's first look at an example of pure states.
Suppose we consider a $d=2$ dimensional quantum system $\hil$, also called
a \emph{qubit}.\index{qubit} We call $\{\ket{0},\ket{1}\}$ the \emph{computational basis}, where \index{computational basis}\index{basis!computational}
$$
\ket{0}= \left(\begin{array}{c} 1 \\ 0\end{array}\right) \mbox{ and }
\ket{1}= \left(\begin{array}{c} 0 \\ 1 \end{array}\right).
$$
Any pure qubit state can then be written
as $\ket{\Psi} = \alpha \ket{0} + \beta\ket{1}$ for some
$\alpha,\beta \in \Complex$ with $|\alpha|^2 + |\beta|^2 = 1$. We take an 
\emph{encoding of '0' or '1'}\index{encoding} in the computational basis to be $\ket{0}$ or $\ket{1}$ respectively, and
use the subscript '+' to refer to an encoding in the computational basis.
An alternative choice of basis would be the \emph{Hadamard basis}\index{basis!Hadamard}\index{Hadamard basis}, given by vectors
$\{\ket{+},\ket{-}\}$, where 
$$
\ket{+} = \frac{1}{\sqrt{2}}(\ket{0} + \ket{1}) \mbox{ and } 
\ket{-} = \frac{1}{\sqrt{2}} (\ket{0} - \ket{1}).
$$
We use '$\times$' to refer to an encoding in the Hadamard basis.
We will often consider systems consisting of $n$ qubits. 
If $\hil$ is a 2-dimensional Hilbert space corresponding
to a single qubit, the system of $n$ qubits is given by the $n$-fold tensor product $\hil^{\otimes n}$ with
dimension $d=2^n$.
A basis for this larger Hilbert space can easily be found by forming the tensor products
of the basis vectors of a single qubit. For example, the computational basis for an $n$-qubit system
is given by the basis vectors $\{\ket{x_1} \otimes \ldots \otimes \ket{x_n}\mid x_j \in \01, j \in [n]\}$ where $[n] = \{1,\ldots,n\}$.
We will often omit the tensor product and use the 
shorthand $\ket{x_1\ldots x_n} = \ket{x_1} \otimes \ldots \otimes \ket{x_n}$.

If $\rho$ is not pure, then $\rho$ is a \emph{mixed state}\index{state!mixed}\index{mixed state} and can be written as 
a \emph{mixture}\index{mixture} of pure states. 
That is, for any state $\rho$ there exist $\lambda_j \geq 0$ with $\sum_j \lambda_j = 1$ and vectors $\ket{\Psi_j}$
such that
$$
\rho = \sum_j \lambda_j \outp{\Psi_j}{\Psi_j}.
$$
Since $\rho$ is Hermitian, we can take $\lambda_j$ and $\ket{\Psi_j}$ to be the
eigenvalues and eigenvectors of $\rho$ respectively. We thus have for any quantum state that
$\Tr(\rho^2) \leq 1$, where equality holds if and only if $\rho$ is a pure state.
We can also consider a mixture of quantum states, pure or mixed. Suppose we have a
physical system whose state $\rho_x$ depends on some value $x \in \mX$ of a classical random variable $X$ 
drawn from $\mX$ according to a probability distribution $P_X$. For anyone who does not know 
the value of $X$ (but does know the distribution $P_X$), 
the state of the system is given as 
$$
\rho = \sum_x P_X(x) \rho_x.
$$
We also call the set $\ens = \{(P_X(x),\rho_x)\mid x \in \mX\}$ an \emph{ensemble},\index{ensemble}
that \emph{gives rise} to the density matrix $\rho$. We generally use the 
common shorthand
$\ens = \{P_X(x),\rho_x\}$. Clearly, for any state $\rho$ we can take its eigendecomposition
as above to find one possible ensemble that gives rise to $\rho$.
With this interpretation in mind, it is now intuitive why 
we wanted $\rho \geq 0$ and $\Tr(\rho)=1$: the first condition ensures that $\rho$ has no negative
eigenvalues and hence all probabilities $\lambda_j$ are non-negative. The second condition ensures
that the resulting distribution in indeed normalized.
We will use $\mS(\hil)$ and $\bop(\hil)$ 
to denote the set of all density matrices and the set of all bounded operators
on a system
$\hil$ respectively.

Let's look at a small example illustrating the concept of mixed quantum states.
The density matrices corresponding to $\ket{0}$ and $\ket{1}$ are 
$\rho_{0+} = \outp{0}{0}$ and $\rho_{1+} = \outp{1}{1}$, and
the density matrices corresponding to $\ket{+}$ and $\ket{-}$ are given by $\rho_{0\times} = \outp{+}{+}$
and $\rho_{1\times} = \outp{-}{-}$.
Let's suppose we are now told that we are given a '0' but encoded in either the computational or Hadamard basis,
each with probability $1/2$. Our quantum state corresponding to this encoding of '0'
is now
$$
\rho_0 = \frac{1}{2}(\rho_{0+} + \rho_{0\times}).
$$
The state corresponding to an encoding of '1' is similarly given by
$$
\rho_1 = \frac{1}{2}(\rho_{1+} + \rho_{1\times}).
$$
It is important to note that the same density matrix can be generated by two different ensembles.
As a simple example, consider the matrix $\rho = (2/3) \outp{0}{0} + (1/3) \outp{1}{1}$. 
Clearly, $\rho \geq 0$ and $\Tr(\rho)=1$ and
thus $\rho$ forms a valid one qubit quantum state. However, 
$\ens_1 = \{(2/3,\ket{0}),(1/3,\ket{1})\}$ and $\ens_2 = \{(1/2,\ket{\phi_0}),(1/2,\ket{\phi_1})\}$
with $\ket{\phi_0} = \sqrt{2/3} \ket{0} + \sqrt{1/3} \ket{1}$ and $\ket{\phi_1} = \sqrt{2/3}\ket{0} - \sqrt{1/3}\ket{1}$
both give rise to $\rho$:
$$
\rho = \frac{2}{3}\outp{0}{0} + \frac{1}{3}\outp{1}{1} = \frac{1}{2} \outp{\phi_0}{\phi_0} + \frac{1}{2}\outp{\phi_1}{\phi_1}.
$$

\bigskip
\noindent
\textbf{\emph{Classical vs.\ Quantum}}\\
Quantum states exhibit an important property known as ``no-cloning'': very much unlike classical states, we cannot
create a copy of an arbitrary quantum state!\index{no-cloning} This is only possible with a small probability. We refer to~\cite{gisin:cloning}
for an excellent overview of known results.

In the following, we call an ensemble \emph{classical} if all states $\rho_x$
commute.\index{classical ensemble}\index{ensemble!classical} This is an interesting special case, we discuss
in more detail below.

\subsection{Multipartite systems}

We frequently need to talk about a quantum state shared by multiple players in a protocol. 
Let $\hil_1,\ldots,\hil_n$ denote the Hilbert spaces corresponding to the quantum systems of players
1 up to $n$. As outlined in the case of multiple qubits above, the joint system $\hil_1 \otimes \ldots \otimes \hil_n$
of all players is formed by taking the tensor product.
For example, suppose that we have only two players, Alice and Bob.
Let $\hil^A$ and $\hil^B$ be the Hilbert spaces corresponding to Alice's and Bob's quantum systems respectively. 
Any \emph{bipartite state}\index{state!bipartite}\index{bipartite state} $\rho^{AB}$ shared by Alice and Bob is a state
living in the joint system $\hil^{A} \otimes \hil^{B}$.
Bipartite states can exhibit an interesting property called entanglement, which we investigate
in Chapter~\ref{chapter:entanglementIntro}. 
In short, if $\ket{\Psi} \in \hil^A \otimes \hil^B$ is a pure state, we say that $\ket{\Psi}$ is \emph{separable} if and only if there exist states $\ket{\Psi^A} \in \hil^A$ and $\ket{\Psi^B} \in \hil^B$ such that $\ket{\Psi} = \ket{\Psi^A} \otimes \ket{\Psi^B}$. A separable pure state is also called a \emph{product state}. A state that is not 
separable is called \emph{entangled}. An example of an entangled pure
state is the so-called EPR-pair
$$
\frac{1}{2}(\ket{00} + \ket{11}).
$$
For mixed states the 
definition is slightly more subtle. Let $\rho \in \mS(\hil^A \otimes \hil^B)$ be a mixed state. Then $\rho$ is 
called a \emph{product state} if there exist $\rho^A \in \hil^A$ and $\rho^B \in \hil^B$ such 
that $\rho = \rho^A \otimes \rho^B$. The state $\rho$ is called \emph{separable}, if there exists an ensemble
$\ens = \{p_j,\ket{\Psi_j}\}$ such that $\ket{\Psi_j} = \ket{\Psi_j^A} \otimes \ket{\Psi_j^B}$ with
$\ket{\Psi_j^A} \in \hil^A$ and $\ket{\Psi_j^B} \in \hil^B$ for all $j$, such
that 
$$
\rho = \sum_j p_j \outp{\Psi_j}{\Psi_j} = \sum_j p_j \outp{\Psi_j^A}{\Psi_j^A} \otimes \outp{\Psi_j^B}{\Psi_j^B}.
$$
Intuitively, if $\rho$ is separable then $\rho$ corresponds to a 
mixture of separable pure states according to
a classical joint probability distribution $\{p_j\}$. We return to such differences in Chapter~\ref{chapter:entanglementIntro}.
From a cryptographic perspective, it is for now merely important to 
note that if the state $\rho^{AB}$ shared between Alice
and Bob is a pure state, then $\rho^{AB}$ is not entangled with any third system $\hil^C$ 
held by Charlie.
That is, $\rho^{AB}$ 
does not depend on any classical random variable $X$ held by Charlie whose value is unknown to Alice and Bob. 
An important consequence is that the outcomes of any measurement (see below) that Alice 
and Bob may perform on $\rho^{AB}$ are therefore independent of $X$, and hence secret with respect to Charlie.

Given a quantum state in a combined, larger, system, what can we say about the state of the individual systems?
For example, given a state $\rho^{AB}$ shared between Alice and Bob, the \emph{reduced state}\index{state!reduced}\index{reduced state} of Alice's system alone is given by
$\rho^A = \Tr_B(\rho^{AB})$, where $\Tr_B$ is the partial trace over Bob's system. 
The partial trace operation $\Tr_B: \bop(\hil^A \otimes \hil^B) \rightarrow \bop(\hil^A)$ is thereby defined
as the unique linear operator that for all $A \in \bop(\hil^A)$ and all $B \in \bop(\hil^B)$ maps
$\Tr_B(A\otimes B) = A\Tr(B)$.
We also say that we \emph{trace out} Bob's system from $\rho^{AB}$ to obtain
$\rho^A$.\index{trace out}
Furthermore,
given any state $\rho^A \in \hil^A$, we can always find a second system $\hil^B$ and a pure
state $\ket{\Psi} \in \hil^A \otimes \hil^B$ such that $\rho^A = \Tr_B(\outp{\Psi}{\Psi})$.
We call $\ket{\Psi}$ a \emph{purification} of $\rho^A$.\index{purification}

\bigskip
\noindent
\textbf{\emph{Classical vs.\ Quantum}}\\
In the quantum world, we encounter a particular effect known as entanglement.
Intuitively, entanglement leads to very strong correlations among Alice and Bob's system, which 
we will examine in detail in Chapter~\ref{chapter:entanglementIntro}. 

\subsection{Quantum operations}

\subsubsection{Unitary evolution}
The evolution of any closed quantum system is described by a \emph{unitary evolution} $U$\index{unitary evolution}
that maps 
$$
\rho \rightarrow U\rho U^{\dagger}.
$$
It is important to note that unitary operations are reversible: We can always apply an
additional unitary $V = U^\dagger$ to retrieve the original state since 
$V(U\rho U^\dagger)V^\dagger = U^\dagger U \rho U^{\dagger} U = \rho$.
In particular, we often make use of the following single qubit unitaries 
known as the Pauli matrices\index{Pauli matrices}
\begin{eqnarray*}
\sigma_x = \left(\begin{array}{cc}
      0 & 1\\
      1 & 0\end{array}\right),
&
\sigma_y = \left(\begin{array}{cc}
      0 & -i\\
      i & 0\end{array}\right),
&
\sigma_z = \left(\begin{array}{cc}
      1 & 0\\
      0 & -1\end{array}\right).
\end{eqnarray*}
Note that $\sigma_y = i \sigma_x \sigma_z$. Furthermore, we also use the Hadamard\index{Hadamard transform}, and the 
K-transform\index{K-transform}\label{ktransform} given by
$$
H = \frac{1}{\sqrt{2}}\left(\begin{array}{cc}
      1 & 1\\
      1 & -1\end{array}\right)
\mbox{ and }
K = \frac{1}{\sqrt{2}}\left(\begin{array}{cc}
      1 & i\\
      i & 1\end{array}\right).
$$
Note that $K = (\id + i\sigma_x)/\sqrt{2}$.

\subsubsection{Measurements}
Besides unitary operations, we can also perform measurements on the quantum state. A \emph{quantum measurement}\index{measurement}
of a state $\rho \in \mS(\hil)$ is a set of operators $\{M_m\}$ acting on $\mS(\hil)$,
satisfying $\sum_m M_m^\dagger M_m = \id$. We will call operators $M_m$
\emph{measurement operators}.\index{measurement!operators}
The probability of obtaining outcome $m$ when measuring
the state $\rho$ is given
by
$$
\Pr[m] = \Tr(M_m^\dagger M_m \rho).
$$
Conditioned on the event that we obtained outcome $m$, the \emph{post-measurement state}\index{post-measurement state}\index{state!post-measurement}\index{measurement!post-measurement state} of the system
is now
$$
\rho_m = \frac{M_m \rho M_m^\dagger}{\Tr(M_m^\dagger M_m \rho)}.
$$
Most measurements \emph{disturb} the quantum state and hence
$\rho_m$ generally differs from $\rho$. We will discuss this effect
in more detail below.
Note that we have $\sum_m \Pr[m] = \Tr\left(\left(\sum_m M_m^\dagger M_m\right)\rho\right) = 1$,
and hence the distribution over outcomes $\{m\}$ is appropriately normalized.

A special case of a quantum measurement is a \emph{projective measurement}\index{measurement!projective}\index{projective measurement}, where all measurement
operators $M_m$ are orthogonal projectors which we write as $P_m = M_m = M_m^\dagger M_m$.
Projective measurements are also described via an 
\emph{observable}\index{observable}\index{measurement!observable} $A = \sum_m m P_m$, where $m \in \Real$.
Note that $A$ is a Hermitian matrix with eigenvalues $\{m\}$. For any given basis $\mB = \{\ket{x_1},\ldots,\ket{x_d}\}$
we speak of \emph{measuring in the basis $\mB$}\index{measurement!in a basis}\index{basis!measurement in}
to indicate that we perform a projective measurement
given by operators $P_k = \outp{x_k}{x_k}$ with $k \in [d]$.

If we are only interested in the measurement outcome, 
but do not care about the post-measurement state, 
it is often simpler to make use of the POVM (positive operator valued measure)\index{POVM}\index{measurement!POVM} formalism. A POVM is a
set of Hermitian operators $\{E_m\}$ such that $\sum_m E_m = \id$ and for all $m$ we have $E_m \geq 0$.
Evidently, from a general measurement we can obtain a POVM by letting $E_m = M_m^\dagger M_m$. We now
have
$$
\Pr[m]= \Tr(E_m \rho).
$$
The advantage of this approach is that we can 
easily solve optimization problems involving probabilities $\Pr[m]$
over the operators $E_m$,
instead of considering the individual operators $M_m$. Since $E_m \geq 0$
such problems can be solved using semidefinite programming, which we describe
in Appendix~\ref{appendix:semidef}.
Finally, it is important to note that quantum measurements
do not always commute: it matters crucially in which order we execute them. Indeed, as we will see later it is this
property that leads to all the interesting quantum effects we will consider.

Let's consider a small example. Suppose we are given a pure quantum state $\ket{\Psi} = \sqrt{2/3} \ket{0} + \sqrt{1/3} \ket{1}$.
When measuring $\ket{\Psi}$ in the computational basis, we perform a measurement determined by operators
$P_0 = \outp{0}{0}$ and $P_1 = \outp{1}{1}$. Evidently, we have
$$
\Pr[0] = \Tr(P_0\outp{\Psi}{\Psi}) = \bra{\Psi}P_0\ket{\Psi} = \frac{2}{3},
$$
and
$$
\Pr[1] = \Tr(P_1\outp{\Psi}{\Psi}) = \bra{\Psi}P_1\ket{\Psi} = \frac{1}{3}.
$$
If we obtained outcome '0', the post-measurement state is given
by
$$
\rho_0 = \frac{P_0 \outp{\Psi}{\Psi} P_0}{\Pr[0]} = \outp{0}{0}.
$$
Similarly, if we obtained outcome '1', the post-measurement state
is
$$
\rho_1 = \frac{P_1 \outp{\Psi}{\Psi} P_1}{\Pr[1]} = \outp{1}{1}.
$$

\subsubsection{Quantum channel}\label{section:invariantMeasure}

The most general way to describe an operation is by means of a CP (completely positive) map\index{CP map}
$\Lambda: \hil^A \rightarrow \hil^B$, where $\hil^A$ and $\hil^B$ denote the in and output systems respectively. We also call $\Lambda$ a \emph{channel}.
Any channel $\Lambda$ can be written as 
$\Lambda(\rho) = \sum_m V_m \rho V_m^\dagger$
where $V_m$ is a linear operator from $\hil^A$ to $\hil^B$, and $\sum_m V_m^\dagger V_m \leq \id$. 
$V_m$ is also referred to as a \emph{Kraus operator}\index{Kraus operator}. 
$\Lambda$
is \emph{trace preserving}\index{quantum channel!trace preserving} if $\sum_m V_m^\dagger V_m = \id$.
Any quantum operation 
can be expressed by means of a CPTP (completely positive trace preserving) map \index{CPTP map}.
We sometimes also refer to such a map as a \emph{superoperator}\index{superoperator}, a 
\emph{quantum channel},\index{quantum channel} 
or
a (measurement) \emph{instrument}\index{instrument}, 
if we think of a POVM with elements $\{V_m\}$.
A channel is called \emph{unital}\index{quantum channel!unital}\index{unital!quantum channel}, if in addition $\sum_m V_m V_m^\dagger = \id$: we then
have $\Lambda(\id) = \id$. 

We give two simple examples.
Consider the unitary evolution $U$ of a state $\rho$: here we have $\Lambda(\rho) = U \rho U^\dagger$.
When we perform our single qubit measurement in the computational basis described above, and ignore
the measurement outcome, we implement the channel $\Lambda(\rho) = P_0 \rho P_0 + P_1 \rho P_1$. Since $P_0$ and $P_1$ form a measurement and
are projectors
we also have that $P_0P_0^\dagger + P_1P_1^\dagger = \id$
and hence the channel is unital.

Any quantum channel can be described by a unitary transformation on the original and an ancilla system, where
the ancilla system is traced out to recover the original operation. More
precisely, given a channel $\Lambda: \hil^A \rightarrow \hil^B$ we can choose a 
Hilbert space $\hil^C$ identical to $\hil^B$, a pure state 
$\hat{\rho} \in \mS(\hil^B \otimes \hil^C)$
and a unitary matrix $U_\Lambda$ acting on $\hil^A \otimes \hil^B \otimes \hil^C$ such that
for any $\rho \in \mS(\hil^A)$ $\Lambda(\rho) = \Tr_{A,C}U_\Lambda(\rho \otimes \hat{\rho})U_\Lambda^\dagger$.
This is all that we need
here, and we refer to~\cite{hayashi:book} for detailed information.

Of particular interest, especially with regard to constructing 
cheat-sensitive\index{cheat-sensitive} protocols, is the 
following statement which specifies which operations leave a given set of states invariant. Clearly,
any cheating party may always perform such operations without being detected. 
It has been shown that
\begin{lemma}(HKL)~\cite{hkl:noiseCommutant}\label{hkl:noiseCommutant}
Let $\Lambda: \hil \rightarrow \hil$ be a unital quantum channel with 
$\Lambda(\rho) = \sum_m V_m \rho V_m^\dagger$, and let $\setS$
be a set of quantum states. Then 
$$
\forall \rho \in \setS, \Lambda(\rho) = \rho \mbox{ if and only if } \forall m \forall \rho \in \setS,  [V_m,\rho]=0.
$$
\end{lemma}

Indeed, the converse direction is easy to see. If we have that for all $m$ and for all $\rho \in \setS$ $[V_m,\rho]=0$,
then $\Lambda(\rho) = \sum_m V_m \rho V_m^\dagger = 
\sum_m V_m V_m^\dagger \rho = \rho$, since $\Lambda$ is unital. If a quantum channel is not of this form, i.e.
it does not leave the state invariant, we also say that it \emph{disturbs}\index{disturb} the state.
The statement above has interesting consequences: consider an ensemble of states $\ens = \{p_x,\rho_x\}$ 
with $\rho_x \in \hil$,
and suppose that there exists a decomposition $\hil = \bigoplus_j \hil_j$ such that for all $x$ we have 
$\rho_x = \sum_j \Pi_j \rho_x \Pi_j$ where $\Pi_j$ is a projector onto $\hil_j$. If we perform the measurement
given by operators $\{\Pi_j\}$ then (ignoring the outcome) the states $\rho_x$ are invariant under such a measurement,
since clearly $[\Pi_j,\rho_x] = 0$ for all $j$ and $x$. The outcome of the measurement tells us which $\hil_j$ we reside in.
However, Lemma~\ref{hkl:noiseCommutant} tells us a lot more: We will see in Chapter~\ref{pistarAlgebra} that if the 
measurement operators from a projective measurement
commute with all the states $\rho_x$, they are in fact of this very form (see also Appendix~\ref{cstar}).
In the following, we call the information about which $\hil_j$ we reside in 
the \emph{classical information of the ensemble} $\ens$.\index{classical information}
Any attempt to gain more information, i.e. by performing measurements which do not satisfy these commutation properties,
necessarily leads to disturbance and can be detected.

An adversary can thus always extract this classical information without affecting the quantum state. 
Looking back at Chapter~\ref{chapter:cryptoIntro}, we can now see that for unital adversary channels we can
define an honest-but-curious\index{honest-but-curious}
player to be honest-but-curious with regard to the classical information, and honest with regard
to the quantum information: he may extract, copy
and memorize the classical information as desired. 
However, if he wants to leave the protocol execution itself unaltered, he cannot perform
any other measurements and must thus be honest on the remaining quantum part of the ensemble.

\bigskip
\noindent
\textbf{\emph{Classical vs.\ Quantum}}\\
Clearly, Lemma~\ref{hkl:noiseCommutant} also tells us that if all the states $\rho_x$ in our ensemble
commute, i.e. the ensemble is classical as defined above, then we can always perform a measurement in their common eigenbasis
``for free''. 
Furthermore, if our ensemble is classical we have $\dim(\hil_j)=1$, i.e. $\hil_j$ 
itself is also classical: it is just a scalar. We thus see that such an ensemble has
no quantum properties: we can extract and copy information at will. 
Informally, we may think of the different
states within the ensemble as different classical probability distributions over their common eigenstates. 
We will return to this idea shortly.

Furthermore, we can look at measurements or observables themselves. Note again from the above that since a quantum
measurement may disturb a state, it matters in which order measurements are executed. That is, quantum operations
do not commute. It is this fact that leads to all the interesting effects we observe: uncertainty relations, 
locking and Bell inequality violations using quantum entanglement are all consequences of the existence of
non-commuting measurements in the quantum world. This lies in stark contrast to the classical world, 
where all our measurement do commute, and we therefore do not encounter such effects.

\section{Distinguishability}\index{distinguishability}\index{state!discrimination}\label{stateDiscriminationVanilla}

\begin{figure}[h]
\begin{center}
\includegraphics[scale=0.6]{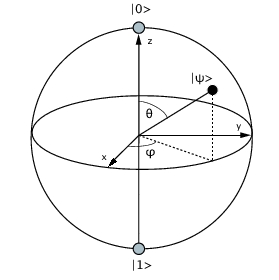}
\caption{Bloch vector
$(r_x,r_y,r_z) = (\cos\psi \sin\theta, \sin\psi \sin\theta, \cos\theta)$}
\label{blochSphere}
\end{center}
\end{figure}

How can we distinguish several quantum states? 
Suppose we are given states $\rho_X$ 
where $X$ is a random variable drawn according to a probability distribution $P_X$ over some finite set $\mX$.
Our goal is now to determine the value of $X$ given an unknown state $\rho \in \{\rho_x\mid x\in \mX\}$.
Cryptographically, this gives an intuitive measure on how well we can guess the value of $X$.
The problem of finding the optimal distinguishing measurement is called \emph{state discrimination},
where optimal refers to finding the measurement that maximizes the probability of successfully
guessing $X$.
For two 
states, the optimal guessing probability is particularly simple to evaluate.
To this end, we first need to introduce the trace distance, and the trace norm:
\begin{definition}\label{def:trdist}\index{trace distance}\index{distance!trace}
The \emph{trace distance} of two states $\rho_0$ and $\rho_1$ is
given by 
$$
D(\rho_0,\rho_1) = \frac{1}{2}||\rho_0 - \rho_1||_1,
$$
where $||A||_1 = \Tr(\sqrt{A^\dagger A})$ is the \emph{trace norm} of $A$.\index{trace norm}
\end{definition}
Alternatively, the trace distance may also be expressed as~\cite{hayashi:book} 
$$
D(\rho_0,\rho_1) = \max_M \Tr(M(\rho_0-\rho_1)),
$$
where the maximization
is taken over all $M \geq 0$.
Indeed, $D$ is really a ``distance'' measure, as it is clearly a metric on the space
of density matrices: We have $D(\rho_0,\rho_1) = 0$ if and only if $\rho_0 = \rho_1$,
and evidently $D(\rho_0,\rho_1) = D(\rho_1,\rho_0)$.
Finally, the triangle inequality holds:
\begin{eqnarray*}
D(\rho_0,\rho_1) &=& \max_M \Tr(M(\rho_0-\rho_1)) = \max_M \left(\Tr(M(\rho_0-\sigma)) + \Tr(M(\sigma - \rho_1))\right)\\
&\leq& D(\rho_0,\sigma) + D(\sigma,\rho_1).
\end{eqnarray*}
When considering single qubits (such as for example in Chapter~\ref{chapter:otNoise}) it is often
intuitive to note that for a single qubit, the trace distance has a 
particularly simple form. Note that $\id$, $\sigma_x$, $\sigma_y$ and $\sigma_z$ form a basis
for the space of $2\times 2$ complex matrices. Since we have $\Tr(\rho)=1$
for any quantum state, we can thus write any single qubit state as
$$
\rho = \frac{\id + \vec{r}\cdot \vec{\sigma}}{2}
= \frac{\id + r_x \sigma_x + r_y \sigma_y + r_z \sigma_z}{2}
$$
where $\vec{\sigma} = (\sigma_x, \sigma_y, \sigma_z)$ and $\vec{r} = (r_x,r_y,r_z)$
is the \emph{Bloch vector} as given in Figure~\ref{blochSphere}.\index{Bloch vector}\index{Bloch sphere}
For $\tau = (\id + \vec{t}\cdot\vec{\sigma})/2$ with $\vec{t} = (t_x,t_y,t_z)$
we then have 
$$
D(\rho,\tau) = \frac{1}{2}||\rho-\tau||_1 = 
\frac{1}{2}\left|\left|\sum_{j \in \{x,y,z\}} (r_j -t_j) \sigma_j\right|\right|_1
= \frac{1}{2}\sqrt{\sum_{j \in \{x,y,z\}} (r_j-t_j)^2},
$$
where we used the fact that all Pauli matrices anti-commute.
Thus, the trace distance between $\rho$ and $\tau$ is exactly half the Euclidean distance
of the corresponding Bloch vectors.

Using the trace distance, we can address the problem of distinguishing \emph{two} quantum 
states:
\begin{theorem}[Helstrom~\cite{helstrom:detection}]\label{helstrom}
  Suppose we are given states $\rho_0$ with probability $q$, and $\rho_1$
  with probability $1-q$. Then the probability to determine whether the state was
  $\rho_0$ and $\rho_1$ is at most
  \[
    p = \frac{1}{2}\left[1 + ||q\rho_0 - (1-q) \rho_1||_1\right].
  \]
  The measurement that achieves $p$ is given by $M_0$, 
  and $M_1 = \id - M_0$, where $M_0$ is the projector onto the positive eigenspace of 
  $q\rho_0 - (1-q)\rho_1$.
\end{theorem}

For $q=1/2$, this gives us $p = 1/2 + D(\rho_0,\rho_1)/2$.
Indeed, it is easy to see why such $M_0$ and $M_1$ form the optimal measurement. Note that here we are only
interested in finding a POVM. To find the optimal POVM we must solve the following optimization
problem for variables $M_0$ and $M_1$:
\begin{sdp}{maximize}{$q \Tr(M_0 \rho_0) + (1-q) \Tr(M_1 \rho_1)$}
&$M_0,M_1 \geq 0$,\\
&$M_0 + M_1 = \id$.
\end{sdp}
We can rewrite our target function as 
\begin{eqnarray*}
q \Tr(M_0 \rho_0) + (1-q) \Tr(M_1 \rho_1) &=&
q \Tr(M_0 \rho_0) + (1-q) \Tr((\id - M_0)\rho_1)\\
&=& \Tr(M_0 (q \rho_0 - (1-q) \rho_1)) + 1 - q\\
&=& \Tr\left(M_0 \left(\sum_{\lambda_j \geq 0} \lambda_j 
\outp{u_j}{u_j}\right)\right)\\
&& + \Tr\left(M_0 \left(\sum_{\lambda_j < 0} \lambda_j \outp{u_j}{u_j}\right)\right) + 1 -q,
\end{eqnarray*}
where $q \rho_0 - (1-q) \rho_1 = \sum_j \lambda_j \outp{u_j}{u_j}$.
Hence, to maximize the above expression, we need to choose $M_0 = \sum_{\lambda_j \geq 0} \outp{u_j}{u_j}$.

Unfortunately, computing the optimal measurement to distinguish more than two states is generally not so
easy. Yuen, Kennedy and Lax~\cite{yuen:maxState} first showed that this problem can be solved using 
semidefinite programming, a technique we describe in
Appendix~\ref{appendix:semidef}. This technique has since been refined to
address other variants such as unambiguous state discrimination where we can output ``don't know'',
but are never allowed to make a mistake~\cite{eldar:sdp}.
Evidently, we can express the optimization problem for any state discrimination problem as
\begin{sdp}{maximize}{$\sum_x P_X(x) \Tr(M_x \rho_x)$}\index{SDP!example}
&$\forall x \in \mX, M_x \geq 0$,\\
&$\sum_{x \in \mX} M_x = \id$.
\end{sdp}
In Chapter~\ref{chapter:pistar}, we will use the above formulation. We also show how to address a variant 
of this problem, where we receive additional classical information after performing the measurement.

Closely related to the trace distance is the notion of fidelity.
\begin{definition}\label{def:fidelity}\index{fidelity}
The \emph{fidelity} of states $\rho$ and $\sigma$ is given
by 
$$
F(\rho,\sigma) = \Tr\sqrt{\rho^{1/2}\sigma\rho^{1/2}}.
$$
\end{definition}
Note that if $\rho = \outp{\Psi}{\Psi}$ is a pure state, this becomes
$$
F(\ket{\Psi},\sigma) = \sqrt{\bra{\Psi}\sigma\ket{\Psi}}.
$$
The fidelity is closely related to the trace distance. In particular, we have
that for any states $\rho$ and $\sigma$
$$
1 - F(\rho,\sigma) \leq D(\rho,\sigma) \leq \sqrt{1-F(\rho,\sigma)^2}.
$$
A proof can be found in~\cite[Section 9.2.3]{nielsen&chuang:qc}.
If $\rho = \outp{\Psi}{\Psi}$ is a pure state, the lower bound can be improved
to 
$$
1 - F(\ket{\Psi},\sigma)^2 \leq D(\ket{\Psi},\sigma).
$$
Many other distance measures of quantum states are known, which may be a more
convenient choice for particular problems. We refer to~\cite{fuchs:diss,hayashi:book} for
an overview.

\bigskip
\noindent
\textbf{\emph{Classical vs.\ Quantum}}\\
Suppose again we are given a classical ensemble of states $\rho$ and $\sigma$. That is, both operators commute and hence
have a common eigenbasis $\{\ket{u_1},\ldots,\ket{u_d}\}$. 
We can thus write $\rho = \sum_j \lambda_j \outp{u_j}{u_j}$ and $\sigma = \sum_j \gamma_j \outp{u_j}{u_j}$,
which allows us to write the trace distance of $\rho$ and $\sigma$ as
$$
D(\rho,\sigma) = \frac{||\sum_j (\lambda_j - \gamma_j)\outp{u_j}{u_j}||_1}{2} = \frac{1}{2} \sum_j |\lambda_j - \gamma_j| = 
D(\lambda_j,\gamma_j),
$$ 
where $D(\lambda_j,\gamma_j)$ is the \emph{classical variational distance}\index{distance!variational}\index{variational distance} between the distributions
$\{\lambda_j\}$ and $\{\gamma_j\}$. Again, we see that there is nothing quantum in this setting. We can view
$\rho$ and $\sigma$ as two different probability distributions over the set $\{\ket{u_j}\}$.
Similarly, it is easy to see that
$$
F(\rho,\sigma) = \Tr\sqrt{\sum_j \lambda_j \gamma_j \outp{u_j}{u_j}} = \sum_j \sqrt{\lambda_j \gamma_j} = F(\lambda_j,\gamma_j),
$$
where $F(\lambda_j,\gamma_j)$ is the \emph{classical fidelity} of the distributions $\{\lambda_j\}$ and $\{\gamma_j\}$.

\section{Information measures}

\subsection{Classical}
We also need the following ways of measuring information. 
Let $X$ be a random variable distributed over a finite set $\setX$ according to 
probability distribution $P_X$. The \emph{Shannon entropy}\index{entropy!Shannon}\index{Shannon entropy} of $X$
is then given
by 
$$
H(X) = - \sum_{x \in \setX} P_X(x) \log P_X(x).
$$
Intuitively, the Shannon entropy measures how much information we gain \emph{on average}
by learning $X$. A complementary view point is that $H(X)$ quantifies the amount of
uncertainty we have about $X$ before the fact. We will also use
$H(P_X)$, if our discussion emphasizes a certain distribution $P_X$.
If $|\setX| = 2$,
we also use the term 
\emph{binary entropy}\index{entropy!binary}\index{binary entropy}
and use the shorthand\label{def:binentropy}
$$
h(p) = - p \log p - (1-p) \log (1-p).
$$
Let $Y$ be a second random variable distributed over a finite set $\setY$ according to
distribution $P_Y$. The \emph{joint entropy}\index{entropy!joint}\index{joint entropy} 
of $X$ and $Y$ can now be expressed as\label{def:jointentropy}
$$
H(X,Y) = - \sum_{x \in \setX, y \in \setY} P_{XY}(x,y) \log P_{XY}(x,y),
$$
where $P_{XY}$ is the joint distribution over $\setX \times \setY$.
Furthermore, we can quantify the uncertainty about $X$ given $Y$ by means
of the \emph{conditional entropy}\index{entropy!conditional}\index{conditional entropy}
\label{def:condentropy}
$$
H(X|Y) = H(X,Y) - H(Y).
$$
To quantify the amount of information $X$ and $Y$ may have in common 
we use the \emph{mutual information}\label{def:mutualinfo}\index{information!mutual}\index{mutual information}
$$
\mI(X,Y) = H(X) + H(Y) - H(X,Y) = H(X) - H(X|Y).
$$
Intuitively the mutual information captures the amount of information
we gain about $X$ by learning $Y$.
The Shannon entropy has many interesting properties, summarized, for example, in~\cite[Theorem 11.3]{nielsen&chuang:qc},
but we do not require them here.
In Chapter~\ref{chapter:locking}, we only need the 
\emph{classical mutual information}\index{classical mutual information}
\index{mutual information!classical} of a 
bipartite\index{bipartite state!mutual information} quantum state $\rho^{AB}$, which is the maximum classical 
mutual information that can be obtained by local measurements $M^A \otimes M^B$
on the state $\rho^{AB}$~\cite{terhal:minfo}:\label{def:classicalmutualinfo}
\begin{equation}
\mI_c(\rho^{AB}) = \max_{M^A \otimes M^B} \mI(A,B),
\end{equation}
where $A$ and $B$ are the random variables corresponding to Alice's and Bob's measurement outcomes respectively.

In a cryptographic setting, the Shannon entropy is not always a desirable measure as it merely captures our
uncertainty about $X$ \emph{on average}. Often, the R{\'e}nyi entropy\index{R{\'e}nyi entropy}\index{entropy!R{\'e}nyi} allows us to make stronger
statements
The \emph{R{\'e}nyi entropy}~\cite{renyi:entropy} of order $\alpha$ is defined as
$$
H_\alpha(X) = \frac{1}{1-\alpha}\log\left[\left(\sum_{x \in \setX} P_X(x)^\alpha\right)^\frac{1}{\alpha-1}\right].
$$
Indeed, the Shannon entropy forms a special case of the R{\'e}nyi entropy by taking the limit $\alpha \rightarrow 1$, i.e.,
$H_1(\cdot) = H(\cdot)$, where we omit the subscript.
Of particular importance is the \emph{min-entropy}\index{entropy!min-entropy}\index{min-entropy}, for $\alpha \rightarrow \infty$:\label{def:minentropy}
$$
H_\infty(X) = - \log\left(\max_{x \in \setX} P_X(x)\right),
$$
and the \emph{collision entropy}\label{def:collision}\index{entropy!collision}\index{collision entropy}
$$
H_2(X) = - \log \sum_{x \in \setX} P_X(x)^2.
$$
We have
$$
\log|\setX| \geq H(X) \geq H_2(X) \geq H_\infty(X).
$$
Intuitively, the min-entropy is determined by the highest peak in the distribution and most closely captures 
the notion of ``guessing'' $x$. Consider the following example:
Let $\setX = \01^n$ and let $x_0 = 0,\ldots,0$ be the all 0 string. Suppose that $P_X(x_0) = 1/2 + 1/(2^{n+1})$
and $P_X(x) = 1/(2^{n+1})$ for $x \neq x_0$, i.e., with probability $1/2$ 
we choose $x_0$ and with probability $1/2$ we choose one string
uniformly at random. Then $H(X) \approx n/2$, whereas $H_\infty(X) = 1$! If $x$ would correspond to an encryption
key used to encrypt an $n$ bit message, 
we would certainly not talk about security if we can guess the key with probability $1/2$! Yet, the Shannon entropy is quite high.
We refer to~\cite{cachin:diss} for an in-depth discussion of security measures in classical cryptography.

\subsection{Quantum}
\label{chapter:covariantIACC}
Similar to the Shannon entropy, the \emph{von Neumann entropy}\index{von Neumann entropy}\index{entropy!von Neumann} of a quantum states $\rho$\index{state!von Neumann entropy}
is given by\label{def:vonneumann}
$$
S(\rho) = - \Tr(\rho \log \rho).
$$
Taking the eigendecomposition of $\rho = \sum_x \lambda_x \outp{x}{x}$ we can also write
$$
S(\rho) = -\sum_x \lambda_x \log \lambda_x,
$$
which corresponds to the Shannon entropy arising from measuring $\rho$ in the basis
given by $\{\outp{x}{x}\}$. We refer to~\cite[Section 11.3]{nielsen&chuang:qc} for the properties
of the von Neumann entropy. 

Here, we will only be concerned with \label{accessInfo}
the \emph{accessible information}~\cite[Eq.\ (9.75)]{peres:book}\index{information!accessible}\index{accessible information}
of an ensemble $\ens = \{p_x, \rho_x\}$ which we 
encounter again in Chapter~\ref{chapter:locking}.\label{def:iacc}
\begin{eqnarray*}
\mI_{acc}(\ens)=
\max_M \left(- \sum_{x} p_{x} \log p_{x} + \right.
\left.\sum_{j} \sum_{x} p_{x} \alpha_j \Tr(M_j \rho_x)
\log \frac{p_{x} \Tr(M_j \rho_x)}{\Tr(M_j\rho)} \right),
\end{eqnarray*}
where $\rho = \sum_x p_x \rho_x$ and the maximization is taken over all POVMs $M = \{M_j\}$. 
It has been shown that we can take all POVM elements to be of rank 1~\cite{davies:access}. However,
maximizing this quantity still remains a hard task~\cite{peres:book}. 
Some upper and lower bounds are known~\cite{fuchs:diss}, but sadly none of them are generally very strong.
The most well-known upper bound is given by
the \emph{Holevo quantity}\index{Holevo quantity}, which is given by\label{def:holevo}
$$
\chi(\rho) = S(\rho) - \sum_x p_x S(\rho_x).
$$
Holevo's theorem~\cite{nielsen&chuang:qc} states that
\begin{equation}\label{iaccupper}
\mI_{acc}(\ens) \leq \chi(\rho).
\end{equation}

\smallskip
\noindent
\textbf{\emph{Classical vs.\ Quantum}}\\
Equality in Eq. (\ref{iaccupper}) is achieved if all states $\rho_x$ have a common eigenbasis (i.e., all $\rho_x$ commute).
Hence, for classical ensembles we do not have a gap between these two quantities. The fact that quantumly
we can obtain such a gap leads to a peculiar effect known as locking classical information in quantum states in Chapter~\ref{chapter:locking}.
However, even if the states $\rho_x$ do not commute, we can still extract the ``classical information'' of the
ensemble: Suppose for all $\rho_x \in \hil$ from our ensemble there exists a decomposition $\hil = \bigoplus_j \hil_j$
such that for all $x$, $\rho_x = \sum_j \Pi_j \rho_x \Pi_j$, where $\Pi_j$ is a projector onto $\hil_j$. That
is, there exists a way to simultaneously block-diagonalize all states. Note that for any measurement
maximizing the accessible information above, we can find an equivalent measurement with measurement
operators $\hat{M} = \sum_j \Pi_j M \Pi_j$, since evidently, $\Tr(\hat{M} \rho_x) = \sum_{j}\Tr(\Pi_j M \Pi_j \rho_x) = \Tr(M \rho_x)$.
Intuitively, this means that we can always first determine which block we are in ``for free'', followed by our original
measurement constrained to this block. 
Note that $[\Pi_j,\rho_x] = 0$ for all $\Pi_j$ and $\rho_x$. Hence, looking back
at Section~\ref{section:invariantMeasure} this is not so surprising:
the measurement leaves our states invariant. 
In general, such commutation relations lead to interesting structural consequences which we examine in more detail
in Appendix~\ref{appendix:cstar} and also exploit in Chapter~\ref{chapter:pistar}.
Finally, it will be useful in Chapter~\ref{chapter:limitations} that the accessible information is additive~\cite{H73,barbara:hiding1}:
For $m$ independent draws 
of an ensemble $\ens$ of separable states (see Chapter~\ref{chapter:entanglementIntro}), i.e., 
we choose $m$ states from $m$ identical ensembles independently, 
we have $\mI_{acc}(\ens^{\otimes m}) = m\mI_{acc}(\ens)$.

\section{Mutually unbiased bases}\label{MUBdef}

In the following chapters, we will be particularly concerned with measurements in \emph{mutually unbiased bases} (MUBs).\index{MUB}\index{mutually unbiased basis}\index{basis!mutually unbiased}\index{basis!MUB}\index{basis!mutually unbiased}
MUBs were initially introduced in the context
of state estimation~\cite{wootters:mub}, but feature in many other problems in quantum information.
The following definition closely follows the one given in~\cite{boykin:mub}.
\begin{definition}[MUBs] \label{def-mub}
Let $\mathcal{B}_1 = \{\ket{\mub^1_1},\ldots,\ket{\mub^1_{d}}\}$ and $\mB_2 =
\{\ket{\mub^2_1},\ldots,\ket{\mub^2_{d}}\}$ be two orthonormal bases in
$\Complex^d$. They are said to be
\emph{mutually unbiased}  if
$|\inp{\mub^1_k}{\mub^2_l}| = 1/\sqrt{d}$, for every $k,l \in[d]$. A set $\{\mathcal{B}_1,\ldots,\mathcal{B}_m\}$ of
orthonormal bases in $\Cn^d$ is called a \emph{set of mutually
unbiased bases} if each pair of bases is mutually unbiased.
\end{definition}

As an example, consider the computational and the Hadamard basis defined above,
and note that we can write $\ket{+} = H\ket{0}$ and $\ket{-} = H\ket{1}$. We then have for $x \in \01^n$
that
$$
|\bra{x}H^{\otimes n}\ket{x}|^2 = \frac{1}{2^n}.
$$
Hence, the computational and the Hadamard basis are mutually unbiased in 
dimension $d=2^n$.

We use $N(d)$ to denote the maximal number of MUBs in dimension $d$.
In any dimension $d$, we have that
$\N(d) \leq d+1$~\cite{boykin:mub}. If $d = p^k$ is a prime power, we have
that $\N(d) = d+1$ and explicit constructions are
known~\cite{boykin:mub,wootters:mub}. If $d = s^2$ is a square,
$\N(d) \geq \MOLS(s)$ where $\MOLS(s)$ denotes the number of mutually orthogonal
$s \times s$ Latin squares~\cite{wocjan:mub}.\index{MOLS}\index{Latin square!mutually orthogonal} In general, we have
$\N(n m) \geq \min\{\N(n),\N(m)\}$ for all $n,m \in \Natural$~\cite{zauner:diss,klappenecker:mubs}.
It is also known that in any dimension, there exists an explicit construction for 3 MUBs~\cite{grassl:mub}.
Unfortunately, not much else is known. For example,
it is still an open problem whether there exists a set of $7$ MUBs in dimension $d=6$.
We say that a unitary $U_t$ \emph{transforms the computational basis into the $t$-th MUB}
$\mB_t = \{\ket{b^t_1},\ldots,\ket{b^t_d}\}$
if for all $k \in [d]$ we have $\ket{b^t_k} = U_t\ket{k}$.
In the next two chapters, we will be particularly concerned with two specific constructions of mutually unbiased bases.
There exists a third construction based on Galois rings~\cite{galois:mub}, which we do not
consider here.

\subsection{Latin squares}\label{latinSquareConstruction}\index{Latin square}\index{MUB!from Latin squares}\index{mutually unbiased basis!from Latin squares}

First, we consider MUBs based on mutually orthogonal Latin squares~\cite{wocjan:mub}.
Informally, an $s \times s$ Latin square over the symbol set $[s]$
is an arrangement
of elements of $[s]$ into an $s \times s$ square such that in each row and each column every element
occurs exactly once. Let $L_{ij}$ denote the entry in a Latin square in row $i$ and column $j$.
Two Latin squares $L$ and $L'$ are called mutually orthogonal if and only if
$\{(L_{i,j},L'_{i,j})|i,j \in [s]\} = \{(u,v)|u,v \in [s]\}$. Intuitively, this means that if we place
one square on top of the other, and look at all pairs generated by the overlaying elements, all possible pairs
occur. An example is given in Figures~\ref{LatinSquareMOLS1} and~\ref{LatinSquareMOLS2} below.
From any $s\times s$ Latin square we can obtain a basis for $\Cn^{s}\otimes \Cn^{s}$.
First, we construct $s$ of the basis vectors from the entries of
the Latin square itself. Let 
$$
\ket{v_{1,\ell}} = \frac{1}{\sqrt{s}} \sum_{i,j\in [s]} E^L_{i,j}(\ell) \ket{i,j},
$$
where $E^L$ is a predicate such that $E^L_{i,j}(\ell) = 1$ if and only if $L_{i,j} = \ell$.
Note that for each $\ell$ we have exactly $s$ pairs $i,j$ such that $E_{i,j}(\ell) = 1$, because
each element of $[s]$ occurs exactly $s$ times in the Latin square.
Secondly, from each such vector we obtain $s-1$ additional vectors by adding successive rows
of an $s \times s$ complex Hadamard matrix $H = (h_{ij})$ as coefficients to obtain the remaining
$\ket{v_{t,j}}$ for $t \in [s]$, where $h_{ij} = \omega^{ij}$ with $i,j \in \{0,\ldots,s-1\}$ and
$\omega = e^{2 \pi i/s}$.
Two additional MUBs can then be obtained in the same way from the two non-Latin squares where
each element occurs for an entire row or column respectively. From each mutually orthogonal Latin square
and these two extra squares which also satisfy the above orthogonality condition, we obtain one basis.
This construction therefore gives $\MOLS(s) + 2$ many MUBs. It is known that if $s = p^k$ is a
prime power itself, we obtain
$p^k+1\approx \sqrt{d}$ MUBs from this construction. Note, however, that there do exist many more
MUBs in prime power dimensions, namely $d+1$. If $s$ is not a prime power, it is merely known
that $\MOLS(s) \geq s^{1/14.8}$~\cite{wocjan:mub}.

\begin{figure}[h]
\begin{minipage}{0.45\textwidth}
\begin{center}
\includegraphics{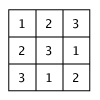}
\caption{Latin Square (LS)}
\label{LatinSquareMOLS1}
\end{center}
\end{minipage}
\begin{minipage}{0.45\textwidth}
\begin{center}
\includegraphics{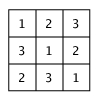}
\caption{Mutually Orthogonal LS}
\label{LatinSquareMOLS2}
\end{center}
\end{minipage}
\end{figure}

As an example, consider the $3 \times 3$ Latin square depicted in Figure~\ref{LatinSquareMOLS1} and
the $3 \times 3$ complex Hadamard matrix\index{complex Hadamard matrix}
$$
H = \left(\begin{array}{ccc}
    1 &1& 1\\
    1 &\omega &\omega^2\\
    1 &\omega^2& \omega
\end{array}\right),
$$
where $\omega = e^{2 \pi i/3}$.
First, we obtain vectors
\begin{eqnarray*}
\ket{v_{1,1}} &=& \frac{1}{\sqrt{3}}(\ket{1,1} + \ket{2,3} + \ket{3,2})\\
\ket{v_{1,2}} &=& \frac{1}{\sqrt{3}}(\ket{1,2} + \ket{2,1} + \ket{3,3})\\
\ket{v_{1,3}} &=& \frac{1}{\sqrt{3}}(\ket{1,3} + \ket{2,2} + \ket{3,1}).
\end{eqnarray*}
With the help of $H$ we obtain 3
additional vectors from the ones above. From the vector $\ket{v_{1,1}}$,
for example, we obtain
\begin{eqnarray*}
\ket{v_{1,1}} &=& \frac{1}{\sqrt{3}}(\ket{1,1} + \ket{2,3} + \ket{3,2})\\
\ket{v_{2,1}} &=& \frac{1}{\sqrt{3}}(\ket{1,1} + \omega \ket{2,3} + \omega^2 \ket{3,2})\\
\ket{v_{3,1}} &=& \frac{1}{\sqrt{3}}(\ket{1,1} + \omega^2 \ket{2,3} + \omega \ket{3,2}).
\end{eqnarray*}
This gives us basis $\mathcal{B} = \{\ket{v_{t,\ell}}|t,\ell \in [s]\}$ for $s = 3$.
The construction of another basis follows in exactly the same way from a mutually orthogonal
Latin square. The fact that two such squares $L$ and $L'$ are mutually orthogonal ensures
that the resulting bases will be mutually unbiased. Indeed, suppose we are given another such basis,
$\mathcal{B'} = \{\ket{u_{t,\ell}}|t,\ell \in [s]\}$ belonging to $L'$. We then have for any $\ell,\ell' \in [s]$ that
$|\inp{u_{1,\ell'}}{v_{1,\ell}}|^2 =
|(1/s) \sum_{i,j\in [s]} E^{L'}_{i,j}(\ell') E^L_{i,j}(\ell)|^2 = 1/s^2$, as there exists exactly only
one pair $\ell,\ell' \in [s]$ such that $E^{L'}_{i,j}(\ell') E^L_{i,j}(\ell) = 1$. Clearly, the same argument
holds for the additional vectors derived from the complex Hadamard matrix.

\subsection{Generalized Pauli matrices}\label{mub:pauliMubs}\index{MUB!from Pauli matrices}\index{mutually unbiased basis!from Pauli matrices}\index{Pauli matrices!generalized}
\index{generalized Pauli matrices}

The second construction we consider is based on the generalized Pauli matrices 
$X_d$ and $Z_d$~\cite{boykin:mub}, defined by their actions on the 
computational basis $C = \{\ket{0},\ldots,\ket{d-1}\}$ as follows: 
\begin{eqnarray*}
X_d\ket{k} &=& \ket{k+1\mod d}\\
Z_d\ket{k} &=& \omega^k\ket{k},~\forall \ket{k} \in C,
\end{eqnarray*}
where $\omega = e^{2 \pi i/d}$.  We say that $\left(X_{d}\right)^{a_1} \left(Z_{d}\right)^{b_1} 
\otimes \cdots \otimes \left(X_{d}\right)^{a_N} \left(Z_{d}\right)^{b_N}$ for 
$a_k,b_k \in \{0,\ldots,d-1\}$ and $k \in [N]$ is a \emph{string of Pauli matrices}.\index{Pauli matrices!strings of}\index{string of Pauli matrices}

If $d$ is a prime, it is known that the $d+1$ MUBs constructed first by
Wootters and Fields~\cite{wootters:mub} can also be obtained as the eigenvectors of the 
matrices $Z_d,X_d,X_dZ_d,X_dZ_d^2,\ldots,X_dZ_d^{d-1}$~\cite{boykin:mub}. If $d = p^k$ is a prime power,
consider all $d^2-1$ possible strings of Pauli matrices excluding the identity and group them
into sets $C_1,\ldots,C_{d+1}$ such that $|C_i| = d - 1$ and $C_i \cap C_j = \{\id\}$ for $i \neq j$ and all
elements of $C_i$ commute. Let $B_i$ be the common eigenbasis of all elements of $C_i$. Then
$B_1,\ldots,B_{d+1}$ are MUBs~\cite{boykin:mub}. A similar result for $d = 2^k$ has also been shown 
in~\cite{lawrence:mub}.
A special case of this construction are the three mutually unbiased bases in dimension $d=2^k$ given by 
the unitaries $\id^{\otimes k}$,$H^{\otimes k}$ and $K^{\otimes k}$ (as defined on page~\pageref{ktransform})
applied to the computational basis.

\section{Conclusion}

We summarized the most important elements of quantum theory that we need here. 
We refer to~\cite{peres:book, nielsen&chuang:qc, hayashi:book} for more information about
each topic. In Chapters~\ref{chapter:uncertainty} and~\ref{chapter:entanglement} we investigate
the two most striking aspects of quantum theory in detail: uncertainty relations and entanglement.
But first, let's examine the case of state discrimination \emph{with} additional post-measurement information.

\chapter[State discrimination with post-measurement information]
{State discrimination\\ with post-measurement information}
\label{stateDiscrimination}
\index{post-measurement
information}\index{state!discrimination with post-measurement information}
\label{pistarChapter}\label{chapter:pistar}

In this chapter, we investigate an extension of the traditional state discrimination problem 
we encountered in Chapter~\ref{stateDiscriminationVanilla}: what if we are given some additional
information \emph{after} the measurement? 
Imagine that you are given a string $x$ encoded in an unknown basis chosen from a known set of bases. 
You may perform any measurement, but you can only store at most $q$ qubits of quantum information afterwards.
Later on, you are told which basis was used. How well can you compute a function
$f$ of $x$, given the initial measurement outcome, the $q$ qubits and the additional basis 
information? 

\section{Introduction}

This question is of central importance for protocols in the bounded quantum 
storage model~\cite{serge:bounded}\index{bounded storage!quantum}, which
we encountered in Chapter~\ref{chapter:cryptoIntro}.
The security of such protocols rests on the realistic assumption that a dishonest
player cannot store more than $q$ qubits for long periods of time. In this model,
even bit commitment and oblivious transfer can be implemented securely which
is otherwise known to be impossible as we saw 
in Chapter~\ref{chapter:cryptoIntro}. 
We formalize this general setting as a state discrimination problem: Here, we are given additional 
information about the state after the measurement or, more generally, after a quantum
memory bound is applied. We prove general bounds on the success probability
for any balanced function. We also show that storing just a \emph{single}
qubit allows you to compute any Boolean function perfectly when two bases are used. However,
we also construct three bases for which you need to keep \emph{all} qubits. 

In general, we consider the following problem:
Take an ensemble of quantum states,
${\cal E} = \{p_{yb},\rho_{yb}\}$, with double indices
$yb \in {\cal Y}\times{\cal B}$, and an integer $q\geq 0$.
Suppose Alice sends Bob the state $\rho_{yb}$, where she alone knows indices $y$ and $b$.
Bob can perform any measurement on his system, 
but afterwards store at most $q$ qubits, and an unlimited amount of classical
information.
Afterwards, Alice tells him $b$. Bob's goal is now
to approximate $y$ as accurately as possible, which means that
he has to make a guess $\hat{Y}$ that maximizes the success probability
$
  P_{\rm succ} = \sum_{yb} p_{yb} \Pr[ \hat{Y} = y | \text{state } \rho_{yb} ].
$
For $|{\cal B}|=1$, i.e.,~no post-measurement information is available, $q$ is irrelevant
and Bob's task is to discriminate among states $\rho_y$. This is the well-known state
discrimination problem, which we encountered in Chapter~\ref{stateDiscriminationVanilla},
a problem studied since the early days of quantum information science.
\begin{figure}
\begin{center}
\includegraphics[scale=1.4]{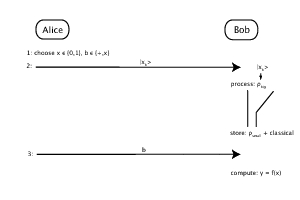}
\caption{Using post-measurement information.}
\label{figure:usingPISTAR}
\end{center}
\end{figure}
A particular case that isolates the aspect
of the timing between measurements and side-information is one where for
each fixed $b$, the states $\rho_{yb}$ are mutually orthogonal:
if Bob knew $b$, he could actually compute $y$ perfectly.
A special case of this problem is depicted in Figure~\ref{figure:usingPISTAR}.
Here, Alice picks a string $x \in_R \01^n$, and a basis $b \in \{+,\times\}$.
She then encodes the string in the chosen basis and sends the resulting state
to Bob.
Bob's goal is now to determine $y = f(x)$ for a fixed function $f$.
The states in this particular problem are thus of the form
$\rho_{yb} = \sum_{x\in f^{-1}(y)} P_{X|B=b}(x) U_b \outp{x}{x} U_b^\dagger$,
for a function $f:{\cal X}\rightarrow{\cal Y}$, and a set of mutually unbiased bases (MUBs) ${\cal B}$, given by the unitaries 
$U_0=\id,U_1,\ldots,U_{|{\cal B}|-1}$ on a Hilbert space
with basis $\{\ket{x}:x\in{\cal X}\}$, where the string $x$ and a basis $b$
are drawn from the distribution $P_{X,B}$. 
We mostly focus on this special case.

This problem also has an interpretation in terms of communication complexity. Suppose
Alice is given $b$, and Bob is given the state $\rho_{yb}$. 
If classical communication is free, what is the minimum number of qubits Bob needs to
send to Alice such that Alice learns $y$? Note that Bob needs to send exactly $q$ qubits 
if and only if there exists a strategy for Bob to compute $y$ in our task, while storing only $q$ qubits.

\subsection{Outline}
In the following, we will close in on our problem in several stages.
First, we briefly recall the case of state discrimination \emph{without} any post-measurement
information in Section~\ref{plainSTAR}. This enables us to draw comparisons later.

Second, in Section~\ref{PISTARnomem} we assume that Bob does receive post-measurement 
information, but has no quantum memory at all, i.e. $q = 0$.
His goal then is to compute $f(x)$ given the classical outcome obtained by
measuring $U_b\ket{x}$ and the later announcement of $b$. 
Clearly, a trivial strategy for Bob is to simply guess the basis, measure to obtain
some string $\hat{x}$ and take $\hat{y} = f(\hat{x})$ as his answer. 
We thus want to find a better strategy. In particular, we will see that
for any number 
of MUBs, any number of function outcomes, and any balanced $f$, Bob has a systematic
advantage over guessing the basis, independent of $|{\cal X}|$.
Furthermore, we show that for \emph{any} Boolean $f$, Bob can succeed
with probability at least $P_{succ} \geq 1/2 + 1/(2\sqrt{2})$ even if he cannot store 
any qubits at all.
The latter result is relevant to the question of whether deterministic privacy
amplification is possible in the protocols of~\cite{serge:bounded}.
Here, Alice uses two MUBs, and secretly chooses a
function from a set of predetermined functions. She later tells
Bob which function he should evaluate, together
with the basis information $b$. 
Is it possible to use a fixed Boolean function
instead? Our result shows that this is not possible.

It is interesting to consider when post-measurement information is useful
for Bob, and how large his advantage is compared to the case where he does not
receive any post-measurement information. To this end, we show how to phrase our problem as a 
semidefinite program (SDP), in the case where Bob has no quantum memory. In Section~\ref{optimalBound},
we examine in detail the specific functions 
XOR and AND, for which we prove optimal bounds on Bob's success probability. In particular, the XOR
on uniformly distributed strings of length $n$ with two or three 
MUBs provides an extreme example of the usefulness of post-measurement information:
We show that for the XOR function
with $n$ odd, $P_{\rm succ} = 1/2 + 1/(2\sqrt{2})$. This is the same as Bob can achieve
\emph{without} the extra basis information. For even $n$, $P_{\rm succ} = 1$
with the additional basis information.
Here, $P_{\rm succ}$ jumps from $3/4$ (without) to certainty (with basis information). 
The advantage that Bob gains can thus be maximal: \emph{without} the post-measurement information, 
he can do no better than guessing the basis. However, \emph{with} it, he can compute $y = f(x)$ 
perfectly. 
For even $n$, this was also observed in~\cite{serge:bounded}.
However, our analysis for odd $n$ shows that the strategy for
even $n$ does \emph{not} work for any linear function as 
claimed in~\cite{serge:bounded}.
It remains an interesting question to find general conditions on 
the ensemble of states that determine how useful post-measurement information 
can be. We return to this question in Chapter~\ref{useofPISTAR}.

Finally, we address the case where Bob does have quantum memory available.
The question we are then interested in is: How large does this memory have
to be so that Bob can compute $y$ perfectly? In Section~\ref{memoryFramework}, 
we derive general conditions that determine
when $q$ qubits are sufficient. Our conditions impose a restriction on the rank of Bob's
measurement operators and require that all such operators commute with the projector onto the support of $\rho_{yb}$,
for all $y$ and $b$. In particular,
we give a general algebraic framework that allows us to 
determine $q$ for any number of bases, functions and outcomes,
in combination with an algorithm given in~\cite{koashi&imoto:operations}. 
In Sections~\ref{limitedInfo2Bases} and~\ref{limitedInfo3Bases}, we 
then consider two specific examples:
First, we show that for \emph{any} Boolean $f$ and \emph{any} two bases,
storing just a \emph{single} qubit is sufficient for Bob to compute $f(x)$
perfectly.
The latter result again has implications to protocols in the bounded quantum storage
model: for all existing protocols, deterministic privacy amplification
is indeed hopeless.
It turns out that part of this specific example also follows from 
known results derived for non-local games as we will discuss below.
Surprisingly, things change dramatically when we are allowed to 
use three bases: 
We show how to construct three bases, such that for \emph{any} balanced $f$ Bob needs
to keep \emph{all} qubits in order to compute $f(x)$ perfectly!

\subsection{Related work}

In Chapter~\ref{stateDiscriminationVanilla}, we already examined the traditional setting
of state discrimination \emph{without} post-measurement information. Some of the tools
we need below have found use in this setting as well.
Many convex optimization problems can be solved using semidefinite programming.
We refer to Appendix~\ref{semidefIntro} for a introduction.
Eldar~\cite{eldar:sdp} and
Eldar, Megretski and Verghese~\cite{eldar:sdpDetector}
used semidefinite programming to solve state discrimination
problems, which is one of the
techniques we also use here. The square-root
measurement~\cite{hausladen:pgm} (also called pretty good measurement)
is an easily constructed measurement to distinguish quantum states,
however, it is only optimal for
very specific sets of states~\cite{eldar:pgm,eldar:symmetric}.
Mochon constructed specific pure state discrimination problems for which the square-root
measurement is optimal~\cite{mochon:pgm}.
We use a variant of the square-root measurement as well.
Furthermore, our problem is related to the task of state
filtering~\cite{bergou:filtering0,bergou:filtering, bergou:filtering2}
and state classification~\cite{wang:classification}.
Here, Bob's goal is to determine whether a given state is
either one specific state or one of several other possible
states, or, more generally, which subset of states a
given state belongs to. Our scenario differs, because
we deal with mixed states and Bob
is allowed to use post-measurement information. Much more is known about pure state
discrimination problems and the case
of unambiguous state discrimination where we are not allowed to make an error.
Since we concentrate on mixed states, we refer
to~\cite{bergou:survey} for an excellent survey on
the extended field of state discrimination.

Regarding state discrimination with post-measurement information,
special instances of the general problem have occurred in the literature
under the heading ``mean king's problem''~\cite{aharonov&englert:meanking,%
klappenecker&roetteler:meanking},
where the stress was on the usefulness of entanglement. Furthermore,
it should be noted that prepare-and-measure quantum key distribution
schemes of the BB84 type also lead to special cases of this problem:
When considering optimal individual attacks, the eavesdropper is faced
with the task of extracting maximal information about the raw key bits, encoded
in an unknown basis, that she learns later during basis reconciliation.

Our result that one qubit of storage suffices for any Boolean function $f$
demonstrates that storing quantum information can give an adversary a great
advantage over storing merely classical information.
It has also been shown in the context of randomness extraction with
respect to a quantum adversary that storing quantum information can sometimes
convey much more power to the adversary~\cite{ronald:qmem}.

\section{Preliminaries}\label{prelimPISTAR}

\subsection{Notation and tools}

We need the following notions. The Bell basis is given by the vectors
$\ket{\Phi^{\pm}} = (\ket{00} \pm \ket{11})/\sqrt{2}$ and $\ket{\Psi^{\pm}}
                  = (\ket{01} \pm \ket{10})/\sqrt{2}$.
Furthermore, let $f^{-1}(y) = \{x \in \mX|f(x) = y\}$.
We say that a function $f$ is balanced if and only if any element
in the image of $f$ is generated by equally many elements in the pre-image of $f$, i.e.
there exists a $k \in \Natural$ such that $\forall y \in \mY: |f^{-1}(y)| = k$.

\subsection{Definitions}\label{STARdef}\label{PISTARdef}
We now give a more formal description of our problem.
Let $\mY$ and $\mB$ be finite sets and
let $P_{YB} = \{p_{yb}\}$ be a probability distribution over 
$\mY \times \mB$. Consider an ensemble 
of quantum states $\ens = \{p_{yb},\rho_{yb}\}$. 
We assume that $\mY$, $\mB$, $\ens$ and $P_{YB}$ 
are known to both Alice and Bob. Suppose 
now that Alice chooses $yb \in \mY \times \mB$ 
according to probability distribution $P_{YB}$, 
and sends $\rho_{yb}$ to Bob. We can then define the tasks:

\begin{definition}\label{def:STAR}\index{STAR}\index{state discrimination}
\emph{State discRimination} ($\text{\emph{STAR}}(\ens)$)
is the following task for Bob.
Given $\rho_{yb}$, determine $y$. He can perform any
measurement on $\rho_{yb}$ immediately upon receipt.
\end{definition}

\begin{definition}\label{def:PISTAR}\index{PI-STAR}\index{post-measurement information}
\index{state discrimination!with post-measurement information}
\emph{State discRimination with Post-measurement Information}
($\text{\emph{PI}}_q\text{\emph{-STAR}}(\ens)$) is 
the following task for Bob.
Given $\rho_{yb}$, determine $y$, where Bob can use the following
sources of information in succession:
\begin{enumerate}
\item \label{item1} First, he can perform any measurement
       on $\rho_{yb}$ immediately upon reception. 
       Afterwards, he can store at most $q$ qubits of quantum
       information about $\rho_{yb}$, and an unlimited amount of
       classical information.
\item \label{item2} After Bob's measurement, Alice announces $b$. 
\item \label{item3} Then, he may perform any measurement
on the remaining $q$ qubits
      depending on $b$ and the measurement outcome obtained in step 1.
\end{enumerate}
\end{definition}

We also say that \emph{Bob succeeds at $\text{\emph{STAR}}(\ens)$ or\index{PI-STAR!to succeed at}\index{STAR!to succeed at}
$\text{\emph{PI}}_q\text{\text{-STAR}}(\ens)$ with probability
$p$} if and only if $p$ is the average success probability 
$p = \sum_{yb} p_{yb} \Pr[ \hat{Y} = y | \text{state } \rho_{yb}]$, 
where $\Pr[\hat{Y} = y|\text{state }\rho_{yb}]$ is the 
probability that Bob correctly determines $y$ given $\rho_{yb}$ in the case of STAR, and 
in addition using information sources \ref{item1}, \ref{item2}
and \ref{item3} in the case of PI-STAR.

Here, we are interested in the following special case: 
Consider a function $f:\mX \rightarrow \mY$ between finite sets, and
a set of mutually unbiased bases $\mB$ as defined in Chapter~\ref{informationIntro}, generated 
by a set of unitaries $U_0,U_1,\ldots,U_{|\mB|-1}$
acting on a Hilbert space with basis \{$\ket{x}\mid x \in \mX$\}.
Take $\ket{\Phi_b^x} = U_b \ket{x}$.
Let $P_X$ and $P_B$ be probability distributions over $\mX$ and $\mB$ respectively.
We assume that $f$, $\mX$, $\mY$, $\mB$, $P_X$, $P_B$, and 
the set of unitaries $\{U_b|b \in \mB\}$
are known to both Alice and Bob. Suppose now that Alice
chooses $x \in \mX$ and  
$b \in \mB$ independently according to probability
distributions $P_X$ and $P_B$ respectively, and sends $\ket{\Phi_b^x}$ to Bob.
Bob's goal is now to compute $y = f(x)$. We thus 
obtain an instance of our problem with states
$\rho_{yb} = \sum_{x \in f^{-1}(y)} P_X(x) \outp{\Phi_b^x}{\Phi_b^x}$. 
We write $\text{STAR}(f)$ and $\text{PI}_q\text{-STAR}(f)$ 
to denote both problems in this special case. We concentrate on the case of 
mutually unbiased bases, as this case is most relevant to our initial goal of analyzing 
protocols for quantum cryptography in the bounded storage model~\cite{serge:bounded}.

Here, we make use of the basis set
$\mB = \{+,\times,\odot\}$, where $\mB_+ = \{\ket{0},\ket{1}\}$ 
is the computational basis,
$\mB_\times = \{ \frac{1}{\sqrt{2}}(\ket{0} + \ket{1}),
                 \frac{1}{\sqrt{2}}(\ket{0} - \ket{1})\}$
is the Hadamard basis, and
$\mB_\odot = \{ \frac{1}{\sqrt{2}}(\ket{0} + i \ket{1}),
                \frac{1}{\sqrt{2}}(\ket{0} - i\ket{1})\}$ is
what we call the K-basis.
The unitaries that give rise to these bases are $U_+ = \id$, $U_\times = H$ and 
$U_\odot =  K$ with $K = (\id+i \sigma_x)/\sqrt{2}$ respectively.
Recall from Chapter~\ref{informationIntro} that the Hadamard matrix is given by 
$H = \frac{1}{\sqrt{2}}(\sigma_x + \sigma_z)$, and that
$\sigma_x$, $\sigma_z$ and $\sigma_y$ are the well-known Pauli matrices.
We generally assume that Bob has no a priori knowledge about the outcome of the function 
and about the value of $b$. This means that
$b$ is chosen uniformly at random from $\mB$, and, in the case of balanced functions, 
that Alice chooses $x$ uniformly at random from $\mX$. More generally,
the distribution is uniform on all $f^{-1}(y)$ and such that each
value $y\in{\cal Y}$ is equally likely.

\subsection{A trivial bound: guessing the basis}

Note that a simple strategy for Bob is to guess the basis, and then measure.
This approach leads to a lower bound on the success probability for 
both $\text{STAR}$ and $\text{PI-STAR}$. In short:
\begin{lemma}\label{guessing}
  Let $P_X(x) = \frac{1}{2^n}$ for all $x \in \01^n$.
  Let $\mB$ denote the set of bases.
  Then for any balanced function $f: \mX \rightarrow \mY$ Bob succeeds at 
  $\text{\emph{STAR}}(f)$ and $\text{\emph{PI}}_0\text{\emph{-STAR}}(f)$ with
  probability at least
  \[
    p_{\text{guess}}=\frac{1}{|\mB|} + \left(1 - \frac{1}{|\mB|}\right) \frac{1}{|\mY|}.
  \]
\end{lemma}
\noindent
Our goal is to beat this bound. We show that for $\text{PI-STAR}$, Bob can indeed 
do much better.

\section{No post-measurement information}\label{noinfo}\label{plainSTAR}

We first consider the standard case of state discrimination. Here, Alice does not
supply Bob with any additional post-measurement information. Instead, Bob's goal is
to compute $y = f(x)$ immediately. This analysis enables us to gain interesting insights
into the usefulness of post-measurement information later.

\subsection{Two simple examples}

We now examine two simple one-qubit examples of a state discrimination problem,
which we make use of later on. Here, Bob's goal is to learn the value of a bit
which has been encoded in two or three mutually unbiased bases while he does
not know which basis has been used.

\begin{lemma}\label{Breitbart}
Let $x \in \01$, $P_X(x) = \frac{1}{2}$ and $f(x) = x$.
Let $\mB =\{+,\times\}$ with $U_+ = \id$ and $U_\times = H$.
Then Bob succeeds at $\text{\emph{STAR}}(f)$ with probability at most
$$
p=\frac{1}{2}+\frac{1}{2\sqrt{2}}.
$$
There exists a strategy for Bob that achieves $p$.
\end{lemma}
\begin{proof}
The probability of success follows from Theorem~\ref{helstrom}
with $\rho_0 =\frac{1}{2}(\outp{0}{0} +H\outp{0}{0}H)$, 
$\rho_1 =\frac{1}{2}(\outp{1}{1} + H\outp{1}{1}H)$ and $q=1/2$.
\end{proof}

\begin{lemma}\label{Kabuki}
Let $x \in \01$, $P_X(x) = \frac{1}{2}$ and $f(x) = x$.
Let $\mB =\{+,\times,\odot\}$ with $U_+ = \id$, $U_\times = H$ and $U_\odot=K$.
Then Bob succeeds at $\text{\emph{STAR}}(f)$ with probability at most
$$
p=\frac{1}{2}+\frac{1}{2\sqrt{3}}.
$$
There exists a strategy for Bob that achieves $p$.
\end{lemma}
\begin{proof}
The proof is identical to that of Lemma~\ref{Breitbart} using
$\rho_0 =\frac{1}{3}(\outp{0}{0} +H\outp{0}{0}H+K\outp{0}{0}K^\dagger)$,
$\rho_1 =\frac{1}{3}(\outp{1}{1} +H\outp{1}{1}H+K\outp{1}{1}K^\dagger)$,
and $q=1/2$.
\end{proof}

\subsection{An upper bound for all Boolean functions}\index{state discrimination!Boolean functions}
\index{STAR!Boolean functions}

We now show that for any Boolean function $f$ and any number of mutually unbiased
bases, the probability that Bob succeeds at $\text{STAR}(f)$ is very limited. 
\begin{theorem}\label{starless0.85}
Let $|\mY|=2$ and let $f$ be a balanced function. Let $\mB$ be a set of 
mutually unbiased bases.
Then Bob succeeds at $\text{\emph{STAR}}(f)$ with probability at most
$$
  p = \frac{1}{2}+\frac{1}{2\sqrt{|\mB|}}.
$$
In particular, for $|\mB|=2$ we obtain $(1 + 1/\sqrt{2})/2 \approx 0.853$;
for $|\mB|=3$, we obtain $(1 + 1/\sqrt{3})/2 \approx 0.789$.
\end{theorem}
\begin{proof}
The probability of success is given by Theorem~\ref{helstrom}
where for $y \in \01$
$$
\rho_y =\frac{1}{2^{n-1}|\mB|} \sum_{b=0}^{|\mB|-1} P_{yb},
$$
with $P_{yb} = \sum_{x \in f^{-1}(y)} U_b \outp{x}{x} U_b^{\dagger}$. 
Using the Cauchy-Schwarz inequality we can show that
\begin{equation}\label{BoundTrDistWithTrofSquare}
\|\rho_0-\rho_1\|_1^2 =  [\Tr(|\rho_0-\rho_1| \id)]^2 \leq \Tr[(\rho_0-\rho_1)^2] \Tr[\id^2]
                       =     2^n\Tr[(\rho_0-\rho_1)^2],
\end{equation}
or
$$
  \|\rho_0-\rho_1\|_1 \leq \sqrt{2^n \Tr[(\rho_0-\rho_1)^2]}.
$$
A simple calculation shows that
$$
\Tr[(\rho_0-\rho_1)^2]=\frac{4}{2^n|\mB|}.
$$
The theorem follows from the previous equation,
together with Theorem~\ref{helstrom} and
Eq.~(\ref{BoundTrDistWithTrofSquare}).
\end{proof}

\subsection{AND function}\index{state discrimination!AND}
\index{STAR!AND}

One of the simplest functions to consider is the AND 
function. Recall, that we always assume that 
Bob has no a priori knowledge about the outcome of the 
function. In the case of the AND, this means 
that we are considering a very specific prior: with 
probability $1/2$ Alice will choose the only
string $x$ for which $\text{AND}(x) = 1$. Without any 
post-measurement information, Bob can already
compute the AND quite well.
\begin{theorem}
Let $P_X(x) = 1/(2(2^n-1))$ for all $x \in \01^n\setminus\{1\ldots1\}$
and $P_X(1\ldots 1) = \frac{1}{2}$.
Let $\mB = \{+,\times\}$ with $U_+ = \id^{\otimes n}$, 
$U_\times = H^{\otimes n}$ and $P_B(+) = P_B(\times) = 1/2$. 
Then Bob succeeds at $\text{\emph{STAR}}(\text{\emph{AND}})$ 
with probability at most
\begin{equation}\label{OptProbSTARAND}
p=\left \{ \begin{array}{ll}
\frac{1}{2}+\frac{1}{2\sqrt{2}} & \textrm{ if } n=1,\\
1-\frac{1}{2(2^n-1)} & \textrm{ if } n\geq 2.
\end{array}
 \right.
\end{equation}
There exists a strategy for Bob that achieves $p$.
\end{theorem}
\begin{proof}
Let $\ket{c_1}=\ket{1}^{\otimes n}$ and $\ket{h_1}=[H\ket{1}]^{\otimes n}$. 
Eq.~(\ref{OptProbSTARAND}) is obtained by substituting
\begin{eqnarray*}
\rho_0&=&\frac{1}{2}\left[\frac{\id-\outp{c_1}{c_1}}{2^n-1}
                         +\frac{\id-\outp{h_1}{h_1}}{2^n-1}\right],\\
\rho_1&=&\frac{\outp{c_1}{c_1}+\outp{h_1}{h_1}}{2},
\end{eqnarray*}
and $q=1/2$ in Theorem \ref{helstrom}.
\end{proof}

In Theorem \ref{PISTARAND}, we show an optimal bound for 
the case that Bob does indeed
receive the extra information. By comparing the previous
equation with Eq.~(\ref{OptProbPISTARAND}) later on, 
we can see that for $n=1$ announcing the basis does not help. 
However, for $n>1$ we will observe an 
improvement of $[2(2^n+2^{n/2}-2)]^{-1}$.

\subsection{XOR function}\index{state discrimination!XOR}
\index{STAR!XOR}

The XOR function provides an example of a Boolean function 
where we observe both the largest advantage 
as well as the smallest advantage in receiving post-measurement 
information: For strings of even length
we show that without the extra information Bob 
can never do better than guessing the basis. For
strings of odd length, however, he can do quite a bit better. 
Interestingly, it turns out that in 
this case the post-measurement information is completely useless to him.
We first investigate how well Bob does at $\text{STAR}(\text{XOR})$ for two bases:

\begin{theorem}\label{xorSTAR2MUBS}\label{evenOddPISTAR}
Let $P_X(x) = \frac{1}{2^n}$ for all $x \in \01^n$.
Let $\mB = \{+,\times\}$ with $U_+ = \id^{\otimes n}$, 
$U_\times = H^{\otimes n}$ and $P_B(+) = P_B(\times) = 1/2$. 
Then Bob succeeds at $\text{\emph{STAR}}(\text{\emph{XOR}})$ 
with probability at most
\[
  p = \begin{cases} \frac{3}{4} & \text{ if } $n$ \text{ is even}, \\
                    \frac{1}{2}\left(1 + \frac{1}{\sqrt{2}}\right) 
                                & \text{ if } $n$ \text{ is odd}.
      \end{cases}
\]
There exists a strategy for Bob that achieves $p$.
\end{theorem}
\begin{proof}
Our proof works by induction on $n$. 
The case of $n=1$ was addressed in Lemma~\ref{Breitbart}.
Now, consider $n=2$: 
Let
$\sigma^{(2)}_0 = \frac{1}{2}(\rho^{(2)}_{0+} + \rho^{(2)}_{0\times})$ and 
$\sigma^{(2)}_1 = \frac{1}{2}(\rho^{(2)}_{1+} + \rho^{(2)}_{1\times})$, where 
$\rho^{(2)}_{0+}$ and $\rho^{(2)}_{1+}$ are defined as 
$\rho^{(n)}_{y b} =
\frac{1}{2^{n-1}} \sum_{x \in \01^n, x \in \text{XOR}^{-1}(y)} U_b\outp{x}{x}U_b^{\dagger}$ with
$y \in \01$ and $b \in \mB = \{+,\times\}$.
A straightforward calculation shows that
$\|\sigma^{(2)}_0 - \sigma^{(2)}_1\|_1 = 1$.

We now show that the trace distance does not change when we go from 
strings of length $n$ to strings of length $n+2$: 
Note that we can write
\begin{eqnarray}\begin{aligned}\label{eq:fromNtoNplus2}
  \rho_{0+}^{(n+2)}      &= \frac{1}{2}(\rho^{(n)}_{0+} \otimes \rho^{(2)}_{0+}
                                      + \rho^{(n)}_{1+} \otimes \rho^{(2)}_{1+})           \\
  \rho_{0\times}^{(n+2)} &= \frac{1}{2}(\rho^{(n)}_{0\times} \otimes \rho^{(2)}_{0\times}
                                      + \rho^{(n)}_{1\times} \otimes \rho^{(2)}_{1\times}) \\
  \rho_{1+}^{(n+2)}      &= \frac{1}{2}(\rho^{(n)}_{0+} \otimes \rho^{(2)}_{1+}
                                      + \rho^{(n)}_{1+} \otimes \rho^{(2)}_{0+})           \\
  \rho_{1\times}^{(n+2)} &= \frac{1}{2}(\rho^{(n)}_{0\times} \otimes \rho^{(1)}_{1\times}
                                      + \rho^{(n)}_{1\times} \otimes \rho^{(2)}_{0\times}).
\end{aligned}\end{eqnarray}
Let $\sigma^{(n)}_0 = \frac{1}{2}(\rho^{(n)}_{0+} + \rho^{(n)}_{0\times})$ and
$\sigma^{(n)}_1 = \frac{1}{2}(\rho^{(n)}_{1+} + \rho^{(n)}_{1\times})$. 
A small calculation shows that 
\begin{equation*}\begin{split}
  \sigma^{(n+2)}_0 - \sigma^{(n+2)}_1 &= \frac{1}{8}\bigl[
        (\rho^{(n)}_{0+} + \rho^{(n)}_{0\times} - \rho^{(n)}_{1+} - \rho^{(n)}_{1\times})
                                              \otimes \outp{\Phi^+}{\Phi^+} \bigr. \\
     &\phantom{===}
      - (\rho^{(n)}_{0+} + \rho^{(n)}_{0\times} - \rho^{(n)}_{1+} - \rho^{(n)}_{1\times}) 
                                              \otimes \outp{\Psi^-}{\Psi^-}        \\
     &\phantom{===}
      + (\rho^{(n)}_{0+} + \rho^{(n)}_{1\times} - \rho^{(n)}_{1+} - \rho^{(n)}_{0\times})
                                              \otimes \outp{\Phi^-}{\Phi^-}        \\
     &\phantom{===}
      - \bigl. (\rho^{(n)}_{0+} + \rho^{(n)}_{1\times} - \rho^{(n)}_{1+} - 
\rho^{(n)}_{0\times})
                                              \otimes \outp{\Psi^+}{\Psi^+} \bigr]
\end{split}\end{equation*}
We then get that
\[
  \|\sigma^{(n+2)}_0 - \sigma^{(n+2)}_1\|_1
     = \frac{1}{2}\left(\|\sigma^{(n)}_0 - \sigma^{(n)}_1\|_1
                   + \|\tilde{\sigma}^{(n)}_0 - \tilde{\sigma}^{(n)}_1\|_1\right),
\]
where $\tilde{\sigma}^{(n)}_0 = \frac{1}{2}(\rho^{(n)}_{0+} + \rho^{(n)}_{1\times})$
and $\tilde{\sigma}^{(n)}_1 = \frac{1}{2}(\rho^{(n)}_{1+} + \rho^{(n)}_{0\times})$. 
Consider the unitary $U = \sigma_x^{\otimes n}$ if $n$ is odd, 
and $U = \sigma_x^{\otimes n-1} \otimes \id$
if $n$ is even. It is easy to verify that 
$\sigma^{(n)}_0 = U\tilde{\sigma}^{(n)}_0U^{\dagger}$ and 
$\sigma^{(n)}_1 = U\tilde{\sigma}^{(n)}_1U^{\dagger}$. We thus have that 
$\|\sigma^{(n)}_0 - \sigma^{(n)}_1\|_1 = \|\tilde{\sigma}^{(n)}_0 - \tilde{\sigma}^{(n)}_1\|_1$ 
and therefore
\[
  \|\sigma^{(n+2)}_0 - \sigma^{(n+2)}_1\|_1 = \|\sigma^{(n)}_0 - \sigma^{(n)}_1\|_1.
\]

It then follows from Helstrom's Theorem~\ref{helstrom} that
the maximum probability to distinguish $\sigma^{(n+2)}_0$
from $\sigma^{(n+2)}_1$ and thus compute the XOR of
the $n+2$ bits is given by 
\[
  \frac{1}{2} + \frac{\|\sigma^{(n)}_0 - \sigma^{(n)}_1\|_1}{4},
\]
which gives the claimed result.
\end{proof}

A similar argument is possible, if we use three mutually unbiased bases.
Intuitively, one might expect Bob's chance of success to drop as we had more bases. 
Interestingly, however, we obtain the same bound of 3/4 if $n$ is even. 

\begin{theorem}\label{xorEvenSTAR3MUBS}
Let $P_X(x) = \frac{1}{2^n}$ for all $x\in\01^n$.
Let $\mB = \{+,\times,\odot\}$ with $U_+ = \id^{\otimes n}$, $U_\times = H^{\otimes n}$,
and $U_\odot=K^{\otimes n}$ with $P_B(+) = P_B(\times) = P_B(\odot) = 1/3$. 
Then Bob succeeds at $\text{\emph{STAR}}(\text{\emph{XOR}})$ with probability at most
\[
  p = \begin{cases} \frac{3}{4} & \text{ if } $n$ \text{ is even}, \\
                    \frac{1}{2}\left(1 + \frac{1}{\sqrt{3}}\right)
                                & \text{ if } $n$ \text{ is odd}.
      \end{cases}
\]
There exists a strategy for Bob that achieves $p$.
\end{theorem}
\begin{proof}
Our proof is very similar to the case of only 2 mutually unbiased bases.
The case of $n=1$ follows from Lemma~\ref{Kabuki}.
This time, we have for $n=2$: 
$\sigma^{(2)}_0 = \frac{1}{3}(\rho^{(2)}_{0+} + \rho^{(2)}_{0\times} + \rho^{(2)}_{0\odot})$ and 
$\sigma^{(2)}_1 = \frac{1}{3}(\rho^{(2)}_{1+} + \rho^{(2)}_{1\times} + \rho^{(2)}_{1\odot})$.
We have $\|\sigma^{(2)}_0 - \sigma^{(2)}_1\|_1 = 1$. 

We again show that the trace distance does not change when we go from 
strings of length $n$ to strings of length $n+2$. 
We use the definitions from Eq. (\ref{eq:fromNtoNplus2}) and let
\begin{eqnarray*}
  \rho_{0\odot}^{(n+2)} 
&=& \frac{1}{2}(\rho^{(n)}_{0\odot} \otimes \rho^{(2)}_{0\odot}
+ \rho^{(n)}_{1\odot} \otimes \rho^{(2)}_{1\odot}),           \\
 \rho_{1\odot}^{(n+2)}      &=& \frac{1}{2}
(\rho^{(n)}_{0\odot} \otimes \rho^{(2)}_{1\odot}
+ \rho^{(n)}_{1\odot} \otimes \rho^{(2)}_{0\odot}).
\end{eqnarray*}
We can compute
\begin{equation*}\begin{split}
  \sigma^{(n+2)}_0 - \sigma^{(n+2)}_1 &= \frac{1}{4}\bigl[
          (\bar{\sigma}^{(n)}_1 - \bar{\sigma}^{(n)}_0) \otimes \outp{\Phi^+}{\Phi^+} \bigr. \\
    &\phantom{===}-(\hat{\sigma}^{(n)}_1 - \hat{\sigma}^{(n)}_0) \otimes \outp{\Psi^-}{\Psi^-}\\
    &\phantom{===}+(\tilde{\sigma}^{(n)}_1 - \tilde{\sigma}^{(n)}_0) \otimes \outp{\Phi^-}{\Phi^-}\\
    &\phantom{===}-\bigl.(\sigma^{(n)}_1 - \sigma^{(n)}_0) \otimes \outp{\Psi^+}{\Psi^+} \bigr],
\end{split}\end{equation*}
where $\bar{\sigma}^{(n)}_1 = (\rho^{(n)}_{0+} + \rho^{(n)}_{0\times} + 
\rho^{(n)}_{1\odot})/3$,
$\bar{\sigma}^{(n)}_0 = (\rho^{(n)}_{1+} + \rho^{(n)}_{1\times} + \rho^{(n)}_{0\odot})/3$,
$\hat{\sigma}^{(n)}_1 = (\rho^{(n)}_{0+} + \rho^{(n)}_{1\times} + \rho^{(n)}_{0\odot})/3$,
$\hat{\sigma}^{(n)}_0 = (\rho^{(n)}_{1+} + \rho^{(n)}_{0\times} + \rho^{(n)}_{1\odot})/3$,
$\tilde{\sigma}^{(n)}_0 = (\rho^{(n)}_{1+} + \rho^{(n)}_{0\times} + \rho^{(n)}_{0\odot})/3$, and
$\tilde{\sigma}^{(n)}_0 = (\rho^{(n)}_{0+} + \rho^{(n)}_{1\times} + \rho^{(n)}_{1\odot})/3$.
Consider the unitaries $\bar{U} = \sigma_y^{\otimes n}$,
$\hat{U} = \sigma_x^{\otimes n}$, and $\tilde{U} = \sigma_z^{\otimes n}$ 
if $n$ is odd, and $\bar{U} = \sigma_y^{\otimes n-1} \otimes \id$,
$\hat{U}= \sigma_x^{\otimes n-1} \otimes \id$, 
and $\tilde{U} = \sigma_z^{\otimes n-1} \otimes \id$ 
if $n$ is even. It is easily verified that
$\sigma^{(n)}_0 = \bar{U}\bar{\sigma}^{(n)}_0\bar{U}^{\dagger}$, 
$\sigma^{(n)}_1 = \bar{U}\bar{\sigma}^{(n)}_1\bar{U}^{\dagger}$, 
$\sigma^{(n)}_0 = \hat{U}\hat{\sigma}^{(n)}_0\hat{U}^{\dagger}$, 
$\sigma^{(n)}_1 = \hat{U}\hat{\sigma}^{(n)}_1\hat{U}^{\dagger}$, 
$\sigma^{(n)}_0 = \tilde{U}\tilde{\sigma}^{(n)}_0\tilde{U}^{\dagger}$, and
$\sigma^{(n)}_1 = \tilde{U}\tilde{\sigma}^{(n)}_1\tilde{U}^{\dagger}$. 
We then get that
\[
  \|\sigma^{(n+2)}_0 - \sigma^{(n+2)}_1\|_1 = \|\sigma^{(n)}_0 - \sigma^{(n)}_1\|_1,
\]
from which the claim follows.
\end{proof}

Surprisingly, if Bob does have some a priori knowledge about the
outcome of the XOR the problem becomes much harder for Bob. 
By expressing the states in the Bell basis and using Helstrom's result,
it is easy to see that if Alice chooses $x \in \01^2$ such that with probability 
$q$, $\text{XOR}(x) = 0$, and with probability $(1-q)$,
$\text{XOR}(x) = 1$, Bob's probability of learning
$\text{XOR}(x)$ correctly is minimized for $q = 1/3$. 
In that case, Bob succeeds with probability
at most $2/3$, which can be achieved  by the trivial 
strategy of ignoring the state he received  and always outputting 1.
This is an explicit example where making a measurement 
does not help in state discrimination. It has previously
been noted by Hunter~\cite{hunter:nouse} that such cases 
can exist in mixed-state discrimination.

\section{Using post-measurement information}\label{haveinfo}\label{PISTARnomem}
\index{post-measurement information!no quantum memory}
\index{PI-STAR!no quantum memory}

We are now ready to advance to the core of our problem. We first consider
the case where Bob does receive post-measurement information, but still has no quantum memory at his disposal. Consider an
instance of $\text{PI}_0\text{-STAR}$ with a function
$f:\mX \rightarrow \mY$ and $m=|\mB|$ bases, and some priors $P_X$
and $P_B$ on the sets $\mX$ and $\mB$. If Bob cannot store any
quantum information, all his nontrivial actions are contained in
the first measurement, which must equip him with possible outputs
$o_i \in \mY$ for each basis $i=1,\ldots,m$. In other words, his
most general strategy is a POVM with $|\mY|^m$ outcomes,
each labeled by the strings $o_1,\ldots,o_{m}$ for $o_i \in \mY$ and $m = |\mB|$. 
Once Alice has announced $b$, Bob outputs $\hat{Y} = o_b$.
Here we first prove a general lower bound on the usefulness of
post-measurement information that beats the guessing bound.
Then, we analyze in detail the AND and the XOR function on $n$ bits.

\subsection{A lower bound for balanced functions}\label{lowerBound}\index{PI-STAR!lower bound}
\index{post-measurement information!lower bound}

We first give a lower bound on Bob's success probability for any 
balanced function and any number of mutually unbiased bases, by constructing an 
explicit measurement that achieves it.
Without loss of generality, we assume in this section that 
$\mB = \{0,\ldots,m-1\}$, as otherwise we could consider a lexicographic ordering of $\mB$. 

\begin{theorem}\label{sipiLower}
Let $f:\mX \rightarrow \mY$ be a balanced function, and let $P_X$ and $P_B$ be the uniform
distributions over $\mX$ and $\mB$ respectively. 
Let the set of unitaries $\{U_b|b \in \mB\}$ give rise to 
$|\mB|$ mutually unbiased bases, and choose an encoding such that 
$\forall x,x'\in \mX: \inp{x}{x'} = \delta_{xx'}$.
Then Bob succeeds at $\text{\emph{PI}}_0\text{\emph{-STAR}}(f)$ with probability at least
\[
  p = p_{\text{guess}} + \begin{cases}
                    \frac{|\mY|-1}{|\mY|(|\mY|+3)} & \text{ if } m = 2,             \\
                    \frac{4(|\mY|^2 - 1)}{3\Y(2 + \Y(\Y + 6))} & \text{ if } m = 3, \\
                    -\frac{2}{2\Y}
                    + \frac{2(\Y + m - 1)}{\Y^2 + 3\Y(m-1) + m^2 - 3m + 2}
                                                               & \text{ if } m \geq 4.
                  \end{cases}
\]
where $p_{\text{guess}}$ is the probability that Bob can achieve by guessing the basis as given
in Lemma~\ref{guessing}. In particular, we always have $p > p_{\text{guess}}$.
\end{theorem}
\begin{proof}
Our proof works by constructing a square-root type 
measurement that achieves the lower bound. As explained above,
Bob's strategy for learning $f(x)$ is to perform a 
measurement with $|\mY|^m$ possible outcomes, 
labeled by the strings $o_1,\ldots,o_{m}$ for $o_i \in \mY$ and $m = |\mB|$. 
Once Alice has announced $b$, Bob outputs $f(x) = o_b$.

Take the projector $P_{yb} = \sum_{x \in f^{-1}(y)} \outp{\Phi^x_b}{\Phi^x_b}$
and $\rho_{yb} = \frac{1}{k} P_{yb}$, where $k = |f^{-1}(y)| = |\mX|/|\mY|$. 
Let $M_{o_1,\ldots,o_m}$
denote the measurement operator corresponding to 
outcome $o_1,\ldots,o_m$. Note that
outcome $o_1,\ldots,o_m$ is the correct outcome for 
input state $\rho_{yb}$ if and only if $o_b = y$. 
We can then write Bob's probability of success as
\[
  \frac{1}{m|\mY|} \sum_{o_1,\ldots,o_m \in \mY} \Tr\left(M_{o_1,\ldots,o_m}
                              \left(\sum_{b \in \mB} \rho_{o_bb}\right)\right).
\]
We make use of the following measurement:
\[
  M_{o_1,\ldots,o_m} = S^{-\frac{1}{2}}\left(\sum_{b\in \mB}
                                               P_{o_bb}\right)^3 S^{-\frac{1}{2}},
  \text{ with }
  S = \sum_{o_1,\ldots,o_m \in \mY} \left(\sum_{b \in \mB} P_{o_bb}\right)^3.
\]
Clearly, we have $\sum_{o_1,\ldots,o_m \in \mY} M_{o_1,\ldots,o_m} = \id$ and 
$\forall o_1,\ldots,o_{m} \in \mY: M_{o_1,\ldots,o_{m}} \geq 0$ by construction and
thus we indeed have a valid measurement. We first show that $S = c_m\id$:
\begin{equation*}\begin{split}
S &= \sum_{o_1,\ldots,o_m \in \mY} \left(\sum_{b \in \mB} P_{o_bb}\right)^3    \\
  =& \sum_{o_1,\ldots,o_m \in \mY} \sum_{b,b',b'' \in \mB} P_{o_bb}P_{o_{b'}b'}
                                                           P_{o_{b''}b''}      \\
  =& \sum_{o_1,\ldots,o_m \in \mY} \left(\sum_{b} P_{o_bb} 
                          + 2 \sum_{bb', b\neq b'} P_{o_bb}P_{o_{b'}b'}\right. \\
  &\phantom{========}
    \left.+ \sum_{bb', b \neq b'} P_{o_bb}P_{o_{b'}b'}P_{o_bb}
                          + \sum_{bb'b'',b \neq b',b \neq b'',b'\neq b''}
                                     P_{o_bb}P_{o_{b'}b'}P_{o_{b''}b''}\right) \\
  =& \bigl[
      m |\mY|^{m-1} + 2 m (m-1)|\mY|^{m-2} + m (m-1) |\mY|^{m-2}
       + m(m-1)(m-2)|\mY|^{m-3}\bar{\delta}_{2m} \bigr] \id,
\end{split}\end{equation*}
where $\bar{\delta}_{2m} = 1 - \delta_{2m}$ and we have used the definition that for any $b$, $P_{o_bb}$ is a projector and 
$\sum_{x \in \mX} \outp{\Phi_b^x}{\Phi_b^x} = \id$ which gives 
$\sum_{o_i \in \mY} P_{o_i b_i}
  = \sum_{o_i \in \mY}\sum_{x \in f^{-1}(y)} \outp{\Phi_b^x}{\Phi_b^x} = \id$.
We can then write Bob's probability of success using this particular measurement as
\[
  \frac{1}{c_m k m |\mY|} \sum_{o_1,\ldots,o_{m} \in \mY} 
                          \Tr\left(\left(\sum_{b \in \mB} P_{o_bb}\right)^4\right).
\]
It remains to evaluate this expression. 
Using the circularity of the trace, we obtain
\begin{equation*}\begin{split}
&\sum_{o_1,\ldots,o_{m} \in \mY}
   \Tr\left(\left(\sum_{b \in \mB} P_{o_bb}\right)^4\right)                             \\
   &= \sum_{o_1,\ldots,o_m \in \mY}
        \Tr\left( \sum_{b}P_{o_bb} + 6 \sum_{bb',b\neq b'} P_{o_bb} P_{o_{b'}b'} \right. \\
   &\phantom{=}
    + 4 \sum_{bb'b'',b\neq b',b \neq b'',b'\neq b''} P_{o_bb}P_{o_{b'}b'}P_{o_{b''}b''}
    + 2 \sum_{bb'b'',b \neq b',b \neq b'',b'\neq b''} P_{o_bb}P_{o_{b'}b'}P_{o_bb}P_{o_{b''}b''}      \\
   &\phantom{=}
    + \left. \sum_{bb'b''\tilde{b},b\neq b',b\neq b'',b \neq \tilde{b},
b'\neq b'',b'\neq \tilde{b},
b''\neq \tilde{b}}
                             P_{o_bb}P_{o_{b'}b'}P_{o_{b''}b''}P_{o_{\tilde{b}}\tilde{b}}
    + \sum_{bb',b \neq b'} P_{o_bb}P_{o_{b'}b'} P_{o_bb} P_{o_{b'}b'} \right)            \\
   &\geq \bigl[ m |\mY|^{m-1} + 6 m (m-1)|\mY|^{m-2} 
                + 6 m(m-1)(m-2)|\mY|^{m-3} \bar{\delta}_{2m} \bigr.                      \\
   &\phantom{==}
    + \bigl. m (m-1)(m-2)(m-3)|\mY|^{t(m-4)}\bar{\delta}_{2m}\bar{\delta}_{3m} \bigr]\Tr(\id)
             + m(m-1)|\mY|^{m-2}k,
\end{split}\end{equation*}
where we have again used the assumption that for any $b$, $P_{o_bb}$ is a projector and 
$\sum_{x \in \mX} \outp{\Phi_b^x}{\Phi_b^x} = \id$ with $\Tr(\id) = |\mX|$.
For the last term we have used the following: 
Note that $\Tr( P_{o_{b}b}P_{o_{b'}b'}) = k^2/|\mX|$, 
because we assumed mutually unbiased bases. 
Let $r = \rank( P_{o_{b}b}P_{o_{b'}b'})$.
Using Cauchy-Schwarz, we can then bound $\Tr((P_{o_bb}P_{o_{b'}b'})^2)
 = \sum_i^r \lambda_i( P_{o_{b}b}P_{o_{b'}b'})^2 
 \geq k^4/(|\mX|^2 r) \geq k^3/|\mX|^2 = k/|\mY|^2$, 
where $\lambda_i(A)$ is the $i$-th eigenvalue of a matrix $A$, 
by noting that $r \leq k$ since $\rank(P_{o_bb}) = \rank(P_{o_b'b'}) = k$.
Putting things together we obtain
\[
p \geq \frac{1}{c_m m}\left[G_m(1) 
        + \left(6 + \frac{1}{|\mY|}\right)G_m(2) + 6 G_m(3) + G_m(4)\right],
\]
where $m = |\mB|$, $c_m = G_m(1) + 3G_m(2) + G_m(3)$
and function $G_m: \Natural \rightarrow \Natural$ defined as
$
G_m(i) = \frac{m!}{(m-i)!} |\mY|^{m-i}\prod_{j=2}^{i-1}\bar{\delta}_{mj}.
$ This expression can be simplified to obtain the claimed result.
\end{proof}

Note that we have only used the assumption that Alice uses mutually unbiased bases
in the very last step to say that 
$\Tr(P_{o_bb}P_{o_{b'}b'}) = k^2/|\mX|$. One could generalize 
our argument to other cases by evaluating 
$\Tr(P_{o_bb}P_{o_{b'}b'})$ approximately. 

In the special case $m=|{\cal Y}|=2$ (i.e.~binary function, with
two bases) we obtain:
\begin{corollary}
  \label{eightyfive-percent}
  Let $f:\01^n \rightarrow \01$ be a balanced function and 
  let $P_X(x) = 2^{-n}$ for all $x\in\01^n$.
  Let $\mB = \01$ with $U_0 = \id^{\otimes n}$, 
  $U_1 = H^{\otimes n}$ and $P_B(0)= P_B(1) = 1/2$.
  Then Bob succeeds at $\text{\emph{PI}}_0\text{\emph{-STAR}}(f)$ with probability 
  $p \geq 0.85$.
\end{corollary}
Observe that this almost attains the upper bound of $\approx .853$
of Lemma~\ref{Breitbart} in the case of no post-measurement information.
In Section~\ref{limitedInfo2Bases} we show that indeed this bound
can always be achieved when post-measurement information is available.

It is perhaps interesting to note that our general bound 
depends only on the number of function values $|\mY|$
and the number of bases $m$. The number of function inputs 
$|\mX|$ itself does not play a direct role. 

\subsection{Optimal bounds for the AND and XOR function}\label{optimalBound}

We now show that for some specific functions, the 
probability of success can even be much larger.
We hereby concentrate on the case where Alice 
uses two or three mutually unbiased bases 
to encode her input. Our proofs thereby lead to explicit 
measurements. In the following, we again assume that Bob
has no a priori knowledge of the function value. 
It turns out that the optimal measurement directly lead us to the essential
idea underlying our algebraic framework of Section~\ref{algebraicFramework}.

\subsubsection{AND function}\index{PI-STAR!AND}
\index{post-measurement information!AND}

\begin{theorem}\label{PISTARAND}
Let $P_X(x) = 1/(2(2^n-1))$ for all $x \in \01^n\setminus\{1\ldots1\}$
and $P_X(1\ldots 1) = \frac{1}{2}$.
Let $\mB = \{+,\times\}$ with $U_+ = \id^{\otimes n}$, 
$U_\times = H^{\otimes n}$ and $P_B(+) = P_B(\times) = 1/2$. 
Then Bob succeeds at $\text{\emph{PI}}_0\text{\emph{-STAR}}(\text{\emph{AND}})$ 
with probability at most
\begin{equation}\label{OptProbPISTARAND}
p=\frac{1}{2}\left[2+\frac{1}{2^n+2^{n/2}-2}-\frac{1}{2^n-1}\right].
\end{equation}
There exists a strategy for Bob that achieves $p$.
\end{theorem}
\begin{proof}
To learn the value of $\text{AND}(x)$, Bob uses the same strategy as in 
Section~\ref{lowerBound}:
he performs a measurement with $4$ possible outcomes, labeled by the strings
$o_+,o_\times$ with $o_+,o_\times \in \01$. 
Once Alice has announced her basis choice $b \in \{+,\times\}$, Bob
outputs $\text{AND}(x) = o_b$. Note that without loss of generality we can assume that Bob's
measurement has only $4$ outcomes, i.e. Bob only stores 2 bits of classical information because
he will only condition his answer on the value of $b$ later on.

Following the approach in the last section, we can write Bob's optimal 
probability of success as a semidefinite program:\index{SDP!example}
\begin{sdp}{maximize}{$\frac{1}{4}\sum_{o_+,o_\times \in \01} 
                       \Tr [b_{o_+ o_\times}M_{o_+o_\times}]$}
&$\forall o_+,o_\times \in \01: M_{o_+o_\times} \geq 0$,\\
&$\sum_{o_+,o_\times \in \01} M_{o_+o_\times} = \id$,
\end{sdp}
where
\begin{eqnarray*}
b_{00}=\rho_{0+}+\rho_{0\times},&
b_{01}=\rho_{0+}+\rho_{1\times},\\
b_{10}=\rho_{1+}+\rho_{0\times},&
b_{11}=\rho_{1+}+\rho_{1\times},
\end{eqnarray*}
with $\forall y \in \01, b \in \{+,\times\}: \rho_{yb} = 
\frac{1}{|AND^{-1}(y)|} \sum_{x \in \text{AND}^{-1}(y)} U_b\outp{x}{x}U_B^{\dagger}$. 
Consider $\hil_2$, the $2$-dimensional Hilbert space spanned by 
$\ket{c_1}\dfdas\ket{1}^{\otimes n}$ and $\ket{h_1}\dfdas\ket{1_\times}^{\otimes n}$.
Let $\ket{c_0}\in \hil_2$ and $\ket{h_0}\in \hil_2$ be the state vectors orthogonal to 
$\ket{c_1}$ and $\ket{h_1}$ respectively. They can be expressed as:
\begin{eqnarray*}
\ket{c_o}&=&\frac{(-1)^{n+1}\ket{c_1}+2^{n/2}\ket{h_1}}{\sqrt{2^n-1}},\\
\ket{h_o}&=&\frac{2^{n/2}\ket{c_1}+(-1)^{n+1}\ket{h_1}}{\sqrt{2^n-1}}.
\end{eqnarray*}
Then $\Pi_\parallel=\outp{c_0}{c_0}+\outp{c_1}{c_1}=\outp{h_0}{h_0}+\outp{h_1}{h_1}$ is 
a projector onto $\hil_2$. Let $\Pi_{\perp}$ be a projector 
onto the orthogonal complement of $\hil_2$.
Note that the $b_{o_+o_\times}$ are all composed of two 
blocks, one supported on $\hil_2$ and the other 
on its orthogonal complement. We can thus write
\begin{eqnarray}\begin{aligned}\label{eq:sepbs}
b_{00}&=\frac{2\Pi_\perp}{2^n-1}&+&\frac{\outp{c_0}{c_0}+\outp{h_0}{h_0}}{2^n-1},\\
b_{01}&=\frac{\Pi_\perp}{2^n-1}&+&\left[\frac{\outp{c_0}{c_0}}{2^n-1}+\outp{h_1}{h_1}\right],\\
b_{10}&=\frac{\Pi_\perp}{2^n-1}&+&\left[\frac{\outp{h_0}{h_0}}{2^n-1}+\outp{c_1}{c_1}\right],\\
b_{11}&=0&+&\outp{c_1}{c_1}+\outp{h_1}{h_1}.
\end{aligned}
\end{eqnarray}
We give an explicit measurement that achieves $p$ and then show
that it is optimal. Take
\begin{eqnarray*}
M_{00}&=\Pi_\perp\\
M_{o_+o_\times}&=\lambda_{o_+o_\times}\outp{\psi_{o_+o_\times}}{\psi_{o_+o_\times}},
\end{eqnarray*}
with 
$\lambda_{01} = \lambda_{10} =(1+\eta)^{-1}$
where 
\begin{eqnarray*}
\eta &=\left|\frac{1-2 \beta^2+ (-1)^{n+1}2\beta\sqrt{1-\beta^2}\sqrt{2^n-1}}{2^{n/2}}\right|,\\
\ket{\psi_{01}}&=\alpha\ket{c_0}+\beta\ket{c_1},\\
\ket{\psi_{10}}&=\alpha\ket{h_0}+\beta\ket{h_1},
\end{eqnarray*}
with $\alpha$ and $\beta$ real and satisfying $\alpha^2+\beta^2=1$. We also 
set $M_{11} = \id - M_{00} - M_{01} - M_{10}$.
We take 
$$
\beta = (-1)^n \frac{1}{\sqrt{2^{2n}+2^{\frac{3}{2}n+1}-2^{\frac{n}{2}+1}}}.
$$
Putting it all together, we thus calculate Bob's probability of success:
$$
p=\frac{1}{2}\left[2+\frac{1}{2^n+2^{n/2}-2}-\frac{1}{2^n-1}\right].
$$

We now show that this is in fact the optimal measurement for Bob. For this
we consider the dual of our semidefinite program above:
\begin{sdp}{minimize}{$\Tr(Q)$}\index{SDP!example}
&$\forall o_+,o_\times \in \01: Q\geq \displaystyle \frac{b_{o_+o_\times}}{4}$.
\end{sdp}
Our goal is now to find a $Q$ such that $p = \Tr(Q)$ and $Q$ is dual feasible. 
We can then conclude from the duality of SDP that $p$ is optimal.
Consider
\begin{eqnarray*}
 Q = &\frac{\Pi_\perp}{2(2^n-1)}
        +\frac{1}{4}\left(\frac{2-{2^{1+n/2}}+{2^{3 n/2}}}{2-3\cdot{2^{n/2}}+{2^{3 n/2}}}\right)
                    (\outp{c_1}{c_1}+ \outp{h_1}{h_1})                                   \\
     &- (-1)^n \frac{1}{4({2^{1-\frac{n}{2}}}+{2^n}-3)}(\outp{c_1}{h_1}+\outp{c_1}{h_1}).
\end{eqnarray*}
Now we only need to show that the $Q$ above satisfies 
the constraints, i.e. $\forall o_+,o_\times \in \01:
Q\geq b_{o_+o_\times}/4$. Let $Q_\perp = \Pi_\perp Q \Pi_\perp$ and 
$Q_\parallel = \Pi_\parallel Q \Pi_\parallel$. 
By taking a look at Eq.\ (\ref{eq:sepbs}) one can easily see 
that $Q_\perp \geq \frac{\Pi_\perp b_{o_+o_\times}\Pi_\perp}{4},$ 
so that it is only left to show that 
$$
  Q_\parallel \geq \frac{\Pi_\parallel b_{o_+o_\times} \Pi_\parallel}{4}, 
                      \textrm{ for }o_+o_\times \in \01, o_+o_\times \neq 00.
$$
These are $2 \times 2$ matrices and this can be done straightforwardly.
We thus have $\Tr(Q) = p$ and the result follows from the duality of semidefinite programming.
\end{proof}

It also follows that if Bob just wants to learn the value of a single bit, he can do no better
than what he could achieve without waiting for Alice's announcement of the basis $b$:
\begin{corollary}\label{breitbartPISTAR}
  Let $x \in \01$, $P_X(x) = \frac{1}{2}$ and $f(x) = x$.
  Let $\mB =\{+,\times\}$ with $U_+ = \id$ and $U_\times = H$.
  Then Bob succeeds at $\text{\emph{PI}}_0\text{\emph{-STAR}}(f)$ with probability at most
$$
p = \frac{1}{2}+\frac{1}{2\sqrt{2}}.
$$
  There exists a strategy for Bob that achieves $p$.
\end{corollary}

The AND function provides an intuitive example of how Bob can compute the value of a
function perfectly by storing just a single qubit.
Consider the measurement with elements $\{\Pi_\parallel,\Pi_\perp\}$ from the previous section.
It is easy to see that the outcome $\perp$ has zero probability if $\text{AND}(x) = 1$.
Thus, if Bob obtains that outcome he can immediately conclude that $\text{AND}(x) = 0$.
If Bob obtains outcome $\parallel$ then the post-measurement states 
live in a $2$-dimensional Hilbert space ($\hil_2$), and can therefore be stored in a single qubit.
Thus, by keeping the remaining state we can calculate 
the AND perfectly once the basis is announced. 
Our proof in Section~\ref{limitedInfo2Bases}, which shows that 
in fact \emph{all} Boolean functions can be
computed perfectly if Bob can store only a single qubit, 
makes use of a very similar effect to
the one we observed here explicitly.

\subsubsection{XOR function}\index{post-measurement information!XOR}\index{PI-STAR!XOR}

We now examine the XOR function. This will be useful in order to gain some insight into
the usefulness of post-measurement information later. For strings of even length, there 
exists a simple strategy for Bob even when three mutually unbiased bases are used.

\begin{theorem}\label{xorEvenPISTAR3MUBS}
Let $n \in \Natural$ be even, and
let $P_X(x) = \frac{1}{2^n}$ for all $x \in \01^n$.
Let $\mB =\{+,\times,\odot\}$ with $U_+ = \id^{\otimes n}$, 
$U_\times = H^{\otimes n}$ and $U_\odot=K^{\otimes n}$, where $K=(\id+i\sigma_x)/\sqrt{2}$. 
Then there is a strategy where Bob succeeds at 
$\text{\emph{PI}}_0\text{\emph{-STAR}}(\text{\emph{XOR}})$ with probability 
$p=1$.
\end{theorem}
\begin{proof}
We first construct Bob's measurement for the first 2 qubits, which allows him
to learn $x_1 \oplus x_2$ with probability 1. Note that the 12 possible states that Alice sends
can be expressed in the Bell basis as follows:
\begin{eqnarray*}
\ket{00} = \frac{1}{\sqrt{2}}(\ket{\Phi^+} + \ket{\Phi^-})
     & H^{\otimes 2}\ket{00} = \frac{1}{\sqrt{2}}(\ket{\Phi^+} + \ket{\Psi^+}) \\
\ket{01} = \frac{1}{\sqrt{2}}(\ket{\Psi^+} + \ket{\Psi^-})
     & H^{\otimes 2}\ket{01} = \frac{1}{\sqrt{2}}(\ket{\Phi^-} + \ket{\Psi^-})\\
\ket{10} = \frac{1}{\sqrt{2}}(\ket{\Psi^+} - \ket{\Psi^-})
     & H^{\otimes 2}\ket{10} = \frac{1}{\sqrt{2}}(\ket{\Phi^-} - \ket{\Psi^-})\\
\ket{11} = \frac{1}{\sqrt{2}}(\ket{\Phi^+} - \ket{\Phi^-})
     & H^{\otimes 2}\ket{11} = \frac{1}{\sqrt{2}}(\ket{\Phi^+} - \ket{\Psi^+})
\end{eqnarray*}
\begin{eqnarray*}
K^{\otimes 2}\ket{00} &=& \frac{1}{\sqrt{2}}(\ket{\Phi^-} +i \ket{\Psi^+})\\
K^{\otimes 2}\ket{01} &=& \frac{1}{\sqrt{2}}(i\ket{\Phi^+} + \ket{\Psi^-})\\
K^{\otimes 2}\ket{10} &=& \frac{1}{\sqrt{2}}(i\ket{\Phi^+} - \ket{\Psi^-})\\
K^{\otimes 2}\ket{11} &=& -\frac{1}{\sqrt{2}}(\ket{\Phi^-} - i\ket{\Psi^+}).
\end{eqnarray*}
Bob now simply measures in the Bell basis and records his 
outcome. If Alice now announces that she used the computational
basis, Bob concludes that $x_1 \oplus x_2 = 0$ if the 
outcome is one of $\ket{\Phi^{\pm}}$ and $x_1 \oplus x_2 = 1$ 
otherwise. If Alice announces she used the Hadamard basis, Bob concludes that
$x_1 \oplus x_2 = 0$ if the outcome was one of 
$\{\ket{\Phi^{+}},\ket{\Psi^{+}}\}$ and $x_1 \oplus x_2 = 1$ 
otherwise. Finally, if Alice announces that she used the 
$\odot$ basis, Bob concludes that $x_1 \oplus x_2 = 0$ 
if the outcome was one of $\{\ket{\Phi^{-}},\ket{\Psi^{+}}\}$
and $x_1 \oplus x_2 = 1$ otherwise.
Bob can thus learn the XOR of two bits with probability 1.
To learn the XOR of the entire string, Bob applies this 
strategy to each two bits individually and then computes
the XOR of all answers. 
\end{proof}

Analogously to the proof of Theorem~\ref{xorEvenPISTAR3MUBS}, we obtain:
\begin{corollary}\label{xorEvenPISTAR2MUBS}
  Let $n \in \Natural$ be even, and
  let $P_X(x) = \frac{1}{2^n}$ for all $x \in \01^n$.
  Let $\mB =\{+,\times\}$ with $U_+ = \id^{\otimes n}$ and $U_\times = H^{\otimes n}$.
  Then there is a strategy where Bob succeeds at 
  $\text{\emph{PI}}_0\text{\emph{-STAR}}(\text{\emph{XOR}})$ with probability 
  $p=1$.
\end{corollary}

Interestingly, there is no equivalent strategy for 
Bob if $n$ is odd. In fact, as we show in the next
section, in this case the post-measurement information 
gives no advantage to Bob at all.

\begin{theorem}\label{xorOddPISTAR2MUBS}
Let $n \in \Natural$ be odd, and
let $P_X(x) = \frac{1}{2^n}$ for all $x \in \01^n$.
Let $\mB = \{+,\times\}$ with $U_+ = \id^{\otimes n}$, 
$U_\times = H^{\otimes n}$ and $P_B(+) = P_B(\times) = 1/2$.
Then Bob succeeds at 
$\text{\emph{PI}}_0\text{\emph{-STAR}}(\text{\emph{XOR}})$ 
with probability at most
\[
  p = \frac{1}{2}\left(1 + \frac{1}{\sqrt{2}}\right).
\]
There exists a strategy for Bob that achieves $p$.
\end{theorem}
\begin{proof}
Similar to the proof of the AND function, we can write Bob's optimal
probability of success as the following semidefinite program in terms
of the length of the input string, $n$:\index{SDP!example}
\begin{sdp}{maximize}{$\frac{1}{4}\sum_{o_+,o_\times \in \01}
                       \Tr [b^{(n)}_{o_+ o_\times}M_{o_+ o_\times}]$}
&$\forall o_+,o_\times \in \01: M_{o_+ o_\times} \geq 0$,\\
&$\sum_{o_+,o_\times \in \01} M_{o_+ o_\times} = \id$,
\end{sdp}
where
\begin{eqnarray*}
b^{(n)}_{o_+ o_\times}&=\rho^{(n)}_{o_+ +}+\rho^{(n)}_{o_\times \times},
\end{eqnarray*}
and $\rho^{(n)}_{o_b b} =
\frac{1}{2^{n-1}} \sum_{x \in \01^n, x \in \text{XOR}^{-1}(o_b)} U_b\outp{x}{x}U_b^{\dagger}$.
The dual can be written as
\begin{sdp}{minimize}{$\frac{1}{4}\Tr(Q^{(n)})$}
&$\forall o_+,o_\times \in \01: Q^{(n)}\geq \displaystyle b^{(n)}_{o_+o_\times}$.
\end{sdp}
Our proof is now by induction on $n$. For $n=1$, let $Q^{(1)} = 2 p\id$.
It is easy to verify that $\forall o_+,o_\times \in \01:
Q^{(1)} \geq b^{(1)}_{o_+o_\times}$ and thus $Q^{(1)}$ is a feasible solution of the dual program.

We now show that for $n + 2$, $Q^{(n+2)} = Q^{(n)} \otimes \frac{1}{4}\id$ is a feasible
solution to the dual for $n+2$, where $Q^{(n)}$ is a solution for
the dual for $n$. Note that the
XOR of all bits in the string can be expressed as the 
XOR of the first $n-2$ bits XORed with the XOR of the last two.
Recall Eq. (\ref{eq:fromNtoNplus2}) and
note that we can write
\begin{align*}
  \rho^{(2)}_{0+} &= \frac{1}{2}(\outp{00}{00} + \outp{11}{11})
               = \frac{1}{2}(\outp{\Phi^+}{\Phi^+} + \outp{\Phi^-}{\Phi^-}) \\
  \rho^{(2)}_{1+} &= \frac{1}{2}(\outp{01}{01} + \outp{10}{10})
               = \frac{1}{2}(\outp{\Psi^+}{\Psi^+} + \outp{\Psi^-}{\Psi^-}).
\end{align*}
It is easy to see that
$\rho^{(2)}_{0\times} = H\rho^{(2)}_{0+}H
                  = \frac{1}{2}(\outp{\Phi^+}{\Phi^+} + \outp{\Psi^+}{\Psi^+})$
and
$\rho^{(2)}_{1\times} = H\rho^{(2)}_{1+}H 
                  = \frac{1}{2}(\outp{\Phi^-}{\Phi^-}  + \outp{\Psi^-}{\Psi^-})$.
By substituting from the above equation we then obtain
\begin{equation*}\begin{split}
  b^{(n+2)}_{00} = \rho^{(n+2)}_{0+} + \rho^{(n+2)}_{0\times}
      &= \frac{1}{4}\bigl( (\rho^{(n)}_{0+} + \rho^{(n)}_{0\times}) \otimes \outp{\Phi^+}{\Phi^+}
                          +(\rho^{(n)}_{0+} + \rho^{(n)}_{1\times}) \otimes \outp{\Phi^-}{\Phi^-}
                    \bigr. \\
      &\phantom{==}
                    \bigl. (\rho^{(n)}_{1+} + \rho^{(n)}_{0\times}) \otimes \outp{\Psi^+}{\Psi^+}
                          +(\rho^{(n)}_{1+} + \rho^{(n)}_{1\times}) \otimes \outp{\Psi^-}{\Psi^-})
                    \bigr) \\
      &\leq \frac{1}{4} Q^{(n)} \otimes \id,
\end{split}\end{equation*}
where we have used the fact that $Q^{(n)}$ is a feasible solution for the dual for $n$ and
that $\outp{\Phi^+}{\Phi^+} + \outp{\Phi^-}{\Phi^-}
       + \outp{\Psi^+}{\Psi^+} + \outp{\Psi^-}{\Psi^-} = \id$.
The argument for $b^{(n+2)}_{01}$, $b^{(n+2)}_{10}$ 
and $b^{(n+2)}_{11}$ is analogous. Thus $Q^{(n+2)}$ satisfies all constraints.

Putting things together, we have for odd $n$ that 
$\Tr(Q^{(n+2)}) = \Tr(Q^{(n)}) = \Tr(Q^{(1)})$ and
since the dual is a minimization problem we know 
that 
$$
p \leq \frac{1}{4}\Tr(Q^{(1)}) = c
$$ 
as claimed.
Clearly, there exists a strategy for Bob that 
achieves $p = c$. He can compute the
XOR of the first $n-1$ bits perfectly, as shown 
in Theorem~\ref{xorEvenPISTAR2MUBS}. 
By Corollary~\ref{breitbartPISTAR}
he can learn the value of the remaining $n$-th 
bit with probability $p=c$.
\end{proof}

We obtain a similar bound for three bases:

\begin{theorem}\label{xorOddPISTAR3MUBS}
Let $n \in \Natural$ be odd, and
let $P_X(x) = \frac{1}{2^n}$ for all $x \in \01^n$.
Let $\mB = \{+,\times,\odot\}$ with $U_+ = \id^{\otimes n}$, $U_\times = H^{\otimes n}$ and
$U_\odot=K^{\otimes n}$, where $K=(\id+i\sigma_x)/\sqrt{2}$, with 
$P_B(+) = P_B(\times) = P_B(\odot) = 1/3$. 
Then Bob succeeds at $\text{\emph{PI}}_0\text{\emph{-STAR}}(\text{\emph{XOR}})$
with probability at most
\[
  p = \frac{1}{2}\left(1 + \frac{1}{\sqrt{3}}\right).
\]
There exists a strategy for Bob that achieves $p$.
\end{theorem}
\begin{proof}
The proof follows the same lines as Theorem \ref{xorOddPISTAR2MUBS}. 
Bob's optimal probability of success is:
\begin{sdp}{maximize}{$\displaystyle \frac{1}{6}\sum_{o_+,o_\times,o_\odot \in \01}
                       \Tr [b^{(n)}_{o_+ o_\times o_\odot}M_{o_+ o_\times o_\odot}]$}
&$\forall o_+,o_\times,o_\odot \in \01 \in \01: M_{o_+ o_\times o_\odot} \geq 0$,\\
&$\displaystyle\sum_{o_+,o_\times,o_\odot \in \01} M_{o_+ o_\times o_\odot} = \id$,
\end{sdp}
where
$$
b^{(n)}_{o_+ o_\times o_\odot}=\sum_{b\in \mB}\rho_{o_b b} ,
$$
and
$$
\rho_{o_b b}=\frac{1}{2^{n-1}}\sum_{x\in XOR(o_b)} U_b\outp{x}{x}U_b^\dagger.
$$
The dual can be written as
\begin{sdp}{minimize}{$\frac{1}{6}\Tr(Q^{(n)})$}
&$\forall o_+,o_\times, o_\odot \in \01: Q^{(n)}\geq \displaystyle b^{(n)}_{o_+ o_\times o_\odot}$.
\end{sdp}
Again, the proof continues by induction on $n$. 
For $n=1$, let $Q^{(1)} = 3 p \id$. It is easy to verify that $\forall o_+,o_\times, o_\odot \in \01:
Q^{(1)} \geq b^{(1)}_{o_+ o_\times o_\odot}$ and thus $Q^{(1)}$ is a feasible solution of the dual program.
The rest of the proof is done exactly in the same 
way as in Theorem \ref{xorOddPISTAR2MUBS} using  that 
\begin{align*}
  \rho^{(2)}_{0\odot} &= \frac{1}{2}(\outp{\Phi^-}{\Phi^-} + \outp{\Psi^+}{\Psi^+}) \\
  \rho^{(2)}_{1\odot} &= \frac{1}{2}(\outp{\Psi^-}{\Psi^-} + \outp{\Phi^+}{\Phi^+}).
\end{align*}
\end{proof}

\section[Using post-measurement information and quantum memory]
{Using post-measurement information\\ and quantum memory}
\index{PI-STAR!quantum memory}
\index{post-measurement information!quantum memory}

\subsection{An algebraic framework for perfect prediction}%
\label{algebraicFramework}
\index{PI-STAR!algebraic framework}
\index{PI-STAR!perfect prediction}
\index{post-measurement information!algebraic framework}
\index{post-measurement information!perfect prediction}
\label{sec:algebra-perfect}\label{memoryFramework}\label{pistarAlgebra}

So far, we had assumed that Bob is not allowed to store any qubits and
can only use the additional post-measurement information to improve his guess.
Now, we investigate the case where he has a certain amount of quantum 
memory at his disposal. 
In particular, we present a general
algebraic approach to determine the minimum dimension $2^q$ of quantum
memory needed to succeed with probability $1$ at an instance of
$\text{PI}_q\text{-STAR}(\ens)$, for any ensemble $\ens = \{p_{yb},\rho_{yb}\}$
as long as the individual states for different values of $y$ 
are mutually orthogonal for a fixed $b$, i.e., 
$\forall y\neq z \in \mY\: \Tr(\rho_{yb},\rho_{zb}) =0$.
In particular, we are looking for an instrument 
consisting of a family of completely positive
maps $\rho \mapsto A \rho A^\dagger$, adding up to a trace preserving
map, such that $\rank(A) \leq 2^q$. This ensures that the post-measurement state
``fits'' into $q$ qubits, and thus takes care of the memory
bound. The fact that after the announcement of $b$
the remaining state $A \rho_{yb} A^\dagger$ gives full information
about $y$ is expressed by demanding orthogonality of the
different post-measurement states:
\begin{equation}
  \label{pms-orth}
  \forall b \in \mB, \forall y\neq z \in \mY\quad
    A \rho_{yb} A^\dagger A \rho_{zb} A^\dagger = 0.
\end{equation}
Note that here we explicitly allow the possibility that, say, $A \rho_{zb} A^\dagger = 0$:
this means that if Bob obtains outcome $A$ and later learns $b$, 
he can exclude the output value $z$.
What Eq.~(\ref{pms-orth}) also implies is that for all states $\ket{\psi}$
and $\ket{\varphi}$ in the support of $\rho_{yb}$ and $\rho_{zb}$, respectively,
one has $A \ketbra{\psi}{\psi} A^\dagger A \ketbra{\varphi}{\varphi} A^\dagger = 0$.
Hence, introducing the support projectors $P_{yb}$ of the $\rho_{yb}$,
we can reformulate Eq.~(\ref{pms-orth}) as
\[
  \forall b \in \mB, \forall y\neq z \in \mY\quad
    A P_{yb} A^\dagger A P_{zb} A^\dagger = 0,
\]
which can equivalently be expressed as
\begin{equation}
  \label{trace-zero}
  \forall b \in \mB, \forall y\neq z \in \mY\quad
    \Tr\bigl( A^\dagger A P_{yb} A^\dagger A P_{zb} \bigr) = 0,
\end{equation}
by noting that $A^\dagger A$ as well as the projectors are positive-semidefinite
operators.
As expected, we see that only the POVM operators
$M=A^\dagger A$ of the instrument play a role in this condition.
Our conditions can therefore also be written as
$M P_{yb} M P_{zb} = 0$.
From this condition, we now derive the following lemma.
 
\begin{lemma}\label{pistarCommutingLemma}
  \label{commutation}
  Bob, using a POVM with operators $\{M_i\}$,
  succeeds at \emph{$\text{PI}_q\text{-STAR}$} with probability $1$,
  if and only if
  \begin{enumerate}
    \item for all $i$, $\rank(M_i) \leq 2^q$,
    \item for all $y\in\mY$ and $b \in \mB$, $[M,P_{yb}] = 0$, where
      $P_{yb}$ is the projection on the support of $\rho_{yb}$.
  \end{enumerate}
\end{lemma}
\begin{proof}
  We first show that these two conditions are necessary. Note that
  only the commutation condition has to be proved.
  Let $M$ be a measurement operator from a POVM succeeding
  with probability $1$. Then, for any $y$, $b$, we have by Eq.~(\ref{trace-zero}) that 
  \[
    \Tr\bigl( MP_{yb}M(\id-P_{yb}) \bigr) = 0,
    \text{ hence }
    \Tr\bigl( MP_{yb}MP_{yb} \bigr) = \Tr\bigl( MP_{yb}M \bigr).
  \]
  Thus, by the positivity of the trace on positive operators, 
  the cyclicity of the trace, and $P_{yb}^2 = P_{yb}$ we have that
  \begin{equation*}\begin{split}
    0 &\leq \Tr\bigl( [M,P_{yb}]^\dagger [M,P_{yb}] \bigr)                              \\
      &=    \Tr\bigl( -(MP_{yb}-P_{yb}M)^2 \bigr)                                       \\
      &=    \Tr\bigl( -MP_{yb}MP_{yb} -P_{yb}MP_{yb}M +P_{yb}M^2P_{yb} +MP_{yb}^2M \bigr)
       =    0.
  \end{split}\end{equation*}
  But that means that the commutator $[M,P_{yb}]$ has to be $0$.

  Sufficiency is easy: since the measurement operators commute with
  the states' support projectors $P_{yb}$, and these are orthogonal
  to each other for fixed $b$, the post-measurement states of these
  projectors, $\propto \sqrt{M}P_{yb}\sqrt{M}$ are also mutually
  orthogonal for fixed $b$. Thus, if Bob learns $b$, he can perform a
  measurement to distinguish the different values of $y$ perfectly.
  The post-measurement states are clearly supported on the support
  of $M$, which can be stored in $q$ qubits. Since Bob's strategy 
  succeeds with probability $1$, it succeeds with probability $1$
  for any states supported in the range of the $P_{yb}$.
\end{proof}

Note that the operators $M$ of the instrument
need not commute with the originally given states $\rho_{yb}$.
Nevertheless, the measurement preserves the orthogonality of $\rho_{yb}$ and $\rho_{zb}$ 
with $y\neq z$ for fixed $b$, i.e., $\Tr(\rho_{yb}\rho_{zb}) = 0$.
Now that we know that the POVM operators of the instrument 
have to commute with all the states' support projectors $P_{yb}$, we can invoke
some well-developed algebraic machinery to find the optimal such instrument.

Looking at Appendix~\ref{cstar}, we see that
$M$ has to come from the commutant
of the operators $P_{yb}$. These themselves generate a $*$-subalgebra
$\mA$ of the full operator algebra $\bop(\hil)$ of
the underlying Hilbert space $\hil$, and the structure of
such algebras and their commutants in finite dimension is well understood.
We know from Theorem~\ref{doubleComm} that 
the Hilbert space ${\cal H}$ has a decomposition (i.e., there is an
isomorphism which we write as an equality)
\begin{equation}
  \label{sum-of-products}
  \hil = \bigoplus_j {\cal J}_j \otimes {\cal K}_j
\end{equation}
into a direct sum of tensor products
such that the $*$-algebra $\mA$
and its commutant algebra
$\Comm(\mA) = \bigl\{ M : \forall P\in{\bop(\hil)}\ [P,M]=0 \bigr\}$
can be written
\begin{align}
  \label{algPISTAR}
  \mA &\isomorph \bigoplus_j {\cal B}({\cal J}_j) \otimes \id_{{\cal K}_j},  \\
  \label{commutantPISTAR}
  \Comm(\mA) &\isomorph \bigoplus_j \id_{{\cal J}_j}     \otimes {\cal B}({\cal K}_j).
\end{align}
 
Koashi and Imoto~\cite{koashi&imoto:operations}, in the context
of finding the quantum operations which leave a set of states
invariant, have described an algorithm to find the commutant $\Comm(\mA)$,
and more precisely the Hilbert space decomposition of Eq. (\ref{sum-of-products}),
of the states $P_{yb}/\Tr P_{yb}$. They show that for this decomposition,
there exist states $\sigma_{j|i}$ on ${\cal J}_j$, 
a conditional probability distribution $\{q_{j|i}\}$, and
states $\omega_j$ on ${\cal K}_j$ which are independent of $i$, such that
we can write them as 
\[
  \forall i\quad \sigma_i = \bigoplus_j q_{j|i}\sigma_{j|i} \otimes \omega_j,
\]
Looking at Eq.~(\ref{commutantPISTAR}), we see that the 
smallest rank operators
$M \in \Comm(\algA)$ are of the form $\id_{{\cal J}_j} \otimes \ketbra{\psi}{\psi}$
for some $j$ and $\ket{\psi}\in{\cal K}_j$, and that they are all admissible.
Since we need a family of operators $M$ that are closed to a POVM (i.e., their sum is equal to the identity), we know that 
all $j$ have to occur. Hence, the minimal quantum memory requirement is
\begin{equation}
  \label{minimal-q}
  \min 2^q = \max_j \dim {\cal J}_j.
\end{equation}
The strategy Bob has to follow is this: For each $j$, pick a basis 
$\{\ket{e_{k|j}}\}$ for
${\cal K}_j$ and measure the POVM
$\{\id_{{\cal J}_j}\otimes\outp{e_{k|j}}{e_{k|j}}\}$,
corresponding to the decomposition
\[
  \hil = \bigoplus_{jk} {\cal J}_j \otimes \outp{e_{k|j}}{e_{k|j}},
\]
which commutes with the $P_{yb}$. For each outcome, he can store the
post-\-measurement state in $q$ qubits [as in Eq.~(\ref{minimal-q})], preserving
the orthogonality of the states for different $y$ but fixed $b$. Once he learns
$b$ he can thus obtain $y$ with certainty.

Of course, carrying out the Koashi-Imoto algorithm may not
be a straightforward task in a given situation. We now consider two explicit examples that one
can understand as two special cases of this general method:
First, we show that in fact \emph{all} 
Boolean functions with two bases (mutually unbiased or not) can be computed perfectly 
when Bob is allowed to store just a single qubit. Second, however, we show that there
exist three bases such that for \emph{any balanced} function, Bob must store \emph{all}
qubits to compute the function perfectly. We also give a recipe how to construct such
bases.

\subsection{Using two bases}\index{PI-STAR!two bases}
\index{post-measurement information!two bases}
\label{limitedInfo2Bases}

For two bases, Bob needs to store only a single qubit to compute any Boolean function
perfectly. As outlined in Section~\ref{sec:algebra-perfect}, 
we need to show that there exists a measurement with the following properties:
First, the post-measurement states of states corresponding 
to strings $x$ such that $f(x)=0$ are 
orthogonal to the post-measurement states of states 
corresponding to strings $y$ such that $f(y)=1$. 
Indeed, if this is true and we keep the post-measurement
state, then after the basis is announced, we 
can distinguish perfectly between both types of states. 
Second, of course, we need that the post-measurement states are 
supported in subspaces of dimension at most $2$.
The following little lemma shows that this is the case for any
Boolean function.
The same statement has been shown independently many times before in a
variety of different contexts. For example, Masanes and also Toner and
Verstraete have shown the same
in the context of non-local games~\cite{masanes:blocks,toner:blocks}.
The key ingredient is also 
present in Bathia's textbook~\cite{bathia:ma}.
Indeed, there is a close connection between the amount of post-measurement information
we require, and the amount of entanglement we need to implement measurements in the setting of non-local
games. We return to this question in Chapter~\ref{entanglementIntro}.

\begin{lemma}\label{directsumdecomposition}
Let $f: \{0,1\}^n\to \{0,1\}$ and $P_{0b}=\sum_{x\in f^{-1}(0)}U_b\outp{x}{x}U_b^\dagger$ 
where $U_0=\id$ and $U_1=U$, then there exists a direct 
sum decomposition of the Hilbert space 
\begin{equation*}
\hil=\bigoplus_{i=1}^m\hil_i, \textrm{ with } \dim \hil_i \leq 2,
\end{equation*}
such that $P_{00}$ and $P_{01}$ can be expressed as
\begin{align*}
  P_{00} &= \sum_{i=1}^m\Pi_i P_{00}\Pi_i, \\
  P_{01} &= \sum_{i=1}^m\Pi_i P_{01}\Pi_i,
\end{align*}
where $\Pi_i$ is the orthogonal projector onto $\hil_i$.
\end{lemma}

\begin{proof}
There exists a basis so that $P_{00}$ and $P_{01}$ can be written as
\begin{eqnarray*}
P_{00}=\left[\begin{array}{@{}cc@{}}\id_{n_0} & 0_{n_0\times n_1} \\
                            0_{n_1\times n_0} & 0_{n_1\times n_1}\end{array}\right],
P_{01}=\left[\begin{array}{@{}cc@{}}A^{00}_{n_0\times n_0} 
                                              & A^{01}_{n_0\times n_1} \\
            (A^{01})^\dagger_{n_1\times n_0} & A^{11}_{n_1\times n_1}\end{array}\right],
\end{eqnarray*}
where $n_{y}=|f^{-1}(y)|$ is the number of 
strings $x$ such that $f(x)=y$, and we have 
specified the dimensions of the matrix blocks for 
clarity. In what follows these dimensions will be 
omitted. We assume without loss of generality that $n_0 \leq n_1$. 
It is easy to check that, since $P_{01}$ is a projector, it must satisfy
\begin{eqnarray}\begin{aligned} \label{P01isaprojector}
A^{00}(\id_{n_0} -A^{00})&=A^{01}{A^{01}}^\dagger,\\
A^{11}(\id_{n_1} -A^{11})&={A^{01}}^\dagger A^{01}.
\end{aligned}\end{eqnarray}
Consider a unitary of the following form
$$
V=\left[\begin{array}{@{}cc@{}}V_0 &0 \\  0 & V_1\end{array}\right],
$$
where $V_0$ and $V_1$ are $n_0 \times n_0$ and $n_1 \times n_1$ unitaries respectively. 
Under such a unitary, $P_{00}$ and $P_{01}$ are transformed to:
\begin{eqnarray}\begin{aligned}\label{decomposed}
VP_{00}V^\dagger &= P_{00},\\
VP_{01}V^\dagger &=\left[\begin{array}{@{}cc@{}} 
                            V_0 A^{00}V_{0}^\dagger & V_0 A^{01} V_1^\dagger \\ 
                            (V_0 A^{01} V_1^\dagger)^\dagger & V_1A^{11} V_1^\dagger
                         \end{array}\right].
\end{aligned}
\end{eqnarray}
We now choose $V_0$ and $V_1$ from the singular value decomposition
(SVD, \cite[Theorem 7.3.5]{horn&johnson:ma}) of
$A^{01} = V_0^\dagger D V_1$ which gives
$$
D= V_0 A^{01} V_1^\dagger = \sum_{k=1}^{n_0} d_k \outp{u_k}{v_k},
$$
where $d_k\geq 0$, $\inp{u_k}{u_l}=\inp{v_k}{v_l}=\delta_{kl}$. 
Since $(A^{01})^\dagger A^{01}$ and 
$A^{01}(A^{01})^\dagger$ are supported in orthogonal 
subspaces, it also holds that $\forall k,l: \inp{u_k}{v_l} = 0$.
Eqs.~(\ref{P01isaprojector}) and (\ref{decomposed}) now give us
\begin{eqnarray*}
V_0 A^{00}V_0^\dagger(\id_{n_0} -V_0 A^{00}V_0^\dagger)
                &=\sum_{k=1}^{n_0} d_k^2 \outp{u_k}{u_k},\\
V_1 A^{11}V_1^\dagger(\id_{n_1} -V_1 A^{11}V_1^\dagger)
                &=\sum_{k=1}^{n_0} d_k^2 \outp{v_k}{v_k}.
\end{eqnarray*}
Suppose for the time being that all the $d_k$ are 
different. Since they are all non-negative, all 
the $d_k^2$ will also be different and it must hold that
\begin{eqnarray*}
V_0 A^{00}V_0^\dagger &=&\sum_{k=1}^{n_0} a^0_k\outp{u_k}{u_k},\\
V_1 A^{11}V_1^\dagger &=&\sum_{k=1}^{n_0} a^1_k \outp{v_k}{v_k}
                         + \sum_{k=n_0+1}^{n_1} a^1_k \outp{\tilde{v}_k}{\tilde{v}_k}
\end{eqnarray*}
for some $a^0_k$, $a^1_k$ and $\ket{\tilde{v}_k}$ with $1\leq k \leq n_1$.. 
Note that we can choose $\ket{\tilde{v}_k}$ such that
$\forall k,k',k\neq k': \inp{\tilde{v}_k}{\tilde{v}_{k'}} = 0$ 
and $\forall k,l: \inp{u_k}{\tilde{v}_l} = 0$. 
We can now express $VP_{01}V^\dagger$ as
\begin{eqnarray*}
&&VP_{01}V^\dagger =\\
&&=\sum_{k=1}^{n_0}\left[ a^0_k \outp{u_k}{u_k}
                           + d_k (\outp{u_k}{v_k} +\outp{v_k}{u_k})
                           + a^1_k \outp{v_k}{v_k}\right]
                   + \sum_{k=n_0+1}^{n_1} a^1_k \outp{\tilde{v}_k}{\tilde{v}_k}.
\end{eqnarray*}
It is now clear that we can choose all
$\hil_k=\spann\{\ket{u_k},\ket{v_k}\}$, and $\hil_{k'}=\spann\{\ket{\tilde{v}_{k'}}\}$
which are orthogonal and together add up to $\hil$.

In the case that all the $d_k$ are not different, there 
is some freedom left in choosing $\ket{u_k}$ and $\ket{v_k}$ 
that still allows us to make $V_0 A^{00}V_0^\dagger$ and 
$V_1 A^{11}V_1^\dagger$ diagonal so that the rest of the 
proof follows in the same way.
\end{proof}

In particular, the previous lemma implies that the post-measurement
states corresponding to strings $x$ for which
$f(x) = 0$ are orthogonal to those corresponding to strings $x$ 
for which $f(x) = 1$, which is expressed in the following lemma.
\begin{lemma}\label{posterior-orth}
Suppose one performs the measurement given by $\{\Pi_i: i\in[m]\}$. 
If the outcome of the measurement is $i$ and the state was $U_b\ket{x}$, 
then the post-measurement state is
$$
\ket{x,i,b}:=\frac{\Pi_iU_b\ket{x}}{\sqrt{\bra{x}U_b^\dagger\Pi_iU_b\ket{x}}}.
$$
The post-measurement states satisfy
\begin{eqnarray*}
\forall x\in f^{-1}(0),~x'\in f^{-1}(1),~i\in [m]: \inp{x,i,b}{x',i,b}=0.
\end{eqnarray*}
\end{lemma}
\begin{proof}
The proof follows straightforwardly from that fact that the $\Pi_i$ commute with 
both $P_{00}$ and $P_{01}$ (which follows from Lemma \ref{directsumdecomposition}).
\end{proof}

Now we are ready to prove the main theorem of this section.

\begin{theorem}\label{onequbitisenough}
Let $|\mY|=|\mB|=2$, then there exists a strategy for Bob
such that he succeeds at $\text{\emph{PI}}_1\text{\emph{-STAR}}(\ens)$ with
probability $p = 1$, for any function $f$ and prior $P_X$ on
${\cal X}$.
\end{theorem}
\begin{proof}
The strategy that Bob uses is the following:
\begin{itemize}
  \item Bob performs the measurement given by  $\{\Pi_i: i\in[m]\}$.
  \item He obtains an outcome $i\in [m]$ and 
    stores the post-measurement state which is supported in 
    the at most two-dimensional subspace $\hil_i$.
  \item After the basis $b\in\{0,1\}$ is announced, he
    measures $\{P_{0b}, P_{1b}\}$ and reports the outcome of this measurement.
\end{itemize}
By Lemma~\ref{posterior-orth} this leads to success probability $1$.
\end{proof}

Our result also gives us a better lower bound for all Boolean functions than 
what we had previously obtained in Section~\ref{sipiLower}. Instead of storing
the qubit, Bob now measures it immediately along the 
lines of Lemma~\ref{Breitbart}. It is not too difficult to convince yourself
that
for one qubit the worst-case post-measurement states to distinguish are in fact 
those in Lemma \ref{Breitbart}.
\begin{corollary} \label{pistarlarger0.85}
  Let $|\mY|=|\mB|=2$, then Bob succeeds at
  $\text{\emph{PI}}_0\text{\emph{-STAR}}(\ens)$ with probability 
  at least $p\geq (1 + 1/\sqrt{2})/2$.
\end{corollary}
In particular, our result implies that for the task 
of constructing Rabin-OT in~\cite{serge:bounded}
it is essential for Alice to choose a random function $f$ from
a larger set, which is initially unknown to Bob.

As a final remark, note that 
the prior distributions do not play any role. Likewise, it is
not actually important that the states $\rho_{yb}$ are proportional to
projectors: 
we only require that for all $b\in \01$, the states $\rho_{0b}$ and $\rho_{1b}$ are orthogonal.

\subsection{Using three bases}\index{PI-STAR!three bases}
\index{post-measurement information!three bases}
\label{limitedInfo3Bases}

We have just shown that Bob can compute any Boolean function perfectly when two bases are used. 
However, we now show that 
for any balanced Boolean function there exist three bases, such that Bob needs to store \emph{all} qubits
in order to compute the function perfectly.
The idea behind our proof is that for a particular choice of three bases, any 
measurement operator that satisfies the conditions set out in Lemma~\ref{commutation} must be proportional 
to the identity. This means that we cannot reduce the number of qubits to be stored by a measurement
and must keep everything. First, we prove the following lemma which we need in our main proof.

\begin{lemma} \label{misproptoid}
Let $M$ be a self-adjoint matrix which is diagonal in two mutually unbiased bases, then $M$ must be 
proportional to the identity.
\end{lemma}
\begin{proof}
Let $\ket{x}$ $\ket{u_x}$ $x\in\{1,\ldots,d\}$ be the two MUBs and let $m_x$ and $m'_x$ be the eigenvalues 
corresponding to $\ket{x}$ and $\ket{u_x}$ respectively, then we can write
$$
M=\sum_{x=1}^d m_x \outp{x}{x}=\sum_{x^\pr=1}^d m'_{x^\pr} \outp{u_{x^\pr}}{u_{x^\pr}}.
$$
From the previous equation, it follows that
$$
 \bra{x}M\ket{x} = m_x=\sum_{{x^\pr}=1}^d m'_{x^\pr} |\inp{u_{x^\pr}}{x}|^2=\frac{1}{d} \Tr M,
$$
which implies the desired result.
\end{proof}

We are now ready to prove the main result of this section.

\begin{theorem}\label{thm:threeSettings}
Let $|\mY|=2$ and $|\mB|=3$, then for any balanced function $f$ and prior $P_X$ on
${\cal X}$ which is uniform on the pre-images $f^{-1}(y)$, there exist three bases 
such that Bob succeeds at $\text{\emph{PI}}_q\text{\emph{-STAR}}(\ens)$ with
probability $p = 1$ if and only if $q=\log d$.
\end{theorem}
\begin{proof}
Let $P_{00}=\sum_{x\in f^{-1}(0)}\outp{x}{x}$, $P_{01}=U_1 P_{00}U_1^\dagger$ and $
P_{02}=U_2 P_{00}U_2^\dagger$. Also, let $s:f^{-1}(0)\to f^{-1}(1)$ be a bijective map, and let $s_x=s(x)$. 
By a reordering of the basis, $P_{00}$, $U_1$ and $U_2$ can be written as
\begin{eqnarray*}
P_{00}=\left[\begin{array}{@{}cc@{}}\id & 0 \\
                            0 & 0\end{array}\right],
U_1=\left[\begin{array}{@{}cc@{}}U_1^{00} 
                                              & U_1^{01} \\
            U_1^{10} & U_1^{11}\end{array}\right],
U_2=\left[\begin{array}{@{}cc@{}}U_2^{00} 
                                              & U_2^{01} \\
            U_2^{10} & U_2^{11}\end{array}\right],
\end{eqnarray*}
where all the blocks are of size $(d/2) \times (d/2)$. $P_{01}$ and $P_{02}$  then take the following form:
\begin{eqnarray*}
P_{01}=\left[\begin{array}{@{}cc@{}}U_1^{00}{U_1^{00}}^\dagger
                                              & U_1^{00}{U_1^{10}}^\dagger \\
            (U_1^{00}{U_1^{10}}^\dagger)^\dagger & U_1^{10}{U_1^{10}}^\dagger\end{array}\right],
P_{02}=\left[\begin{array}{@{}cc@{}}U_2^{00}{U_2^{00}}^\dagger
                                              & U_2^{00}{U_2^{10}}^\dagger \\
            (U_2^{00}{U_2^{10}}^\dagger)^\dagger & U_2^{10}{U_2^{10}}^\dagger\end{array}\right].
\end{eqnarray*}

It follows from Lemma~\ref{commutation}, that we only have to prove that 
$[M,P_{00}]=[M,P_{01}]=[M,P_{02}]=0$ implies that $M$ must be proportional to the identity. Write
\begin{eqnarray*}
M=\left[\begin{array}{@{}cc@{}}M^{00} 
                                              & M^{01} \\
            (M^{01})^\dagger & M^{11}\end{array}\right].
\end{eqnarray*}
Commutation with $P_{00}$ implies $M^{01}=0$. Commutation with $P_{01}$ and $P_{02}$ implies
\begin{eqnarray}
\label{condsforM1}
{[}M^{00},U_1^{00}{U_1^{00}}^\dagger{]}&=&{[}M^{00},U_2^{00}{U_2^{00}}^\dagger{]}=0,\\
\label{condsforM2}
[M^{11},U_1^{10}{U_1^{10}}^\dagger]&=&[M^{11},U_2^{10}{U_2^{10}}^\dagger]=0,\\
\label{last1Cond}
M^{00}(U_1^{00}{U_1^{10}}^\dagger)&=&(U_1^{00}{U_1^{10}}^\dagger)M^{11},\\
\label{last2Cond}
M^{00}(U_2^{00}{U_2^{10}}^\dagger)&=&(U_2^{00}{U_2^{10}}^\dagger)M^{11}.
\end{eqnarray}
We choose $U_1$ and $U_2$ in the following way:
\begin{eqnarray*}
U_1 &=&\sum_{x\in f^{-1}(0)}\left[a_x (\outp{x}{x}+\outp{s_x}{s_x}) +\sqrt{1-a_x^2}
(\outp{x}{s_x}-\outp{s_x}{x})\right],\\
U_2 &=&\sum_{x\in f^{-1}(0)}\left[a_x (\outp{u_x}{u_x}+\outp{v_x}{v_x}) +
\sqrt{1-a_x^2}(\outp{u_x}{v_x}-\outp{v_x}{u_x})\right],
\end{eqnarray*}
with $a_x\in [0,1]$, satisfying $a_x=a_{x^\pr}$ if and only if $x=x^\pr$. 
Furthermore, choose $\ket{u_x}$ and $\ket{v_x}$ such that 
\begin{eqnarray*}
\forall x,x^\pr \in f^{-1}(0),~ \inp{x}{v_{x^\pr}}=\inp{s_x}{u_{x^\pr}}=0,~
|\inp{x}{u_{x^\pr}}|^2=|\inp{s_x}{v_{x^\pr}}|^2=2/d.
\end{eqnarray*}

\noindent
With this choice for $U_1$ and $U_2$ we have that
\begin{eqnarray*}
U_1^{00}{U_1^{00}}^\dagger&=&\sum_{x\in f^{-1}(0)} a_x^2 \outp{x}{x},\\
U_2^{00}{U_2^{00}}^\dagger&=&\sum_{x\in f^{-1}(0)} a_x^2 \outp{u_x}{u_x},
\end{eqnarray*}
i.e., $\{\ket{x}\}$ and $\{\ket{u_x}\}$ form an eigenbasis for $U_1^{00}{U_1^{00}}^\dagger$ and 
$U_2^{00}{U_2^{00}}^\dagger$ respectively. Furthermore, since all the $a_x^2$ are different, 
the eigenbases are unique. Now, using Eq.\ (\ref{condsforM1}), we see that $M^{00}$ must commute 
with both $U_1^{00}{U_1^{00}}^\dagger$  and $U_2^{00}{U_2^{00}}^\dagger$, and since their eigenbases 
are unique, it must be true that $M^{00}$ is diagonal in both  $\{\ket{x}\}$ and $\{\ket{u_x}\}$. Using the 
result of Lemma \ref{misproptoid} it follows that $M^{00}=m_0 \id_{d/2}$. In exactly the same way 
we can prove that $M^{11}=m_1 \id_{d/2}$ using Eq.\ (\ref{condsforM2}). 
It remains to prove that $m_0=m_1$, which follows directly from either Eq.\ (\ref{last1Cond}) or Eq.\ (\ref{last2Cond}).
\end{proof}

From our proof it is clear how to construct $U_1$ and $U_2$. 
For $P_{00}$ as defined above, we could choose vectors
of the form $\ket{x} = \ket{0}\ket{\hat{x}}$
and $\ket{s_x} = \ket{1}\ket{\hat{x}}$ where $\hat{x} \in \01^{n-1}$ to
construct $U_1$.
For $U_2$ we could then pick $\ket{u_x} = \ket{0}H^{\otimes n-1}\ket{\hat{x}}$ and analogously $\ket{v_x} = \ket{1}H^{\otimes n-1}\ket{\hat{x}}$. As we will see in Chapter~\ref{entanglementIntro}, 
our example shows that for non-local games we cannot hope to prove
a statement analogous to~\cite{masanes:blocks,toner:blocks} for three
measurement settings where each measurement has two outcomes.

Note, however, that whereas
we know that for such unitaries Bob must store all qubits in order to compute the value of the function
perfectly, it remains unclear how close he can get to computing 
the function perfectly when storing fewer qubits. In particular,
he can always choose two of the three bases, and employ the strategy outlined in the previous section:
he stores the one qubit that allows him to succeed with probability $1$ for two of the bases. If he gets the third basis 
then he just flips a coin. In this case, he is correct with probability $2/3 + 1/(3 \cdot 2) = 5/6$ 
for a balanced function and a uniform prior.
It remains an important open question to address the approximate case.

\section{Conclusion}
\label{conclusionPISTAR}
We have introduced a new state discrimination problem, motivated by
cryptography: discrimination with extra information about the state
after the measurement, or, more generally, after a quantum memory bound
applies.  We have left most general questions open, but we found fairly complete
results in the case of guessing $y = f(x)$ with mutually unbiased encodings.

We have shown that storing just a single qubit allows Bob to succeed at
PI-STAR perfectly for \emph{any} Boolean function and any two bases.
In contrast, we showed how to construct \emph{three} bases such that
Bob needs to store \emph{all} qubits in order to compute the function
perfectly.
We have also given an explicit strategy for two functions, namely the AND 
and the XOR. More generally, it would be interesting to determine,
how many qubits Bob needs to store to compute $f(x)$ 
perfectly for any function $f: \mX \rightarrow \mY$
in terms of the number of outputs $|\mY|$ and the number of bases $|\mB|$.
It should be clear that the algebraic techniques of
Section~\ref{sec:algebra-perfect} allow us to answer
these questions for any given function in principle. However, so far, we 
have not been able to obtain explicit structures for wider
classes of functions.
Our results imply that in existing protocols in the bounded quantum storage 
model~\cite{serge:bounded} we cannot restrict ourselves to a single fixed function $f$ to perform privacy amplification.
Note that our algebraic framework can also address the question of 
using more than one function, where
$f$ is also announced after the memory bound applies~\cite{serge:bounded}:
we merely obtain a larger problem. Yet, it is again difficult to determine 
a general bound.

In the 
important case of two mutually unbiased bases and 
balanced functions, we have shown (Theorem \ref{starless0.85} 
and Corollary \ref{pistarlarger0.85}) that there exists a clear separation 
between the case where Bob gets the post-measurement information (PI-STAR) 
and when he does not (STAR).  Namely, for any such function, Bob's
optimal success probability is never larger
than $(1 + 1/\sqrt{2})/2 \approx 0.853$ for STAR and always 
at least as large as the same number for PI-STAR.

In some cases the gap between STAR and PI-STAR can be 
more dramatic. The XOR function on strings of even
length with two mutually unbiased bases is one of 
these cases. We have shown that in this case the advantage can be 
maximal. Namely, \emph{without} the extra information 
Bob can never do better than guessing the 
basis, \emph{with} it however, he can compute the 
value of the function perfectly. This contrasts with the 
XOR function on strings of odd length, where the optimal success probabilities 
of STAR and PI-STAR are both $(1 + 1/\sqrt{2})/2$ 
and the post-measurement information is completely useless for Bob. 
It would be interesting to see, how large the
gap between STAR and PI-STAR can be for any function 
$f: \01^n \rightarrow \01^k$ where $k > 2$.
We return to this
question in Chapter~\ref{useofPISTAR}. 

It would also be nice to show a general lower bound for non-balanced
functions or a non-uniform prior. As the example for $3$ bases
showed, the uniform prior is not necessarily the one that leads to the largest gap, and
thus the prior can play an important role.
Another generalization would be to consider functions 
of the form $f: [d]^n \rightarrow [d]^k$. 

We now turn our attention to uncertainty relations. These will play an important
role in locking in Chapter~\ref{chapter:locking}. In the problem of locking, we also distinguish
measurement \emph{with} basis information, analogous to our $\text{PI}_q\text{-STAR}$
with $q=n$, and \emph{without}
corresponding to $\text{PI}_0\text{-STAR}$.
So far, our objective has been to obtain an accurate guess of a value, e.g. $y=f(x)$.
In Chapter~\ref{chapter:locking}, we are interested in a slightly different problem: How can we maximize
the classical mutual information? In particular, can we use mutually unbiased
bases to obtain locking effects?

\chapter{Uncertainty relations}\label{uncertaintyChapter}\index{uncertainty}\label{chapter:uncertainty}\index{uncertainty relation}

Uncertainty relations lie at the very core of quantum mechanics. Intuitively, they quantify how much we can
learn about different properties of a quantum system simultaneously. 
Some properties lead to very strong uncertainty relations: 
if we decide to learn one, we remain entirely ignorant about the others. 
But what characterizes such properties? In this chapter, we first investigate whether choosing our
measurements to be mutually unbiased bases allows us to obtain strong uncertainty relations. Sadly, it
turns out that mutual unbiasedness is not sufficient. Instead, we need to consider anti-commuting measurements.

\section{Introduction}

Heisenberg first realized that quantum mechanics leads to uncertainty relations for conjugate observables
such as position and momentum~\cite{heisenberg:uncertainty}. Uncertainty relations are probably best known
in the form given by Robertson~\cite{robinson:uncertainty}, who extended Heisenberg's result to any two observables
$A$ and $B$. Robertson's relation states that if we prepare many copies of the state $\ket{\psi}$, and measure
each copy individually using either $A$ or $B$, we have\index{uncertainty relation!Heisenberg}
$$
\Delta A \Delta B \geq \frac{1}{2} |\bra{\psi}[A,B]\ket{\psi}|
$$
where $\Delta X = \sqrt{\bra{\psi}X^2\ket{\psi} - \bra{\psi}X\ket{\psi}^2}$ for $X \in \{A,B\}$ is
the standard deviation resulting from measuring $\ket{\psi}$ with observable $X$.
Recall from Chapter~\ref{classicalTheoriesChapter}, that classically we always have $[A,B] = 0$, and
there is no such limiting lower bound. 
Hence, uncertainty relations are another characteristic that sets apart quantum theory.
The consequences are rather striking: even if we had a perfect measurement apparatus, we are nevertheless
limited! 

Entropic uncertainty relations are an alternative way to state Heisenberg's uncertainty principle.
They are frequently a more useful characterization, because the ``uncertainty'' is lower
bounded by a quantity that does not depend on the state to be measured~\cite{deutsch:uncertainty,kraus:entropy}.
Recently, entropic uncertainty relations have gained importance in the context of quantum cryptography
in the bounded storage model, where proving the security of such protocols ultimately
reduces to establishing such relations~\cite{serge:new}. Proving new entropic uncertainty relations
could thus give rise to new protocols. Intuitively, it is clear that uncertainty relations have a significant
impact on what kind of protocols we can obtain in the quantum settings. Recall 
the cryptographic task of oblivious transfer
from Chapter~\ref{chapter:cryptoIntro}:
the receiver should be able to extract information about one particular property of a system, but should learn 
as little as possible about all other properties. It is clear that, without placing any additional
restrictions on the receiver, uncertainty relations intuitively quantify how well we are able to implement
such a primitive. 

\emph{Entropic} uncertainty relations were first introduced by Bialynicki-Birula and Mycielski~\cite{bm:uncertainty}. 
For our purposes, we will be interested in uncertainty relations in the form put forward by
Deutsch~\cite{deutsch:uncertainty}. Following a conjecture by Kraus~\cite{kraus:entropy}, Maassen and Uffink~\cite{maassen:entropy}
have shown that if we measure the state $\ket{\psi}$ with observables $A$ and $B$ determined
by the bases $\mA = \{\ket{a_1},\ldots,\ket{a_d}\}$ and $\mB = \{\ket{b_1},\ldots,\ket{b_d}\}$ respectively, we have
$$
\frac{1}{2}\left( H(\mA|\ket{\psi})+H(\mB|\ket{\psi})\right)\geq -\log c(\mA,\mB),
$$
where $c(\mA,\mB)=\max \left \{|\inp{a}{b}|\mid \ket{a} \in \mA, \ket{b} \in \mB\right\}$, and
$$
H(\mX|\ket{\psi}) =  - \sum_{i = 1}^{d} |\inp{\psi}{x_i}|^2 \log |\inp{\psi}{x_i}|^2
$$ is the Shannon entropy~\cite{shannon:info} arising from 
measuring the state $\ket{\psi}$ in the basis $\mX = \{\ket{x_1},\ldots,\ket{x_d}\}$.
In fact, Maassen and Uffink provide a more general statement which also leads to uncertainty
relations for higher order R{\'e}nyi entropies. Such relations have also been shown by Bialynicki-Birula~\cite{bia:uncertaintyRenyi}
for special sets of observables. Note that the above relation achieves our initial goal: the lower bound no 
longer depends on the state $\ket{\psi}$, but only on $A$ and $B$ itself.
What is the strongest possible relation we could obtain? That is, which choices of $\mA$ and $\mB$ maximize 
$- \log c(\mA,\mB)$? It is not hard to see that choosing $\mA$ and $\mB$
to be mutually unbiased (see Section~\ref{MUBdef}) 
provides us with a lower bound of $(\log d)/2$ which is
the strongest possible uncertainty relation: 
If we have no entropy for one of the bases, then the entropy for the other
bases must be maximal. For example, in case of a one qubit system of $d=2$
choosing $\mA = \{\ket{0},\ket{1}\}$ and $\mB = \{\ket{+},\ket{-}\}$ to be the computational and the Hadamard basis respectively, we obtain a lower bound
of $1/2$. 

Can we derive a similar relation for measurements using three or more observables? Surprisingly, 
very little is known for a larger number of measurement settings~\cite{azarchs:entropy}. 
Sanchez-Ruiz~\cite{sanchez:entropy,sanchez:entropy2} (using results of Larsen~\cite{larsen:entropy}) has
shown that for measurements using all $d+1$ mutually unbiased bases, we can obtain strong uncertainty relations. 
Here, we provide an elementary proof of his result in dimension $d=2^n$.
Given the fact that mutually unbiased bases seem to be a good choice if we use only two or $d+1$ measurement
settings, it may be tempting to conclude that choosing our measurements to be mutually unbiased always gives
us good uncertainty relations for which the lower bound is as large as possible.
Numerical results for MUBs in prime dimensions up to 29 indicate that
MUBs may indeed be a good choice~\cite{terhal:locking}. However, we show that merely being
mutually unbiased is not sufficient to obtain strong uncertainty relations. 
To this end, we prove tight entropic uncertainty relations for measurements in a large number of 
mutually unbiased bases (MUBs) in square dimensions. In particular, we consider any MUBs derived
from mutually orthogonal Latin squares~\cite{wocjan:mub}, and \emph{any} set of MUBs 
obtained from the set of unitaries of the form $\{U \otimes U^*\}$, where $\{U\}$ 
gives rise to a
set of MUBs in dimension $s$ when applied to the basis elements of the computational basis. 
For any $s$, there are at most $s+1$ such MUBs in a Hilbert space of dimension $d=s^2$: recall from Section~\ref{MUBdef} that we can have at most $s+1$ MUBs in a space of dimension $s$.
Let $\mathbb{B}$ be the set of MUBs coming from one of these two constructions. 
We prove that for any subset $\mathbb{T} \subseteq \mathbb{B}$ of these bases we
have 
$$ 
\min_{\ket{\psi}} \sum_{\mB \in \mathbb{T}} H(\mB|\ket{\psi}) = \frac{|\mathbb{T}|}{2} \log d.
$$

Our result shows that one needs to be careful to 
think of ``maximally incompatible'' measurements
as being necessarily mutually unbiased. When we take entropic uncertainty relations as our measure of
``incompatibility'', mutually unbiased measurements are not always the most incompatible when
considering more than two observables. In particular, it has been shown \cite{winter:randomizing} that if we choose approximately
$(\log d)^4$ bases uniformly at random, then with high probability
$\min_{\ket{\psi}} (1/|\mathbb{T}|) \sum_{\mB \in \mathbb{T}} H(\mB|\ket{\psi}) \geq \log d -3$. This means that
there exist $(\log d)^4$ bases for which this sum of entropies is very large, i.e., measurements in such bases
are very incompatible. However, we show that when $d$ is large, there exist $\sqrt{d}$ 
mutually unbiased bases that are much less incompatible according to this measure.
When considering entropic uncertainty relations as a measure of ``incompatibility'', we must therefore
look for different properties for the bases to define incompatible measurements.

Luckily, we are able to obtain maximally strong uncertainty relations for two-outcome measurements for
\emph{anti-commuting} observables. In particular, we obtain for $\Gamma_1,\ldots,\Gamma_{K}$ with $\{\Gamma_i,\Gamma_j\} = 0$
that 
$$
\min_\rho \frac{1}{K} \sum_{j=1}^K H(\Gamma_j|\rho) = 1 - \frac{1}{K},
$$
where $H(\Gamma_j|\rho) = - \sum_{b \in \01}\Tr(\Gamma_j^b \rho) \log \Tr(\Gamma_j^b \rho)$
and $\Gamma_j^0$, $\Gamma_j^1$ are projectors onto the positive and negative eigenspace of
$\Gamma_j$ respectively.
Thus, if we have zero entropy for one of the terms, we must have maximal 
entropy for all others.
For the collision entropy
we obtain something slightly suboptimal
$$
\min_\rho \frac{1}{K} \sum_{j=1}^K H_2(\Gamma_j,\rho) \approx 1 - \frac{\log e}{K}
$$
for large $K$, where $H_2(\Gamma_j|\rho) = - \log \sum_{b \in \01} \Tr(\Gamma_j^b\rho)^2$.
Especially our second uncertainty relation is of interest for cryptographic applications.

\section{Limitations of mutually unbiased bases}\label{mubUncertainty}

We first prove tight entropic uncertainty for measurements in MUBs in square dimensions. 
We need the result of Maassen and Uffink~\cite{maassen:entropy} mentioned above:
\begin{theorem}[Maassen and Uffink]\index{uncertainty relation!Maassen and Uffink}
\index{uncertainty relation!for two bases}
\label{eq:maasenuffinkbound}
Let $\mB_1$ and $\mB_2$ be two orthonormal basis in a Hilbert space of dimension $d$. Then
for all pure states $\ket{\psi}$
$$
\frac{1}{2}\left( H(\mB_1|\ket{\psi})+H(\mB_2|\ket{\psi})\right)\geq -\log c(\mB_1,\mB_2),
$$
where $c(\mB_1,\mB_2)=\max \left \{|\inp{b_1}{b_2}|\mid \ket{b_1} \in \mB_1,\ket{b_2}\in \mB_2\right \}$.
\end{theorem}
The case when $\mB_1$ and $\mB_2$ are MUBs is of special interest for us. More generally, when one has a set of MUBs a trivial application of Theorem~\ref{eq:maasenuffinkbound} leads to the following corollary also noted in~\cite{azarchs:entropy}.
\begin{corollary}\label{MUderived}
\label{eq:maassenuffinkij}
Let $\mathbb{B}=\{\mB_1,\ldots,\mB_m\}$ be a set of MUBs in a Hilbert space of dimension $d$. Then
$$
\frac{1}{m} \sum_{t=1}^m H(\mB_t|\ket{\psi})\geq \frac{\log d}{2}.
$$
\end{corollary}
\begin{proof}
Using Theorem~\ref{eq:maasenuffinkbound}, one gets that for any pair of MUBs $\mB_t$ and $\mB_{t'}$ with $t\neq t'$
$$
\frac{1}{2}\left[ H(\mB_t|\psi)+H(\mB_{t'}|\psi)\right]\geq \frac{\log d}{2}.
$$
Adding up the resulting equation for all pairs $t\neq t'$ we get the desired result.
\end{proof}

We now show that this bound can in fact be tight for a large set of MUBs. 

\subsection{MUBs in square dimensions}
Corollary \ref{MUderived} gives  a lower bound on the average of the entropies of a set of MUBs. But how good is this bound?
We show that the bound is indeed tight when we consider product MUBs in a Hilbert space of square dimension.
\begin{theorem}\label{squareThm}\index{uncertainty relation!for MUBs in square dimensions}
Let $\mathbb{B}=\{\mB_1,\ldots,\mB_m\}$ with $m\geq 2$ be a set of MUBs in a Hilbert space $\hil$ of dimension $s$. Let $U_t$ be the unitary operator that transforms the computational basis to $\mB_t$. 
Then $\mathbb{V}=\{\mV_1,\ldots,\mV_m\}$, where 
$$
\mV_t=\left \{U_t\ket{k}\otimes U_t^* \ket{l}\mid k,l\in[s] \right \},
$$
is a set of MUBs in $\hil \otimes \hil$, and it holds that
$$
\min_{\ket{\psi}} \frac{1}{m} \sum_{t=1}^m H(\mV_t|\ket{\psi})= \frac{\log d}{2},
$$
where $d=\dim(\hil \otimes \hil)=s^2$. 
\end{theorem}
\begin{proof}
It is easy to check that $\mathbb{V}$ is indeed a set of MUBs. Our proof works by constructing a state $\ket{\psi}$
that achieves the bound in Corollary~\ref{MUderived}.
It is easy to see that the maximally entangled state
$$
\ket{\psi}= \frac{1}{\sqrt{s}}\sum_{k=1}^{s}\ket{kk},
$$
satisfies $U\otimes U^*\ket{\psi}=\ket{\psi}$ for any $U\in \U(d)$. Indeed,
\begin{eqnarray*}
\bra{\psi}U\otimes U^*\ket{\psi}&=&\frac{1}{s}\sum_{k,l=1}^{s} \bra{k}U\ket{l}\bra{k}U^*\ket{l}\\
						&=&\frac{1}{s}\sum_{k,l=1}^{s} \bra{k}U\ket{l}\bra{l}U^\dagger\ket{k}\\
						&=&\frac{1}{s}\Tr UU^\dagger=1. 
\end{eqnarray*}
Therefore, for any $t\in[m]$ we have that
\begin{eqnarray*}
H(\mV_t|\ket{\psi})	&=&-\sum_{kl}|\bra{kl}U_t\otimes U_t^*\ket{\psi}|^2\log|\bra{kl}U_t\otimes U_t^*\ket{\psi}|^2\\
				&=&-\sum_{kl}|\inp{kl}{\psi}|^2\log|\inp{kl}{\psi}|^2\\
				&=&\log s=\frac{\log d}{2}.
\end{eqnarray*}
Taking the average of the previous equation over all $t$ we obtain the result.
\end{proof}

\subsection{MUBs based on Latin squares}

We now consider mutually unbiased bases based on Latin squares~\cite{wocjan:mub}
as described in Section~\ref{latinSquareConstruction}. Our proof again follows by providing a state
that achieves the bound in Corollary~\ref{MUderived}, which turns out to have
a very simple form.
\begin{lemma}\label{LSentropy}\index{uncertainty relation!for Latin square MUBs}
Let $\mathbb{B}=\{\mB_1,\ldots,\mB_m\}$ with $m \geq 2$ be any set of MUBs in a Hilbert space of dimension $d=s^2$ constructed
on the basis of Latin squares. Then
$$
\min_{\ket{\psi}} \frac{1}{m} \sum_{\mB\in\mathbb{B}} H(\mB|\ket{\psi}) = \frac{\log d}{2}.
$$
\end{lemma}
\begin{proof}
Consider the state $\ket{\psi} = \ket{1,1}$ and fix a basis
$\mB_t = \{\ket{v^t_{i,j}}|i,j \in [s]\} \in \mathbb{B}$
coming from a Latin square.
It is easy
to see that there exists exactly one $j \in [s]$ such that $\inp{v^t_{1,j}}{1,1} = 1/\sqrt{s}$. Namely this
will be the $j \in [s]$ at position $(1,1)$ in the Latin square. Fix this $j$. For any other
$\ell \in [s], \ell \neq j$, we have $\inp{v^t_{1,\ell}}{1,1} = 0$. But this means that
there exist exactly $s$ vectors in $\mB$ such that $|\inp{v^t_{i,j}}{1,1}|^2 = 1/s$, namely
exactly the $s$ vectors derived
from $\ket{v^t_{1,j}}$ via the Hadamard matrix. The same argument holds for any such basis $\mB \in \mathbb{T}$.
We get
\begin{eqnarray*}
\sum_{\mB \in \mathbb{T}} H(\mB|\ket{1,1}) &=& \sum_{\mB \in \mathbb{T}}
\sum_{i,j \in [s]} |\inp{v^t_{i,j}}{1,1}|^2 \log |\inp{v^t_{i,j}}{1,1}|^2\\
&=& |\mathbb{T}| s \frac{1}{s} \log \frac{1}{s}\\
&=& |\mathbb{T}| \frac{\log d}{2}.
\end{eqnarray*}
The result then follows directly from Corollary~\ref{MUderived}.
\end{proof}

\subsection{Using a full set of MUBs} 

We now provide an alternative proof of an entropic uncertainty relation
for a full set of mutually unbiased bases. This has previously been proved
in~\cite{sanchez:entropy,sanchez:entropy2}. We already provided an alternative proof using
the fact that the set of all mutually unbiased bases forms a 2-design~\cite{wehner06c}.
Here, we provide a new alternative proof for dimension $d=2^n$ which has the advantage that it neither requires the 
introduction of 2-designs, nor the results of~\cite{larsen:entropy} that were used in
the previous proof by Sanchez-Ruiz~\cite{sanchez:entropy,sanchez:entropy2}. Instead, our proof 
is extremely simple: After choosing a convenient parametrization of quantum states, 
the statement follows immediately using only elementary Fourier analysis.

For the parametrization,
we first introduce a basis for the space of $2^n \times 2^n$ matrices 
with the help of mutually unbiased bases. Recall from Section~\ref{MUBdef}
that in dimension $2^n$, we can find exactly $2^n+1$ MUBs.
\begin{lemma}\index{basis!MUB}
Consider the Hermitian matrices
$$
S^j_b = \sum_{x \in \01^n} (-1)^{j \cdot x} \outp{x_b}{x_b},
$$
for $b \in [d+1]$,
$j \in [d-1]$ and for all $x,x' \in \01^n$ and
$b\neq b' \in [d+1]$ we have $|\inp{x_b}{x'_{b'}}|^2 = 1/d$. Then 
the set $\{\id\} \cup \{S^j_b\mid b \in [d+1], j \in [d-1]\}$ forms
a basis for the space of $d \times d$ matrices, where for all $j$ and $b$, $S^j_b$ is traceless and $(S^j_b)^2 = \id$.
\end{lemma}
\begin{proof}
First, note that we have $(d+1)(d-1) + 1 = d^2$ matrices. We now show that
they are all orthogonal. Note that
$$
\Tr(S^j_b) = \sum_{x\in\01^n} (-1)^{j \cdot x} = 0,
$$
since $j \neq 0$, and hence $S^j_b$ is traceless. Hence $\Tr(\id S^j_b) = 0$. Furthermore,
\begin{equation}\label{orthogonalBasis}
\Tr(S^j_b S^{j'}_{b'}) = \sum_{x,x' \in \01^n} (-1)^{j \cdot x} (-1)^{j' \cdot x'} |\inp{x_b}{x'_{b'}}|^2.
\end{equation}
For $b\neq b'$, Eq.~(\ref{orthogonalBasis}) gives us $\Tr(S^j_b S^{j'}_{b'}) = (1/d) \left(\sum_x (-1)^{j \cdot x}\right)
\left(\sum_{x'} (-1)^{j' \cdot x'}\right) = 0$, since $j,j' \neq 0$. For $b=b'$, but $j \neq j'$, we get
$\Tr(S^j_b S^{j'}_{b'})= \sum_x (-1)^{(j \oplus j') \cdot x} = 0$ since $j \oplus j' \neq 0$.

Finally, $\left(S^j_b\right)^2 = \sum_{xx'} (-1)^{j \cdot x}(-1)^{j \cdot x'} \outp{x_b}{x_b}\outp{x'_b}{x'_b} = \id$.
\end{proof}

Since $\{\id,S^j_b\}$ form a basis for the $d \times d$ matrices, we can thus express the state $\rho$ of a $d$-dimensional
system as
$$
\rho = \frac{1}{d}\left(\id + \sum_{b \in [d+1]}\sum_{j \in [d-1]} s^j_b S^j_b\right),
$$
for some coefficients $s^j_b \in \Real$. It is now easy to see that
\begin{lemma}\label{MUBpurity}
Let $\rho$ be a pure state parametrized as above. Then 
$$
\sum_{b \in [d+1]}\sum_{j \in [d-1]} (s^j_b)^2 = d-1.
$$
\end{lemma}
\begin{proof}
If $\rho$ is a pure state, we have $\Tr(\rho^2) = 1$. Hence
\begin{eqnarray*}
\Tr(\rho^2) &=& \frac{1}{d^2}\left(\Tr(\id) + \sum_{b \in [d+1]}\sum_{j \in [d-1]} (s^j_b)^2 \Tr(\id)\right)\\
&=&\frac{1}{d}\left(1 + \sum_b \sum_j (s^j_b)^2\right) = 1,
\end{eqnarray*}
from which the claim follows.
\end{proof}
Suppose now that we are given a set of $d+1$ MUBs $\mB_1,\ldots,\mB_{d+1}$
with $\mB_b = \{\ket{x_b}\mid x \in \01^n\}$. Then the following simple observation lies at the core of our proof:
\begin{lemma}\label{MUBexpansion}
Let $\ket{x_b}$ be the $x$-th basis vector of the $b$-th MUB. Then for any state $\rho$
$$
\Tr(\outp{x_b}{x_b} \rho) = \frac{1}{d}\left(1 + \sum_{j \in [d-1]} (-1)^{j \cdot x} s^j_{b}\right).
$$
\end{lemma}
\begin{proof}
We have
$$
\Tr(\outp{x_b}{x_b}\rho) = \frac{1}{d}\left(\Tr(\outp{x_b}{x_b}) + \sum_{b',j} s^j_{b'} \Tr(S^j_{b'} \outp{x_b}{x_b})\right)
$$
Suppose $b \neq b'$. Then $\Tr(S^j_{b'} \outp{x_b}{x_b}) = (1/d) \sum_{x'} (-1)^{j \cdot x'} = 0$, since $j \neq 0$.
Suppose $b = b'$. Then $\Tr(S^j_{b'} \outp{x_b}{x_b}) = \sum_{x'} (-1)^{j \cdot x'} |\inp{x_b}{x'_b}|^2 = (-1)^{j \cdot x}$, 
from which the claim follows.
\end{proof}

We are now ready to prove an entropic uncertainty relation for $N$ mutually unbiased bases.
\begin{theorem}\label{MUBmany}\index{uncertainty relation!for all MUBs}
Let $\mS = \{\mB_1,\ldots,\mB_N\}$ be a set of mutually unbiased bases. Then
$$
\frac{1}{N}\sum_{b \in [N]} H_2(\mB_b,\ket{\Psi}) \geq - \log \frac{N + d - 1}{d N}.
$$
\end{theorem}
\begin{proof}
First, note that we can define functions $f_b(j) = s^j_b$ for $j \in [d-1]$ and $f_b(0) = s^0_b = 1$. 
Then $\hat{f}_b(x) = (1/\sqrt{d})(\sum_{j\in\{0,\ldots,d-1\}} (-1)^{j \cdot x} s^j_b)$ is the Fourier
transform of $f_b$ and $(1/\sqrt{d}) \hat{f}_b(x) = \Tr(\outp{x_b}{x_b}\rho)$ by Lemma~\ref{MUBexpansion}.
Thus
\begin{eqnarray*}
\frac{1}{N}\sum_{b \in [N]} H_2(\mB_b,\ket{\Psi}) &=& -\frac{1}{N}\sum_{b \in [N]} \log \sum_{x \in \01^n} 
|\inp{x_b}{\Psi}|^4\\
&\geq& - \log \frac{1}{d N} \sum_b \sum_x \hat{f}_b(x)^2\\
&=& - \log \frac{1}{d N} \sum_b (1 + \sum_j (s^j_b)^2)\\
&=& - \log \frac{1}{d N} (N + d - 1),
\end{eqnarray*}
where the first inequality follows from Jensen's inequality and the concavity of $\log$. The next equality
follows from Parseval's equality, and the last follows from the fact that $\ket{\Psi}$ is a pure state
and Lemma~\ref{MUBpurity}.
\end{proof}

\begin{corollary}
Let $\mS = \{\mB_1,\ldots,\mB_N\}$ be a set of mutually unbiased bases. Then
$$
\frac{1}{N}\sum_{b \in [N]} H(\mB_b|\ket{\Psi}) \geq - \log \frac{N + d - 1}{d N}.
$$
In particular, for a full set of $N = d+1$ MUBs we have $(1/N) \sum_b H(\mB_b|\ket{\Psi}) \geq \log((d+1)/2)$.
\end{corollary}
\begin{proof}
This follows immediately from Theorem~\ref{MUBmany} and the fact that $H(\cdot) \geq H_2(\cdot)$.
\end{proof}

It is interesting to note that this bound is the same that arises from interpolating between 
the results of Sanchez-Ruiz~\cite{sanchez:entropy,sanchez:entropy2} and Maassen and Uffink~\cite{maassen:entropy}
as was done by Azarchs~\cite{azarchs:entropy}.

\section{Good uncertainty relations}
\label{cliffordUncertainty}\label{goodUncertainty}\index{Clifford algebra}\index{Clifford algebra!uncertainty relation}\index{uncertainty relation!for anti-commuting observables}\index{uncertainty relation!Clifford algebra}

As we saw, merely choosing our measurements to be mutually unbiased is not sufficient
to obtain good uncertainty relations. 
However, we now investigate measurements using \emph{anti-commuting}\index{anti-commuting observables}\index{observable!anti-commuting}
observables for which we do obtain maximally strong uncertainty relations!
In particular, we consider the matrices $\Gamma_1,\ldots,\Gamma_{2n}$, satisfying
the anti-commutation
relations
\begin{equation}\label{antiComm}
\Gamma_i\Gamma_j = - \Gamma_j \Gamma_i, \mbox{   } \Gamma_i^2 = \id
\end{equation}
for all $i,j \in [2n]$. 
Such operators $\Gamma_1,\ldots,\Gamma_{2n}$ form generators for the Clifford
algebra, which we explain in more detail in Appendix~\ref{appendix:clifford}.

Intuitively, these operators have a property that is very similar to being mutually unbiased:
Recall from Appendix~\ref{appendix:clifford} that we can write for all $j \in [2n]$ 
$$
\Gamma_j = \Gamma_j^0 - \Gamma_j^1,
$$
where $\Gamma_j^0$ and $\Gamma_j^1$ are projectors onto the positive and negative eigenspace
of $\Gamma_j$ respectively. We also have that for all $i,j \in [2n]$ with $i \neq j$
$$
\Tr(\Gamma_i\Gamma_j) = \frac{1}{2}\Tr(\Gamma_i \Gamma_j + \Gamma_j \Gamma_i)=0.
$$
Hence the positive and negative eigenspaces of such operators
are similarly mutually unbiased as bases can be: from
$$
\Tr(\Gamma_i\Gamma_j^0) = \Tr(\Gamma_i\Gamma_j^1),
$$
we immediately see that if we would pick a vector lying in the positive or negative eigenspace of $\Gamma_j$
and perform a measurement with $\Gamma_i$, the probability to obtain outcome $\Gamma_i^0$ or outcome
$\Gamma_i^1$ must be the same.
Thus, one might intuitively hope to obtain good uncertainty relations
for measurements using such operators. We now show that this is indeed the case.

\subsection{Preliminaries}
Before we can turn to proving our uncertainty relations, we recall a few simple observations
from Appendix~\ref{appendix:clifford}.
The operators $\Gamma_1,\ldots,\Gamma_{2n}$ have a unique (up to unitary) representation in terms
of the matrices
\begin{align*}
  \Gamma_{2j-1} &= \sigma_y^{\otimes(j-1)} \otimes \sigma_x \otimes \id^{\otimes(n-j)}, \\
  \Gamma_{2j}   &= \sigma_y^{\otimes(j-1)} \otimes \sigma_z \otimes \id^{\otimes(n-j)},
\end{align*}
for $j=1,\ldots,n$.  We now fix this representation.
The product $\Gamma_0 := i\Gamma_1\Gamma_2 \cdots \Gamma_{2n}$ is also called the
pseudo-scalar.
A particularly useful fact is that the collection of operators
\[\begin{split}
  \id           & \\
  \Gamma_j     & \phantom{===} (1\leq j\leq 2n) \\
  \Gamma_{jk}  &= i\Gamma_j\Gamma_k \ (1\leq j < k \leq 2n) \\
  \Gamma_{jk\ell} &= \Gamma_j\Gamma_k\Gamma_\ell \ (1\leq j < k < \ell \leq 2n) \\
    \vdots     &\\
  \Gamma_{12\ldots (2n)} &= \Gamma_0
\end{split}
\]
forms an orthogonal basis for the $d \times d$ complex matrices for $d = 2^n$, where
in the definition of the above operators we introduce a factor of
$i$ to all with an even number of indices to make the whole set a
basis for the Hermitian operators with real valued coefficients.
Hence we can write every state $\rho \in \hil$ as
\begin{equation}
  \label{eq:op-basis}
  \rho = \frac{1}{d}\left(\id + \sum_j g_j\Gamma_j + \sum_{j<k} g_{jk}\Gamma_{jk}
                              + \ldots + g_0 \Gamma_0 \right).
\end{equation}
The real valued
coefficients $(g_1,\ldots,g_{2n})$ in this expansion are called ``vector'' components,
the ones belonging to higher degree products of $\Gamma$'s are
``tensor'' or ``k-vector'' components.

\begin{figure}[h]
\begin{center}
\includegraphics[scale=0.9]{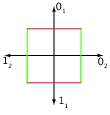}
\caption{2n = 2-cube}
\label{fig2Cube}
\end{center}
\end{figure}

\begin{figure}[h]
\begin{center}
\includegraphics[scale=0.9]{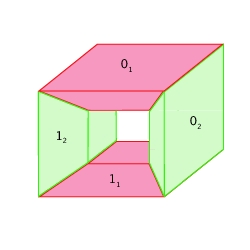}
\caption{2n = 4-cube}
\label{fig4Cube}
\end{center}
\end{figure}

Recall that we may think of the operators
$\Gamma_1,\ldots,\Gamma_{2n}$ as the basis vectors of a $2n$-dimensional real vector
space. Essentially, we can then think of the positive and negative eigenspace of such operators 
as the positive and negative direction of the basis vectors. We can visualize the $2n$ basis
vectors with the help of a $2n$-dimensional hypercube. Each basis vector determines
two opposing faces of the hypercube\footnote{Note that the face of an $2n$-dimensional hypercube is a $2n-1$ dimensional
hypercube itself.},
where we can think of the two faces as corresponding to the positive and negative eigenspace of each operator 
as illustrated in Figures~\ref{fig2Cube} and~\ref{fig4Cube}.

Finally, recall that within the Clifford algebra two vectors are orthogonal
if and only if they anti-commute. Hence, if we transform the generating set
of $\Gamma_j$ linearly,
\[
  \Gamma_k' = \sum_{j} T_{jk}\Gamma_j,
\]
the set $\{ \Gamma_1',\ldots,\Gamma_{2n}'\}$ satisfies the anti-commutation
relations if and only if $(T_{jk})_{jk}$ is an orthogonal matrix. In that case there
exists a matching unitary $U(T)$ of ${\cal H}$ which transforms the operator
basis as
$$
  \Gamma_j' = U(T) \Gamma_j U(T)^\dagger.
$$
We thus have an $\mbox{O}(2n)$ symmetry of the generating set $\Gamma_1,\ldots,\Gamma_{2n}$.
Indeed, this can be extended to a $\mbox{SO}(2n+1)$ symmetry by viewing $\Gamma_0$ as an additional
''vector'': It is not difficult to see that $\Gamma_0$ anti-commutes with $\Gamma_1,\ldots,\Gamma_{2n}$. We are thus
free to remove one of these operators from the generating set and replace it with $\Gamma_0$
to obtain a new set of generators. Evidently, we may also view these as basis vectors.
This observation forms the basis of the following little lemma, which allows
us to prove our uncertainty relations:
\begin{lemma}\label{reduceGrades}
The linear map $\mathbb{P}$ taking $\rho$ as
in Eq.~(\ref{eq:op-basis}) to
\begin{equation}
  \label{eq:projection}
  \mathbb{P}(\rho) := \frac{1}{d}\left( \1 + \sum_{j=0}^{2n} g_j\Gamma_j \right)
\end{equation}
is positive.
I.e., if $\rho$ is a state, then so is $\mathbb{P}(\rho)$,
and in this case $\sum_{j=0}^{2n} g_j^2 \leq 1$.
Conversely, if $\sum_{j=0}^{2n} g_j^2 \leq 1$, then
\[
  \sigma = \frac{1}{d}\left( \1 + \sum_{j=0}^{2n} g_j\Gamma_j \right)
\]
is positive semidefinite, hence a state.
\end{lemma}
\begin{proof}
First, we show that there exists a unitary $U$ such that
$\rho' = U\rho U^\dagger$ has no pseudo-scalar $\Gamma_0$, and only one nonzero vector component,
say at $\Gamma_1$.
Hence, our goal is to find the transformation $U$ that rotates $g = \sum_{j=0}^{2n} g_j \Gamma_j$
to the vector $b = \sqrt{\ell} \Gamma_1$, where we let $\ell := \sum_{j=0}^{2n} g_j^2 = (g_1')^2$. Finding such a transformation
for only the first $2n$ generators can easily be achieved, as we saw in Appendix~\ref{appendix:clifford}.
The challenge is thus to include $\Gamma_0$. To this
end we perform three individual operations: First, we rotate $g' = \sum_{j=1}^{2n} g_j \Gamma_j$ onto the vector
$b' = \sqrt{\ell'} \Gamma_1$ with $\ell' := \sum_{j=1}^{2n} g_j^2$. Second, we exchange $\Gamma_2$ and $\Gamma_0$.
And finally we rotate the vector $g'' = \sqrt{\ell'} \Gamma_1 + g_0 \Gamma_2$ onto the vector $b = \sqrt{\ell}\Gamma_1$.

First,  we rotate $g' = \sum_{j=1}^{2n} g_j \Gamma_j$ onto the vector
$b' = \sqrt{\ell'} \Gamma_1$: This is exactly analogous to the transformation constructed in Appendix~\ref{appendix:clifford}.
Consider the vector
$\hat{g} = \frac{1}{\sqrt{\ell'}} g'$ .
We have $\hat{g}^2 = |\hat{g}|^2 \1 = \1$ and
thus the vector is of length 1.
Let $m = \hat{g}+\Gamma_1$ denote the vector lying in the plane
spanned by $\Gamma_1$ and $\hat{g}$ located exactly halfway between $\Gamma_1$ and $\hat{g}$.
Let $\hat{m} = c (\hat{g}+\Gamma_1)$ with
$c = 1/\sqrt{2(1+g_1/\sqrt{\ell'})}$. It is easy to verify that $\hat{m}^2 = \1$ and hence the
vector $\hat{m}$ has length 1. To rotate the vector $g'$ onto the vector $b'$, we now need to first reflect
$g'$ around the plane perpendicular to $\hat{m}$, and then around the plane perpendicular to $\Gamma_1$.
Hence, we now define $R = \Gamma_1\hat{m}$.
Evidently, $R$ is unitary since $RR^{\dagger} = R^{\dagger}R = \1$.
First of all, note that
\begin{eqnarray*}
Rg' &=& \Gamma_1\hat{m}g'\\
& =& c \Gamma_1 \left(\frac{1}{\sqrt{\ell'}} g' + \Gamma_1\right)g' \\
&=& c \left(\Gamma_1\frac{g'^2}{\sqrt{\ell'}} + \Gamma_1^2 g'\right)\\
&=& c \sqrt{\ell'}\left(\Gamma_1 + \frac{1}{\sqrt{\ell'}}g'\right)\\
&=&\sqrt{\ell'}\hat{m}.
\end{eqnarray*}
Hence,
$$
Rg'R^\dagger = \sqrt{\ell'}\hat{m}\hat{m}\Gamma_1 = \sqrt{\ell'}\Gamma_1 = b',
$$
as desired.
Using the geometry of the Clifford algebra, one can see that $k$-vectors remain $k$-vectors when transformed with the
rotation $R$ (see Appendix~\ref{appendix:clifford}). Similarly, it is easy to see that $\Gamma_0$ is untouched by the operation $R$
\begin{eqnarray*}
R \Gamma_0 R^\dagger = \Gamma_0 R R^\dagger = \Gamma_0,
\end{eqnarray*}
since $\{\Gamma_0,\Gamma_j\}=0$ for all $j \in \{1,\ldots,2n\}$.
We can thus conclude that
\begin{eqnarray*}
R \rho R^\dagger = \frac{1}{d}\left(\1 + \sqrt{\ell'} \Gamma_1 + g_0 \Gamma_0 + \sum_{j<k} g'_{jk} \Gamma_{jk}+ \ldots\right),
\end{eqnarray*}
for some coefficients $g'_{jk}$ and similar for the terms involving higher products. 

Second, we exchange $\Gamma_2$ and $\Gamma_0$: To this end, recall that
$\Gamma_2,\ldots,\Gamma_{2n},\Gamma_0$ is also a generating set for the Clifford algebra. Hence, we can now
view $\Gamma_0$ itself as a vector with respect to the new generators.
To exchange $\Gamma_0$ and $\Gamma_2$, we now simply rotate $\Gamma_0$ onto $\Gamma_2$. Essentially,
this corresponds to a rotation about 90 degrees in the plane spanned by vectors $\Gamma_0$ and $\Gamma_2$.
Consider the vector $n = \Gamma_0 + \Gamma_2$ located exactly halfway between both vectors.
Let $\hat{n} = n/\sqrt{2}$ be the normalized vector. Let $R' = \Gamma_2 \hat{n}$. A small calculation analogous to
the above shows that
$$
R' \Gamma_0 R{'^\dagger} = \Gamma_2\mbox{ and } R'\Gamma_2 R^{'\dagger} = - \Gamma_0.
$$
We also have that $\Gamma_1$, $\Gamma_3,\ldots,\Gamma_{2n}$ are untouched by the operation: for $j\neq 0$
and $j \neq 2$, we have that
$$
R' \Gamma_j R^{'\dagger} = \Gamma_j,
$$
since $\{\Gamma_0,\Gamma_j\} = \{\Gamma_2,\Gamma_j\} = 0$. How does $R'$ affect the $k$-vectors in terms
of the original generators $\Gamma_1,\ldots,\Gamma_{2n}$? Using the anti-commutation relations and the definition
of $\Gamma_0$ it is easy to convince yourself that all $k$-vectors are mapped to $k'$-vectors with $k' \geq 2$ (except for $\Gamma_0$ itself). Hence,
the coefficient of $\Gamma_1$ remains untouched.
We can thus conclude that
\begin{eqnarray*}
R' R\rho R^\dagger R^{'\dagger} = \frac{1}{d}\left(\1 + \sqrt{\ell'} \Gamma_1 + g_0 \Gamma_2 + \sum_{j<k} g''_{jk} \Gamma_{jk} + \ldots\right),
\end{eqnarray*}
for some coefficients $g''_{jk}$ and so on.

Finally, we now rotate the vector $g'' = \sqrt{\ell'} \Gamma_1 + g_0 \Gamma_2$ onto the vector $b$.
Note that $(g'')^2 = (\ell + g_0^2) \1 = \ell \1$. Let $\hat{g}'' = g''/\sqrt{\ell}$ be the normalized
vector. Our rotation is derived exactly analogous to the first step: Let $k = \hat{g}'' + \Gamma_1$, and
let $\hat{k} = k/\sqrt{2(1+\sqrt{\ell'}/\sqrt{\ell})}$. Let $R'' = \Gamma_1\hat{k}$. A simple calculation analogous
to the above shows that
$$
R''g''R^{''\dagger} = \sqrt{\ell} \Gamma_1,
$$
as desired. Again, we have $R'' \Gamma_k R''^{\dagger} = \Gamma_k$ for $k \neq 1$ and $k \neq 2$. Furthermore,
$k$-vectors remain $k$-vectors under the actions of $R''$~\cite{doran:book}.
Summarizing, we obtain
\begin{eqnarray*}
R'' R' R\rho R^\dagger R^{'\dagger} R^{''\dagger}= \frac{1}{d}\left(\1 + \sqrt{\ell} \Gamma_1 + \sum_{j<k} g'''_{jk} \Gamma_{jk}+ \ldots\right),
\end{eqnarray*}
for some coefficients $g'''_{jk}$ and so on. Thus, we can take $U = R'' R' R$ to 
arrive at a new, simpler looking, state
\[\begin{split}
  \rho' &= U \rho U^\dagger \\
        &= \frac{1}{d}\left( \1 + g_1'\Gamma_1 + \sum_{j<k} g'''_{jk}\Gamma_{jk}
                                                   + \ldots + 0\,\Gamma_0 \right),
\end{split}\]
for some $g'''_{jk}$, etc.

Similarly, there exist of course orthogonal transformations $F_j$ that take
$\Gamma_k$ to $(-1)^{\delta_{jk}}\Gamma_k$. Such transformations
flip the sign of a chosen Clifford generator. In a similar way to the above,
it is easy to see that $F_j = \Gamma_0 \Gamma_j$ fulfills this task:
we rotate $\Gamma_j$ by 90 degrees in the plane given by $\Gamma_0$ and
$\Gamma_j$ as in the example we examined in Appendix~\ref{appendix:clifford}.
Now, consider
$$
  \rho'' = \frac{1}{2}\left(\rho' + F_j \rho' F_j^\dagger\right),
$$
for $j>1$.
Clearly, if $\rho'$ was a state, $\rho''$ is a state as well.
Note that we no longer have terms involving $\Gamma_j$ in 
the basis expansion:
Note that if we flip the sign of precisely those
terms that have an index $j$ (i.e., they have a factor $\Gamma_j$
in the definition of the operator basis), and then the
coefficients cancel with those of $\rho'$.

We now iterate this map through $j=2,3,\ldots, 2n$, and
we are left with a final state $\hat{\rho}$ of the form
\[
  \hat{\rho} = \frac{1}{d}\left( \1 + g_1'\Gamma_1 \right).
\]
By applying $U^\dagger = (R'' R' R)^\dagger$ from above, we now transform $\hat{\rho}$ to
$U^\dagger \hat{\rho} U = \mathbb{P}(\rho)$, which is the first
part of the lemma.

Looking at $\hat{\rho}$ once more,
we see that it can be positive semidefinite only if $g_1' \leq 1$,
i.e., $\sum_{j=0}^{2n} g_j^2 \leq 1$.
Evidently, $\Tr(\hat{\rho}) = 1$ and hence $\hat{\rho}$ is a state.

Conversely, if $\sum_{j=0}^{2n} g_j^2 \leq 1$, then the (Hermitian)
operator $A = \sum_j g_j \Gamma_j$ has the property
\[
  A^2 = \sum_{jk} g_j g_k \Gamma_j\Gamma_k
      = \sum_j g_j^2 \1 \leq \1,
\]
i.e. $-\1 \leq A \leq \1$, so $\sigma = \frac{1}{d}(\1+A) \geq 0$.
\end{proof}

\subsection{A meta-uncertainty relation}\index{uncertainty relation!meta}
We now first use the above tools to prove a ``meta''-uncertainty relation, from which
we will then derive two new entropic uncertainty relations.
Evidently, we have immediately from the above that
\begin{lemma}\label{metaUR}
Let $\rho \in \hil$ with $\dim\hil = 2^n$ be a quantum state,
and consider $K \leq 2n+1$ anti-commuting observables $\Gamma_j$.
Then,
$$
  \sum_{j=0}^{K-1} \bigl( \Tr(\rho\Gamma_j) \bigr)^2
     \leq \sum_{j=0}^{2n} \bigl( \Tr(\rho\Gamma_j) \bigr)^2
	= \sum_{j=0}^{2n} g_j^2
     \leq 1.
$$
\end{lemma}
Our result is essentially a generalization of the Bloch sphere picture to higher dimensions:
For $n=1$ ($d=2$) the state is parametrized by
$\rho = \frac{1}{2}(\1 + g_1 \Gamma_1 + g_2 \Gamma_2 + g_0 \Gamma_0)$
where $\Gamma_1 = X$, $\Gamma_2 = Z$ and $\Gamma_0 = Y$ are the familiar Pauli matrices.
Lemma~\ref{metaUR} tells
us that $g_0^2 + g_1^2 + g_2^2 \leq 1$, i.e., the state must lie inside the Bloch sphere (see Figure~\ref{blochSphere}).
Our result may be of independent interest, since it is often hard to find conditions on the coefficients
$g_1,g_2,\ldots$ such that $\rho$ is a state.

Notice that the $g_j = \Tr(\rho\Gamma_j)$ are directly interpreted
as the expectations of the observables $\Gamma_j$. Indeed, $g_j$ is
precisely the bias of the $\pm1$-variable $\Gamma_j$:
\[
  \Pr[ \Gamma_j = 1 | \rho ] = \frac{1+g_j}{2}.
\]
Hence, we can interpret Lemma~\ref{metaUR} as a form of uncertainty relation
between the observables $\Gamma_j$: if one or more of the observables have a large
bias (i.e., they are more precisely defined), this limits the bias of
the other observables (i.e., they are closer to uniformly distributed).

\subsection{Entropic uncertainty relations}

It turns out that Lemma~\ref{metaUR} has strong consequences for the R\'{e}nyi and von Neumann
entropic averages
$$
  \frac{1}{K} \sum_{j=0}^{K-1} H_\alpha\left(\Gamma_j|\rho\right),
$$
where $H_\alpha(\Gamma_j|\rho)$ is the R{\'e}nyi entropy at $\alpha$
of the probability distribution arising from measuring the state $\rho$ with
observable $\Gamma_j$.
The minima over all states $\rho$ of such expressions
can be interpreted as giving entropic uncertainty relations, as we shall now do
for $\alpha=2$ (the collision entropy) and $\alpha=1$ (the Shannon entropy).

\begin{theorem}
Let $\dim\hil = 2^n$,
and consider $K \leq 2n+1$ anti-commuting observables as defined in Eq.~(\ref{antiComm}).
Then,
$$ 
\min_{\rho} \frac{1}{K} \sum_{j=0}^{K-1} H_2\left(\Gamma_j|\rho\right)
      = 1 - \log\left( 1+\frac{1}{K} \right)
      \sim 1 - \frac{\log e}{K},
$$
where $H_2(\Gamma_j|\rho) = - \log \sum_{b \in \{0,1\}} \Tr(\Gamma_j^b \rho)^2$,
and the minimization is taken over all states $\rho$.
The latter holds asymptotically for large $K$.
\end{theorem}
\begin{proof}
Using the fact that $\Gamma_j^b = (\1 + (-1)^b \Gamma_j)/2$ we
can first rewrite
\begin{equation*}\begin{split}
  \frac{1}{K} \sum_{j=0}^{K-1} H_2\left(\Gamma_j|\rho\right)
      &=    - \frac{1}{K} \sum_{j=0}^{K-1} \log \left[\frac{1}{2}\left(1 + \Tr(\rho\Gamma_j)^2\right)\right]  \\
      &\geq - \log \left( \frac{1}{2K} \sum_{j=0}^{K-1} \left(1 + g_j^2\right) \right) \\
      &\geq 1 - \log\left(1 + \frac{1}{K}\right),
\end{split}\end{equation*}
where the first inequality follows from Jensen's inequality and the concavity of the log,
and the second from Lemma~\ref{metaUR}.
Clearly, the minimum is attained if
all $g_j = \Tr(\rho\Gamma_j) = \sqrt{\frac{1}{K}}$.
It follows from Lemma~\ref{reduceGrades} that our inequality is tight.
Via the Taylor expansion of $\log\left(1 + \frac{1}{K}\right)$ we obtain
the asymptotic result for large $K$.
\end{proof}

For the Shannon entropy ($\alpha=1$) we obtain something even nicer:
\begin{theorem}
Let $\dim\hil = 2^n$,
and consider $K \leq 2n+1$ anti-commuting observables as defined in Eq.~(\ref{antiComm}).
Then,
$$
\min_{\rho} \frac{1}{K} \sum_{j=0}^{K-1} H(\Gamma_j|\rho) = 1 - \frac{1}{K},
$$
where $H(\Gamma_j|\rho) = - \sum_{b\in \{0,1\}} \Tr(\Gamma_j^b \rho) \log \Tr(\Gamma_j^b \rho)$,
and the minimization is taken over all states $\rho$.
\end{theorem}
\begin{proof}
To see this, note that by rewriting our objective as above,
we observe that we need to minimize the expression
\[
  \frac{1}{K} \sum_{j=0}^{K-1} H\left( \frac{1 \pm \sqrt{t_j}}{2} \right),
\]
subject to $\sum_j t_j \leq 1$ and $t_j \geq 0$, via the identification
$t_j = (\Tr(\rho\Gamma_j))^2$. An elementary calculation
shows that
the function $f(t) = H\left( \frac{1 \pm \sqrt{t}}{2} \right)$ is
concave in $t\in[0,1]$: 
\[
  f'(t) = \frac{1}{4\ln 2}\frac{1}{\sqrt{t}} \bigl( \ln(1-\sqrt{t})-\ln(1+\sqrt{t}) \bigr),
\]
and so
\[
  f''(t) = \frac{1}{8\ln 2}\frac{1}{t^{3/2}} \left( \ln\frac{1+\sqrt{t}}{1-\sqrt{t}}
                                                            - \frac{2\sqrt{t}}{1-t} \right).
\]
Since we are only interested in the sign of the second derivative, we ignore the
(positive) factors in front of the bracket, and are done if we can show that
\[\begin{split}
  g(t) &:= \ln \frac{1+\sqrt{t}}{1-\sqrt{t}} - \frac{2\sqrt{t}}{1-t} \\
       &=  \ln(1+\sqrt{t}) + \frac{1}{1+\sqrt{t}}
           - \ln(1-\sqrt{t}) - \frac{1}{1-\sqrt{t}}
\end{split}\]
is non-positive for $0 \leq t \leq 1$. Substituting $s = 1-\sqrt{t}$, which
is also between $0$ and $1$, we rewrite this as
\[
  h(s) = -\ln s - \frac{1}{s} + \ln(2-s) + \frac{1}{2-s},
\] 
which has derivative
\[
  h'(s) = (1-s)\left( \frac{1}{s^2} - \frac{1}{(2-s)^2}\right),
\] 
and this is clearly positive for $0<s<1$. In other words, $h$ increases
from its value at $s=0$ (where it is $h(0)=-\infty$) to its value at $s=1$
(where it is $h(1)=0$), so indeed $h(s) \leq 0$ for all $0\leq s\leq 1$.
Consequently, also $f''(t) \leq 0$ for $0\leq t \leq 1$.

Hence, by Jensen's inequality, the minimum
is attained with one of the $t_j$
being $1$ and the others $0$,
giving just the lower bound of $1-\frac{1}{K}$.
\end{proof}

We have shown that anti-commuting Clifford observables obey the
strongest possible uncertainty relation for the von Neumann entropy.
It is interesting that in the process of the proof, however,
we have found three uncertainty type inequalities
(the sum of squares bound, the bound on $H_2$, and finally the bound on
$H_1$), and all three have a different structure of attaining the
limit. The sum of squares bound can be achieved in every direction
(meaning for every tuple satisfying the bound we get one attaining
it by multiplying all components by some appropriate factor),
the $H_2$ expression requires all components to be equal,
while the $H_1$ expression demands exactly the opposite.

\section{Conclusion}
We showed that merely choosing our measurements to be mutually unbiased does not lead to strong uncertainty relations.
However, we were able to identify another property which does lead to optimal entropic uncertainty relations
for two outcome measurements! 
\emph{Anti-commuting} Clifford observables obey the
strongest possible uncertainty relation for the von Neumann entropy: if we have no uncertainty for one
of the measurements, we have maximum uncertainty for all others.
We also obtain a slightly suboptimal uncertainty relation for the collision entropy
which is strong enough for all cryptographic purposes. Indeed, one could use our entropic uncertainty relation
in the bounded quantum storage setting to construct, for example, 1-K oblivious transfer protocols
analogous to~\cite{serge:new}. Here, instead of encoding a single bit into either the computational or Hadamard
basis, which gives us a 1-2 OT, we now encode a single bit into the positive or negative eigenspace of each
of these $K$ operators. It is clear from the representation of such operators discussed earlier, that such 
an encoding can be done experimentally as easily as encoding a single bit into three mutually unbiased basis
given by $\sigma_x$, $\sigma_y$, $\sigma_z$. Indeed, our construction can be seen as a direct extension of such
an encoding: we obtain the uncertainty relation for the three MUBs previously proved by 
Sanchez~\cite{sanchez:entropy,sanchez:entropy2} as a special case of our analysis for $K=3$.
It is perhaps interesting to note that the same operators also play a prominent role in the setting
of non-local games as discussed in Chapter~\ref{tsirelsonConstruction}.

Sadly, strong uncertainty relations for measurements with more than two outcomes remain inaccessible to us. 
It has been shown~\cite{serge:personal} that uncertainty relations for more outcomes can be obtained via
a coding argument from uncertainty relations as we construct them here. Yet, these are far from optimal.
A natural choice would be to consider the generators of a generalized Clifford algebra, yet such an
algebra does not have such nice symmetry properties which enabled us to implement operations on the
vector components above. It remains an exciting open question whether such operators form a good
generalization, or whether we must continue our search for new properties.

\chapter{Locking classical information}\label{lockingChapter}\index{locking}\label{chapter:locking}

Locking classical correlations in quantum states~\cite{terhal:locking} is an exciting feature
of quantum information,
intricately related to entropic uncertainty relations. 
In this chapter, we will investigate whether good locking effects can be obtained using mutually
unbiased bases.

\section{Introduction}

Consider a two-party protocol with one or more rounds of communication. Intuitively, 
one would expect that in each round the amount of correlation between
the two parties cannot increase by much more than the amount of data transmitted.
For example, transmitting  $2\ell$ classical bits or $\ell$ qubits (and using superdense coding) 
should not increase the amount of correlation by more than $2\ell$ bits,
no matter what the initial state of the two-party system was. This intuition
is accurate when we take the classical mutual information $\mI_c$ as
our correlation measure, and require all communication to be classical.
However, when quantum communication was possible at some point
during the protocol, everything changes: there exist two-party mixed quantum 
states, such that transmitting just a 
single extra bit of classical communication can result in an arbitrarily large 
increase in $\mI_c$~\cite{terhal:locking}. 
The magnitude of this increase thereby
only depends on the dimension of the initial mixed state. Since then similar locking
effects have been observed, also for other correlation
measures~\cite{m:locking,karol:locking}. Such effects play a role in very different
scenarios: they have been used to explain physical phenomena related to black holes~\cite{jsmo:locking}, but
they are also important in crypto\-gra\-phic applications such as quantum key distribution~\cite{rk:locking}
and quantum bit string commitment that we will encounter in Chapter~\ref{chapter:qsc}. We are thus interested in
determining how exactly we can obtain locking effects, and how dramatic they can be.

\subsection{A locking protocol}\index{locking!protocol}
The correlation measure considered here, is the classical mutual
information of a bipartite quantum state $\rho_{AB}$, which is the maximum classical 
mutual information that can be obtained by local measurements 
$M_A \otimes M_B$
on the state $\rho_{AB}$ (see Chapter~\ref{informationChapter}):
\begin{equation}\label{mutualInfo}
\mI_c(\rho_{AB}) = \max_{M_A \otimes M_B} \mI(A,B).
\end{equation}
Recall from Chapter~\ref{chapter:informationIntro} that
the mutual information is defined as 
$\mI(A,B) = H(P_A) + H(P_B) - H(P_{AB})$
where $H$ is the Shannon entropy. $P_A$, $P_B$, and $P_{AB}$ are
the probability distributions corresponding to the individual and joint outcomes
of measuring the state $\rho_{AB}$ with $M_A \otimes M_B$. 
The mutual information between $A$ and $B$ is a measure of the information
that $B$ contains about $A$.
This measure of correlation is of particular relevance for quantum bit string commitments in Chapter~\ref{chapter:qsc}.
Furthermore, the first locking effect was observed for this quantity 
in the following protocol 
between two parties: Alice (A) and Bob (B). 
Let $\mathbb{B} = \{\mB_1,\ldots,\mB_m\}$ with $\mB_t = \{\ket{b^t_1},\ldots,\ket{b^t_d}\}$
be a set of $m$ MUBs in $\Complex^d$.
Alice picks an
element $k \in \{1,\ldots,d\}$ and a basis $\mB_t \in \mathbb{B}$ uniformly at random. She then
sends $\ket{b^t_k}$ to Bob, while keeping $t$ secret. 
Such a protocol gives 
rise to the joint state
$$
\rho_{AB} = \frac{1}{m d} 
\sum_{k=1}^{d} \sum_{t=1}^{m} (\outp{k}{k} \otimes \outp{t}{t})_A
\otimes (\outp{b_k^t}{b^t_k})_B.
$$

\begin{figure}[h]
\begin{center} 
\includegraphics[scale=1.3]{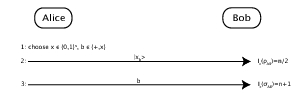}
\caption{A locking protocol for 2 bases.}
\label{lockingFig}
\end{center}
\end{figure}

Clearly, if Alice told her basis choice $t$ to Bob, 
he could measure in the right basis and obtain the correct $k$. 
Alice and Bob would then share $\log d + \log m$ bits of correlation,
which is also their mutual information $\mI_c(\sigma_{AB})$,
where $\sigma_{AB}$ is the state obtained from $\rho_{AB}$ after the announcement of $t$. 
But, how large is $\mI_c(\rho_{AB})$,  when Alice does \emph{not} announce $t$ to Bob? 
It was shown~\cite{terhal:locking} that in dimension $d=2^n$, using the two MUBs given by 
the unitaries $U_+= \id^{\otimes n}$ and $U_\times = H^{\otimes n}$ applied to the computational basis
we have $\mI_c(\rho_{AB}) = (1/2) \log d$ (see Figure~\ref{lockingFig}, where $\ket{x_b} = U_b \ket{x}$). This means 
that the single bit of basis information Alice transmits to Bob
``unlocks'' $(1/2) \log d$ bits: \emph{without} this bit, the mutual information
is $(1/2) \log d$, but \emph{with} this bit it is $\log d + 1$.
To get a good locking protocol,
we want to use only a small number of bases, i.e., $m$ should be as small as possible, while 
at the same time forcing $\mI_c(\rho_{AB})$ to be as low as possible.
That is, we want $\log m/(\log d - \mI_c(\rho_{AB}))$ to be small. 

It is also known that if Alice
and Bob randomly choose a large set of unitaries from the Haar
measure to construct $\mathbb{B}$, then $\mI_c(\rho_{AB})$
can be brought down to a small constant~\cite{winter:randomizing}.
However, no explicit constructions with more than two bases are known 
that give good locking effects. Based on numerical studies for
spaces of prime dimension $3 \leq d \leq 30$, one might hope that adding
a third MUB would strengthen the locking effect 
and give $\mI_c(\rho_{AB}) \approx (1/3) \log d$~\cite{terhal:locking}.

Here, however, we show that this intuition fails us. We 
prove that for three MUBs given by
$\id^{\otimes n}$, $H^{\otimes n}$, and $K^{\otimes n}$ where $K = (\id + i\sigma_x)/\sqrt{2}$ and 
dimension $d=2^n$ for some even integer $n$, we have 
\begin{equation}\label{minf}
\mI_c(\rho_{AB}) = \frac{1}{2} \log d,
\end{equation}
the same locking effect as with two MUBs. 
We also show that for any subset of the MUBs based on Latin squares and the MUBs in square
dimensions based on generalized Pauli matrices~\cite{boykin:mub}, we again obtain Eq.~(\ref{minf}), i.e., using
two or all $\sqrt{d}$ of them makes no difference at all!
Finally, we show that for any set of MUBs $\mathbb{B}$ based on generalized Pauli matrices in \emph{any} dimension,
$\mI_c(\rho_{AB}) = \log d - \min_{\ket{\phi}}(1/|\mathbb{B}|)\sum_{\mB \in \mathbb{B}}H(\mB|\ket{\phi})$, i.e., it 
is enough to determine a bound on the entropic uncertainty relation to determine the 
strength of the locking effect.
Although bounds for general MUBs still elude us, our results show
that merely choosing the bases to be mutually unbiased is not sufficient and we must look
elsewhere to find bases which provide good locking. 

\subsection{Locking and uncertainty relations}\index{locking!uncertainty relations}

We first explain the connection between locking
and entropic uncertainty relations. In particular, we will see that for MUBs based on generalized
Pauli matrices, we only need to look at such uncertainty relations to determine the exact strength 
of the locking effect. 

In order to determine how large the locking effect is for some set
of mutually unbiased bases $\mathbb{B}$,
and the shared state
\begin{equation}\label{rhoAB}
\rho_{AB} =
\sum_{t=1}^{|\mathbb{B}|}
\sum_{k=1}^{d} p_{t,k}
(\outp{k}{k} \otimes \outp{t}{t})_A
\otimes (\outp{b^t_k}{b^t_k})_B,
\end{equation}
we must find the value of $\mI_c(\rho_{AB})$ or at least a good upper bound.
That is, we must find a POVM
$M_A \otimes M_B$ that maximizes Eq.\ (\ref{mutualInfo}).
Here, $\{p_{t,k}\}$ is a probability distribution over $\mathbb{B} \times [d]$.
It has been shown in~\cite{terhal:locking} that we can restrict ourselves to taking
$M_A$ to be the local measurement determined by the projectors $\{\outp{k}{k} \otimes \outp{t}{t}\}$.
It is also known that we can limit ourselves to take the measurement $M_B$ consisting of rank one
elements $\{\alpha_i \outp{\Phi_i}{\Phi_i}\}$ only~\cite{davies:access}, where $\alpha_i \geq 0$ and $\ket{\Phi_i}$
is normalized.
Maximizing over $M_B$ then corresponds to maximizing Bob's accessible information as defined in Chapter~\ref{informationChapter} for
the ensemble $\ens = \{p_{k,t},\outp{b^t_k}{b^t_k}\}$
\begin{equation}
\begin{aligned}\label{accessible}
&\mI_{acc}(\ens)=\\
&\max_{M_B} \left(- \sum_{k,t} p_{k,t} \log p_{k,t} + \right.
\left.\sum_{i} \sum_{k,t} p_{k,t} \alpha_i \bra{\Phi_i}\rho_{k,t}\ket{\Phi_i}
\log \frac{p_{k,t} \bra{\Phi_i}\rho_{k,t}\ket{\Phi_i}}{\bra{\Phi_i}\mu\ket{\Phi_i}} \right),
\end{aligned}
\end{equation}
where $\mu = \sum_{k,t} p_{k,t} \rho_{k,t}$ and $\rho_{k,t} = \outp{b^t_k}{b^t_k}$. 
Therefore, we have $\mI_c(\rho_{AB}) = \mI_{acc}(\ens)$.
As we saw in Chapter~\ref{chapter:informationIntro}, maximizing the accessible information
is often a very hard task. Nevertheless, for our choice of MUBs, the problem will turn out to be quite easy 
in the end.

\section{Locking using mutually unbiased bases}\index{locking!mutually unbiased bases}
\subsection{An example}

We now determine how well we can lock information using specific sets of mutually unbiased bases.
We first consider a very simple example with only three MUBs that provides the intuition behind
the remainder of our proof. The three MUBs we consider now are generated by the unitaries
$\id$, $H$ and $K = (\id + i\sigma_x)/\sqrt{2}$ when applied to the computational basis.
For this small example, we also investigate the role of the prior over the bases and the encoded
basis elements. It turns out that this does not affect the strength of the locking effect positively, i.e.,
we do not obtain a stronger locking affect using a non-uniform prior. 
Actually, it is possible
to show the same for encodings in many other bases. However, we 
do not consider this case in full generality as to not obscure our main line of argument.

\begin{lemma}\label{3mubLocking}
Let $U_1=\id^{\otimes n}$,$U_2 = H^{\otimes n}$, and $U_3 = K^{\otimes n}$, 
and take $k \in \{0,1\}^n$ where $n$ is an even integer.
Let $\{p_t\}$ with $t \in [3]$ be a probability distribution over
the set $\mS = \{U_1,U_2,U_3\}$. Suppose that $p_1,p_2,p_3 \leq 1/2$ and let $\{p_{t,k}\}$ with
$p_{t,k} = p_t/d$ be the joint distribution over $\mS \times \{0,1\}^n$.
Consider the ensemble $\ens = \{ p_t \frac{1}{d},U_t \outp{k}{k}U_t^\dagger\}$, then
$$
\mI_{acc}(\ens) = \frac{n}{2}.
$$ 
If, on the other hand, there exists a $t \in [3]$ such that
$p_t > 1/2$, then $\mI_{acc}(\ens) > n/2$.
\end{lemma}
\begin{proof}
We first give an explicit measurement strategy and then prove a matching upper bound
on $\mI_{acc}$.
Consider the Bell basis vectors $\ket{\Gamma_{00}} = (\ket{00} + \ket{11})/\sqrt{2}$, 
$\ket{\Gamma_{01}} = (\ket{00} - \ket{11})/\sqrt{2}$, $\ket{\Gamma_{10}} = (\ket{01} + \ket{10})/\sqrt{2}$,
and $\ket{\Gamma_{11}} = (\ket{01} - \ket{10})/\sqrt{2}$. Note that we can write
for the computational basis
\begin{eqnarray*}
\ket{00} &=& \frac{1}{\sqrt{2}}(\ket{\Gamma_{00}} + \ket{\Gamma_{01}}),\\
\ket{01} &=& \frac{1}{\sqrt{2}}(\ket{\Gamma_{10}} + \ket{\Gamma_{11}}),\\
\ket{10} &=& \frac{1}{\sqrt{2}}(\ket{\Gamma_{10}} - \ket{\Gamma_{11}}),\\
\ket{11} &=& \frac{1}{\sqrt{2}}(\ket{\Gamma_{00}} - \ket{\Gamma_{01}}). 
\end{eqnarray*}
The crucial fact to note is that if we fix some $k_1,k_2$, then 
there exist exactly two Bell basis vectors $\ket{\Gamma_{i_1i_2}}$ such that 
$|\inp{\Gamma_{i_1i_2}}{k_1,k_2}|^2 = 1/2$. For the remaining two basis vectors
the inner product with $\ket{k_1,k_2}$ will be zero. 
A simple calculation shows that we can express the two-qubit basis states of
the other two mutually unbiased bases analogously: for each two qubit basis state
there are exactly two Bell basis vectors such that the 
inner product is zero and for the other two the inner product squared is $1/2$.

We now take the measurement given by $\{\outp{\Gamma_i}{\Gamma_i}\}$ with 
$\ket{\Gamma_i} = \ket{\Gamma_{i_1i_2}} \otimes \ldots \otimes \ket{\Gamma_{i_{n-1}i_{n}}}$ for the binary
expansion of $i = i_1i_2\ldots i_n$. Fix a $k = k_1k_2\ldots k_n$. By the above argument,
there exist exactly $2^{n/2}$ strings $i \in \01^n$ such that 
$|\inp{\Gamma_i}{k}|^2 = 1/2^{n/2}$. Putting everything together, Eq.\ (\ref{accessible})
now gives us for any prior distribution $\{p_{t,k}\}$ that
\begin{equation}\label{Ibell}
-\sum_i \bra{\Gamma_i}\mu\ket{\Gamma_i} \log \bra{\Gamma_i}\mu\ket{\Gamma_i} - \frac{n}{2} \leq \mI_{acc}(\ens).
\end{equation}
For our particular distribution we have $\mu = \id/d$ and thus 
$$
\frac{n}{2} \leq \mI_{acc}(\ens).
$$

We now prove a matching upper bound that shows that our measurement is optimal.
For our distribution, we can rewrite Eq.\ (\ref{accessible}) for the POVM 
given by $\{\alpha_i \outp{\Phi_i}{\Phi_i}\}$ to\label{rewriteIACC}

\begin{eqnarray*}
\mI_{acc}(\ens) &=& \max_M \left(\log d + 
\sum_i \frac{\alpha_i}{d} \sum_{k,t} p_{t} |\bra{\Phi_i}U_t\ket{k}|^2 \log |\bra{\Phi_i}U_t\ket{k}|^2 \right)\\
&=& \max_M \left(\log d -  \sum_i \frac{\alpha_i}{d} \sum_{t} p_t H(\mB_t|\ket{\Phi_i}) \right),
\end{eqnarray*}
for the bases $\mB_t = \{U_t\ket{k}\mid k \in \01^n\}$.

It follows from Corollary~\ref{MUderived} that $\forall i\in \{0,1\}^n$ and
$p_1,p_2,p_3\leq 1/2$
\begin{eqnarray*}
(1/2-p_1) [H(\mB_2|\ket{\Phi_i}) + H(\mB_3|\ket{\Phi_i}) ]&+&\\
(1/2-p_2) [H(\mB_1|\ket{\Phi_i}) + H(\mB_3|\ket{\Phi_i})]&+&\\
(1/2-p_3)[H(\mB_1|\ket{\Phi_i})+H(\mB_2|\ket{\Phi_i})]&& \geq n/2,
\end{eqnarray*}
where we used the fact that $p_1+p_2+p_3 = 1$.
Reordering the terms we now get
$\sum_{t=1}^3 p_{t} H(\mB_t|\ket{\Phi_i})\geq n/2.$
Putting things together and using the fact that $\sum_i \alpha_i=d$, we obtain
$$
\mI_{acc}(\ens) \leq \frac{n}{2},
$$
from which the result follows.

If, on the other hand, there exists a $t \in [3]$ such that $p_t > 1/2$, then 
by measuring in the basis $\mB_t$ we obtain $\mI_{acc}(\ens) \geq p_t n > n/2$, since the entropy will be 0 for basis $\mB_t$ and we have 
$\sum_t p_t = 1$.
\end{proof}

Above, we have only considered a non-uniform prior over the set
of \emph{bases}. 
In Chapter~\ref{pistarChapter}, we observed that when we want to 
guess the XOR of a string of length $2$ encoded in one (unknown to us) 
of these three bases,
the uniform prior on the strings is not the one that gives the smallest
probability of success. This might lead one to think that a similar
phenomenon could be observed in the present setting, i.e., that one
might obtain better locking with three basis for a non-uniform prior on
the strings. In what follows, however, we show that this is not the case.

Let $p_t=\sum_{k} p_{k,t}$ be the marginal distribution on the basis,
then the difference in Bob's knowledge between receiving only the 
quantum state and receiving the quantum state \emph{and} the basis information,
where we will ignore the basis information itself, is given by 
\begin{eqnarray*}
 \Delta(p_{k,t})=H(p_{k,t})-\mI_{acc}(\ens) -H(p_t),
\end{eqnarray*}
Consider the post-measurement state
$\nu=\sum_i \bra{\Gamma_i}\mu\ket{\Gamma_i}\ket{\Gamma_i}\bra{\Gamma_i}$. Using Eq.~(\ref{Ibell}) we obtain
\begin{eqnarray} \label{gap1}
 \Delta(p_{k,t})\leq H(p_{k,t})-S(\nu)+n/2 -H(p_t),
\end{eqnarray}
where $S$ is the von Neumann entropy. Consider the state
\begin{eqnarray*}
\rho_{12} =
\sum_{k=1}^{d} \sum_{t=1}^{3} p_{k,t}(\outp{t}{t})_1
\otimes (U_t \outp{k}{k} U_t^{\dagger})_2,
\end{eqnarray*}
for which we have that
\begin{eqnarray*}
S(\rho_{12})=H(p_{k,t}) &\leq S(\rho_1) +S(\rho_2)\\
                        & = H(p_t) +S(\mu)\\
                        &\leq H(p_t)+S(\nu).
\end{eqnarray*}
Using Eq.~(\ref{gap1}) and the previous equation we get
\begin{eqnarray*}
 \Delta(p_{k,t})\leq n/2,
\end{eqnarray*}
for any prior distribution. This bound is saturated by the uniform prior
and therefore we conclude that the uniform prior results in the largest
gap possible. 

\subsection{MUBs from generalized Pauli matrices}

We now consider MUBs based on the generalized Pauli matrices $X_d$ and $Z_d$ as described
in Chapter~\ref{mub:pauliMubs}. We consider a uniform prior over the elements of each basis and the set of bases.
Choosing a non-uniform prior does not lead to a better locking effect.

\begin{lemma}\label{equiv}
Let $\mathbb{B}=\{\mB_1,\ldots,\mB_{m}\}$ be any set of MUBs constructed on the basis of generalized Pauli matrices
in a Hilbert space of prime power dimension $d = p^N$.
Consider the ensemble $\ens = \{ \frac{1}{d m},\outp{b^t_k}{b^t_k}\}$. Then
$$
\mI_{acc}(\ens) = \log d - 
\frac{1}{m} \min_{\ket{\psi}} \sum_{\mB_t \in \mathbb{B}} H(\mB_t|\ket{\psi}).
$$
\end{lemma}
\begin{proof}
We can rewrite Eq.\ (\ref{accessible}) for a POVM $M_B$ 
of the form $\{\alpha_i \outp{\Phi_i}{\Phi_i}\}$ as
\begin{eqnarray*}
\mI_{acc}(\ens) &=& \max_{M_B} \left(\log d + 
\sum_i \frac{\alpha_i}{d m} \sum_{k,t} |\inp{\Phi_i}{b^t_k}|^2 \log |\inp{\Phi_i}{b^t_k}|^2 \right)\\
&=& \max_{M_B} \left(\log d -  \sum_i \frac{\alpha_i}{d} \sum_{t} p_{t} H(\mB_t|\ket{\Phi_i}) \right).
\end{eqnarray*}
For convenience, we split up the index $i$ into $i = a,b$ with $a = a_1,\ldots,a_N$ and $b=b_1,\ldots,b_N$, 
where $a_\ell,b_\ell \in \{0,\ldots,p-1\}$ in the following.

We first show 
that applying generalized Pauli matrices to the basis vectors of a MUB merely permutes those vectors.
\begin{claim}\label{lemma:pauliCovariance}
Let $\mB_t = \{\ket{b^t_1},\ldots,\ket{b^t_d}\}$ be a basis based on generalized Pauli matrices 
(Chapter~\ref{mub:pauliMubs}) with
$d = p^N$. Then $\forall a,b \in \{0,\ldots,p-1\}^N, \forall k \in [d]$ we have that $\exists k' \in [d],$ such 
that $\ket{b^{t}_{k'}} = X_d^{a_1}Z_d^{b_1} \otimes \ldots 
\otimes X_d^{a_N}Z_d^{b_N}\ket{b^t_k}$.
\end{claim}
\begin{proof}
Let $\Tau_p^i$ for $i \in \{0,1,2,3\}$ denote the generalized Pauli's 
$\Tau_p^0 = \id_p$, 
$\Tau_p^1 = X_p$, 
$\Tau_p^3 = Z_p$, and
$\Tau_p^2 = X_p Z_p$. Note that $X_p^uZ_p^v = \omega^{uv} Z_p^v X_p^u$,
where $\omega = e^{2\pi i/p}$.
Furthermore, define
$
\Tau_p^{i,(x)} = \id^{\otimes (x - 1)} \otimes \Tau_p^{i} \otimes \id^{N-x}
$
to be the Pauli operator $\Tau_p^i$ applied to the $x$-th qupit.
Recall from Section~\ref{mub:pauliMubs} that there exist sets of
Pauli operators $C_t$ such that the basis $\mB_t$ is the unique simultaneous eigenbasis 
of the set of operators in $C_t$, i.e., for all $k \in [d]$ and $f,g \in [N]$,
$\ket{b^t_k} \in \mB_t$ and $c_{f,g}^t \in C_t$, we have
$c_{f,g}^t \ket{b^t_k}=\lambda_{k,f,g}^t \ket{b^t_k} \textrm{ for some value }\lambda^t_{k,f,g}$.
Note that any vector $\ket{v}$ that satisfies this equation
is proportional to a vector in $\mB_t$. To prove
that any application of one of the generalized Paulis merely permutes the vectors in $\mB_t$
is therefore equivalent to proving that $\Tau^{i,(x)}_{p}
\ket{b^t_k}$ are eigenvectors of $c_{f,g}^t$ for any $f,g \in [k]$ and $i \in \{1,
3\}$. This can be seen as follows: Note that $c_{f,g}^t=\bigotimes_{n=1}^N
\left(\Tau^{1, (n)}_{p}\right)^{f_N} \left(\Tau^{3,(n)}_{p}\right)^{g_N}$ 
for $f = (f_1,\ldots,f_N)$ and $g=(g_1, \ldots, g_N)$ 
with $f_N,g_N \in \{0,\ldots,p-1\}$~\cite{boykin:mub}. A calculation then shows that
$$
c_{f,g}^t \Tau^{i,(x)}_p \ket{b^t_k}= \tau_{f_x,g_x, i} \lambda_{k,f,g}^t \Tau^{i,(x)}_{p} \ket{b^t_k},
$$
where  $\tau_{f_x,g_x, i}=\omega^{g_x}$ for $i = 1$ and 
$\tau_{f_x,g_x,i}=\omega^{-f_x}$ for $i = 3$. Thus
$\Tau^{i,(x)}_{p} \ket{b^t_k}$ is an eigenvector of $c^t_{f,g}$ for
all $t, f, g$ and $i$, which proves our claim.
\end{proof}

Suppose we are given $\ket{\psi}$ that minimizes 
$\sum_{\mB_t \in \mathbb{T}} H(\mB_t|\ket{\psi})$. 
We can then construct a full POVM with $d^2$ elements by taking
$\{\frac{1}{d}\outp{\Phi_{ab}}{\Phi_{ab}}\}$ with $\ket{\Phi_{ab}} = (X_d^{a_1}Z_d^{b_1} \otimes \ldots 
\otimes X_d^{a_N}Z_d^{b_N})^\dagger\ket{\psi}$. However, it follows from our claim
above that $\forall a,b,k, \exists k'$ such that $|\inp{\Phi_{ab}}{b^t_k}|^2 = |\inp{\psi}{b^{t}_{k'}}|^2$, 
and thus
$H(\mB_t|\ket{\psi}) = H(\mB_t|\ket{\Phi_{ab}})$ from which the result follows.
\end{proof}

Determining the strength of the locking effects for such MUBs is thus equivalent to proving bounds on 
entropic uncertainty relations. We thus obtain as a corollary of 
Theorem~\ref{squareThm} and Lemma~\ref{equiv}, that, for dimensions which are the square of a prime power 
(i.e. $d = p^{2N}$), using any product MUBs based on generalized Paulis does not give us any better
locking than just using 2 MUBs. 
\begin{corollary}\label{pauliLocking}
Let $\mathbb{S}=\{\mS_1,\ldots,\mS_{m}\}$ with $m \geq 2$ be any set of MUBs constructed on the basis of generalized Pauli matrices
in a Hilbert space of prime (power) dimension $s = p^N$. 
Define $U_t$ as the unitary that transforms the computational basis
into the $t$-th MUB, i.e., $\mS_t = \{U_t\ket{1},\ldots,U_t\ket{s}\}$. 
Let $\mathbb{B} = \{\mB_1,\ldots,\mB_{m}\}$ be the set of product MUBs with
$\mB_t = \{U_t \otimes U_t^* \ket{1},\ldots,U_t \otimes U_t^*\ket{d}\}$ in dimension $d=s^2$.
Consider the ensemble $\ens = \{ \frac{1}{d m},\outp{b^t_k}{b^t_k}\}$. Then
$$
\mI_{acc}(\ens) = \frac{\log d}{2}.
$$
\end{corollary}
\begin{proof}
The claim follows from Theorem~\ref{squareThm} and the proof of Lemma~\ref{equiv}, by constructing
a similar measurement formed from vectors $\ket{\hat{\Phi}_{\hat{a}\hat{b}}} = K_{a^1b^1} 
\otimes K_{a^2b^2}^* \ket{\psi}$
with $\hat{a} = a^1a^2$ and $\hat{b} = b^1b^2$, where $a^1,a^2$ and $b^1,b^2$ are
defined like $a$ and $b$ in the proof of Lemma~\ref{equiv}, and $K_{ab} = (X_d^{a_1}Z_d^{b_1}\otimes\ldots\otimes X_d^{a_N}Z^{b_N}_d)^\dagger$
from above.
\end{proof}

The simple example we considered above is in fact a special case of Corollary~\ref{pauliLocking}. 
It shows that if the vector that minimizes the sum of entropies has certain symmetries, 
the resulting POVM can even be much simpler. For example, the Bell states are vectors which
such symmetries.

\subsection{MUBs from Latin squares}

At first glance, one might think that maybe the product MUBs based on generalized Paulis are not well suited
for locking just because of their product form. Perhaps MUBs with entangled basis vectors do not exhibit this problem?
Let's examine how well MUBs based on Latin squares can lock classical information in a quantum state.
All such MUBs are highly entangled, with the exception of the two extra MUBs based on non-Latin squares.
Surprisingly, it turns out, however, that \emph{any} set of at least two MUBs based on Latin squares, does equally well
at locking as using just 2 such MUBs. Thus such MUBs perform equally ``badly'', i.e., we cannot improve the strength of 
the locking effect by using more MUBs of this type. 
   
\begin{lemma}\label{LSlocking}
Let $\mathbb{B}=\{\mB_1,\ldots,\mB_m\}$ with $m \geq 2$ be any set of MUBs in a Hilbert space of dimension $d=s^2$ constructed
on the basis of Latin squares.
Consider the ensemble $\ens = \{ \frac{1}{d m},\outp{b^t_k}{b^t_k}\}$. Then
$$
\mI_{acc}(\ens) = \frac{\log d}{2}.
$$ 
\end{lemma}
\begin{proof}
Note that we can again rewrite $\mI_{acc}(\ens)$ as in the proof of Lemma~\ref{equiv}. Consider 
the simple measurement in the computational basis $\{\outp{i,j}{i,j}\mid i,j \in [s]\}$. The result
then follows by the same
argument as in Lemma~\ref{LSentropy}.
\end{proof}

\begin{figure}[h]\label{subSquareMeasure}
\begin{center}
\includegraphics{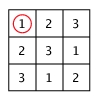}
\caption{Measurement for $\ket{1,1}$.}
\end{center}
\end{figure}

Intuitively, our measurement outputs one sub-square of the Latin square used to construct the MUBs as depicted in Figure~\ref{subSquareMeasure}. 
As we saw in the construction of MUBs based on Latin squares in Chapter~\ref{latinSquareConstruction},
each entry ``occurs'' in exactly $\sqrt{d} = s$ MUBs.

\section{Conclusion}

We have shown tight bounds on locking for specific sets of mutually unbiased bases. 
Surprisingly, it turns out that using more mutually unbiased basis does not always lead to a better locking effect. 
It is interesting to consider what may make these bases so special. The example of three MUBs considered in Lemma~\ref{3mubLocking}
may provide a clue. These three bases are given by the common eigenbases of $\{\sigma_x \otimes \sigma_x, \sigma_x \otimes \id,
\id \otimes \sigma_x\}$, $\{\sigma_z \otimes \sigma_z, \sigma_z \otimes \id, \id \otimes \sigma_z\}$ and $\{\sigma_y \otimes \sigma_y,
\sigma_y \otimes \id, \id \otimes \sigma_y\}$ respectively~\cite{boykin:mub}. However, $\sigma_x \otimes \sigma_x$, $\sigma_z \otimes \sigma_z$ and
$\sigma_y \otimes \sigma_y$ commute and thus also share a common eigenbasis, namely the Bell basis. This is exactly the basis we will use as our
measurement. For all MUBs based on generalized Pauli matrices, the MUBs in prime power dimensions are given as the common eigenbasis
of similar sets consisting of strings of Paulis. It would be interesting to determine the strength of the locking effect on the
basis of the commutation relations of elements of \emph{different} sets. 
Furthermore, perhaps it is possible to obtain good locking from a subset of such MUBs
where none of the elements from different sets commute.

It is also worth noting that the numerical results of~\cite{terhal:locking}
indicate that at least in dimension $p$ using more than three bases does indeed lead to a stronger
locking effect. It would be interesting to know, whether the strength of the locking effect
depends not only on the number of bases, but also on the dimension of the system in question.

Whereas general bounds still elude us, we have shown that merely choosing mutually unbiased bases is not sufficient to obtain 
good locking effects. We thus have to look for different properties. Sadly, whereas we were able to obtain good uncertainty relations
in Chapter~\ref{goodUncertainty}, the same approach does not work here: To obtain good locking we must not only find good uncertainty
relations, but also find a way to encode many bits using only a small number of encodings. 

\part{Entanglement}\label{entanglementPart}

\chapter{Introduction}\label{entanglementIntroChapter}\index{entanglement}\label{chapter:entanglementIntro}\label{chapter:entanglement}\label{entanglementIntro}

Entanglement is possibly the most intriguing element of quantum theory. It plays a crucial role
in quantum algorithms, quantum cryptography and the understanding of quantum mechanics itself. 
It enables us to perform quantum teleportation, as well as superdense coding~\cite{nielsen&chuang:qc}.
In this part, we investigate one particular aspect of quantum
entanglement: the violation of Bell-inequalities, and their
implications for classical protocols.
But first, let's take a brief look at the history of entanglement, and introduce the essential ingredients 
we need later.

\section{Introduction}

In 1935, Einstein, Podolsky and Rosen (EPR)\index{EPR}
identified one of the striking consequences of what latter became
known as entanglement. In their seminal article~\cite{epr:original} 
''Can Quantum Mechanical Description of Physical Reality Be Considered Complete?'' the authors define
``elements of reality'' as follows: \begin{quote}
\emph{If, without in any way disturbing a system, we can predict with certainty (i.e.\ with probability equal to unity)
the value of a physical quantity, then there exists an element of physical reality\index{physical reality} 
corresponding to this physical
quantity.}
\end{quote} 
EPR call a theory that satisfies this condition \emph{complete}\index{complete}. 
They put forward the now famous EPR-Paradox\index{EPR!Paradox},
here stated informally using discrete variables as put forward by Bohm~\cite{peres:book}. EPR assume that
if we have a state shared between two spatially separated systems, Alice and Bob, that do not interact at the time of a
measurement, 
\begin{quote}
\emph{no real change can take place in the second system as a consequence of anything that may be done
to the first system.}
\end{quote}
That means that Alice and Bob cannot use the shared state itself to transmit information.
We will also refer to this as the \emph{no-signaling} condition.
Now consider the shared state
\begin{equation}\label{EPRstate}\index{EPR!state}\index{state!EPR pair}\index{EPR!pair}
\ket{\Psi} = \frac{1}{\sqrt{2}}(\underbrace{\ket{0}}_{Alice}\underbrace{\ket{0}}_{Bob} + 
\underbrace{\ket{1}}_{Alice}\underbrace{\ket{1}}_{Bob}) = 
\frac{1}{\sqrt{2}}(\underbrace{\ket{+}}_{Alice}\underbrace{\ket{+}}_{Bob} + 
\underbrace{\ket{-}}_{Alice}\underbrace{\ket{-}}_{Bob}).
\end{equation}
Suppose that we measure Alice's system in the computational basis to obtain outcome $c_A$.
Note that we can now predict the outcome of a measurement of Bob's system in the computational
basis with certainty: $c_B = c_A$, without having disturbed Bob's system in any way. Thus $c_B$ is an 
``element of physical reality''. However, we might as well have measured Alice's system in the Hadamard basis
to obtain outcome $h_A$. Likewise, we can now predict with certainty the outcome of measuring Bob's system
in the Hadamard basis, $h_B = h_A$, again without causing any disturbance to the second system. Thus
$h_B$ should also be an ``element of physical reality''. But as we saw in 
Chapter~\ref{chapter:uncertainty}, quantum mechanics
forbids us to assign exact values to both $c_B$ and $h_B$ simultaneously, as measurements in the computational and
Hadamard basis are non-commutative. Indeed, in Chapter~\ref{mubUncertainty},
we saw that these two measurements give the strongest entropic uncertainty relation for two measurements. EPR
thus conclude 
\begin{quote}
\emph{that the quantum mechanical description of reality given by 
the wave function is not complete.}
\end{quote}
EPR's article spurred a flurry of discussion that continues up to the present
day. Shortly after the publication of their article, Schr\"odinger
published two papers in which he coined the
term entanglement (German: Verschr\"ankung)~\cite{schroedinger:eprEnglish, schroedinger:eprGerman} and investigated this
phenomenon which he described as ``not one, but rather \emph{the} characteristic trait of quantum mechanics, the one that enforces its entire departure from classical lines of thought''~\cite{schroedinger:eprEnglish}. One point of discussion in the ensuing years was whether the fact that quantum mechanics is not complete, means that there might 
exist a more detailed description of nature which \emph{is} complete. 
Even though, these more detailed
descriptions
also called ``hidden variables''\index{hidden variables} 
had remained inaccessible to us so far: a better theory and better technology might enable us to learn them. 
Thus quantum mechanical observations would merely appear to be 
probabilistic in the absence of our knowledge of such hidden variables.

\subsection{Bell's inequality}
This idea was put to rest by Bell~\cite{bell:epr}
in 1964, when he proposed conditions that \emph{any} classical theory, i.e.\ any theory based on local hidden variables, has to satisfy, and which can be verified experimentally. 
These conditions are known as \emph{Bell inequalities}\index{Bell inequality}. 
Intuitively, Bell inequalities measure the strength of non-local correlations attainable
in any classical theory. 
Non-local correlations arise as the result of measurements performed on a quantum system
shared between two spatially separated parties. Imagine two parties, Alice and Bob,
who are given access to a shared quantum state $\ket{\Psi}$, but cannot communicate. In the simplest case, 
each of them is able to perform one of two possible measurements. Every measurement has two possible
outcomes labeled $\pm 1$.
Alice and Bob now measure $\ket{\Psi}$ using an independently chosen measurement setting and record
their outcomes. In order to obtain an accurate estimate for the correlation between their measurement
settings and the measurement outcomes, they perform this experiment independently 
many times using an identically
prepared state $\ket{\Psi}$ in each round.
\begin{figure}[h]
\includegraphics[scale=0.9]{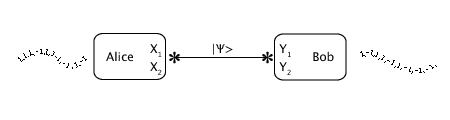}
\caption{Alice and Bob measure many copies of $\ket{\Psi}$}
\end{figure}

Both classical and quantum theories impose limits on the strength of non-local correlations.
In particular, both should not violate the non-signaling condition of special relativity as put
forward by EPR above. That is,
the local choice of the measurement setting does not allow Alice and Bob to transmit information.
Limits on the strength of correlations which are possible in the framework of any
\emph{classical} theory are the Bell inequalities.
The best known Bell inequality is the
Clauser, Horne, Shimony and Holt (CHSH)\index{CHSH inequality}
\index{Bell inequality!CHSH} 
inequality~\cite{chsh:nonlocal}
\begin{equation}\label{CHSHinequality}
\langle CHSH \rangle_c = |\ev{X_1}{Y_1} + \ev{X_1}{Y_2} + \ev{X_2}{Y_1} - \ev{X_2}{Y_2}| \leq 2,
\end{equation}
where $X_1, X_2$ and $Y_1, Y_2$ are the observables representing the 
measurement settings of Alice and Bob respectively and we use
$\ev{X_i}{Y_j} = \bra{\Psi}X_i \otimes Y_j\ket{\Psi}$
to denote the mean value of $X_i$ and $Y_j$.
Quantum mechanics allows for a violation of the CHSH inequality, and is thus indeed non-classical: If we
take the shared state $\ket{\Psi} = (\ket{00} + \ket{11})/\sqrt{2}$ and let $X_1 = \sigma_x$, $X_2 = \sigma_z$, 
$Y_1 = (\sigma_x + \sigma_z)/2$, and $Y_2 = (\sigma_x - \sigma_z)/2$ we obtain
$$
\langle CHSH \rangle_q = |\ev{X_1}{Y_1} + \ev{X_1}{Y_2} + \ev{X_2}{Y_1} - \ev{X_2}{Y_2}| = 2\sqrt{2}.
$$
Most importantly, this violation can be experimentally verified allowing us to test the validity of the theory.
The first such tests were performed by Clauser~\cite{ch:experiment}
and Aspect, Dalibard, Grangier, and
Roger~\cite{aspect:test1, aspect:test2}. Over the years these tests have been 
refined considerably, ruling out
many loopholes present in the initial experiments such as for example detector inefficiency~\cite{bellDetectors}. Yet, no
conclusive test has been achieved so far. 
Unfortunately, such experimental concerns are outside the scope of this thesis, and we merely point to
an overview of such issues~\cite{aspect:overview}.

\subsection{Tsirelson's bound}

Curiously, even quantum mechanics itself still limits the strength of non-local correlations. 
Tsirelson's bound\index{Tsirelson!bound}~\cite{tsirel:original} says that for quantum mechanics
$$
|\ev{X_1}{Y_1} + \ev{X_1}{Y_2} + \ev{X_2}{Y_1} - \ev{X_2}{Y_2}| \leq 2\sqrt{2},
$$
and thus the above measurements are optimal. We provide a simple proof
of this fact in Chapter~\ref{chapter:optimalStrategies}. It is interesting to consider what
would happen if quantum mechanics allowed for more powerful non-local correlations.
To this end, it is convenient to rewrite the CHSH inequality from Eq.~(\ref{CHSHinequality})
in the form
$$
\sum_{x,y \in \01} \Pr [a_x \oplus b_y = x \cdot y] \leq 3.
$$
Here, $x \in \01$ and $y \in \01$ denote the choice of Alice's and
Bob's measurement, $a_x \in \01$ and $b_y \in \01$ the
respective binary outcomes, and $\oplus$ addition modulo $2$ (see Section~\ref{nonlocalGames} for
details).
In this form, quantum mechanics allows a violation up to the 
maximal value of $2 + \sqrt{2}$.
Since special relativity would even
allow a violation of Tsirelson's bound, Popescu and
Rohrlich~\cite{popescu:nonlocal,popescu:nonlocal2,popescu:nonlocal3}
raised the question why nature is not more 'non-local'? That is,
why does quantum mechanics not allow for a stronger violation of
the CHSH inequality up to the maximal value of 4? To gain more
insight into this question, they constructed a toy-theory based on
non-local boxes. Each such box takes inputs $x,y \in \01$ from
Alice and Bob respectively and always outputs measurement outcomes
$a_x$,$b_y$ such that $x\cdot y = a_x \oplus b_y$. Alice
and Bob still cannot use this box to transmit any information.
However, since for all $x$ and $y$,
$\Pr[a_x \oplus b_y = x \cdot y] = 1$, the above sum equals 4 and thus
non-local boxes lead to a maximum violation of the CHSH
inequality.

Van Dam~\cite{wim:nonlocal,wim:thesis}
has shown that having access
to such non-local boxes allows Alice and Bob to perform any
kind of distributed computation by transmitting only a \emph{single} bit
of information. This is even true for slightly less perfect boxes
achieving weaker correlations~\cite{falk:nonlocal}. In~\cite{wehner05b}, we showed
that given any non-local boxes, Alice and Bob could perform bit commitment
and oblivious transfer, which is otherwise known to be impossible.
Thus, such cryptographic principles are in principle compatible with the theory
of non-signaling: non-signaling itself does not prevent us from implementing them.

Looking back to the uncertainty relations in Chapter~\ref{uncertaintyChapter}, which
rest at the heart of the EPR paradox, we might suspect that the violation of the CHSH inequality
likewise depends on the commutation relations between the local measurements of Alice and Bob.
Indeed, it has been shown by Landau~\cite{landau:bell}, and Khalfin and Tsirelson~\cite{tsirel:obscure}, there exists a state $\ket{\Psi}$ such that
$$
|\ev{X_1}{Y_1} + \ev{X_1}{Y_2} + \ev{X_2}{Y_1} - \ev{X_2}{Y_2}| = 2\sqrt{1 + 4\norm{[X_1^0,X_2^0][Y_1^0,Y_2^0]}},
$$
for any $X_1 = X_1^0 - X_1^1$, $X_2 = X_2^0 - X_2^1$ and $Y_1 = Y_1^0 - Y_1^1$, $Y_2 = Y_2^0 - Y_2^1$, where we use the superscripts '0' and '1' to denote
the projectors onto the positive and negative eigenspace respectively. 
Thus, given any observables $X_1, X_2$ and $Y_1, Y_2$, the CHSH inequality is violated
if and only if $[X_1^0,X_2^0][Y_1^0,Y_2^0] \neq 0$. 

\section{Setting the stage}

\subsection{Entangled states}
The state given in Eq.~(\ref{EPRstate}) is just one possible example of an entangled state. Recall from
Chapter~\ref{informationIntro} that if $\ket{\Psi} \in \hil^A \otimes \hil^B$ is a pure state, we say that $\ket{\Psi}$ is \emph{separable} if and only if there exist states $\ket{\Psi^A} \in \hil^A$ and $\ket{\Psi^B} \in \hil^B$ such that $\ket{\Psi} = \ket{\Psi^A} \otimes \ket{\Psi^B}$. A separable pure state is also called a \emph{product state}. A state that is not 
separable is called \emph{entangled}. For mixed states the 
definition is slightly more subtle. Let $\rho \in \mS(\hil^A \otimes \hil^B)$ be a mixed state. Then $\rho$ is 
called a \emph{product state} if there exist $\rho^A \in \mS(\hil^A)$ and $\rho^B \in \mS(\hil^B)$ such 
that $\rho = \rho^A \otimes \rho^B$. The state $\rho$ is called \emph{separable}, if there exists an ensemble
$\ens = \{p_j,\outp{\Psi_j}{\Psi_j}\}$ such that $\ket{\Psi_j} = \ket{\Psi_j^A} \otimes \ket{\Psi_j^B}$ with
$\ket{\Psi_j^A} \in \hil^A$ and $\ket{\Psi_j^B} \in \hil^B$ for all $j$, such
that 
$$
\rho = \sum_j p_j \outp{\Psi_j}{\Psi_j} = \sum_j p_j \outp{\Psi_j^A}{\Psi_j^A} \otimes \outp{\Psi_j^B}{\Psi_j^B}.
$$
Intuitively, if $\rho$ is separable then $\rho$ corresponds to a mixture of separable pure 
states according to a joint probability distribution $\{p_j\}$, a purely classical form of correlation. 
Given a description of a mixed state $\rho$ it is an NP-hard problem to decide whether $\rho$ is 
separable~\cite{gurvits:nphard}. However, many criteria 
and approximation algorithms have been 
proposed~\cite{andrew:sep1,andrew:sep2,andrew:sep3,lawrence:sep,lawrence:sep2}. 
It is 
an interesting question to determine the maximal violation of a given Bell-inequality for a fixed 
state $\rho$~\cite{andrew:states}.
Here, we only concern ourselves with \emph{maximal} violations of Bell inequalities, and 
refer to~\cite{lawrence:thesis} for an overview of the separability problem. Generally, the maximal violation is obtained
by using the maximally entangled state. However, there are cases for which the maximal violation is achieved
by a non maximally entangled state~\cite{gisin:nonmaxEnt}.
Note that we can never observe a Bell inequality violation for a separable state: it is no more than a classical
mixture of separable pure states. 
On the other hand, any two-qubit \emph{pure} state that is entangled violates the CHSH inequality~\cite{gisin:pureStateViolates}.
However, not all entangled \emph{mixed} states violate the CHSH inequality! A counterexample was given by Werner~\cite{werner:state}
\index{Werner state} with the so-called Werner-state\index{state!Werner}
$$
\rho_W = p \frac{2}{d^2+d}P_{sym} + (1-p) \frac{2}{d^2-d}P_{asym},
$$
where $P_{sym}$
and $P_{asym}$
are projectors onto the symmetric and the anti-symmetric subspace respectively.
For $p \geq 1/2$ this state is separable, but it is entangled for $p < 1/2$. Yet, the CHSH inequality is not
violated.
A lot of work has been done to quantify the amount of entanglement in quantum states, and we refer 
to~\cite{barbara:thesis, eisert:thesis, matthias:thesis} for an overview. 

\subsection{Other Bell inequalities}\index{Bell inequalities}

The CHSH inequality we encountered above is by no means the only Bell inequality. Recall that
non-local correlations arise as the result of measurements performed on a quantum system
shared between two spatially separated parties. Let $x$ and $y$ be the variables corresponding to Alice and Bob's 
choice of measurement. Let $a$ and $b$ denote the corresponding outcomes\footnote{For simplicity, we assume that the set of possible outcomes is the same for each setting.}. Let $\Pr[a,b|x,y]$ be the probability of obtaining
outcomes $a, b$ given settings $x,y$. What values are allowed for $\Pr[a,b|x,y]$? Clearly, we want
that for all $x,y,a,b$ we have that $\Pr[a,b|x,y] \geq 0$ and $\sum_{a,b} \Pr[a,b|x,y] = 1$. From the 
no-signaling
condition we furthermore obtain that the marginals obey $\Pr[a|x] = \Pr[a|x,y] = \sum_{b} \Pr[a,b|x,y]$ and 
likewise
for $\Pr[b|y]$, i.e.\ the probability of Alice's measurement outcome is independent of Bob's choice of measurement setting, and vice versa.
For $n$ players, who each perform one of $N$ measurements with $k$ possible outcomes, we have $(Nk)^n$ such probabilities
to assign, giving us a $(Nk)^n$ dimensional vector. To find all Bell inequalities, we now look for inequalities 
that bound the classically accessible region (a convex polytope) for such assignments. It is clear that we
can find a huge number of such inequalities. Of course, often the most interesting inequalities are the ones
that are satisfied only classically, but where we can find a better quantum strategy. Much work has been
done to identify such inequalities, and we refer to~\cite{werner:overview} for an excellent overview.
In the following chapters, we are interested in the following related question: Given an inequality, what is
the optimal quantum measurement strategy that maximizes the inequality?

\subsection{Non-local games}\label{nonlocalGames}\index{games!non-local}\index{non-local games}

It is often convenient to view Bell experiments as a game between two, or more, distant players, who cooperate
against a special party. We call this special party the \emph{verifier}\index{verifier}.
In a two player game with players Alice and Bob, the verifier picks two questions, say $s_1$ and $s_2$, and hands them to Alice and Bob respectively, who
now need to decide answers $a_1$ and $a_2$. To this end, they may agree on any strategy beforehand, but can no longer communicate once the game starts. The verifier then decides according to a fixed set of public rules, whether 
Alice and Bob win by giving answers $a_1,a_2$ to questions $s_1,s_2$.
In a quantum game, Alice and Bob may perform measurements on a shared entangled
state to determine their answers. We can thus think of the questions as measurement settings and the 
answers as measurement outcomes. 

More formally, we consider games among $N$ players $P_1,\ldots,P_N$. Let $S_1,\ldots,S_N$ and
$A_1,\ldots,A_N$ be finite sets corresponding to the possible questions and answers respectively.
Let $\pi$ be a probability distribution on $S_1 \times \ldots \times S_N$, and let $V$ be a predicate on
$A_1 \times \ldots \times A_N \times S_1 \times \ldots \times S_N$. Then $G = G(V,\pi)$ is the following
$N$-player cooperative game: A set of questions $(s_1,\ldots,s_N) \in S_1 \times \ldots \times S_N$ is chosen
at random according to the probability distribution $\pi$. Player $P_j$ receives question $s_j$, and then responds 
with answer $a_j \in A_j$. The players win if and only if $V(a_1,\ldots,a_N,s_1,\ldots,s_N) = 1$. 
We write $V(a_1,\ldots,a_N|s_1,\ldots,s_N) = V(a_1,\ldots,a_N,s_1,\ldots,s_N)$ to emphasize
the fact that $a_1,\ldots,a_N$ are the answers given questions $s_1,\ldots,s_N$. 
\begin{figure}[h]
\begin{center}
\includegraphics[scale=0.65]{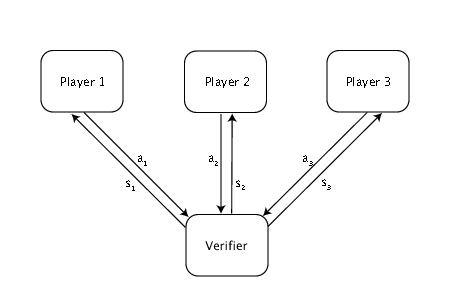}
\caption{Multiplayer non-local games.}
\end{center}
\end{figure}

The \emph{value}
of the game $\omega(G)$ is the probability that the players win the game, maximized over all
possible strategies. We use $\omega_c(G)$ and $\omega_q(G)$ to differentiate between the value of the
game in the classical and quantum case respectively. Classically, $\omega_c(G)$ can always be attained
by a deterministic strategy~\cite{cleve:nonlocal}. We can thus write
\begin{equation}\label{classicalGameValue}
\omega_c(G) = \max_{f_1,\ldots,f_N} \sum_{s_1,\ldots,s_N} \pi(s_1,\ldots,s_N) 
V(f_1(s_1),\ldots,f_N(s_N)|s_1,\ldots,s_N),
\end{equation}
where the maximization is taken over all functions $f_j: S_j \rightarrow A_j$
that determine answers $a_j = f_j(s_j)$.

Quantumly, the strategy of the players consists of their choice of measurements and
shared entangled state. Let $\ket{\Psi}$ denote the players' choice of state, and
let $X_{s_j}^{[j]} = \{X_{s_j}^{a_j,[j]}\mid a_j \in A_j\}$ denote the POVM of player $P_j$ for 
question $s_j \in S_j$. Here, we always assume that the underlying Hilbert space
is finite-\-dimensional. The value of the quantum game is then
\begin{equation}\label{quantumGameValue}
\omega_q(G) = \max_{X^{[1]},\ldots,X^{[N]}} \sum_{s_1,\ldots,s_N} \pi(s_1,\ldots,s_N) \sum_{a_1,\ldots,a_N}
\bra{\Psi} X_{s_1}^{a_1,[1]} \otimes \ldots \otimes X_{s_N}^{a_N,[N]} \ket{\Psi},
\end{equation}
where the maximization is taken over all POVMS $X_{s_j}^{[j]}$ for all $j \in [N]$ and $s_j \in S_j$. In the following, 
we say that a set of measurement operators \emph{achieves} $p$, if 
$$
p = \sum_{s_1,\ldots,s_N} \pi(s_1,\ldots,s_N) \sum_{a_1,\ldots,a_N}
\bra{\Psi} X_{s_1}^{a_1,[1]} \otimes \ldots \otimes X_{s_N}^{a_N,[N]} \ket{\Psi}.
$$
Of particular
relevance in the next chapters is a special class of two-player games known as XOR-games~\cite{cleve:nonlocal}:\index{games!XOR}
Here, $N=2$ and we assume that $A_1 = A_2 = \01$. The two players $P_1$ (Alice) and $P_2$ (Bob) each 
have only two possible measurement outcomes. Furthermore, the winning condition only depends on the XOR
of answers $a_1$ and $a_2$ and thus we write $V(c|s_1,s_2)$ with $c = a_1 \oplus a_2$. It can be shown~\cite{cleve:nonlocal}
that the optimal POVM in this case consists only of projectors. We can thus write $X_{s_1}^{[1]}$ and 
$X_{s_2}^{[2]}$ as observables
with two eigenvalues: $X_{s_1}^{[1]} = X_{s_1}^{0,[1]} - X_{s_1}^{1,[1]}$ and 
$X_{s_2}^{[2]} = X_{s_2}^{0,[2]} - X_{s_2}^{1,[2]}$ where $s_1 \in S_1$
and $s_2 \in S_2$. A small calculation using the fact that $X_{s_1}^{0,[1]} + X_{s_1}^{1,[1]} = \id$ and
$X_{s_2}^{0,[2]} + X_{s_2}^{1,[2]} = \id$ shows that we can rewrite the optimal value of a quantum XOR-game as
\begin{align}\label{xorGameValue}
\omega_q(G) =&\\
& \max_{X^{[1]},X^{[2]}} \frac{1}{2} \sum_{s_1,s_2} \pi(s_1,s_2) \sum_{c \in \01} V(c|s_1,s_2) 
(1 + (-1)^c \bra{\Psi} X_{s_1}^{[1]} \otimes X_{s_2}^{[2]} \ket{\Psi}).
\end{align}
From the above, we can see that XOR-games correspond to correlation inequalities with two-outcome
measurements. We will see in Chapter~\ref{optimalStrategiesChapter} that this reformulation enables us to determine
the optimal measurements for such XOR-games in a very simple manner. Indeed, the CHSH inequality can be rephrased as 
a simple quantum XOR-game. Here, Alice and Bob win if and only if given questions $s_1,s_2$ they return answers
$a_1,a_2$ such that $s_1 \cdot s_2 = a_1 \oplus a_2$, i.e. we have $V(c|s_1,s_2) = 1$ if and only if $s_1 \cdot s_2 = c$.
Recalling Eq.~(\ref{CHSHinequality}) we can write
$$
\omega(CHSH) = \frac{1}{2} \left(1 + \frac{\langle CHSH \rangle}{4}\right),
$$
from which we obtain $\omega_q(CHSH) = 1/2 + 1/(2\sqrt{2})$ vs. $\omega_c(CHSH) = 3/4$.

\section{Observations}

In the following chapters, we are concerned with finding the optimal quantum measurement strategies for 
Bell inequalities. To this end, we first make a few simple observations that help us understand the 
structural properties of our problem. In particular, this also enables us to understand the relation
between Bell inequalities and the problem of post-measurement information in Chapter~\ref{piRelation}.
We then present a theorem by Tsirelson~\cite{tsirel:original, tsirel:separated} that plays a crucial role
in the subsequent chapters.

\subsection{Simple structural observations}

Suppose we are given a set of measurements for Alice and Bob and a shared state $\rho$.
Can we reduce the dimension of Alice and Bob's measurements operators and
the thereby amount of entanglement they need? As we saw in Chapter~\ref{informationChapter},
we can often simplify our problem by identifying its classical and quantum part. Indeed, this is
also the case here.
\begin{lemma}\label{lemma:reducingGames}\label{simpleIsEnough}
Let $\hil = \hil^A \otimes \hil^B$ and let $\setA = \{\aoq \in \bop(\hil^A)\}$ and $\setB = \{\boq \in \bop(\hil^B)\}$
be the set of Alice and Bob's measurement operators respectively. 
Let $\rho \in \mS(\hil)$ be the
state shared by Alice and Bob. Suppose that for such operators we have
$$
q = \sum_{s \in S,t \in T} \pi(s,t) \sum_{a \in A,b \in B} V(a,b|s,t)\Tr(\aoq \otimes \boq \rho).
$$
Then there exist measurement operators
$\tilde{\setA} = \{\tildeaoq\}$ and 
$\tilde{\setB} = \{\tildeboq\}$
and a state $\tilde{\rho}$ such
$$
q \leq \sum_{s \in S,t \in T} \pi(s,t) \sum_{a \in A,b \in B} V(a,b|s,t)\Tr(\tildeaoq \otimes \tildeboq \tilde{\rho}).
$$
and the $C^*$-algebra generated by $\tilde{\setA}$ and $\tilde{\setB}$ is simple.
\end{lemma}
\begin{proof}
Let $\algA = \langle \setA \rangle$ and $\algB = \langle \setB \rangle$.
If $\algA$ and $\algB$ are simple, we are done. If not, we know from Lemma~\ref{finiteMeansSemisimple}
and Lemma~\ref{isomorphBop}
that there exists a decomposition 
$\hil^A \otimes \hil^B = \bigoplus_{jk} \hil^A_j \otimes \hil^B_k$.
Consider $\Tr((M^A \otimes M^B)\rho)$, where $M^A \otimes M^B \in \algA \otimes \algB$.
It follows from the above that
$M^A \otimes M^B = \bigoplus_{jk} (\Pi^A_j \otimes \Pi^B_k) M^A \otimes M^B 
(\Pi^A_j \otimes \Pi^B_k)$,
where $\Pi^A_j$ and $\Pi^B_k$ are projectors onto $\hil^A_j$ and $\hil^B_k$ respectively.
Let $\hat{\rho} = \bigoplus_{jk} (\Pi^A_j \otimes \Pi^B_k) \rho 
(\Pi^A_j \otimes \Pi^B_k)$.
Clearly,
\begin{eqnarray*}
\Tr((M^A \otimes M^B)\hat{\rho}) &=& \sum_{jk}
\Tr\left((\Pi^A_j \otimes \Pi^B_k) M^A \otimes M^B (\Pi^A_j \otimes \Pi^B_k) \rho\right)\\
&=&
\Tr((M^A \otimes M^B)\rho).
\end{eqnarray*}
The statement now follows immediately by convexity:
Alice and Bob can now measure $\rho$
using $\{\Pi^A_j \otimes \Pi^B_k\}$ and record the classical outcomes $j, k$. The new measurements
are then $\tilde{A}_{s,j}^a
= \Pi^A_j \aoq \Pi^A_j$ and $\tilde{B}_{t,k}^b = \Pi^B_k \boq \Pi^B_k$
on state $\tilde{\rho}_{jk} = (\Pi^A_j \otimes \Pi^B_k)\rho
(\Pi^A_j \otimes \Pi^B_k)/\Tr((\Pi^A_j \otimes \Pi^B_k)\rho)$.
By construction, $\tilde{\algA}_j = \{\tilde{A}_{s,j}^a\}$ and 
$\tilde{\algB}_k = \{\tilde{B}_{t,k}^b\}$ are simple.

Let $q_{jk}$ denote the probability that we obtain outcomes $j, k$, and let 
$$
r_{jk} = 
\sum_{s \in S,t \in T} \pi(s,t) \sum_{a \in A,b \in B} V(a,b|s,t)\Tr(\tilde{A}^{a}_{s,j} \otimes \tilde{B}^{b}_{t,j}
\tilde{\rho}_{jk}).
$$
Then $q = \sum_{jk} q_{jk} r_{jk} \leq \max_{jk} r_{jk}$.
Let $u,v$ be such that $r_{u,v} = \max_{jk} r_{jk}$.
Hence, we can skip the initial measurement
and instead use measurements $\tildeaoq= \tilde{A}_{s,u}^a$, $\tildeboq = \tilde{B}_{t,v}^b$
and state $\tilde{\rho} = \tilde{\rho}_{u,v}$.
\end{proof}

It also follows immediately from the above proof that
\begin{corollary}
$\dim(\tilde{\rho}) \leq \dim(\hil^A_u)\dim(\hil^B_v)$
\end{corollary}

We can thus assume without loss of generality, that
the algebra generated by Alice and Bob's optimal measurements is always simple.
We also immediately see why we can simulate the quantum measurement
classically if Alice or Bob's measurements commute locally. Indeed, the above proof 
tells us how to construct the appropriate 
classical strategy:
\begin{corollary}\label{classicalStrategyCorollary}
Let $\hil = \hil^A \otimes \hil^B$ and let $\setA = \{\aoq \in \bop(\hil^A)\}$ and $\setB = \{\boq \in \bop(\hil^B)\}$
be the set of Alice and Bob's measurement operators respectively. 
Let $\rho \in \mS(\hil)$ be the
state shared by Alice and Bob.
Let $p$ be the value of the non-local game achieved using these measurements.
Suppose that for all $s,s'$,and $a,a'$ we have that $[X_s^a,X_{s'}^{a'}] = 0$ 
(or for all $t,t'$, $b,b'$ $[Y_t^b,Y_{t'}^{b'}] = 0$). Then there exists a classical
strategy for Alice and Bob that achieves $p$. 
\end{corollary}
\begin{proof}
Our conditions imply that either $\algA$ or $\algB$ is abelian. Suppose wlog that $\algA$ is abelian. 
Hence, by the above proof we have $\max_j \dim(\hil^A_j) = 1$.
Again, Alice and Bob perform the measurements determined by $\Pi^A_j$ and $\Pi^B_k$ and record
their outcomes $j,k$. Since $\dim(\hil^A_j) = 1$, Alice's post-measurement state is in fact classical, 
and we have no further entanglement between Alice and Bob.
\end{proof}

To violate a Bell inequality, Alice and Bob must thus use measurements which do not
commute locally.
However, since Alice and Bob are spatially separated, we can
write Alice and Bob's measurement operators as 
$X = \hat{X} \otimes \id$ and $Y = \id \otimes \hat{Y}$ respectively as for any
$\rho$ we can write $\Tr(\rho (X \otimes Y)) = \Tr(\rho (\hat{X} \otimes \id)(\id \otimes \hat{Y}))$. Thus
$[X,Y] = 0$. Thus from a bipartite structure we obtain certain commutation relations. 
How about the converse? As it turns out, in any finite-dimensional $C^*$-algebra\footnote{or indeed any
Type-I von Neumann algebra}, these two notions are equivalent: From commutation we immediately obtain a bipartite
structure! We encounter this well-known, rather beautiful observation in 
Appendix~\ref{cstar}.

\subsection{Vectorizing measurements}
\label{tsirelsonConstruction}

In Chapter~\ref{optimalStrategies}, we show how to obtain the optimal measurements for
any bipartite correlation inequality. At first sight, this may appear to be a daunting problem: We must
simultaneously maximize Eq.~(\ref{xorGameValue}) over the state $\rho$ as well as measurement operators
of the form $X \otimes Y$, a problem which is clearly not convex. Yet, the following brilliant observation
by Tsirelson~\cite{tsirel:original, tsirel:separated} greatly simplifies our problem.

\begin{theorem}[Tsirelson]\label{tsirel}\index{Tsirelson!vectors}\index{Tsirelson!construction}
Let $X_1,\ldots,X_n$ and $Y_1,\ldots,Y_m$ be observables with eigenvalues in the interval $[-1,1]$.
Then for any state $\ket{\Psi} \in \hil^A \otimes \hil^B$ and for all $s \in [n]$, $t \in [m]$, 
there exist real unit vectors $x_1,\ldots,x_n$,$y_1,\ldots,y_m \in \Real^{n+m}$ such that
$$
\bra{\Psi}X_s \otimes Y_t\ket{\Psi} = x_s \cdot y_t,
$$
where $x_s \cdot y_t$ is the standard inner product.
Conversely, let $x_s,y_t \in \Real^{N}$ be real unit vectors.
Let $\ket{\Psi} \in \hil^A \otimes \hil^B$ be any maximally entangled state where
$\dim(\hil^A) = \dim(\hil^B) = 2^{\floor{N/2}}$. Then for all $s \in [n]$, $t \in [m]$
there exist observables $X_s$ on $\hil^A$ and $Y_t$ on $\hil^B$ with eigenvalues in $\{-1,1\}$
such that
$$
x_s \cdot y_t = \bra{\Psi}X_s \otimes Y_t\ket{\Psi}.
$$
\end{theorem}

In fact, by limiting ourselves onto the space spanned by the vectors $x_1,\ldots,x_n$ or $y_1,\ldots,y_m$, we could
further decrease the dimension of the vectors to $N = \min\{n,m\}$~\cite{tsirel:separated}.
The result was proven by Tsirelson in a more general form for any finite-dimensional
$C^*$-algebra. 
Here, we do not consider this more abstract argument, but instead simply sketch how to obtain the vectors
and state how to find the corresponding
measurement operators in turn~\cite{tsirel:hadron}. 
To find vectors $x_s$ and $y_t$, we merely need to consider the vectors
$$
x_s = X_s \otimes \id \ket{\Psi} \mbox{ and } y_t = \id \otimes Y_t \ket{\Psi},
$$
where may take the vectors to be real~\cite{tsirel:original}.
Recall that we are only interested in the inner products.
But clearly we can then bound the dimension of our vectors as the number of our
vectors is strictly limited and thus cannot span a space of dimension larger than
$N$. 

To construct observables corresponding to a given set of vectors, consider the generators of
a Clifford algebra
$\Gamma_1,\ldots,\Gamma_N$ with $N$ even\footnote{If $N$ is odd,
we obtain one additional element from $\Gamma_0$.}
that we already encountered in Section~\ref{cliffordUncertainty}, i.e., we have that
for all $j \neq k \in [N]$, $\{\Gamma_j, \Gamma_k\} = 0$ and $\Gamma_j^2 = \id$. Note that
we also have $\Tr(\Gamma_j\Gamma_k) = \delta_{jk}$ as the two matrices anti-commute. 
Consider two vectors $x_s, y_t \in \Real^N$
with $x_s = (x_s^1,\ldots,x_s^N)$ and $y_t = (y_t^1,\ldots,y_t^N)$.
Define $X_s = \sum_{j \in [N]} x_s^j \Gamma_j^T$ and $Y_t = \sum_{j \in [N]} y_t^j \Gamma_j$ and
let $\ket{\Psi} = (1/\sqrt{d}) \sum_k \ket{k}\ket{k}$ with $d = 2^{\floor{N/2}}$ be the maximally entangled state. We then have
$$
\bra{\Psi}X_s \otimes Y_t\ket{\Psi} = \frac{1}{d}\sum_{jk}x_s^j y_t^k\Tr(\Gamma_j \Gamma_k) = 
\frac{1}{d}\sum_j x_s^j y_t^k \Tr(\id) = x_s \cdot y_t.
$$
Note that in principle we could have chosen any set of orthogonal operators $\Gamma_1,\ldots,\Gamma_N$ to obtain
the stated equality. However,
we obtain from their anti-commutation that
$$
X_s^2 = 
\sum_{jk}x_s^j x_s^k \Gamma_j \Gamma_k = 
\frac{1}{2}\sum_{jk}x_s^j x_s^k \{\Gamma_j,\Gamma_k\} = 
\sum_j (x_s^j)^2 \id = \id,
$$
since $||x_s|| = 1$. Hence, $X_s$ has eigenvalues in $\{-1,1\}$ as desired.
Curiously, $\Gamma_1,\ldots,\Gamma_N$ were also the 
right choice of operators to obtain good uncertainty relations in Chapter~\ref{goodUncertainty}.

\section{The use of post-measurement information}\index{post-measurement information!gap}\label{piRelation}\label{useofPISTAR}\index{PI-STAR!gap}

Looking back to Chapter~\ref{pistarChapter}, we see that we have already encountered the same structure in the context of
post-measurement information. Recall that there our goal was to determine $y$ given some 
$\rho_{yb} \in \{\rho_{yb}\mid y \in \mY \mbox{ and } b \in \mB\}$ after receiving additional post-measurement 
information $b$. In particular, 
as we explain in more detail in Chapter~\ref{dimensionalityChapter} 
we see that the question of how much post-measurement information is required is the same as the following: given a set of observables, how large does our quantum state have to be in order to implement the resulting non-local game? However, we can further exploit the relationship between these two problems to prove a gap between the optimal success probability in the setting of state discrimination (STAR)\index{STAR} and the setting of state discrimination \emph{with} post-measurement information (PI-STAR)\index{PI-STAR}. In particular, we show that for some problems, if we can succeed 
perfectly in the setting of PI-STAR without keeping any qubits at all, our success at STAR can in fact be bounded by 
a Bell-type inequality! Of course, PI-STAR itself is not a non-local problem.
However, as we saw in Appendix~\ref{cstar}, the commutation relations which are necessary for Bob to succeed at PI-STAR perfectly in Lemma~\ref{pistarCommutingLemma}, 
do induce a bipartite structure. We now exploit the structural similarity of the two problems.

\begin{figure}[h]
\begin{minipage}{0.45\textwidth}
\begin{center}
\includegraphics[scale=0.8]{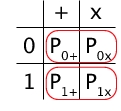}
\caption{Original problem}
\label{starProblem1Figure}
\end{center}
\end{minipage}
\begin{minipage}{0.45\textwidth}
\begin{center}
\includegraphics[scale=0.8]{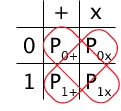}
\caption{Derived problem}
\label{starProblem2Figure}
\end{center}
\end{minipage}
\end{figure}

We first consider the very simple case of two bases and a Boolean function. Here, it turns out that we can
bound the value of the STAR problems using the CHSH inequality. We do this by showing a bound on the average
of two equivalent STAR problems, illustrated in Figures~\ref{starProblem1Figure}
and~\ref{starProblem2Figure}. 
The XOR function considered in Chapter~\ref{pistarChapter} is an example
of such a problem. Below we construct a generalization
of the CHSH inequality which allows us to make more general statements. We state our result in the notation
introduced in Chapter~\ref{pistarChapter}.
For simplicity, we use indices $+$ and $\times$ to denote two \emph{arbitrary} bases
and use the notation $\mbox{STAR}(\rho_0,\ldots,\rho_{n-1})$ to refer to a state discrimination problem
between $n$ different states. 

\begin{lemma}\label{pistarGapLemma}\index{STAR}\index{PI-STAR}
Let $P_X(x) = 1/2^n$ for all $x \in \01^n$ and let $f: \01^n \rightarrow \01$ be any
Boolean function. Let $\mB = \{+,\times\}$ denote a set of two bases, and suppose
there exists a unitary $U$ such that $\rho_{0+} = U\rho_{0+}U^{\dagger}$, $\rho_{1+} =
U \rho_{1+}U^{\dagger}$, $\rho_{1\times} = U\rho_{0\times}U^{\dagger}$ and $\rho_{0\times} = U\rho_{1\times}U^{\dagger}$.
Suppose Bob succeeds at $\mbox{PI-STAR}_0(f)$ with probability $p=1$. Then 
he succeeds at $\mbox{STAR}(\rho_0,\rho_1)$
with probability at most 3/4, where $\rho_0 = (\rho_{0+} + \rho_{0\times})/2$,
$\rho_1 = (\rho_{1+} + \rho_{1\times})/2$.
\end{lemma}
\begin{proof}
Let $P_{0+}$, $P_{1+}$, $P_{0\times}$ and $P_{1\times}$ be projectors onto
the support of $\rho_{0+}$, $\rho_{1+}$, $\rho_{0\times}$ and $\rho_{1\times}$ respectively.
Suppose that Bob succeeds with probability $p$ at $\mbox{STAR}(\rho_0,\rho_1)$. Then there exists
a strategy for Alice and Bob to succeed at the CHSH game with probability $p$, where Alice's
measurements are given by $\{P_{0+},P_{1+}\}$ and $\{P_{0\times},P_{1\times}\}$: 

Let $\hat{\rho}_0 = (\rho_{0+} + \rho_{1\times})/2$ and $\hat{\rho}_1 = (\rho_{1+} + \rho_{0\times})/2$.
Note that since there exists such a $U$, we have that Bob succeeds at $\mbox{STAR}(\hat{\rho}_0,\hat{\rho}_1
)$
with probability $p$ as well. Suppose that Alice and Bob share the maximally entangled state
$\ket{\Psi_{AB}}^{\otimes n}$ with $\ket{\Psi_{AB}} = (\ket{00} + \ket{11})/\sqrt{2}$. With probability $1/2
$
Alice chooses measurement setting $x=0$ and then her measurement is given by $\{P_{0+},P_{1+}\}$. Let $a$ denote
her measurement outcome. Bob's system is now in the state $\rho_{a+}$. Similarly, with probability $1/2$ Alice
sets $x=1$ and measures $\{P_{0\times},P_{1\times}\}$, which leaves Bob's system in the state $\rho_{a\times}$.
Let $y$ denote Bob's measurement setting. The CHSH game now requires Bob to obtain a measurement outcome $b$
such that $x \cdot y = a \oplus b$. Thus, for $y=0$, Bob always tries to obtain $b = a$ which means he
wants to solve $\mbox{STAR}(\rho_0,\rho_1)$. For $y=1$, Bob tries to obtain $b=a$ for $x=0$ but $b = 1-a$ for
$x=1$, i.e., he wants to solve $\mbox{STAR}(\hat{\rho}_0,\hat{\rho}_1)$. Since Bob chooses $y \in \01$ uniformly 
at random, we obtain the stated result.

Now suppose that Bob succeeds at $\mbox{PI-STAR}_0(f)$ with probability $p=1$. We know from Lemma~\ref{pistarCommutingLemma}
that for all $y,y' \in \mY$ and $b,b' \in \mB$ we have $[P_{yb},P_{y'b'}] = 0$ where $P_{yb}$ is a projector
onto the support of $\rho_{yb}$. Now, suppose that on the contrary
he succeeds at $\mbox{STAR}(\rho_0,\rho_1)$ with probability greater than $3/4$. Then we know from the
above argument that there exists a strategy for Alice and Bob to succeed at the CHSH game with probability 
greater than $3/4$ where Alice measures two \emph{commuting} observables, which contradicts
Corollary~\ref{classicalStrategyCorollary}.
\end{proof}

It may appear unrealistic to assume that the two STAR problems are essentially equal. Note however, that
this is indeed the case in the example of the XOR function and two mutually unbiased bases (e.g. computational and
Hadamard). The unitary here is just $U = \sigma_z \otimes \id^{\otimes n-1}$, as $\sigma_z$ acts as a bit-flip in the Hadamard
basis, but leaves the computational basis invariant. We saw in the proof of Theorem~\ref{evenOddPISTAR} that
such unitaries exist for any choice of two bases from the computational, Hadamard and K-basis. Indeed,
for the XOR function on a string of length $n$ with $n$ even we saw that for STAR the optimal probability
is $p = 3/4$, whereas for PI-STAR we obtained $p=1$, as expected. 

To generalize this approach, we need to consider more complicated inequalities. In general, there are many possibilities
for such inequalities, and one should choose an inequality that reflects the equivalences of possible STAR problems:
For example, for the XOR function the CHSH inequality is a good choice as we could identify $U = \sigma_z\otimes \id^{\otimes n-1}$
to give us an equivalence between the two problems. Of course, we would like to ensure that for one of Bob's measurement
settings he needs to solve the original STAR problem where Bob's goal
is to determine Alice's measurement outcome. 
At the same time, we would like to minimize the number of
possibly inequivalent additional STAR problems created in a similar proof, i.e. we would like to find an inequality 
where Bob has only a small number of measurement settings. As an example, we consider the following easy way to extend
the CHSH inequality. Here, we assume that Alice has equally many measurement outcomes as she has measurement settings. In the language of PI-STAR that means
we have wlog $A = \mY = S = \mB$. We fix the number of Bob's measurement settings to 2, but allow an 
arbitrarily large number of settings $|S| = |\mB|$ for Alice. Wlog we use $T = \01$ and $S = \{0,\ldots,|\mB|-1\}$.
We now define the predicate $V$ with the help of the function $\tau^s_t$ for $s \in S$ and $t \in T$.
Let $\tau^s_0(y) = y$ for all $s \in S$ and let $\tau^0_1(y) = y$ and $\tau^s_1(y) = \sigma^s(y)$ for all $s \in \{1,\ldots,|\mB|-1\}$, where $\sigma = (1,\ldots,|\mB|-1)$ is the cyclic permutation. We now define the inequality as a non-local game with predicate $V(a,b|s,t) = 1$ if and 
only if $b = \tau^s_t(a)$. Intuitively, this means that if Bob chooses setting $t=0$ he is required to solve
the original STAR problem, where he tries to guess Alice's measurement outcome. For the setting $t=1$ he has to solve the problem where the values of $y$ are shifted
depending on the basis. Note that the CHSH inequality is a special case of this inequality. Recall from Section~\ref{nonlocalGames} that the optimal value of a classical game can always be attained by a deterministic strategy. 
Let $f_A: S \rightarrow \mY$ and $f_B: T \rightarrow \mY$ denote the functions implementing this strategy for Alice
and Bob respectively. Looking at Eq.~(\ref{classicalGameValue}) 
we see that we can write 
$$
\omega_c(G) = \max \frac{1}{2|\mB|} \sum_{t,s} [\tau^s_t(f_A(s)) = f_B(t)],
$$ 
where
$[x=y]=1$ if and only if $x=y$.
It is easy to see that for a uniform choice of
Alice and Bob's measurements, the best thing Bob can do is answer $f_B(t) = x$ for all $t$ where we choose
any fixed $x \in \mY$ and let $x = f_A(s)$ for all $s \in S$, i.e.\ Alice and Bob agree on a particular outcome
which will always be their answer regardless of their setting. For $t=0$ this means that Bob is always correct,
and for $t=1$ he will be correct if Alice obtained $a=0$. We then have $\omega_c(G) = (|\mB| + 1)/(2|\mB|)$.
For the CHSH case, this gives us $\omega_c(G) = 3/4$ as expected. It is now possible to make a similar 
statement
then in Lemma~\ref{pistarGapLemma} for a bigger PI-STAR problem.

The connection to Bell inequalities helped us understand the case where there exists a clear gap between 
the two problems. Here, post-measurement information was extremely helpful to us. However, as we saw in Chapter~\ref{pistarChapter} 
there do exist cases where post-measurement information is entirely useless: we can do equally well without it, if we cannot
store any quantum information. Interestingly, in the example of the XOR function on an odd number of input bits this happens
exactly when the corresponding states correspond to a measurement that maximally violates CHSH. We have thus 
reached an extremal point of our problem. Is it possible to find conditions on a set of states which determine when post-measurement information is indeed useful?

\section{Conclusion}

As we saw, entanglement is an inherent aspect of quantum theory. We can experimentally violate Bell's inequality, 
because we can indeed measure non-commuting observables. The existence of such violations is, next to uncertainty relations and locking,
another consequence of the existence of non-commuting measurements within quantum theory.
This illustrates their close link to uncertainty relations, locking and even post-measurement information 
we encountered in the preceding chapters. In essence, in all these tasks we are faced with exactly the same 
problem: what are the consequences of non-commuting measurements? And how can we find maximally ``incompatible'' 
measurements? 

In the following chapters, we examine entanglement from a variety of viewpoints. In Chapter~\ref{optimalStrategiesChapter}, we first consider Bell inequalities, and show how to find upper bounds on their violation in a quantum setting. Our approach allows us to find the optimal measurements for any bipartite correlation inequality with two-outcome measurements in a very easy manner. We then consider more general multipartite inequalities. Sadly, our method does not easily apply for more general inequalities. In fact, it is not even clear how large our optimization problem would have to be. We therefore consider a related problem in Chapter~\ref{dimensionalityChapter}: Given a probability distribution over measurement outcomes, how large a state do we need to implement such a strategy? We prove a very weak lower bound on the dimension on the 
resulting state for a very restricted class of games. 
Finally, we consider the effects that entanglement has on classical protocols in Chapter~\ref{interactiveProofsChapter}. To this end we examine interactive proof systems where the two provers are allowed to share entanglement. Surprisingly, it turns out that two such provers can be simulated by just a single quantum prover.

\chapter{Finding optimal quantum strategies}\label{optimalStrategiesChapter}
\label{optimalStrategies}\label{chapter:optimalStrategies}
\index{quantum strategy}\index{games!optimal strategies}

In the previous chapter, we encountered the CHSH inequality and its generalizations in the guise of
quantum games. Tsirelson has proven an upper bound on the CHSH inequality that can be achieved using a quantum 
strategy. But how can we prove upper bounds for more general inequalities? Or actually, how can we find the optimal
measurement strategy? In this chapter, we answer these questions for a restricted class of
inequalities by presenting a method that yields the optimal strategy for any 
two-player correlation inequality with 
$n$ measurement settings and two measurement outcomes, i.e. an XOR-game.

\section{Introduction}

Optimal strategies for generalized inequalities not only have applications
in computer science with regard to interactive proof systems, but may also be important to ensure security in 
cryptographic protocols. From a physical perspective finding such bounds may also be helpful.
As Braunstein and Caves~\cite{braunstein:inequ} have shown, it is interesting to consider
inequalities based on many measurement settings, in particular, the chained CHSH inequality in Eq.~\ref{chainedCHSH} below:
Here, the gap between the classical and the quantum bound 
is larger than for the original CHSH inequality 
with only two measurement settings. This can be helpful in real experiments
that inevitably include noise, as this inequality
leads to a larger gap achieved by 
the optimal classical and the quantum strategy, and may thus lead to a better test. 
However, determining bounds on the correlations that \emph{quantum} theory allows remains a difficult 
problem~\cite{massar:tsirel}. All Tsirelson-type bounds are known for 
correlation inequalities with two measurement settings and two outcomes for both Alice 
and Bob~\cite{tsirel:hadron}. Landau~\cite{landau:compat} has taken a step towards finding Tsirelson-type bounds
by considering when two-party correlations of two measurement settings for both Alice and Bob can
be realized using quantum measurements.
Filipp and Svozil~\cite{filipp:bell} have considered the case of
three measurement settings analytically and conducted numerical studies for a larger number of settings.
Werner and Wolf~\cite{werner:bellBound} also considered obtaining Tsirelson-type bounds for two-outcome 
measurements for multiple parties and studied the case of three and four settings explicitly. However, their method is 
hard to apply to general inequalities.
Finally, Buhrman and Massar have shown a bound for a generalized CHSH inequality using
three measurement settings with three outcomes each~\cite{massar:tsirel}. It is not known whether this
bound can be attained.

Our approach is based on semidefinite programming in combination with Tsirelson's seminal  
results~\cite{tsirel:original,tsirel:separated,tsirel:hadron} as outlined in Section~\ref{tsirelsonConstruction}. See Appendix~\ref{appendix:semidef} for a brief introduction to semidefinite programming.
It is very easy to apply and gives tight bounds as we can find the optimal measurements explicitly.
Let $X$ and $Y$ be Alice's and Bob's observables, and let $\ket{\Psi}$ be a state shared by Alice and Bob.
The key benefit we derive from Tsirelson's construction\index{Tsirelson!construction} is that it saves us from the need to maximize over all
states $\ket{\Psi}$ and observables. Instead, we can replace any terms of the form 
$\bra{\Psi}X \otimes Y\ket{\Psi}$ with the inner product of two real unit vectors $x\cdot y$, and
then maximize over all such vectors instead.
Our method is thereby similar to methods used in computer science for the two-way partitioning problem~\cite{boyd:book} 
and the approximation algorithm for MAXCUT by Goemans and Williamson~\cite{goemans:maxcut}. 
Semidefinite programming allows for an efficient way to approximate Tsirelson's 
bounds for any CHSH-type inequalities numerically. However, it can also be used to prove Tsirelson type bounds
analytically. As an illustration, we first give an alternative proof of Tsirelson's original bound using
semidefinite programming. We then prove a new Tsirelson-type bound for the following 
generalized CHSH inequality~\cite{peres:book,braunstein:inequ}. Classically, it can be shown that
\begin{equation}\label{chainedCHSH}
|\sum_{i = 1}^n \ev{X_i}{Y_i} + \sum_{i = 1}^{n-1} \ev{X_{i+1}}{Y_i} - \ev{X_1}{Y_n}| \leq 2n - 2.
\end{equation}
Here, we show that for quantum mechanics
$$
|\sum_{i = 1}^n \ev{X_i}{Y_i} + \sum_{i = 1}^{n-1} \ev{X_{i+1}}{Y_i} - \ev{X_1}{Y_n}| 
\leq 2 n\cos\left(\frac{\pi}{2n}\right),
$$
where $\{X_1,\ldots,X_n\}$ and $\{Y_1,\ldots,Y_n\}$ are observables with eigenvalues $\pm 1$
employed by Alice and Bob respectively, corresponding to their $n$ possible measurement settings.
It is well known that this bound can be achieved~\cite{peres:book,braunstein:inequ} for a specific 
set of measurement settings if Alice and Bob share a singlet state. Here, we show that this bound
is indeed optimal for \emph{any} state $\ket{\Psi}$ and choice of measurement settings. This method
generalizes to other CHSH inequalities, for example, the inequality 
considered by Gisin~\cite{gisin:chsh}. 

\section{A simple example: Tsirelson's bound}\label{tsirelBoundSection}\index{Tsirelson!bound}

To illustrate our approach we first give a detailed proof of Tsirelson's bound 
using semidefinite programming. This proof is more complicated than Tsirelson's original 
proof. However, it serves as a good introduction to the following section.
Let $X_1,X_2$ and $Y_1,Y_2$ denote the observables with eigenvalues $\pm 1$
used by Alice and Bob respectively. Our goal is now to show an upper bound for\index{CHSH inequality}
$$
|\ev{X_1}{Y_1} + \ev{X_1}{Y_2} + \ev{X_2}{Y_1} - \ev{X_2}{Y_2}|.
$$
From Theorem~\ref{tsirel} we know that there exist
real unit vectors $x_s,y_t \in \Real^4$ such that for all $s,t \in \01$
$\ev{X_s}{Y_t} = x_s \cdot y_t$. 
In order to find Tsirelson's bound, we thus want to solve the following
problem: 
maximize $x_1 \cdot y_1 + x_1 \cdot y_2 + x_2 \cdot y_1 - x_2 \cdot y_2$, 
subject to $\norm{x_1} = \norm{x_2} = \norm{y_1} = \norm{y_2} = 1$. Note that we can
drop the absolute value since any set of vectors maximizing the above equation, simultaneously
leads to a set of vectors minimizing it by taking $-y_1,-y_2$ instead.
We now phrase this as a semidefinite program. Let $G = [g_{ij}]$ be the 
Gram matrix of the vectors $\{x_1,x_2,y_1,y_2\} \subseteq \Real^4$ with respect to the inner product:
$$
G = 
\left(\begin{array}{cccc}
x_1 \cdot x_1 & x_1 \cdot x_2 & x_1 \cdot y_1 & x_1 \cdot y_2 \\
x_2 \cdot x_1 & x_2 \cdot x_2 & x_2 \cdot y_1 & x_2 \cdot y_2\\
y_1 \cdot x_1 & y_1 \cdot x_2 & y_1 \cdot y_2 & y_1 \cdot y_2\\
y_2 \cdot x_1 & y_2 \cdot x_2 & y_2 \cdot y_1 & y_2 \cdot y_2
\end{array}\right).
$$
$G$ can thus be written as $G = B^TB$ where the columns of $B$ are the vectors
$\{x_1,x_2,y_1,y_2\}$. By~\cite[Theorem 7.2.11]{horn&johnson:ma} we can write $G = B^TB$ if
and only if $G$ is positive semidefinite.
We thus impose the constraint that $G \geq 0$.
To make sure that we obtain unit vectors, we add the constraint that
all diagonal entries of $G$ must be equal to $1$.
Define
$$
W = \left(\begin{array}{cccc} 0 &0& 1& 1\\
0  & 0 & 1 & -1\\
1 & 1 & 0 & 0\\
1 & -1 & 0 & 0
\end{array}\right).
$$
Note that the choice of order of the vectors in $B$ is not unique, however, a different order
only leads to a different $W$ and does not change our argument.
We can now
rephrase our optimization problem as the following SDP:
\begin{sdp}{maximize}{$\frac{1}{2}\Tr(GW)$}
&$G \geq 0$
and $\forall i, g_{ii} = 1$
\end{sdp}
We can then write for the Lagrangian 
$$
L(G,\lambda) = \frac{1}{2}\Tr(GW) - \Tr(\diag(\lambda)(G - I)),
$$
where $\lambda = (\lambda_1, \lambda_2, \lambda_3,\lambda_4)$.
The dual function is then
\begin{eqnarray*}
g(\lambda) &=& \sup_G \Tr\left(G\left(\frac{1}{2}W - \diag(\lambda)\right)\right) + \Tr(\diag(\lambda))\\
           &=& \left\{\begin{array}{ll}
                       \Tr(\diag(\lambda)) & \mbox{if } \frac{1}{2}W - \diag(\lambda) \preceq 0\\[0.5mm]
                       \infty & \mbox{otherwise}
                      \end{array} \right.
\end{eqnarray*}
We then obtain the following dual formulation of the SDP
\begin{sdp}{minimize}{$\Tr(\diag(\lambda))$}
&$-\frac{1}{2}W + \diag(\lambda) \geq 0$
\end{sdp}
Let $p'$ and $d'$ denote optimal values for the primal and Lagrange dual 
problem respectively. From weak duality it follows that $d' \geq p'$. 
For our example, it is not difficult to see that this is indeed true as
we show in Appendix~\ref{appendix:semidef}.

In order to prove Tsirelson's bound, we now exhibit an optimal solution for both the primal
and dual problem and then show that the value of the primal problem equals the value of the dual problem.
The optimal solution is well 
known~\cite{tsirel:original,tsirel:separated,peres:book}.
Alternatively, we could easily guess the optimal solution based on numerical 
optimization by a small program for
Matlab\footnote{See {http://www.cwi.nl/\~{}wehner/tsirel/} for the Matlab example code.} and the package 
SeDuMi~\cite{sedumi} for semidefinite programming. Consider the following solution for the
primal problem
$$
G' = \left( \begin{array}{cccc}
              1 & 0 & \frac{1}{\sqrt{2}} & \frac{1}{\sqrt{2}}\\
              0 & 1 & \frac{1}{\sqrt{2}} & -\frac{1}{\sqrt{2}}\\
              \frac{1}{\sqrt{2}} & \frac{1}{\sqrt{2}} & 1 & 0\\
              \frac{1}{\sqrt{2}} & - \frac{1}{\sqrt{2}} & 0 & 1\\
            \end{array}
     \right),
$$
which gives rise to the primal value $p' = \frac{1}{2}\Tr(G'W) = 2\sqrt{2}$.
Note that $G' \geq 0$ since all its eigenvalues are non-negative~\cite[Theorem 7.2.1]{horn&johnson:ma},
and all its diagonal entries are 1. Thus all constraints are satisfied.
The lower left quadrant of $G'$ is in fact the same as the well known 
correlation matrix for 2 observables~\cite[Equation 3.16]{tsirel:hadron}.
Next, consider the following solution for the dual problem
$$
\lambda' = \frac{1}{\sqrt{2}} \left(1,1,1,1\right).
$$
The dual value is then $d' = \Tr(\diag(\lambda')) = 2\sqrt{2}$.
Because $-W + \diag(\lambda') \geq 0$, $\lambda'$ satisfies the constraint.
Since $p' = d'$, $G'$ and $\lambda'$ are in fact optimal solutions for
the primal and dual respectively. We can thus conclude that
$$
|\ev{X_1}{Y_1} + \ev{X_1}{Y_2} + \ev{X_2}{Y_1} - \ev{X_2}{Y_2}| \leq 2\sqrt{2},
$$
which is Tsirelson's bound~\cite{tsirel:original}. By Theorem~\ref{tsirel}, this
bound is achievable.

\section{The generalized CHSH inequality}\label{generalSemiDef}\index{CHSH inequality!generalized}

We now show how to obtain bounds for inequalities based on more than 2 observables
for both Alice and Bob. In particular, we prove a bound for the chained
CHSH inequality 
for the quantum case.
It is well known~\cite{peres:book} that it is possible to choose observables
$X_1,\ldots,X_n$ and $Y_1,\ldots,Y_n$, and the maximally entangled state, 
such that
$$
|\sum_{i = 1}^n \ev{X_i}{Y_i} + \sum_{i = 1}^{n-1} \ev{X_{i+1}}{Y_i} - \ev{X_1}{Y_n}|
= 2 n \cos\left(\frac{\pi}{2n}\right).
$$
We now show that this is optimal. Our proof is similar to the last section.
However, it is more difficult to show feasibility for
all $n$.
\begin{theorem}
Let $\rho \in \mA \otimes \mB$ be an arbitrary state, where $\mA$ and $\mB$
denote the Hilbert spaces of Alice and Bob. Let $X_1,\ldots,X_n$ and $Y_1,\ldots,Y_n$ 
be observables with eigenvalues $\pm 1$ on $\mA$ and $\mB$ respectively. 
Then 
$$
|\sum_{i = 1}^n \ev{X_i}{Y_i} + \sum_{i = 1}^{n-1} \ev{X_{i+1}}{Y_i} - \ev{X_1}{Y_n}| 
\leq 2 n\cos\left(\frac{\pi}{2n}\right),
$$
\end{theorem}

\begin{proof}
By Theorem~\ref{tsirel}, our goal is to find the maximum value for
$
x_1 \cdot y_1 + x_2 \cdot y_1 + x_2 \cdot y_2 + x_3 \cdot y_2 + \ldots
+ x_n \cdot y_n - x_1 \cdot y_n,
$
for real unit vectors $x_1,\ldots,x_n,y_1,\ldots,y_n \in \Real^{2n}$. 
As above we can drop the absolute value. 
Let $G = [g_{ij}]$ be the Gram matrix of the vectors 
$\{x_1,\ldots,x_n,y_1,\ldots,y_n\} \subseteq \Real^{2n}$.
As before, we can thus write $G = B^TB$, where the 
columns of $B$ are the vectors $\{x_1,\ldots,x_n,y_1,\ldots,y_n\}$, if
and only if $G\geq 0$.
To ensure we obtain unit vectors, we again demand that 
all diagonal entries of $G$ equal $1$. 
Define $n \times n$ matrix $A$ and $2n \times 2n$ matrix $W$ by

\begin{eqnarray*}
A = \left(\begin{array}{ccccc}
            1 & 1 & 0 & \ldots &  0\\
            0 & 1 & 1 &  &  \vdots \\
            \vdots &  & \ddots & \ddots &  0\\
            0  &  & &               1 & 1\\        
            -1 & 0 & \ldots &  0 & 1\\
          \end{array}
    \right),\mbox{~}
W = \left(\begin{array}{cc} 
            0 & A^{\dagger}\\
            A & 0 
          \end{array}
     \right).
\end{eqnarray*}
We can now phrase our maximization problem as the following SDP:
\begin{sdp}{maximize}{$\frac{1}{2}\Tr(GW)$}
&$G \geq 0$
and $\forall i, g_{ii} = 1$
\end{sdp}
Analogous to the previous section,
the dual SDP is then:
\begin{sdp}{minimize}{$\Tr(\diag(\lambda))$}
&$-\frac{1}{2}W + \diag(\lambda) \geq 0$
\end{sdp}
Let $p'$ and $d'$ denote optimal values for the primal and dual 
problem respectively. As before, $d' \geq p'$. 

\paragraph{Primal}
We now show
that the vectors suggested in~\cite{peres:book} are optimal. For $k \in [n]$, choose 
unit vectors $x_k,y_k \in \Real^{2n}$ to be of the form
\begin{eqnarray*}
x_k &=& (\cos(\phi_k), \sin(\phi_k),0,\ldots,0),\\
y_k &=& (\cos(\psi_k), \sin(\psi_k),0,\ldots,0),
\end{eqnarray*}
where $\phi_k = \frac{\pi}{2n}(2k - 2)$ and $\psi_k = \frac{\pi}{2n}(2k - 1)$.
The angle between $x_k$ and $y_k$ is given by $\psi_k - \phi_k = \frac{\pi}{2n}$
and thus $x_k \cdot y_k = \cos\left(\frac{\pi}{2n}\right)$.
The angle between $x_{k+1}$ and $y_k$ is $\phi_{k+1} - \psi_k = \frac{\pi}{2n}$
and thus $x_{k+1} \cdot y_k = \cos\left(\frac{\pi}{2n}\right)$. 
Finally, the angle between
$-x_1$ and
$y_n$
is $\pi - \psi_n = \frac{\pi}{2n}$ and so $-x_1 \cdot y_n = \cos\left(\frac{\pi}{2n}\right)$.
The value of our primal problem is thus given by
$$
p' = \sum_{k=1}^n x_k \cdot y_k + \sum_{k=1}^{n-1} x_{k+1} \cdot y_k - x_1 \cdot y_{n}
= 2n \cos\left(\frac{\pi}{2n}\right).
$$
Let $G'$ be the Gram matrix constructed from all vectors $x_k,y_k$ as described earlier.
Note that our constraints are satisfied: $\forall i: g_{ii} = 1$ and $G' \geq 0$, because
$G'$ is symmetric and of the form $G' = B^TB$.

\paragraph{Dual}
Now consider the $2n$-dimensional vector
$$
\lambda' = \cos\left(\frac{\pi}{2n}\right) \left(1,\ldots,1\right).
$$
In order to show that this is a feasible solution to the dual problem, we have to prove
that $-\frac{1}{2}W + \diag(\lambda') \geq 0$ and thus the
constraint is satisfied. 
To this end, we first show that
\begin{claim}\label{aClaim}
The eigenvalues of $A$ are given by
$\gamma_s = 1 + e^{i\pi(2s+1)/n}$ with $s = 0,\ldots,n-1$.
\end{claim}
\begin{proof}
Note that if the lower left corner of $A$ were $1$, $A$ would
be a circulant matrix~\cite{circulant}, i.e.~each row of $A$ 
is constructed by taking the previous row and shifting it one place to the right.
We can use ideas from
circulant matrices to guess eigenvalues $\gamma_s$ with eigenvectors
$$
u_s = (\rho_s^{n-1},\rho_s^{n-2},\rho_s^{n-3},\ldots,\rho_s^1,\rho_s^0),
$$
where $\rho_s = e^{-i\pi(2s+1)/n}$ and $s = 0,\ldots,n-1$. 
By definition, $u = (u_1,u_2,\ldots,u_n)$ is an eigenvector of $A$ with 
eigenvalue $\gamma$ if and only if $Au = \gamma u$. Here, $Au = \gamma u$
if and only if 
\begin{eqnarray*}
(i)&& 
\forall j \in \{1,\ldots,n-1\}:
u_j + u_{j+1} = \gamma u_j,\\
(ii)&& -u_1 + u_n = \gamma u_n.
\end{eqnarray*}
Since for any $j \in \{1,\ldots,n-1\}$
\begin{eqnarray*}
u_j + u_{j+1} 
&=&
\rho_s^{n-j} + \rho_s^{n-j-1} = 
\\
&=&
e^{-i(n-j)\pi(2s+1)/n}(1+e^{i\pi(2s+1)/n}) = \\
&=&
\rho_s^{n-j} \gamma_s = \gamma_s u_j,
\end{eqnarray*}
(i) is satisfied.
Furthermore 
(ii) is satisfied, since
\begin{eqnarray*}
-u_1 + u_n &=& - \rho_s^{n-1} + \rho_s^0 = \\
&=& 
- e^{-i \pi(2s+1)} e^{i\pi(2s+1)/n} + 1 = \\
&=&
1 + e^{i\pi(2s+1)/n} = \\
&=&
\gamma_s \rho_s^0 = \gamma_s u_n.
\end{eqnarray*}
\end{proof}

\begin{claim}\label{biggestVal}
The largest eigenvalue of $W$ is given by $\gamma = 2\cos\left(\frac{\pi}{2n}\right)$.
\end{claim}
\begin{proof}
By~\cite[Theorem 7.3.7]{horn&johnson:ma}, the eigenvalues of $W$ are given
by the singular values of $A$ and their negatives. It follows from
Claim~\ref{aClaim} 
that the singular values of $A$ are
$$
\sigma_s = \sqrt{\gamma_s \gamma_s^*} = 
\sqrt{2 + 2\cos\left(\frac{\pi(2s+1)}{n}\right)}.
$$
Considering the shape of the cosine function, it is easy to see that the largest 
singular value of $A$ is given by $\sqrt{2 + 2\cos(\pi/n)}
= \sqrt{4 \cos^2(\pi/(2n))}$,
the largest eigenvalue of $W$ is 
$\sqrt{2 + 2\cos(\pi/n)} =  2 \cos(\pi/(2n))$.
\end{proof}

Since $-\frac{1}{2}W$ and $\diag(\lambda')$ are both Hermitian, Weyl's 
theorem~\cite[Theorem 4.3.1]{horn&johnson:ma} implies that 
$$
\gamma_{min}\left(-\frac{1}{2}W + \diag(\lambda')\right)
\geq
\gamma_{min}\left(-\frac{1}{2}W\right) + \gamma_{min}\left(\diag(\lambda')\right),
$$
where $\gamma_{min}(M)$ is the smallest eigenvalue of a matrix $M$. It then follows 
from the fact that $\diag(\lambda')$ is diagonal and 
Claim~\ref{biggestVal} 
that
$$
\gamma_{min}\left(-\frac{1}{2}W + \diag(\lambda')\right)
 \geq 
-\frac{1}{2}\left(2\cos\left(\frac{\pi}{2n}\right)\right) + \cos\left(\frac{\pi}{2n}\right)
= 
0.
$$
Thus $-\frac{1}{2}W + \diag(\lambda') \geq 0$
and $\lambda'$ is a feasible solution to the dual problem. The value of the
dual problem is then
$$
d' = \Tr(\diag(\lambda')) = 2 n \cos\left(\frac{\pi}{2n}\right).
$$
Because $p' = d'$, $G'$ and $\lambda'$ are optimal solutions for
the primal and dual respectively,
which completes our proof.
\end{proof}

Note that for the primal problem we are effectively dealing with $2$-dimensional 
vectors, $x_k,y_k$. As we saw in Section~\ref{tsirelsonConstruction}, it follows from
Tsirelson's construction~\cite{tsirel:hadron} that in this case we just need a single EPR pair such that we can
find observables that achieve this bound. In fact, these vectors just determine the measurement directions
as given in~\cite{peres:book}.

\begin{figure}[h]
\begin{center}
\includegraphics[scale=0.7]{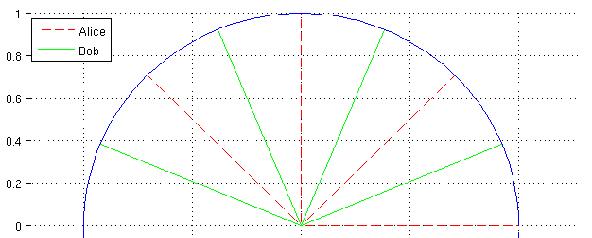}
\caption{Optimal vectors for $n=4$ obtained numerically using Matlab.}
\end{center}
\label{vectorsgenCHSH}
\end{figure}

\section{General approach and its applications}

\subsection{General approach}
Our approach can easily be generalized to other correlation inequalities.
For another inequality, we merely use a different matrix $A$ in $W$. For example, for Gisin's CHSH 
inequality~\cite{gisin:chsh}, $A$ is the matrix with 1's in the upper left half and on the diagonal, and
-1's in the lower right part. Otherwise our approach stays exactly the same, and thus we do not consider 
this case here. Numerical results provided by our Matlab example code suggest that Gisin's observables
are optimal. Given the framework of semidefinite programming, the only difficulty in proving bounds for 
other inequalities is to determine the eigenvalues of the corresponding $A$, a simple matrix. All bounds
found this way are tight, as we can always implement the resulting strategy using a maximally entangled
state as shown in Section~\ref{tsirelsonConstruction}.

With respect to finding numerical bounds, we see that the optimal strategy can be found in time exponential 
in the number of measurement settings: The size of the vectors scales exponentially with the number of settings, however,
we can fortunately find the optimal vectors in time polynomial in the length of the vectors using well-known
algorithms for semidefinite programming~\cite{boyd:book}.

\subsection{Applications}
In Chapter~\ref{interactiveProofsChapter}, we will see that the mere existence of such a semidefinite program has implications for
the computational complexity of interactive proof systems with entanglement. 
Cleve, H{\o}yer, Toner and
Watrous~\cite{cleve:nonlocal}
have also remarked during their presentation at CCC'04 that Tsirelson's constructions leads to an approach by semidefinite
programming in the context of multiple interactive proof systems with entanglement, but never gave
an explicit argument. 

The above semidefinite program has also been used to prove results about compositions of quantum games, in particular, parallel 
repetitions of quantum XOR-games~\cite{falk:paralell}. One particular type of composition studied by Cleve, Slofstra, Unger and
Upadhyay~\cite{falk:paralell} is the
XOR-composition of non-local games. For example, an XOR-composition of a CHSH game is a new game where Alice and Bob each 
have $n$ inputs $x_1,\ldots,x_n$ and $y_1,\ldots,y_n$ with $x_j,y_j \in \01$ and must give answers $a$ and $b$ such that
$a \oplus b = \bigoplus x_j\cdot y_j$. In terms of our semidefinite program, this is indeed easy to analyze. The matrix
defining the game is now given by
\begin{eqnarray*}
A = \left(\begin{array}{cc}
    1 & 1\\
    1 & -1
\end{array}\right)&
W = \left(\begin{array}{cc}
    0 & A^{\otimes n}\\
    A^{\otimes n} & 0
\end{array}\right).
\end{eqnarray*}
Note that the eigenvalues of $W$ are given by $\pm \sqrt{\gamma(A)\gamma(A)^*}$ where $\gamma(A) = \pm (\sqrt{2})^n$ is an eigenvalue 
of $A^{\otimes n}$. Consider the matrix $G = \id + W/(\sqrt{2})^n$. Clearly, $W/(\sqrt{2})^n$ has eigenvalues $\pm 1$
so we have $G \geq 0$. Thus $G$ is a valid solution to our primal problem, for which we obtain $p = \Tr(GW)/2 = (2\sqrt{2})^k$.
Consider $\lambda = (1,\ldots,1)((\sqrt{2})^n/2)$. Clearly, it is a valid solution to our dual problem
as $-W/2 + \diag(\lambda) \geq 0$, again using Weyl's theorem. This gives for our dual problem $d = \Tr(\diag(\lambda)) = (2 \sqrt{2})^n = p$ and thus our primal solution is optimal. For more general problems, such a composition may be more complicated
as the dual solution is not immediately related to the eigenvalues of $W$. Nevertheless, it can be readily evaluated using
Schur's complement trick~\cite{falk:paralell}. By rewriting, one can then relate such compositions to the questions of parallel repetition: Given multiple runs of the game, does there exist a better quantum measurement than executing the optimal strategy of each round many times? 
It is very interesting that this is in fact not true for XOR-games~\cite{falk:paralell}. However, there exist inequalities
and specific quantum states for which collective measurements are better. Such 
examples can be found in the works of Peres~\cite{peres:paralell} and
Liang and Doherty~\cite{andrew:paralell}. Sadly, our approach fails here.

\section{Conclusion}

We have provided a simple method to obtain the optimal measurements for any bipartite correlation inequality, i.e. any two-player XOR game. Our method easily allows us to obtain bounds using numerical analysis, but also suits itself to construct analytical proofs as demonstrated by our examples.
However, the above discussion immediately highlights the shortcomings of our approach. How can we find the optimal strategies for more generalized inequalities, where we have more than two players or a non-correlation inequality? Or more than two measurement outcomes? How can we find the optimal strategy for a fixed quantum state that is given to Alice and Bob? 
To address more than two measurement outcomes, we can rescale the observables such that they 
have eigenvalues in the interval $[-1,1]$. Indeed, examining Tsirelson's proof, it is easy to see
that we could achieve the same by demanding that the vectors have a length proportional to the number of settings. However, it is clear that the converse of Tsirelson's theorem that allows us to 
construct measurement from the vectors can no longer hold. Indeed, any matrix $M = \sum_j m_j X_j$ that can be written as a sum of anti-commuting matrices $X_1,\ldots,X_n$ with $X_j^2 = \id$ and $\sum_j m_j^2 = 1$ must have eigenvalues $\pm 1$ itself since $M^2 = \sum_{jk} m_j m_k X_j X_k = \left(\sum_j m_j^2\right)\id = \id$.

Since the completion of this work, exciting progress has been made to answer the above questions. 
Liang and Doherty~\cite{andrew:states} have shown how to obtain lower and upper bounds on the optimal strategy achievable using
a fixed quantum state using semidefinite programming relaxations. Kempe, Kobayashi, Matsumoto, Toner and Vidick~\cite{julia:mip} have since shown that there exist three-player games for which the optimal quantum strategy cannot be computed using a semidefinite program that is exponential in the number of measurement settings unless P=NP. Finally, Navascu{\'e}s, Pironio and Ac{\'i}n~\cite{acin:bell} have shown how to obtain bounds for general two-party inequalities with more measurement outcomes using semidefinite programming, inspired by Landau~\cite{landau:compat}. Their beautiful approach used successive hierarchies of semidefinite programs to obtain better and better bounds. In their approach, they consider whether a given distribution over outcomes can be obtained using a quantum strategy. Sadly, it does not give a general method to construct actual measurements and thus show that an obtained bound is tight. A similar result obtained
using an approach that is essentially dual to~\cite{acin:bell} has been
obtained in~\cite{andrew:qmp}, which also proves a convergence result
for such a hierarchy.

One of the difficulties we face when trying to find tight bounds for more general inequalities is to determine how large our optimization problem has to be. But even if we are given some distribution over possible outcomes, how can we decide how large our
system has to be in order to implement a quantum strategy? In general, this is a tricky problem which we will consider in the next chapter.

\chapter{Bounding entanglement in NL-games}
\label{boundingEntanglementChapter}\label{dimensionalityChapter}\label{chapter:bounding}\label{chapter:boundingEntanglement}

In the previous chapter, we provided a simple method to determine the optimal quantum strategy for
two-outcome XOR games. However, when trying to find the optimal strategies for
more general games, we are faced with a fundamental issue: 
How large do we have to choose our state and measurements such that
we can achieve the optimal quantum value?

\section{Introduction}

Determining an upper bound and the amount of entanglement we need, given the description of the game alone,
turns out to be a tricky problem in the general case. Hence, we address an intermediate problem: 
Given the description of a non-local game and associated probabilities, how large a state do Alice and Bob 
need to implement such a strategy? 
Navascu{\'e}s, Pironio and Ac{\'i}n~\cite{acin:bell} 
and also~\cite{andrew:qmp} have shown how to obtain upper bounds for the violation
of more general quantum games using multiple hierarchies of semidefinite programs. 
However, their method does not provide us with an explicit strategy, and it remains unclear how many levels of
the hierarchy we need to consider in order to obtain a tight bound. Yet, from their method we can obtain
a probability distribution over measurement outcomes. Using our approach, we can then determine an extremely weak lower bound
on the dimension of the quantum state we would need in order to implement a corresponding quantum strategy.

The idea behind our approach is to transform a non-local game
into a random access code. A random access code is an encoding of a string into a quantum
state such that we can retrieve at least one entry of our choice from this string with some probability.
Intuitively, Alice's measurements will create an encoding. Bob's choice of measurement then determines
which bit of this ``encoding'' he wants to retrieve. We prove a general lower bound for any independent one-to-one
non-local game among $n$ players, where a one-to-one non-local game is a game where for each possible
measurement setting there exists exactly one correct measurement outcome. In particular, we show that
in any one-to-one non-local game where player $P_j$ obtains the correct outcome $a \in A_j$ for any
measurement setting $s \in S_j$ with probability $p$
the dimension $d$ of player $P_j$'s
state obeys
$$
d \geq 2^{(\log|A_j|- H(p)-(1-p)\log(|A_j|-1))|S_j|}.
$$ 
Even though our bound is very weak, and the class of games very restricted, we are hopeful that our
approach may lead to stronger results in the future.
Finally, we discuss how we could obtain upper bounds from the description of the non-local game alone without resorting to probability distributions.

\section{Preliminaries}

Before we can prove our lower bound, we first introduce the notion of a random access code. For our purposes,
we need to generalize the existing results on random access codes. We use $M(\rho)$ to denote the random variable corresponding to the outcome 
of a measurement $M$ on a state $\rho$. We also use $A^n$ to denote an $n$-element string where each element is 
chosen from an alphabet $A$.
We will also use the notation $\vec{s}_{-j}$ to denote the string
$\vec{s} = (s_1,\ldots,s_n)$ without the element $s_j$.

\subsection{Random access codes}\index{random access code}\index{RAC}

A quantum $(n,m,p)$-random access code (RAC)~\cite{nayak:original,nayak:rac} over a binary 
alphabet is an encoding of an 
$n$-bit string $x$ into an $m$-qubit state $\rho_x$ such that for any $i \in [n]$ we can retrieve $x_i$ from $\rho_x$ with
probability $p$. Note that we are only interested in retrieving a single bit of the original string $x$ from
$\rho_x$. In general, it is unlikely we will be able to retrieve more than a single bit. For such codes the
following lower bound has been shown~\cite[Theorem 2.3]{nayak:rac}, where it is assumed that the original strings
$x$ are chosen uniformly at random:

\begin{theorem}[Nayak]\label{nayak}
Any $(n,m,p)$-random access code has $m \geq (1-H(p))n$. 
\end{theorem}

In the following, we make use of a generalization of random access codes to larger alphabets. 
We also need two additional generalizations:
First, we also want to obtain a bound on such a RAC encoding if the string $x$ is chosen from a slightly more general, 
possibly non-uniform, distribution. Let $P_{X_t}$ be a probability distribution over $\Sigma$ and let
$P_X = P_{X_1} \times \ldots \times P_{X_n}$ be a probability distribution 
over $\Sigma^n$. That is, a particular string $x$ is chosen with probability 
$P_X(x) = \Pi_{t=1}^n P_{X_t}(x_t)$. Note that we assume that the individual entries
of $x$ are chosen independently. 

Second, we allow for unbalanced random access codes, where each entry of the string $x$ may have a different
probability of being decoded correctly. We define
\begin{definition}\index{random access code!unbalanced}\index{URAC}
An $(n,m,(p_1,\ldots,p_n))_{|\Sigma|}$-unbalanced random access code (URAC) over a finite alphabet $\Sigma$ is an encoding of an 
$n$-element string $x \in \Sigma^n$ into an $m$-qubit state $\rho_x$ such that for any $t \in [n]$, there exists
a measurement $M_t$ with outcomes $\Sigma$ such that for all $x \in \Sigma^n$ we have $\Pr[M_t(\rho_x) = x_t] \geq p_t$.
\end{definition}

Fortunately, it is straightforward to extend the analysis of 
Nayak~\cite{nayak:rac} to this setting. We extend the proof by Nayak as opposed
to other known proofs of this lower bound in order to deal with unbalanced random access codes more easily.
\begin{lemma}\label{genRACS}
Let $P_X = P_{X_1} \times \ldots \times P_{X_n}$ be a probability distribution 
over $\Sigma^n$.
Then any $(n,m,(p_1,\ldots,p_n))_{|\Sigma|}$-unbalanced random access code 
has 
$$
m \geq \sum_{t=1}^{n} H(X_t) - H(p_t) - (1-p_t)\log(|\Sigma|-1)
$$
where $X_t$ is a random variable chosen from $\Sigma$ according to the probability distribution
$P_{X_t}$.
\end{lemma}
\begin{proof}
The proof follows along the same lines as Lemma 4.1 and Claim 4.6 of~\cite{nayak:rac}. We state the adaption 
for clarity:

We first consider decoding a single element. Let $\sigma_a$ with $a \in \Sigma$ be density matrices, 
and let $P$ be a probability distribution over $\Sigma$. Define $\sigma = \sum_{a \in \Sigma} P(a) \sigma_a$.
Let $M$ be a measurement with outcomes $\Sigma$ that given any state $\sigma_a$ gives the correct 
outcome $a$ with average probability $p$. Let $X$ be a random variable over $\Sigma$ chosen according to probability
distribution $P$, and let $Z$ be a random variable over $\Sigma$ corresponding to the 
outcome of the measurement. It now follows from Fano's inequality (see for example~\cite[Theorem 2.2]{hayashi:book}) that 
$\mI(X,Z) = H(X) - H(X|Z) \geq H(X) - H(p) - (1-p)\log(|\Sigma|-1)$.
Using Holevo's bound, we then have $S(\sigma) \geq \sum_{a \in \Sigma} P(a) S(\sigma_{a}) + 
H(X) - H(p) - (1-p)\log(|\Sigma|-1)$.

We now consider an entire string $x$ encoded as a state $\rho_x$. Consider $k$ with $n \geq k \geq 0$ and
define $\rho_y = \sum_{z \in \Sigma^{n-k}} q_z \rho_{zy}$ with $q_z = \Pi_{j=n-k}^n P_{X_j}(z_j)$
where we used indices $z = z_n,\ldots,z_{n-k}$ and $P_{X_j}$ to denote the probability
distribution over $\Sigma$ according to which the $j$-th entry was encoded.
We now claim that $S(\rho_y) \geq \sum_{a \in \Sigma} P_{X_{n-k}}(a) S(\rho_{ay})
+ H(X_{n-k}) - H(p_{n-k}) - (1-p_{n-k})\log(|\Sigma| - 1)$. 
The proof follows by downward induction over $k$: Consider $n=k$, 
clearly $S(\rho_y) \geq 0$ and the claim is valid. Now suppose our claim holds for $k+1$. Note that
we have $\rho_y = \sum_{a \in \Sigma} P_{X_{n-k}}(a) \rho_{ay}$. Note that strings encoded by the
density matrices $\rho_{ay}$
only differ by one element $a \in \Sigma$. We can therefore distinguish them with probability $p_{n-k}$.
From the above discussion we have that 
$S(\rho_y) \geq \sum_{a \in \Sigma} P_{X_{n-k}}(a) S(\rho_{ay})
+ H(X_{n-k}) - H(p_{n-k}) - (1-p_{n-k})\log(|\Sigma| - 1)$. 

Using the inductive hypothesis, letting $y$ be
the empty string and using the fact that $S(\rho) \leq \log d = m$ then completes the proof.
\end{proof}

\subsection{Non-local games and state discrimination}

For our purpose, we need to think of non-local games as a special form of state discrimination. When each subset of 
players performs a measurement on their part of the state, they effectively
prepare a certain state on the system of the remaining players.
Let $\chi_{\vec{a}_{-j}}^{\vec{s}_{-j}}$ denote the state of Player $P_j$ if the remaining players chose
measurement settings $\vec{s}_{-j}$ and obtained outcomes $\vec{a}_{-j}$. Note that the probability that player
$P_j$ holds $\chi_{\vec{a}_{-j}}^{\vec{s}_{-j}}$ is $\Pr[\vec{a}_{-j},
\vec{s}_{-j}] = 
\Pr[\vec{a}_{-j}|\vec{s}_{-j}] \Pi_{\ell = 1, \ell \neq j}^N \pi_\ell(s_\ell)$.
Define the state
$$
\zeta^{s_j}_{a_j} = \frac{1}{q^{s_j}_{a_j}} \left(\sum_{\vec{s}_{-j}} \sum_{\vec{a}_{-j}} V(\vec{a}|\vec{s}) 
\Pr[\vec{a}_{-j}|\vec{s}_{-j}] \chi_{\vec{a}_{-j}}^{\vec{s}_{-j}} \right)
$$
where $q^{s_j}_{a_j} = \sum_{\vec{s}_{-j}}\sum_{\vec{a}_{-j}} V(\vec{a}|\vec{s})
\Pr[\vec{a}_{-j}|\vec{s}_{-j}]$ to ensure normalization. We call a game \emph{independent}, if
the sets of probabilities
$\{q^{u}_{a_j} \mid a_j \in A_j\}$ and $\{q^{v}_{a_k}\mid a_k \in A_j\}$
are uncorrelated for all measurement settings $u,v \in S_j$ with $u \neq v$.
Note that $q^{s_j}_{a_j}$ is the probability
that player $P_j$ holds state $\zeta^{s_j}_{a_j}$, and that $\sum_{a_j \in A_j} q^{s_j}_{a_j} = 1$
since the game is one-to-one.
If player $P_j$ now chooses measurement setting $s_j$ he is effectively trying to solve a state discrimination 
problem, given the ensemble $\{q^{s_j}_{a_j}, \zeta^{s_j}_{a_j}| a_j \in A_j\}$.

Note that we already encountered this viewpoint in Chapter~\ref{piRelation}. Consider the simple case of the
CHSH game. Here, Alice (Player 1) and Bob (Player 2) had 
to give answers $a_1$ and $a_2$ for settings $s_1$ and $s_2$ such that
$s_1 \cdot s_2 = a_1 \oplus a_2$. Let $\zeta_{a_1}^{s_1}$ denote Bob's state if Alice chose
measurement setting $s_1$ and obtained outcome $a_1$.
If Bob chooses setting $s_2 = 0$, he has to solve the state discrimination problem
described by Figure~\ref{starProblem1Figure}: he must answer $a_2=a_1$, and hence his goal is to learn $a_1$. 
That is, he must solve the state discrimination problem given by $\rho_0 = (\zeta_{0}^{0} + \zeta_{0}^{1})/2$
and $\rho_1 = (\zeta_{1}^{0} + \zeta_{1}^{1})/2$. For $s_2=1$, he has to solve the problem given by 
Figure~\ref{starProblem2Figure}: For $s_1=0$, he must answer $a_2=a_1$, but for $s_1=1$ he must answer $a_2 \neq a_1$.
Hence, he must solve the state discrimination problem given by $\tilde{\rho}_0 = (\zeta_{0}^{0} + \zeta_{1}^{1})/2$
and $\tilde{\rho}_1 = (\zeta_{1}^{0} + \zeta_{0}^{1})/2$.

\section{A lower bound}

We now show how to obtain a random access encoding 
from a one-to-one non-local game. This enables us to find a lower bound on the dimension of the quantum
state necessary for any player $P_j$ to implement particular non-local strategies. Recall that 
we are trying to give a bound given all parameters of the game. In particular, we are given the probabilities
$\Pr[\vec{a}_{-j}|\vec{s}_{-j}]$ that the remaining players obtain outcomes $\vec{a}_{-j}$ 
for their measurement settings $\vec{s}_{-j}$, as well as the value of the game. 
Note that we do not need to know an actual state and measurement strategy for the players.
We just want to give a lower bound for a chosen set of parameters, whether these can be obtained or not. 

\begin{theorem}\label{conversion}
Any one-to-one independent non-local game where player $P_j$ obtains the correct outcome $a_j \in A_j$ for
measurement setting $s_j \in S_j$ with probability $p_{s_j}$ for all 
$\vec{s}_{-j} \in 
S_1\times\ldots\times S_{j-1}\times S_{j+1}\times\ldots\times S_N$ and 
$\vec{a}_{-j} \in 
A_1\times\ldots\times A_{j-1}\times A_{j+1} \times\ldots\times A_N$
is a $(|S_j|,m,(p_1,\ldots,p_{|S_j|}))_{|A_j|}$-unbalanced random access code. 
\end{theorem}
\begin{proof}
To encode a string, the other players choose measurement settings $\vec{s}_{-j}$
and measure their part of the state as in the non-local game
to obtain outcomes $\vec{a}_{-j}$. Note that the string is chosen randomly
by the measurement.
Since our game was one-to-one we can define a 
function 
$$
g(\vec{s}_{-j},\vec{a}_{-j}) = 
f_{1}(\vec{s}_{-j},\vec{a}_{-j}),\ldots,f_{|S_j|}(\vec{s}_{-j},\vec{a}_{-j}).
$$
Let $x = g(\vec{s}_{-j},\vec{a}_{-j})$ be the encoded string and note that 
$\rho_x = \chi_{\vec{a}_{-j}}^{\vec{s}_{-j}}$. We have $P_{X_t}(c) = q^{s_j}_c$,
since our game is one-to-one. Since our game is independent, we have that $P_X$ is a product
distribution.
To retrieve the $t$-th entry of $x$, player $P_j$ then has to distinguish 
$\zeta_{a_j}^{s_j}$ as in the non-local game which he can do with probability $p_{s_j}$ by assumption.
\end{proof}

Now that we can obtain a random access code from a non-local game, we can easily give a lower bound on
the dimension of the state from a lower bound of the size of the random access code. It follows immediately
from Theorem~\ref{conversion} and Lemma~\ref{genRACS} that

\begin{corollary}
In any one-to-one independent non-local game where player $P_j$ obtains the correct outcome $a \in A_j$ for
measurement setting $s \in S_j$ with probability $p_{s}$ for all 
measurement settings 
$\vec{s}_{-j} \in 
S_1\times\ldots\times S_{j-1}\times S_{j+1}\times\ldots\times S_N$ 
and outcomes
$\vec{a}_{-j} \in 
A_1\times\ldots\times A_{j-1}\times A_{j+1} \times\ldots\times A_N$ 
of the other players,
the dimension $d$ of player $P_j$'s
state obeys
$$
d \geq 2^{\sum_{t=1}^{|S_j|}H(X_t)-H(p_t)-(1-p_t)\log(|A_j|-1)},
$$ 
where $X_t$ is a random variable chosen from $A_j$ where $\Pr[X_t = a] = q^{t}_a$.
\end{corollary}

For almost all known games, we can obtain a simplified bound as each player will choose
a measurement setting uniformly at random. Likewise, in most cases we can assume that the probability
that the players obtain certain outcomes is also uniform. Indeed, if we do not know a particular measurement
strategy for a given game, we can find a bound if we assume that the distribution over the outcomes
given the choice of measurement settings is uniform. In this case, we also assume that the
probability of giving the correct answer is the same for each possible choice of measurement settings
and is equal to the value of the game. We then obtain

\begin{corollary}\label{corr:boundDim}
In any one-to-one independent non-local game where player $P_j$ obtains the correct outcome $a \in A_j$ for any
measurement setting $s \in S_j$ with probability $p$ where
$q^t_a = 1/|A_j|$ for all $t \in S_j$ and measurement settings $\vec{s}_{-j} \in 
S_1\times\ldots\times S_{j-1}\times S_{j+1}\times\ldots\times S_N$ and 
outcomes
$\vec{a}_{-j} \in 
A_1\times\ldots\times A_{j-1}\times A_{j+1} \times\ldots\times A_N$ 
of the other players,
the dimension $d$ of player $P_j$'s
state obeys
$$
d \geq 2^{(\log(|A_j|)- H(p)-(1-p)\log(|A_j|-1))|S_j|}.
$$ 
\end{corollary}

Note that if we are willing to assume that the optimal value of the game is achieved when the players share
a maximally entangled state, we can improve this bound to $d \geq \max_j 2^{(\log(|A_j|)- H(p)-(1-p)\log(|A_j|-1))|S_j|}$.

Let's look at a small example
which illustrates the proof. Consider the CHSH inequality. Here, we have only two players, Alice (Player 1) and 
Bob (Player 2).
Bob's goal is to obtain an outcome $a_2$ such that $s_1 \cdot s_1 = a_1 + a_2 \mod 2$. 
This means we define the function $g(s_1,a_1) = x$ as $g(0,0) = 0,0$, $g(1,0) = 1,1$, $g(0,1) = 1,0$ and 
$g(1,1) = 0,1$. For the lower bound we do not need to consider a
specific encoding, however, for the well-known CHSH state and measurements we would have an encoding of $\rho_{00} = \outp{0}{0}$, $\rho_{01} = \outp{-}{-}$, 
$\rho_{10} = \outp{+}{+}$, and $\rho_{11} = \outp{1}{1}$ and $q_{x_1}^1 = q_{x_2}^2 = 1/2$ for all $x_1,x_2 \in \01$.
How many qubits does Bob need to use if he wants to give the correct answers with probability $p = 1/2 + 1/(2\sqrt{2})$? 
Since everything is uniform we obtain $\log d \geq (1 - H(p))2 \approx 0.8$, i.e., Bob needs to keep at least 
one qubit.

Our bound contains a tradeoff between the probability $p$ of giving the correct answer, the number of measurement settings, and the number of possible outcomes. Clearly, our bound will only be good, if the number of measurement 
settings is large. It is also clear that it performs badly as $p$ approaches $1/2$ and $|A_j|$ is large, and thus for most cases our bound will be 
very unsatisfactory. The following figures illustrate
the tradeoff between the different parameters of Corollary~\ref{corr:boundDim}. 

\begin{figure}[h]
\begin{center}
\includegraphics[scale=0.3]{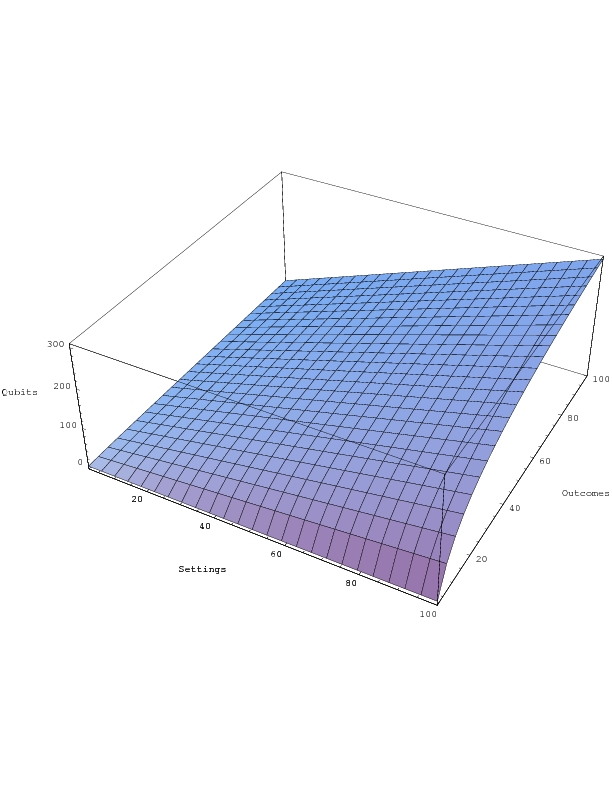}
\caption{Tradeoff for $p=0.6$.}
\label{figure:fixedPQubits}
\end{center}
\end{figure}
\begin{figure}[h]
\begin{center}
\includegraphics[scale=0.3]{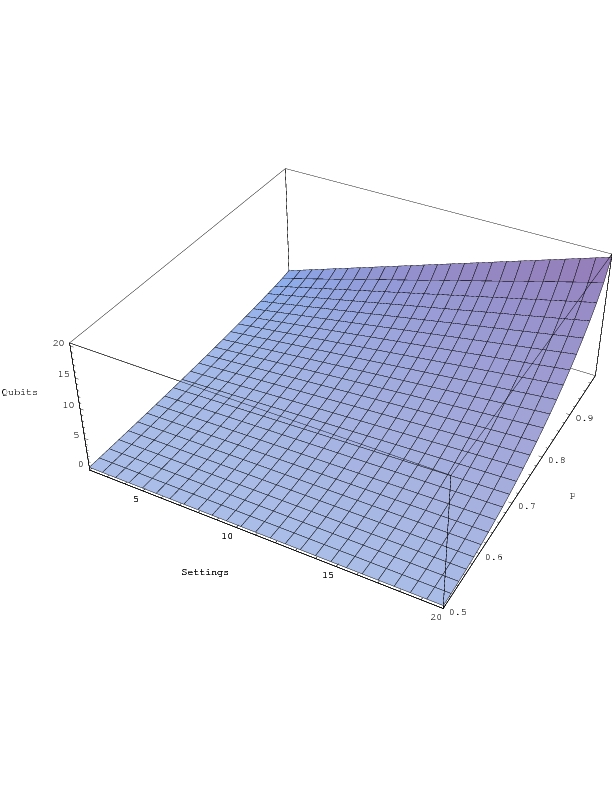}
\caption{Tradeoff for 2 outcomes.}
\label{figure:fixedOutcomes}
\end{center}
\end{figure}

\section{Upper bounds}

Ideally, we would find an upper bound on the amount of entanglement we need purely from the description of the game alone. 
Clearly, Tsirelson's construction from Chapter~\ref{tsirelsonConstruction} tells us that for any XOR game the local 
dimension of Alice's and Bob's system is $d \leq 2^{N/2}$, where $N$ is the number of measurement settings. 
Similarly to XOR games, we can consider mod $k$-games. Here, Alice and Bob have to give answers $a_1$, $a_2$
given questions $s_1$, $s_1$ such that $f(s_1,s_2) = a_1 + a_2 \mod k$ for 
some function $f: S_1 \times S_2 \rightarrow \{0,\ldots,k-1\}$. 
One may hope that for mod $k$-games, similarly than for XOR-games, the following
holds:
\begin{conjecture}
For any mod $k$-game, the dimension of Alice's and Bob's systems obeys $d \leq k^{N/2}$, where
$N$ is the number of measurement settings for Alice and Bob.
\end{conjecture}

An alternative approach to bounding the dimension would be to consider how far we can reduce the size
of an existing state and observables using Lemma~\ref{lemma:reducingGames}. Suppose that Alice has only two measurement
settings $X_0 = X_0^0 - X_0^1$ and $X_1 = X_1^0 - X_1^1$ with $X_0^0 + X_0^1 = \id$ and $X_1^0 + X_1^1 = \id$.
We know from Lemma~\ref{directsumdecomposition} that there exist projectors $\Pi_j$ such that
we can decompose $X_s$ as $X_s = \sum_j \Pi_j X_s^a \Pi_j$ for 
$s,b \in {0,1}$, where $\rank(\Pi_j) \leq 2$. Hence, we can immediately conclude from Lemma~\ref{lemma:reducingGames}
that if Alice only measures two possible observables with two outcomes each, the dimension of her state
does not need to exceed $d = 2$. This has previously been proved by Masanes~\cite{masanes:blocks}.
Could we prove something similar for three measurement settings? Sadly, Theorem~\ref{thm:threeSettings} tells
us that this is not possible! There do exist three measurements for which no such decomposition exists.
It is not hard to see that the question of how large Alice's entangled state has to be given a specific set of measurement operators is 
essentially equivalent to
the question of how many qubits we need to store in the problem of post-measurement information to achieve
perfect success. In both settings we are interested in reducing the dimension
by finding a way to block-diagonalize the matrices.

\section{Conclusion}

Bounding the amount of entanglement that we need to implement the optimal strategy in non-local games remains a tricky problem. We have given a simple lower bound on the amount of entanglement necessary for an extremely restricted class of games. The CHSH game forms an instance of such a game. Even though our bound is very weak, and the class of games quite restricted, we are hopeful that our approach may lead to stronger statements in the future.
We also showed how our earlier considerations and Tsirelson's construction led to an upper bound for the specific case of XOR-games. Sadly, better bounds still elude us so far.

\chapter{Interactive Proof Systems}
\label{interactiveProofsChapter}\index{interactive proof system}\label{chapter:interactiveProofs}

As we saw in the past chapters, two spatially separated parties, Alice and Bob, can use entanglement to obtain 
correlations that are impossible to achieve classically, without any additional communication. However, there do
exist classical systems whose strength, or security, indeed depends crucially on the fact that specific parties
cannot communicate during the course of the protocol. How are such systems affected by the presence of entanglement?
Can Alice and Bob use their shared entanglement to gain a significant advantage? Here, we study interactive proof systems
which are a specific case of such a classical system. Surprisingly, it turns out that the space-like separation is
lost alltogether and we can simulate two classical parties with just a single quantum one.

\section{Introduction}

\subsection{Classical interactive proof systems}

Before getting to the heart of the matter, we first need to take a closer look at interactive proof systems.
Classical interactive proof systems have received considerable 
attention~\cite{babai:ipNexp,benor:ip,cai:ip,feige:twoProverOneRound,lapidot:ip,feige:mip}
since their introduction by Babai~\cite{babai:ip} and
Goldwasser, Micali and Rackoff~\cite{goldwasser:ip} in 1985.
An interactive proof system takes the form of a protocol
of one or more rounds between two parties, a verifier and a prover. Whereas the prover 
is computationally unbounded, the verifier is limited to probabilistic polynomial time. Both 
the prover and the verifier have access to a common input string $x$. The goal of the prover is 
to convince the verifier that $x$ belongs to a pre-specified language $L$. 
The verifier's aim, on the other hand, is to determine whether the prover's claim is indeed valid. 
In each round, the verifier sends a polynomial (in $x$) size query to the prover, who returns a 
polynomial size answer. At the end of the protocol, the verifier decides to accept, and 
conclude $x \in L$, or reject based on the messages exchanged and his own private randomness.
A language has an interactive proof if there exists a verifier $V$
and a prover $P$ such that: If $x \in L$, the prover can always convince $V$ to accept.
If $x \notin L$, no strategy of the prover can convince $V$ to accept with non-negligible probability.
IP\index{IP} denotes the class of languages having an interactive proof system. 
Watrous~\cite{watrous:qip} first considered the notion of \emph{quantum} interactive proof systems.
Here, the prover has unbounded quantum computational power whereas the verifier is restricted
to quantum polynomial time. In addition, the two parties can now exchange quantum messages.
QIP is the class of languages having a quantum interactive proof system.
Classically, it is known that
$\mbox{IP} = \mbox{PSPACE}$~\cite{shamir:ipPspace,shen:ipPspace},
where $\mbox{PSPACE}$ is the class of languages decidable using only polynomial space.
For the quantum case, it has been shown that 
$\mbox{PSPACE} \subseteq \mbox{QIP} \subseteq \mbox{EXP}$~\cite{watrous:qip,kitaev&watrous:qip}.\index{QIP}
If, in addition, the verifier is given polynomial size quantum advice, the resulting class
$\mbox{QIP}/\mbox{qpoly}$ contains all languages~\cite{raz:quantumPCP}.
Let $\mbox{QIP}(k)$ denote the class where the prover and verifier are restricted to exchanging $k$ messages. 
It is known that $\mbox{QIP} = \mbox{QIP}(3)$~\cite{kitaev&watrous:qip} and
$\mbox{QIP}(1) \subseteq \mbox{PP}$~\cite{vyalyi:qma,mariott:qip},
where $\mbox{PP}$ is the class of all problems solvable by a probabilistic machine in polynomial time.
We refer to~\cite{mariott:qip} for an overview of the extensive work 
done on $\mbox{QIP}(1)$, also known as $\mbox{QMA}$.
Very little is known about 
$\mbox{QIP}(2)$ and its relation to either $\mbox{PP}$ or $\mbox{PSPACE}$. 

In multiple-prover interactive proof systems the verifier can interact with multiple, computationally
unbounded provers. Before the protocol starts, the provers are allowed to
agree on a joint strategy, however they can no longer communicate during the execution of the protocol. 
Let MIP\index{MIP} denote the class of languages having a \emph{multiple}-prover interactive proof system.
Here, we are especially interested in two-prover interactive proof systems as introduced
by Ben-Or, Goldwasser, Kilian and Widgerson~\cite{benor:ip}. 
Babai, Fortnow and Lund~\cite{babai:ipNexp}, and 
Feige and Lov\'asz~\cite{feige:mip} have shown that a language is in NEXP if and only if 
it has a two-prover one-round proof system, i.e., MIP[2] = NEXP. Feige and Lov\'asz have also shown that
a system using more than two-provers is thus no more powerful than a system with only two provers, i.e., MIP[2] = MIP.
Let $\oplus\mbox{MIP}[2]$\index{MIP[2]}
denote the restricted class where the verifier's output is a function of 
the XOR of two binary answers. Even for such
a system $\oplus \mbox{MIP}[2] = \mbox{NEXP}$, for certain
soundness and completeness parameters~\cite{cleve:nonlocal}. Classical multiple-prover interactive 
proof systems are thus more powerful than classical proof systems based on a single prover, assuming
$\mbox{PSPACE} \neq \mbox{NEXP}$. 

\subsection{Quantum multi-prover interactive proof systems}

Given the advent of quantum computing, one can also consider quantum interactive proof systems with \emph{multiple} provers.
These can be grouped into two categories:
First, one can consider provers and a verifier that are quantum themselves and can exchange quantum messages.
Kobayashi and Matsumoto have considered such \emph{quantum} multiple-prover interactive
proof systems which form an extension of quantum single prover interactive proof systems as described
above. 
Let $\mbox{QMIP}$ denote the resulting class. In particular, they showed that
$\mbox{QMIP} = \mbox{NEXP}$ if the provers do \emph{not} share quantum entanglement~\cite{kobayashi:mip}. 
If the provers share at most polynomially many entangled qubits the resulting class
is contained in $\mbox{NEXP}$~\cite{kobayashi:mip}.

Secondly, one can consider proof systems where all communication remains classical, but the provers
can share any entangled state as part of their strategy on which they are allowed to perform
arbitrary measurements.
Cleve, H{\o}yer, Toner and Watrous~\cite{cleve:nonlocal} have raised the question whether 
a \emph{classical} two-prover system is weakened in such a setting.
We write $\mbox{MIP}^*$ if the provers share entanglement.
The authors provide a number of examples which demonstrate that the soundness condition of 
a classical proof system can be compromised, i.e. the interactive proof system is 
weakened, when entanglement is used.
In their paper, it is proved that 
$\oplus\mbox{MIP}^*[2] \subseteq \mbox{NEXP}$.\index{$\oplus\mbox{MIP}^*$} Later, the same authors also showed 
that $\oplus\mbox{MIP}^*[2] \subseteq \mbox{EXP}$ using 
semidefinite programming~\cite{cleve:nonlocalTalk}. Entanglement thus clearly weakens an
interactive proof system, assuming $\mbox{EXP} \neq \mbox{NEXP}$. 

Intuitively, entanglement
allows the provers to coordinate their answers, even though they cannot use it to communicate.
By measuring the shared entangled state the provers can generate correlations which they can use 
to deceive the verifier. Tsirelson~\cite{tsirel:original,tsirel:separated} has shown that even quantum mechanics
limits the strength of such correlations, as we saw in Chapter~\ref{chapter:entanglementIntro}.
Recall that Popescu and Roehrlich~\cite{popescu:nonlocal,popescu:nonlocal2,popescu:nonlocal3} have raised 
the question why nature imposes such limits. To this end, they constructed a 
toy-theory based on non-local boxes~\cite{popescu:nonlocal,wim:thesis}, which are 
hypothetical ``machines'' generating
correlations stronger than possible in nature. In their full generalization, non-local boxes
can give rise to any type of correlation as long as they cannot be used to signal.
Preda~\cite{preda:talk} showed that sharing non-local boxes allows two provers
to coordinate their answers perfectly and obtained $\oplus\mbox{MIP}_{\mbox{\tiny{NL}}} = \mbox{PSPACE}$, 
where we write $\oplus\mbox{MIP}_{\mbox{\tiny{NL}}}$\index{$\oplus\mbox{MIP}_{\mbox{\tiny{NL}}}$} to indicate that the two provers share non-local boxes.

Kitaev and Watrous~\cite{kitaev&watrous:qip} mention that it is unlikely that a single-prover 
\emph{quantum} interactive proof system can simulate multiple classical provers, because then 
from $\mbox{QIP} \subseteq \mbox{EXP}$ and $\mbox{MIP} = \mbox{NEXP}$ it follows that
$\mbox{EXP} = \mbox{NEXP}$.

Surprisingly, it turns out that when the provers are allowed to share entanglement
it can be possible to simulate two such classical provers by one quantum prover. This indicates that
entanglement among provers truly leads to a weaker proof system. In particular, we show
that a two-prover one-round interactive proof system where the verifier computes the XOR
of two binary answers and the provers are allowed to share an arbitrary entangled state,
can be simulated by a single quantum interactive proof system with two 
messages: $\oplus\mbox{MIP}^*[2] \subseteq \mbox{QIP(2)}$\index{QIP(2)}. Since very little is known about
$\mbox{QIP}(2)$ so far~\cite{kitaev&watrous:qip}, we hope that our result may help shed
some light on its relation to $\mbox{PP}$ or $\mbox{PSPACE}$.
Our result also leads to a proof that $\oplus\mbox{MIP}^*[2] \subseteq \mbox{EXP}$.

\section{Proof systems and non-local games}

\subsection{Non-local games}

\begin{figure}[h]
\begin{center}
\includegraphics[scale=0.9]{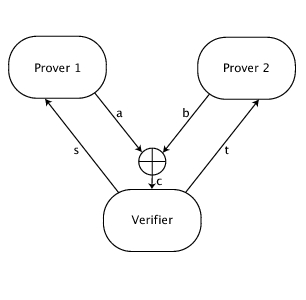}
\caption{A one-round XOR proof system.}
\label{xorProofSystemFigure}
\end{center}
\end{figure}

For our proof, it is necessary to link interactive proof systems to non-local games, as we
described in Chapter~\ref{nonlocalGames}. Since we consider only two parties, we omit 
unnecessary indices and use separate letters to refer to the sets of possible questions and answers.
We briefly recap our setup, summarized in Figure~\ref{xorProofSystemFigure}:
Let $S$, $T$, $A$ and $B$ be finite sets, and $\pi$ a probability distribution on $S \times T$.
Let $V$ be a predicate on $S \times T \times A \times B$. Then $G = G(V,\pi)$ is the following
two-person cooperative game\footnote{Players 1 and 2 collaborate against the verifier}: 
A pair of questions $(s,t) \in S \times T$ is chosen at random
according to the probability distribution $\pi$. Then $s$ is sent to player 1, henceforth called
Alice, and $t$ to player 2, which we call Bob. Upon receiving $s$, Alice has to reply
with an answer $a \in A$. Likewise, Bob has to reply to question $t$ with an answer $b \in B$.
They win if $V(s,t,a,b) = 1$ and lose otherwise. Alice and Bob may agree on any kind of strategy
beforehand, but they are no longer allowed to communicate once they have received questions $s$ and $t$.
The value $\omega(G)$ of a game $G$ is the maximum probability that Alice and Bob win the game.
We write $V(a,b|s,t)$ instead of $V(s,t,a,b)$ to emphasize
the fact that $a$ and $b$ are answers given questions $s$ and $t$. 

Here, we are particularly interested in non-local games. Alice and Bob are allowed to
share an arbitrary entangled state $\ket{\Psi}$ to help them win the game. Let $\hil^A$
and $\hil^B$ denote the Hilbert spaces of Alice and Bob respectively.
The state 
$\ket{\Psi} \in \hil^A \otimes \hil^B$ is
part of the quantum strategy that Alice and Bob can agree on beforehand. This means that for each game, 
Alice and Bob can choose a specific $\ket{\Psi}$ to maximize their chance of success.
In addition,
Alice and Bob can agree on quantum measurements. For each $s \in S$, Alice has a projective measurement described by 
$
\{X_s^a\mid a \in A\}
$ on $\hil^A$. For each $t \in T$, Bob has a projective measurement described
by 
$
\{Y_t^b\mid b \in B\}
$ 
on $\hil^B$.
For questions $(s,t) \in S \times T$, Alice 
performs the measurement corresponding to $s$ on her part of $\ket{\Psi}$ which gives her outcome 
$a$. Likewise, Bob performs the measurement corresponding to $t$ on his part of $\ket{\Psi}$ with outcome $b$.
Both send their outcome, $a$ and $b$, back to the verifier. The probability that Alice and Bob answer $(a,b) \in A \times B$ is then given by 
$$
\bra{\Psi}X_s^a \otimes Y_t^b\ket{\Psi}.
$$
The probability that Alice and Bob win the game is now given by
\begin{equation}\label{AliceandBobWin}
\Pr[\mbox{Alice and Bob win}] = \sum_{s,t} \pi(s,t) \sum_{a,b} V(a,b|s,t) \bra{\Psi}X_s^a \otimes Y_t^b\ket{\Psi}.
\end{equation}

The \emph{quantum value} $\omega_q(G)$ of a game $G$ is the maximum probability over all possible 
quantum strategies that Alice and Bob win. Recall that \emph{XOR game} is a game where the value of $V$ 
only depends on $c = a \oplus b$ and not on $a$ and $b$ independently. For XOR games we 
write $V(c|s,t)$ instead of $V(a,b|s,t)$. 
Here, we are only interested in the case that $a \in \01$ and $b \in \01$ and XOR games. Alice and Bob's
measurements are then described by $\{X_s^0,X_s^1\}$ for $s \in S$ and $\{Y_t^0,Y_t^1\}$ for $t \in T$ respectively. 
Note that $X_s^0 + X_s^1 = \id$
and $Y_t^0 + Y_t^0 = \id$ and thus these measurements can be
expressed in the form of observables $X_s$ and $Y_t$ with eigenvalues $\pm 1$: 
$X_s = X_s^0 - X_s^1$ and $Y_t =  Y_t^0 - Y_t^1$. 
Recall from Chapter~\ref{tsirelsonConstruction} that Tsirelson~\cite{tsirel:original,tsirel:separated} has shown that 
for any $\ket{\Psi} \in \hil^A \otimes \hil^B$ 
there exists real unit vectors $x_s,y_t \in \Real^N$ with $N = |S| + |T|$ such 
that $\bra{\Psi}X_s \otimes Y_t\ket{\Psi} = \inp{x_s}{y_t}$. 
It is then easy to see from Eq.~(\ref{AliceandBobWin}) that for XOR games we can express the maximum winning probability 
as
\begin{equation}\label{maxQuantumGame}\index{games!XOR, winning probability}
\omega_q(G) = \max_{x_s,y_t} \frac{1}{2}\sum_{s,t} \pi(s,t) \sum_{c} V(c|s,t) \left(1 + (-1)^{c} \inp{x_s}{y_t}\right),
\end{equation}
where the maximization is taken over all unit vectors $x_s,y_t \in \Real^N$.

\subsection{Multiple classical provers}
It is well known~\cite{cleve:nonlocal,feige:mip}, that two-prover one-round interactive 
proof systems with classical communication can be modeled as (non-local) games. Here, Alice and Bob 
take the role of the two 
provers. The verifier now poses questions $s$ and $t$, and evaluates the resulting answers.
A proof system associates with each string $x$ a game $G_x$, where $\omega(G_x)$ determines the 
probability that the verifier accepts
(and thus concludes $x \in L$). The string $x$, and thus the nature of the game $G_x$ is known to both
the verifier and the provers. Ideally, for all $x \in L$ the value of $\omega(G_x)$ is close to one,
and for $x \notin L$ the value of $\omega(G_x)$ is close to zero. It is possible to extend the game
model for MIP[2] to use a randomized predicate for the acceptance predicate $V$. This corresponds to $V$
taking an extra input string chosen at random by the verifier. However, known applications of MIP[2] proof
systems do not require this extension~\cite{feige:stateOfArt}. Our argument in Section~\ref{simulatingProver} 
can easily be extended to deal with randomized predicates. Since $V$ is not a randomized 
predicate in~\cite{cleve:nonlocal}, we follow this approach.

We concentrate on proof systems involving two provers, one round of communication,
and single-bit answers. The provers are computationally unbounded, but limited by the laws of
quantum physics. However, the verifier is probabilistic polynomial time bounded.
As defined by Cleve, H{\o}yer, Toner and Watrous~\cite{cleve:nonlocal},
\begin{definition}\index{$\oplus\mbox{MIP}[2]$}
For $0 \leq s < c \leq 1$, let $\oplus\mbox{MIP}_{c,s}[2]$ denote the class of all languages
$L$ recognized by a classical two-prover interactive proof system of the following form:
\begin{itemize}
\item They operate in one round, each prover sends a single bit in response to the verifier's question,
and the verifier's decision is a function of the parity of those two bits.
\item If $x \in L$ then there exists a strategy for the provers for which the 
probability that the verifier accepts is at least $c$ (the \emph{completeness} probability).
\item If $x \notin L$ then, whatever strategy the two provers follow, the probability that the verifier 
accepts is at
most $s$ (the \emph{soundness} probability).
\end{itemize}
\end{definition}

\begin{definition}
\index{$\oplus{\mbox{MIP}}^*[2]$}
For $0 \leq s < c \leq 1$, let $\oplus\mbox{MIP}^*_{c,s}[2]$ denote the class corresponding
to a modified version of the previous definition: all communication remains classical, but 
the provers may share prior quantum entanglement, which may depend on $x$, and perform 
quantum measurements.
\end{definition}
We generally omit indices $c$, $s$, unless they are explicitly relevant.

In Chapter~\ref{optimalStrategiesChapter}, we discussed how to find the optimal strategies for XOR-games.
In particular, we saw that we can determine the optimal value of $\omega_q(G_x)$ in time exponential 
in $\min(|S|,|T|)$ using semidefinite programming. This implies immediately that 
$\oplus\mbox{MIP}^* \subseteq \mbox{EXP}$, as was shown by 
Cleve, H{\o}yer, Toner and Watrous~\cite{cleve:nonlocal} during their presentation at CCC'04.
Here, we show something stronger, namely that $\oplus\mbox{MIP}^* \subseteq \mbox{QIP(2)}$.

\subsection{A single quantum prover}

Instead of two classical provers, we can also consider a system consisting of a single quantum prover $P_q$ 
and a quantum polynomial time verifier $V_q$ as defined by Watrous~\cite{watrous:qip}. Again, the quantum 
prover $P_q$ is computationally unbounded, however, he is limited by the laws of quantum physics.
The verifier and the prover can communicate over a quantum channel. In this thesis, we are only interested
in one round quantum interactive proof systems: the verifier sends a single quantum message to the prover,
who responds with a quantum answer. We here express the definition of $\mbox{QIP}(2)$~\cite{watrous:qip} in 
a form similar to the definition of $\oplus\mbox{MIP}^*$:

\begin{definition}\index{QIP(2)}
Let $\mbox{QIP}(2,c,s)$ denote the class of all languages
$L$ recognized by a quantum one-prover one-round interactive proof system of the following form:
\begin{itemize}
\item If $x \in L$ then there exists a strategy for the quantum prover for which the probability
that the verifier accepts is at least $c$.
\item If $x \notin L$ then, whatever strategy the quantum prover follows, the probability that
the quantum verifier accepts is at most $s$.
\end{itemize}
\end{definition}

\section{Simulating two classical provers with one quantum prover}\label{simulatingProver}

We now show that an interactive proof system where the verifier bases his decision only on the XOR of
two binary answers is in fact no more powerful than a system based on a single quantum prover.
The main idea behind our proof is to combine two classical queries into one quantum query,
and thereby simulate the classical proof system with a single quantum prover. Similar
techniques have been used to prove results about classical locally decodable codes~\cite{kerenidis&wolf:qldc,wehner04a}.
Recall that the two provers can use an arbitrary entangled state as part of their strategy. 

For our proof we make use of the fact that we can write the optimal value of the game as in Eq.~(\ref{maxQuantumGame}).

\begin{theorem}\label{xormipINqip}
\index{$\oplus{\mbox{MIP}}^*[2]$}
For all $s$ and $c$ such that $0 \leq s < c \leq 1$,
$\oplus\mbox{\emph{MIP}}_{c,s}^*[2] \subseteq \mbox{\emph{QIP}}(2,c,s)$.
\end{theorem}
\begin{proof}
Let $L \in \oplus\mbox{MIP}_{c,s}^*[2]$ and let $V_{e}$ be a verifier witnessing this fact.
Let $P_{e}^1$ (Alice) and $P_{e}^2$ (Bob) denote the two provers sharing entanglement.
Fix an input string $x$. As mentioned above, interactive proof systems can be modeled
as games indexed by the string $x$. It is therefore sufficient to show that there exists 
a verifier $V_{q}$ and a quantum prover $P_q$, such that $\omega_{sim}(G_x) = \omega_q(G_x)$, where
$\omega_{sim}(G_x)$ is the value of the simulated game. 

Let $s$,$t$ be the questions that $V_{e}$ sends to the two provers $P_{e}^1$ and $P_{e}^2$ 
in the original game. 
The new verifier $V_{q}$ now constructs the following state in $\mV \otimes \mM$
$$
\ket{\Phi_{init}} = \frac{1}{\sqrt{2}}(\underbrace{\ket{0}}_{\mV}\underbrace{\ket{0}\ket{s}}_{\mM} + 
\underbrace{\ket{1}}_{\mV}\underbrace{\ket{1}\ket{t}}_{\mM}),
$$
and sends register $\mM$ to the single quantum 
prover $P_{q}$.

We first consider the honest strategy of the prover. 
Let $a$ and $b$ denote the answers of the two classical provers to questions $s$ and $t$ 
respectively. The quantum prover now transforms the state to
$$
\ket{\Phi_{honest}} = \frac{1}{\sqrt{2}}((-1)^a \underbrace{\ket{0}}_{\mV}\underbrace{\ket{0}\ket{s}}_{\mM} + 
(-1)^b \underbrace{\ket{1}}_{\mV}\underbrace{\ket{1}\ket{t}}_{\mM}),
$$
and returns register $\mM$ back to the verifier. 
The verifier $V_{q}$ now performs
a measurement on $\mV \otimes \mM$ described by the following projectors
\begin{eqnarray*}
P_0 &=& \outp{\Psi^+_{st}}{\Psi^+_{st}} \otimes I\\
P_1 &=& \outp{\Psi^-_{st}}{\Psi^-_{st}} \otimes I\\
P_{reject} &=& I - P_0 - P_1,
\end{eqnarray*}
where $\ket{\Psi^{\pm}_{st}} = (\ket{0}\ket{0}\ket{s} \pm \ket{1}\ket{1}\ket{t})/\sqrt{2}$.
If he obtains outcome ``reject'', he immediately aborts and concludes that the quantum
prover is cheating. If he obtains outcome $m \in \01$, the verifier concludes
that $c = a \oplus b = m$. Note that $\Pr[m = a \oplus b|s,t] = 
\bra{\Phi_{honest}} P_{a \oplus b} \ket{\Phi_{honest}} = 1$, so the verifier can reconstruct
the answer perfectly. 

We now consider the case of a dishonest prover.  
In order to convince the verifier, the prover applies a transformation
on $\mM \otimes \mP$ and send register $\mM$ back to the verifier. We show that for any
such transformation the value of the resulting game is at most $\omega_q(G_x)$: Note that the 
state of the total system in $\mV \otimes \mM \otimes \mP$ can now be described as
$$
\ket{\Phi_{dish}} = \frac{1}{\sqrt{2}}(\ket{0}\ket{\phi_s} + \ket{1}\ket{\phi_t})
$$
where $\ket{\phi_s} = \sum_{u \in S' \cup T'} \ket{u} \ket{\alpha_u^s}$ and 
$\ket{\phi_t} = \sum_{v \in S' \cup T'} \ket{v} \ket{\beta_v^t}$ with
$S' = \{ 0s | s \in S\}$ and $T' = \{ 1t | t \in T\}$. Any transformation
employed by the prover can be described this way.
We now have that
\begin{eqnarray}\label{crude1}
\Pr[m = 0|s,t] = \bra{\Phi_{dish}}P_0\ket{\Phi_{dish}} &=& \frac{1}{4}(\inp{\alpha_s^s}{\alpha_s^s}
+ \inp{\beta_t^t}{\beta^t_t}) + \frac{1}{2}\Re(\inp{\alpha_s^s}{\beta_t^t})\\\label{crude2}
\Pr[m = 1|s,t] = \bra{\Phi_{dish}}P_1\ket{\Phi_{dish}} &=& \frac{1}{4}(\inp{\alpha_s^s}{\alpha_s^s}
+ \inp{\beta_t^t}{\beta^t_t}) - \frac{1}{2}\Re(\inp{\alpha_s^s}{\beta_t^t})
\end{eqnarray}
The probability that the prover wins is given by
$$
\Pr[\mbox{Prover wins}] = \sum_{s,t} \pi(s,t) \sum_{c \in \01} V(c|s,t) \Pr[m = c|s,t].
$$
The prover will try to maximize his chance of success by maximizing 
$\Pr[m = 0|s,t]$ or $\Pr[m = 1|s,t]$. We can therefore restrict ourselves to 
considering real unit vectors for which $\inp{\alpha_s^s}{\alpha_s^s} = 1$ and $\inp{\beta_t^t}{\beta_t^t} = 1$,
as the dimension of our vectors is directly determined by their number.
Hence, we may also assume that $\ket{\alpha_s^{s'}} = 0$ iff $s \neq s'$ and $\ket{\beta_t^{t'}} = 0$ iff $t \neq t'$:
any other strategy can lead to rejection and thus to a lower probability of success.
By substituting into Eqs.~(\ref{crude1}) and~(\ref{crude2}), it follows that the probability 
that the quantum prover wins the game (when he avoids rejection) is
\begin{equation}\label{prob}
\frac{1}{2} \sum_{s,t,c} \pi(s,t) V(c|s,t) (1 + (-1)^c \inp{\alpha_s^s}{\beta_t^t}).
\end{equation}
In order to convince the verifier, the prover's goal is to choose real 
vectors $\ket{\alpha_s^s}$ and $\ket{\beta_t^t}$ which maximize Eq.~(\ref{prob}).
Since in $\ket{\phi_s}$ and $\ket{\phi_t}$ we sum over $|S'| + |T'| = |S| + |T|$ 
elements respectively, the dimension of $\mP$ need not exceed $N = |S| + |T|$.
Thus, it is sufficient to restrict the maximization to vectors in $\Real^{|S| + |T|}$. 
Given Eq.~(\ref{prob}), we thus have 
$$
\omega_{sim}(G_x) = \max_{\alpha_s^s,\beta_t^t} \frac{1}{2} \sum_{s,t,c} \pi(s,t) 
V(c|s,t) (1 + (-1)^c \inp{\alpha_s^s}{\beta_t^t}),
$$
where the maximization is taken over vectors $\{\alpha_s^s \in \Real^N: s \in S\}$,
and $\{\beta_t^t \in \Real^N: t \in T\}$.
However, we know from Eq.~(\ref{maxQuantumGame}) that
$$
\omega_{q}(G_x) = \max_{x_s,y_t} \frac{1}{2} \sum_{s,t,c} \pi(s,t) V(c|s,t) (1 + (-1)^c \inp{x_s}{y_t})
$$
where the maximization is taken over unit vectors $\{x_s \in \Real^N: s \in S\}$ and 
$\{y_t \in \Real^N: t \in T\}$. We thus have
$$
\omega_{sim}(G_x) = \omega_q(G_x)
$$
which completes our proof.
\end{proof}

\begin{corollary}\label{plusmipINexp}
For all $s$ and $c$ such that $0 \leq s < c \leq 1$, $\oplus\mbox{\emph{MIP}}_{c,s}^*[2] \subseteq \mbox{\emph{EXP}}$.
\end{corollary}
\begin{proof}
This follows directly from Theorem~\ref{xormipINqip} and the result that 
$\mbox{QIP(2)} \subseteq \mbox{EXP}$~\cite{kitaev&watrous:qip}.
\end{proof}

\section{Conclusion}

As we have shown, the strength of classical systems can be weakened considerably in the presence of
entanglement. In our example above, we showed that the systems can be weakened so much that all space-like
separation is lost: we saw that two classical parties with entanglement are as powerful as
a single quantum party. 

It would be interesting to show that this result also holds for a proof system where the 
verifier is not restricted to computing the XOR of both answers, but some other Boolean function. 
However, the approach based on vectors from Tsirelson's results does not work for binary games.
Whereas it is easy to construct a single quantum query which allows the verifier to compute an arbitrary 
function of the two binary answers with some advantage, it thus remains unclear how the value of the 
resulting game is related to the value of a binary game.
Furthermore, mere classical tricks trying to obtain the value of a binary function from XOR itself seem to confer
extra cheating power to the provers.  

Examples of non-local games with longer answers~\cite{cleve:nonlocal}, such as the Kochen-Specker or the 
Magic Square game, seem to make it even easier for the provers to cheat by taking advantage of 
their entangled state. Furthermore, existing proofs that $\mbox{MIP}=\mbox{NEXP}$ break down if
the provers share entanglement. It is therefore an open question whether $\mbox{MIP}^* = \mbox{NEXP}$ or,
$\mbox{MIP}^* \subseteq \mbox{EXP}$. 

As described, non-locality experiments between two space-like separated observers, Alice and Bob, can be cast in the form 
of non-local games. For example, the experiment based on the well known CHSH inequality~\cite{chsh:nonlocal}, 
is a non-local game with binary answers of which the verifier computes the XOR~\cite{cleve:nonlocal}. Our 
result implies that this non-local game can be simulated in superposition by a single prover/observer:
Any strategy that Alice and Bob might employ in the non-local game can be mirrored by the single prover
in the constructed ``superposition game'', and also vice versa, due to Tsirelson's 
constructions~\cite{tsirel:original,tsirel:separated} mentioned earlier. 
This means that the ``superposition game''\index{superposition game}\index{games!superposition} corresponding to the non-local CHSH game is in fact 
limited by Tsirelson's inequality~\cite{tsirel:original}, even though it itself has no non-local character.
Whereas this may be purely coincidental, it would be interesting to know its physical interpretation, if any.
Perhaps it may be interesting to ask whether Tsirelson-type inequalities have any consequences on
local computations in general, beyond the scope of these very limited games.

\part{Consequences for Crytography}

\chapter{Limitations}\label{chapter:qsc}\index{bit commitment!limitations}\label{chapter:sc}\index{string commitment}\label{chapter:limitations}

Finally, we turn our attention to cryptographic protocols directly. As we saw in Chapter~\ref{chapter:cryptoIntro},
it is impossible to implement bit commitment even in the quantum setting!
In the face of the negative results, what can we still hope to achieve? 

\section{Introduction}

Here, we consider the task of committing to an entire string of
$n$ bits at once when both the honest player and the adversary have
unbounded resources. Since perfect bit commitment is impossible,
perfect bit string commitment is clearly impossible as well. Curiously, however,
we can still make interesting statements in the quantum setting, if we 
give both Alice and Bob a limited ability to cheat. That is, we allow Alice 
to change her mind about the committed string within certain limited
parameters, and allow Bob to gain some information about the committed
string. It turns out that it matters crucially how we measure Bob's information gain.

First, we introduce a
framework for the classification of bit string commitments in terms
of the length $n$ of the string, Alice's ability to cheat on at most
$a$ bits and Bob's ability to acquire at most $b$ bits of
information before the reveal phase. We say that Alice can cheat on
$a$ bits if she can reveal up to $2^a$ strings successfully. Bob's
security definition is crucial to our investigation: If $b$
determines a bound on his probability to guess Alice's string, then
we prove that $a + b$ is at least $n$. This implies that the trivial
protocol, where Alice's commitment consists of sending $b$ bits of
her string to Bob, is optimal. If, however, $b$ is a bound on the
accessible information that the quantum states contain about Alice's
string, then we show that non-trivial schemes exist. More precisely,
we construct schemes with $a=4\log n+O(1)$ and $b=4$. This is
impossible classically. We also present a simple, implementable, protocol,
that achieves $a=1$ and $b=n/2$. This protocol can furthermore be made
cheat-sensitive.
Quantum commitments of strings have previously been considered by
Kent~\cite{kent:sc}, who pointed out that in the quantum world
useful bit string commitments could be possible despite the no-go
theorem for bit commitment. 
His scenario differs significantly from
ours and imposes an additional constraint, which is not present in
our work: Alice does not commit to a superposition of strings.

\section{Preliminaries}

\subsection{Definitions}

We first formalize the notion of quantum string commitments in a quantum setting. 

\begin{definition}\label{securityDef}\index{QBSC}
\index{commitment!quantum bit string}\index{quantum bit string commitment}
\emph{An $(n,a,b)$-\emph{Quantum Bit String Commitment (QBSC)} is a
quantum communication protocol between two parties, Alice (the
committer) and Bob (the receiver), which consists of two phases:
\begin{itemize}
\item (Commit Phase) Assume that both parties are honest.
Alice chooses a string $x \in \01^n$ with probability $p_x$.
Alice and Bob communicate and at the end Bob holds state $\rho_x$.
\item (Reveal Phase)
If both parties are honest, Alice and Bob communicate and at the end
Bob outputs $x$. Bob accepts.
\end{itemize}
We have the following two security requirements:
\begin{itemize}
\item (Concealing)
If Alice is honest, then for any strategy of Bob
$$
\sum_{x \in \01^n}p^B_{x|x} \leq 2^{b},
$$ 
where
$p^B_{x|x}$ is the probability that Bob correctly guesses $x$ before
the reveal phase.
\item (Binding)
If Bob is honest, then for any strategy of Alice 
$$
\sum_{x \in \01^n} p^A_x \leq 2^a,
$$ 
where $p^A_x$ is the probability that Alice
successfully reveals $x$ (Bob accepts the opening of $x$).
\end{itemize}}
\end{definition}

Bob thereby accepts
the opening of a string $x$, if he performs a test depending on the individual
protocol to check Alice's honesty and concludes that she was indeed honest.
Note that quantumly, Alice can always
commit to a superposition of different strings without being
detected. Thus even for a perfectly binding bit string commitment
(i.e. $a=0$) we only demand that $\sum_{x \in \{0,1\}^n}p^A_x \leq 1$,
whereas classically one wants that $p^A_{x'}=\delta_{x, x'}$.
Note that our concealing definition reflects Bob's a priori
knowledge about $x$. We choose an a priori uniform distribution
(i.e. $p_x = 2^{-n}$) for $(n,a,b)$-QBSCs, which naturally comes from the
fact that we consider $n$-bit strings. A generalization to any
$(P_X,a,b)$-QBSC where $P_X$ is an arbitrary distribution is
possible but omitted in order not to obscure our main line of
argument.

Instead of Bob's guessing probability, one can take any information
measure $B$ to express the security against Bob. In general, we consider an
$(n,a,b)$-$\mbox{QBSC}_B$ where the new Concealing-condition reads
\begin{itemize}
\item (General Concealing) If Alice is honest, then
for any ensemble $\ens = \{p_x,\rho_x\}$ that Bob can obtain
by a cheating strategy $B(\ens) \leq b$.
\end{itemize}
Later, we will show that for $B$ being the
\emph{accessible information},
non-trivial protocols, i.e. protocols
with $a+b \ll n$, do exist. Recall that the accessible information was defined 
in Section~\ref{accessInfo} as
$\mI_{acc}(\ens) = \max_{M} I(X,Y)$, where $P_X$ is the prior
distribution of the random variable $X$, $Y$ is the random variable
of the outcome of Bob's measurement on $\ens$, and the
maximization is taken over all measurements $M$.

\subsection{Model}\index{two-party protocols}
We work in the model of two-party non-relativistic quantum protocols of
Yao \cite{yao:otFromBc}, simplified by Lo and
Chau~\cite{lo&chau:bitcom} which is usually adopted in this context.
Here, any two-party quantum protocol can be regarded as a pair of
quantum machines (Alice and Bob), interacting through a quantum channel.
Consider the product of
three Hilbert spaces $\hilbert_A$, $\hilbert_B$ and $\hilbert_C$ of
bounded dimensions, representing the Hilbert spaces of Alice's and
Bob's machines and the channel, respectively. Without loss of
generality, we assume that each machine is initially in a specified
pure state. Alice and Bob perform a number of rounds of
communication over the channel. Each such round can be modeled as a
unitary transformation on $\mathcal{H}_A \otimes \mathcal{H}_C$ and
$\mathcal{H}_B \otimes \mathcal{H}_C$ respectively.
Since the protocol is known to both Alice and Bob, they know
the set of possible unitary transformations used in the protocol.
We assume that Alice and Bob are in
possession of both a quantum computer and a quantum storage device.
This enables them to add ancillae to the quantum machine and use
reversible unitary operations to replace measurements. 
The state of this ancilla can then be read off only
at the end of the protocol, and
by doing so,
Alice and Bob can effectively delay any measurements until the very end.
The resulting protocol will be equivalent to the original 
and thus we can limit ourselves
to protocols where both parties only measure at the
very end. Moreover, any classical computation or communication that
may occur can be simulated by a quantum computer.

\subsection{Tools}

We now gather the essential ingredients for our proof.
First, we now show that every $(n,a,b)$-$\mbox{QBSC}$ is an
$(n,a,b)$-$\mbox{QBSC}_\xi$. The security measure $\xi(\ens)$ is
defined by
\begin{equation} \label{xi}
\xi (\ens) := n- H_2(\rho_{AB}|\rho),
\end{equation}
where $\rho_{AB} = \sum_x p_x \outp{x}{x} \otimes \rho_x$ and $\rho
= \sum_x p_x \rho_x$ are only dependent on the ensemble $\ens=\{p_x,
\rho_x\}$. $H_2(\cdot|\cdot)$ is an entropic quantity defined
in~\cite{renato:diss}\index{$H_2(\cdot\mid \cdot)$} 
$$
H_2(\rho_{AB}|\rho) := -\log\Tr 
\left(\left[\id \otimes \rho^{-\frac{1}{2}})\rho_{AB}\right]^2\right).
$$\label{def:qcollision}
Interestingly, this quantity is directly
connected to Bob's maximal average probability of successfully guessing
the string:
\begin{lemma} \label{guessing-lemma} 
Bob's maximal average probability of successfully guessing
the committed string,~i.e. $\sup_M \sum_x p_x p^{B, M}_{x|x}$ where $M = \{M_x\}$
ranges over all measurements and $p^{B, M}_{y|x} = \Tr(M_y\rho_x)$ is the conditional
probability of outputting $y$ given $\rho_x$, obeys
$$
\sup_M \sum_x p_x p^{B, M}_{x|x} \geq 2^{- H_2(\rho_{AB}|\rho)}.
$$
\end{lemma} 
\begin{proof}
By definition, the maximum average guessing
probability is lower bounded by the average guessing probability for
a particular measurement strategy. We choose the \emph{square-root
measurement} which has operators 
$$
M_x = p_x \rho^{-\frac{1}{2}}
\rho_x \rho^{-\frac{1}{2}}.
$$ 
We use $p^B_{x|x} = \Tr(M_x \rho_x)$ to denote the
probability that Bob guesses $x$ given $\rho_x$, hence
\begin{eqnarray*}
\log\sum_x p_x p^{B,\max}_{x|x}
&\geq& \log\sum_x p_x^2 \Tr(\rho^{-\frac{1}{2}} 
\rho_x \rho^{-\frac{1}{2}} \rho_x)\\
&=&\log\Tr\left(\sum_x p_x^2 \outp{x}{x} \otimes \rho^{-\frac{1}{2}}
\rho_x \rho^{-\frac{1}{2}} \rho_x\right)\\
&=& \log\Tr \left(\left[(\id \otimes \rho^{-\frac{1}{2}})\rho_{AB}\right]^2\right)\\
&=& - H_2(\rho_{AB}|\rho)
\end{eqnarray*}
\end{proof}

Related estimates were derived in~\cite{barnumknill}.

\label{PAused}
Furthermore, we make use of the following theorem, known as \emph{privacy
amplification against a quantum adversary}\index{privacy amplification}
with two-universal hash functions\index{two-universal hash function},
which we state in a form that is most convenient for our purposes in this chapter.
 A class $\setF$ of functions $f: \01^n
\rightarrow \01^\ell$ is thereby called \emph{two-universal} 
if for all $x\neq y \in
\01^n$ and for uniformly at random chosen $f \in \setF$ we have
$\Pr[f(x) = f(y)] \leq 2^{-\ell}$. For example, the set of all affine
functions\footnote{Geometrically, an affine function is 
a linear function plus a translation}.
 from $\01^n$ to $\01^\ell$ is two-universal~\cite{CarWeg79}.  
The following
theorem expresses how hash functions can decrease Bob's knowledge about a random variable
when he holds some quantum information.
In our case, Bob will hold
some quantum memory and privacy amplification is used to find Alice's
attack.
\begin{theorem}[Th.~5.5.1 in~\cite{renato:diss} (see
also~\cite{KoMaRe05})]\label{theorem:renato}\index{privacy amplification}
 Let $\mathcal{G}$ be a
class of two-universal hash functions
from $\01^n$ to $\01^s$. Application of $g \in
\mathcal{G}$ to the random variable $X$ maps the ensemble
$\ens=\{p_x, \rho_x\}$ to $\ens_g=\{q^g_y, \sigma^g_y\}$ with
probabilities $q^g_y = \sum_{x \in g^{-1}(y)} p_x$ and quantum
states $\sigma^g_y = \sum_{x \in g^{-1}(y)} p_x \rho_x$. Then 
\begin{equation}
\label{eq-renner-koenig} \frac{1}{|\mathcal{G}|} \sum_{g \in
\mathcal{G}} d(\ens_g)  \leq
\frac{1}{2}2^{-\frac{1}{2}[H_2(\rho_{AB}|\rho) - s]},
\end{equation}
where
$d(\ens) := D\big(\sum_x p_x \proj{x} \otimes \rho_x,
\id/2^n\otimes \rho\big)$ (and similarly for $d(\ens_g)$).
\end{theorem}

Finally, the following reasoning that is used to prove the impossibility of
quantum bit commitment~\cite{lo&chau:bitcom, mayers:trouble} will be
essential:\index{bit commitment!impossibility}\index{impossibility!bit commitment}
Suppose $\rho_0$ and $\rho_1$ are density operators that correspond to the state
of Bob's system if Alice committed 
a ``0'' or a ``1'' respectively. Let
$\ket{\phi_0}$ and $\ket{\phi_1}$ be the corresponding purifications
on the joint system of Alice and Bob: Alice holds the purification of $\rho_0$ and $\rho_1$. 
If $\rho_0$ equals $\rho_1$
then Alice can find a local unitary transformation $U$ that Alice
can apply to her part of the system such that
$\ket{\phi_1} = U \otimes \id \ket{\phi_0}$. This enables Alice to change the total
state from $\ket{\phi_0}$ to $\ket{\phi_1}$ and thus cheat using entanglement! This
reasoning also holds in an approximate
sense~\cite{mayers:trouble}, here used in the following
form:
\begin{lemma}\label{lemma:mayers}
Let $D(\rho_0, \rho_1)\leq \epsilon$ and assume that the
bit-commitment protocol is error-free if both parties are honest. Then there exists a
method for Alice to cheat such that the probability of successfully
revealing a $0$ during the reveal phase, given that she honestly committed herself
to a $1$ during
the commit phase, is at least
$1-\sqrt{2\epsilon}$.
\end{lemma}
\begin{proof}
$D(\rho_0, \rho_1) \leq \epsilon$ implies 
$\max_{U}| \bra{\phi_0} U \otimes \id\ket{\phi_1}| \geq 1 - \eps$ by Uhlmann's theorem~\cite{Uhlmann:1976}. 
Here, $\ket{\phi_0}$ and $\ket{\phi_1}$
correspond to the joint states after the commit phase if Alice committed to a '0'
or '1' respectively where
the maximization
ranges over all unitaries $U$ on Alice's (i.e. the purification)
side. Let $\ket{\psi_0} = U \otimes \id \ket{\phi_1}$ for a $U$
achieving the maximization, be the state that Alice prepares by applying
$U$ to the state 
on her side when she wants to reveal a '1', given a prior
honest commitment to '0'.
We then have
\begin{eqnarray*}
D(\proj{\phi_0},
\proj{\psi_0})&=&\sqrt{1-|\braket{\phi_0}{\psi_0}|^2}\\
&\leq& \sqrt{1-(1-\epsilon)^2}\\
&\leq& \sqrt{2\epsilon}.
\end{eqnarray*}
If Bob is honest,
the reveal phase can be regarded as a measurement resulting
in a distribution $P_Y$ (or $P_Z$) if $\ket{\phi_0}$ (or $\ket{\psi_0}$)
was the state before the reveal phase. The random variables $Y$ and
$Z$ can take values $\01$ (corresponding to the opened bit) or the value `reject (r)'. Since the trace
distance does not increase under measurements, $D(P_Y, P_Z)
\leq D(\proj{\phi_0}, \proj{\psi_0}) \leq \sqrt{2\epsilon}$.
Hence $\frac{1}{2}(|P_Y(0)-P_Z(0)| +
|P_Y(1)-P_Z(1)|+|P_Y(r)-P_Z(r)|)  \leq \sqrt{2\epsilon}$. Since
$\ket{\phi_0}$ corresponds to Alice's honest commitment to $0$ we
have $P_Y(0)=1$, $P_Y(1)=P_Y(r)=0$ and hence $P_Z(0)\geq
1-\sqrt{2\epsilon}$. 
\end{proof}

\section[Impossibility of quantum string commitments]
{Impossibility\\ of quantum string commitments}
\index{bit string commitment!impossibility}

As we saw above, any $(n,a,b)$-QBSC is also
an $(n,a,b)$-$\mbox{QBSC}_\xi$ with the security measure $\xi(\ens)$
defined in Eq.~(\ref{xi}). To prove our impossibility result we now
prove that an
$(n,a,b)$-$\mbox{QBSC}_\xi$ can only exist for values $a$, $b$ and
$n$ obeying $a + b + c \geq n$, where $c$ is a small constant
independent of $a$, $b$ and $n$. This in turn implies the
impossibility of an $(n,a,b)$-QBSC for such parameters. Finally, we
show that if we execute the protocol many times in parallel, the
protocol can only be
secure if $a + b \geq n$. 

The intuition behind our proof is simple: To cheat, Alice first chooses
a two-universal hash function $g$. She then commits to a superposition
of all strings for which $g(x) = y$ for a specific $y$. We now know from
the privacy amplification theorem above, that even though Bob may gain
some knowledge about $x$, he is entirely ignorant about $y$. But then 
Alice can change her mind and reveal a string from a different set of strings for which
$g(x) = y'$ with $y \neq y'$ as we saw above! The following figure
illustrates this idea.

\begin{figure}[h]
\begin{center}
\includegraphics{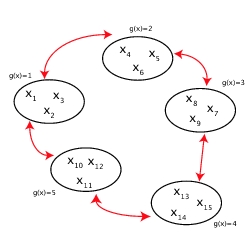}
\caption{Moving from a set of string with $g(x) = y$ to a set of strings with
$g(x) = (y \mod 5) + 1$.}
\end{center}
\end{figure}

\begin{theorem} \label{nogo}\index{impossibility!bit string commitment}\index{QBSC!impossibility}
$(n,a,b)$-$\mbox{QBSC}_\xi$ schemes
with $a+b+c < n$ do not exist,
where $c = 5 \log 5 - 4\approx 7.61$ is a constant.
\end{theorem}
\begin{proof}
Consider an $(n,a,b)$-$\mbox{QBSC}_\xi$ and the case
where both Alice and Bob are honest. Alice committed to
$x$. We denote the joint state of the system Alice-Bob-Channel
$\hilbert_A \otimes \hilbert_B \otimes \hilbert_C$ after the commit
phase by $\ket{\phi_x}$ for input state $\ket{x}$.
Let $\rho_x$ be Bob's reduced density matrix, and let
$\ens=\{p_x, \rho_x\}$
where $p_x=2^{-n}$.

Assuming that Bob is honest, we will give a cheating strategy for
Alice in the case where $a+b+5 \log 5 - 4 < n$. The strategy will
depend on the two-universal hash function $g:\mathcal{X} = \01^n \rightarrow
\mathcal{Y} = \01^{n -m}$, for appropriately chosen $m$. Alice
picks a $y\in \mathcal{Y}$ and constructs the state $ (\sum_{x \in
g^{-1}(y)}\ket{x}\ket{x})/\sqrt{|g^{-1}(y)|}$. She then gives the
second half of this state as input to the protocol and stays honest
for the rest of the commit phase. The joint state of Alice and Bob
at the end of the commit phase is thus $\ket{\psi^g_y} =
(\sum_{x \in g^{-1}(y)}\ket{x}\ket{\phi_x})/\sqrt{|g^{-1}(y)|}$. The
reduced states on Bob's side are
$\sigma^g_y = \frac{1}{q^g_y}\sum_{x \in g^{-1}(y)} p_x \rho_x$
with probability $q^g_y = \sum_{x \in g^{-1}(y)} p_x$. We
denote this ensemble by $\ens_g$. Let 
$\sigma^g = \sum_y q^g_y \sigma^g_y$.

We now apply Theorem~\ref{theorem:renato} with $s=n-m$ and
$\xi(\ens)\leq b$ and obtain $ \frac{1}{|\mathcal{G}|} \sum_{g \in
\mathcal{G}} d(\ens_g) \leq \eps $
where $\eps=\frac{1}{2}2^{-\frac{1}{2}(m-b)}$. Hence, there is at
least one $g$ such that $d(\ens_g) \leq \eps$. Intuitively, this
means that Bob knows only very little about the value of $g(x)$.
This $g$ defines Alice's cheating strategy. It is straightforward to verify that
$d(\ens_g) \leq \eps$ implies
\begin{equation}\label{sum_eq}
2^{-(n-m)} \sum_{y \in \mathcal{Y}} D(\sigma^g, \sigma^g_y) \leq 2\eps.
\end{equation}
We therefore assume without loss of generality that Alice
chooses $y_0\in \setY$ with $D(\sigma^g_{y_0}, \sigma^g) \leq
2\eps$.

We first observe that the probability to successfully reveal some $x$ in
$g^{-1}(y)$ given $\ket{\psi^g_y}$ is one\footnote{Alice learns
$x$, but can't pick it: she committed to a superposition and $x$ is
chosen randomly by measurement.}. We say that Alice reveals $y$
if she reveals an $x$ such that $y=g(x)$. We then also have that
the probability for Alice to reveal
$y$ given $\ket{\psi^g_y}$
successfully is one. Let $\tilde{p}_x$
and $\tilde{q}^g_y$ denote the probabilities to successfully reveal
$x$ and $y$ respectively and $\tilde{p}^g_{x|y}$ be the conditional
probability to successfully reveal $x$, given $y$. We have
$$
\sum_x \tilde{p}_x =\sum_y \tilde{q}^g_y
\sum_{x \in g^{-1}(y)}\tilde{p}^g_{x|y} \geq \sum_y \tilde{q}^g_y,
$$
where the inequality follows from our observation above.

As in the impossibility proof of bit commitment, Alice can now
transform $\ket{\psi^g_{y_0}}$ approximately
into $\ket{\psi^g_y}$ if $\sigma^g_{y_0}$ is sufficiently close to
$\sigma^g_y$ by using only local transformations on her part. 
Indeed, Lemma~\ref{lemma:mayers} tells us how to bound the
probability of revealing $y$, given that the state was really
$\ket{\psi_{y_0}}$. Since this reasoning applies to all $y$, on
average, we have
\begin{eqnarray*}
\sum_y \tilde{q}^g_y &\geq& \sum_y \left(1- \sqrt{2}
 \sqrt{D(\sigma^g_{y_0}, \sigma^g_{y})}\right)\\
&\geq& 2^{n-m} - 
2^{n-m} 
\sqrt{2} 
\sqrt{2^{m-n} \sum_y D(\sigma^g_{y_0},\sigma^g_{y})}\\
&\geq& 2^{n-m}\left(1 - 
\sqrt{2}
\sqrt{2^{m-n}
\left(\sum_y D(\sigma^g_{y_0},\sigma^g) + D(\sigma^g,\sigma^g_{y})\right)}\right)\\
&\geq& 2^{n-m}(1 - 2 \sqrt{2\eps}),
\end{eqnarray*} where the first inequality follows from
Lemma~\ref{lemma:mayers}, the second from Jensen's inequality and
the concavity of the square root function, the third from the
triangle inequality and the fourth from Eq.~(\ref{sum_eq}) and
$D(\sigma^g_{y_0}, \sigma^g) \leq 2\eps$. Recall that to be
secure against Alice, we require $2^a \geq 2^{n-m}(1- 2
\sqrt{2\eps})$. We insert
$\eps=\frac{1}{2}2^{-\frac{1}{2}(m-b)}$, define $m=b+\gamma$ and
take the logarithm on both sides to get 
\begin{equation} 
\label{eq-delta} a+b
+\delta \geq n,
\end{equation}
where $\delta = \gamma -\log (1-2^{-\gamma/4
+1})$. Keeping in mind that $1-2^{-\gamma/4 +1}>0$ (or equivalently
$\gamma>4$), we find that the minimum value of $\delta$ for which
Eq.~(\ref{eq-delta}) is satisfied is $\delta=5 \log 5 -4$ and
arises from $\gamma = 4 (\log 5 - 1)$. Thus,
no $(n,a,b)$-$\mbox{QBSC}_\xi$ with $a+b+5 \log 5 - 4 < n$ exists.
\end{proof}

It follows immediately that the same restriction holds for an
$(n,a,b)$-QBSC:
\begin{corollary}
$(n,a,b)$-$\mbox{QBSC}$ schemes, 
with $a+b+c < n$ do not exist,
where $c = 5 \log 5 - 4\approx 7.61$ is a constant.
\end{corollary}
\begin{proof}
For the uniform
distribution $p_x = 2^{-n}$, we have from the concealing
condition that $\sum_x p^B_{x|x} \leq 2^{b}$, which by
Lemma~\ref{guessing-lemma} implies $\xi(\ens) \leq b$. Thus, a
$(n,a,b)$-QBSC is an $(n,a,b)$-$\mbox{QBSC}_\xi$ from which the
result follows.
\end{proof}

Since the constant $c$ does not depend on $a$, $b$ and $n$, multiple
parallel executions of the protocol can only be secure if $a+b\geq
n$. This follows by considering $m$ parallel executions of the protocol
as a single execution with a string of length $m n$. 
\begin{corollary} Let $P$ be an $(n,a,b)$-QBSC with $P^m$ an
$(mn,ma,mb)$-QBSC. Then $n < a + b + c/m$. In particular, no $(n, a,
b)$-QBSC with $a+b<n$ can be executed securely an arbitrary number
of times in parallel. 
\end{corollary} 
Thus, we can indeed hope to do no better than the trivial protocol. 
It follows directly from~\cite{kitaev:super} that
the results in this section also hold in the presence of
superselection rules, where, very informally, quantum actions are restricted
to act only on certain subspaces of a larger Hilbert space.

\section{Possibility}\index{QBSC!possibility}
\index{bit string commitment!possibility}
Surprisingly, if one is willing to
measure Bob's ability to learn $x$ using a weaker measure of
information, the accessible information,
non-trivial protocols become possible. These protocols are based on
a discovery known as ``locking of classical information in quantum
states'' which we already encountered in Chapter~\ref{chapter:locking}.


The protocol, which we call LOCKCOM($n$, $\setU$), uses this effect
and is specified by a set $\setU = \{U_1,\ldots,U_{|\setU|}\}$ of
unitaries. We have

\begin{protocol}{LOCKCOM($n, \mU$)}{}\index{LOCKCOM}\index{locking!application}
\item Commit phase: Alice has the string $x \in \01^n$ and randomly chooses
$r \in \{1,\ldots,|\setU|\}$. She sends the state $U_r \ket{x}$ to
Bob, where $U_r \in \setU$.
\item Reveal phase: Alice announces $r$ and $x$. Bob applies
$U_r^\dagger$ and measures in the computational basis
to obtain $x'$. He accepts if and only if $x'=x$.
\end{protocol}
\smallskip
We now first show that our protocol is secure with respect to Definition~\ref{securityDef}
if Alice is dishonest. Note that our proof only depends on the number of unitaries used, 
and is independent of a concrete instantiation of the protocol.
\begin{lemma}\label{Alice-lemma}
For any LOCKCOM$(n,\setU)$ protocol the security against a dishonest 
Alice is bounded by $2^a \leq |\setU|$,
\end{lemma}
\begin{proof}
Let $\tilde{p}_x$ denote the probability that Alice reveals $x$ successfully. Then,
$\tilde{p}_x \leq \sum_r \tilde{p}_{x,r}$, where $\tilde{p}_{x,r}$
is the probability that $x$ is accepted by Bob when the reveal
information was $r$. Let $\rho$ denote the state of Bob's system.
Summation over $x$ now yields 
\begin{eqnarray*}
\sum_x \tilde{p}_x &\leq&  \sum_{x, r}\tilde{p}_{x,r} \\
&=& \sum_{x,r} \Tr \proj{x} U_r^\dagger \rho U_r \\
&=& \sum_r \Tr \rho = 2^a.
\end{eqnarray*}
\end{proof}

In order to examine security against a dishonest Bob, we have to consider the actual form
of the unitaries. We first show that there do indeed exist interesting protocols. Secondly, 
we present a simple, implementable, protocol.
To see that interesting protocols can exist, let Alice choose a set of $O(n^4)$ unitaries independently
according to the Haar measure (approximately discretized) and announce the resulting set $\setU$
to Bob. They then perform LOCKCOM($n,\setU$). 
Following the work of~\cite{winter:randomizing}, we now show that this variant is secure
against Bob with high probability. That is, there exist $O(n^4)$ unitaries that bring Bob's 
accessible information down to a
constant: $\I_{acc}(\ens) \leq 4$:
\begin{theorem} \label{theorem-LOCKCOM}
For $n \geq 3$, there exist $(n, 4\log n +O(1), 4)$-$
\mbox{QBSC}_{\I_{acc}}$ protocols.
\end{theorem}
\begin{proof}
Let $\setU_{ran}$ denote the set of $m$ randomly chosen bases
and consider the LOCKCOM($n,a,b$) scheme using unitaries $\setU = \setU_{ran}$.
Security against Alice is again given by Lemma~\ref{Alice-lemma}. We now need
to show that this choice of unitaries achieves the desired locking effect and
thus security against Bob.
Again, let $d=2^n$ denote the dimension. As we saw in Section~\ref{rewriteIACC}
we have that
$$
\mI_{acc}(\ens)\leq \log d +\max_{\ket{\phi}} \sum_j \frac{1}{m} H(X_j),
$$
where $X_j$ denotes the outcome of the measurement of $\ket{\phi}$ in basis
$j$ and the maximum is taken over all pure states $\ket{\phi}$.
According to \cite[Appendix B]{winter:randomizing} there is a
constant $C'>0$ such that
\begin{eqnarray*}
\Pr [\inf_\phi \frac{1}{m} \sum_{j=1}^m H(X_j) &\leq& (1-\eps) \log d -3 ]\\
&\leq& \left( \frac{10}{\eps} \right)^{2d}
2^{ -m \left(\frac{\eps C' d}{2(\log d)^2} -1\right)
},
\end{eqnarray*}
for $d \geq 7$ and $\eps \leq 2/5$. Set $\eps =\frac{1}{\log d}$.
The RHS of the above equation then decreases provided that $m > \frac{8}{C'} (\log d)^4$.
Thus with $d=2^n$ and $\log m=4 \log n + O(1)$,
the accessible information of the ensemble corresponding to the
commitment is then $\mI_{acc}(\ens) \leq \log d -
(1-\eps) \log d +3 = \eps\log d +3=4$ for our choice
of $\eps$.
\end{proof}

Unfortunately, the protocol is inefficient both in terms of
computation and communication. It remains open to find an efficient
constructive scheme with those parameters.

In contrast, for only two bases, an efficient construction exists
and uses the identity and the Hadamard transform as unitaries. For this
case, the security of the standard LOCKCOM protocol follows immediately from the locking
arguments of Chapter~\ref{chapter:locking}.
It has been shown that this protocol can be made cheat-sensitive~\cite{matthias:thesis}.

\begin{theorem} \label{two-bases}
LOCKCOM($n,1,n/2$) using $\mU =
\{\id^{\otimes n}, H^{\otimes n}\}$ is a $(n, 1, n/2)$-$\mbox{QBSC}_{\I_{acc}}$
protocol.
\end{theorem}
\begin{proof}
The result follows immediately from Lemma $\ref{Alice-lemma}$ and the fact that by Corollary~\ref{pauliLocking} 
$\mI_{acc}(\ens) \leq n/2$ for Bob.
\end{proof}

We can thus obtain non-trivial protocols by exploiting the locking
effects discussed in Chapter~\ref{chapter:locking}. Note, however, that
the security parameters are very weak. Indeed, if Alice uses only two
possible bases chosen with equal probability then 
Bob is always able to obtain the encoded string with probability at least
$1/2$: he simply guesses the basis and performs the corresponding
measurement. 

\section{Conclusion}
We have introduced a framework for
quantum commitments to a string of bits. Even if we consider 
string commitments that are weaker than bit commitments, 
no non-trivial protocols can exist if we choose a very strong measure of security. A property of
quantum states known as \emph{locking}, however, allowed us to
propose meaningful protocols for a much weaker security demand.
One could extend our method to the case of weak secure function evaluation
as was done for the original bit commitment protocol in~\cite{lo:insecurity}.
After completion of our work, Jain~\cite{jain:qbsc} has also shown using a different
method that $\mbox{QBSC}_\chi$ protocols with
$a + 16b + 31 < n$ cannot exist.

A drawback of weakening the security requirement is that LOCKCOM
protocols are not necessarily composable. Thus, if LOCKCOM is
used as a sub-protocol in a larger protocol, the security of the
resulting scheme has to be evaluated on a case by case basis.
However, LOCKCOM protocols are secure when
executed in parallel. This is a consequence of the definition of
Alice's security parameter and the additivity of the accessible
information (see Chapter~\ref{chapter:informationIntro}), and sufficient for many cryptographic
purposes.

However, two important open questions remain: First, how can we construct
efficient protocols using more than two bases? It may be tempting to
conclude that we could simply use a larger number of mutually unbiased bases,
such as given by the identity and Hadamard transform. Yet, as we saw in Chapter~\ref{chapter:uncertainty}
using more mutually unbiased bases does not necessarily lead to a better
locking effect and thus better string commitment protocols. Finally,
are there any novel applications for this weak quantum string commitment?

Fortunately, it turns out that we can implement protocol with very strong security parameters if we are willing to introduce additional assumptions.
We now show how to obtain oblivious transfer from the assumption
that qubits are affected by noise during storage.

\renewcommand{\appendixtocname}{Appendix}
\noappendicestocpagenum
\addappheadtotoc

\chapter{Possibilities: Exploiting storage errors}\label{storageErrorsChapter}\label{goCryptoChapter}\label{chapter:noise}\label{sec:cryptoFromNoise}\label{OTfromNoise}\label{chapter:otNoise}\label{chapter:storageNoise}\index{bounded storage!due to noise}\index{noisy storage}

Given the negative results from the last chapter, what can we still hope to achieve?
Fortunately, the situation is not quite as bleak if we are taking
advantage of the technical limitation that
quantum storage is necessarily noisy. Here, the very problem that still
prevents us from implementing
a quantum computer can actually be turned to our advantage! As we saw in Chapter~\ref{chapter:cryptoIntro}
the primitive of oblivious transfer allows us to implement essentially all 
cryptographic protocols among two mutually distrustful players,
and hence we
focus on this primitive.

\section{Introduction}

As outlined in Chapter~\ref{chapter:cryptoIntro}, 
it was recently shown that secure OT is possible when the
receiver Bob has a limited amount of quantum\index{bounded storage!quantum}
memory~\cite{serge:bounded,serge:new} at his disposal. Within this
`bounded-quantum-storage model' OT can be implemented securely as\index{OT}
\index{oblivious transfer}
long as a dishonest receiver Bob can store at most $n/4-O(1)$ qubits
coherently, where $n$ is the number of qubits transmitted from Alice
to Bob. The problem with this approach is that it assumes an
explicit limit on the physical number of qubits (or more precisely,
the rank of the adversary's
quantum state). However, at present we do not know of any practical
physical situation which enforces such a limit for quantum
information. On the other hand it is a fact that currently and in
the near-future storing photonic qubits is noisy. We therefore
propose an alternative model of \emph{noisy-quantum storage}\index{noisy quantum storage model} inspired by
present-day physical implementations: We require no explicit memory
bound, but we assume that any qubit that is placed into quantum
storage undergoes a certain amount of noise. Here,
we take
the 1-2 OT protocol from~\cite{serge:new} as our starting point, and
analyze it in this model.
This simple 1-2 OT protocol can be implemented using photonic qubits (using
polarization or phase-encoding) with standard BB84 quantum key\index{BB84}
distribution~\cite{bb84, GRTZ:qkd_review} hardware, only with different classical
post-processing.

Our adversary model
\index{adversary model}
is that of collective attacks (in analogy with collective eavesdropping attacks in the
quantum key distribution setting). 
More precisely:
\begin{itemize}
\item Bob may choose to (partially) measure (a subset of)
his qubits immediately upon reception using an error-free {\em product} measurement.
\item Bob may store each incoming qubit, or post-measurement state from a prior partial measurement, separately 
and wait until he gets
additional information from Alice (at Step~3 in Protocol~1). 
\item Once he obtained the additional information he may perform
an arbitrary coherent measurement on 
his stored qubits and stored classical data.
\end{itemize}

We assume that a qubit $q_i$ undergoes some noise while in 
storage, where we denote the combined channel given by Bob's initial (partial) measurement, followed by the noise by super-operator $\mS_i$.
The source of noise can be due to the transfer of qubit onto a different physical
carrier, such as an atomic ensemble or atomic state for example, or into an error-correcting code with
fidelity less than 1. In addition, the (encoded) qubit will undergo noise once it has been
transferred into `storage'. Hence, the quantum operation $\mS_i$
in any real world setting will necessarily include some form of noise. Note
that such noise is typically much larger than the noise experienced
by honest players who only need to make immediate complete
measurements in the BB84 basis.

First of all, we show that for any initial measurement by Bob, and
any noisy superoperator $\mS_i$ the 1-2 OT protocol is secure if the
honest players can perform {\em perfect} noise-free quantum
operations. As an explicit example, we consider 
depolarizing
noise for which reduce the set of optimal attacks to two simple
ones: measure in the so-called Breidbart basis or let the qubits
undergo depolarizing noise. This allows us to obtain an explicit
tradeoff between the amount of noise in storage and the security
of the protocol.

In a real implementation using photonic qubits the execution of the
protocol by the honest players is imperfect: their quantum
operations can be inaccurate or noisy, weak laser pulses instead of
single photon sources are used and qubits undergo decoherence in
transmission. Note, however, that unlike in QKD, we also want
to execute such protocols over very short distances (for example in
banking applications) such that the depolarization rate
during transmission in free-space is very low.
Our practical 1-2 OT-protocol is a small modification of the perfect
protocol, so that we can separately deal with erasure errors (i.e.
photon loss) and the rate of these errors does not affect the
security of the protocol. We then show for this practical protocol
how one can derive trade-offs between the amount of storage noise,
the amount of noise for the operations performed by the honest
players, and the security of the protocol.
At the end, we discuss the issue of analyzing fully
coherent attacks for our protocol. Indeed, there is 
a close relation between the 1-2 OT
protocol and BB84 quantum key distribution.

Our security analysis can in principle be carried over to obtain a
secure identification scheme in the noisy-quantum-storage model
analogous to~\cite{DFSS:secureid}. This scheme achieves
password-based identification and is of particular practical
relevance as it can be used for banking applications.

\subsection{Related work}
Precursors of the idea of basing the security of 1-2 OT on storage-noise
are already present in~\cite{crepeau:practicalOT} which laid the
foundations for the protocol in~\cite{serge:new}, but no rigorous
analysis was carried through in that paper. Furthermore, it was
pointed out in~\cite{chris:diss,DFSS08journal} how the original
bounded-quantum-storage analysis applies in the case of noise levels
which are so large that the rank of a dishonest player's quantum
storage is reduced to $n/4$. In contrast, we are able to give an
explicit security tradeoff even for small amounts of noise.
We furthermore note that our security proof is not exploiting the
noise in the communication channel (which has been done in the
classical setting to achieve cryptographic tasks, see
e.g.~\cite{crepeau:weakened,crepeau:efficientOT,CMW04}), but is solely based on the
fact that the dishonest receiver's quantum storage is
noisy. 
Another technical limitation has been considered in
\cite{salvail:physical} where a bit-commitment scheme was shown
secure under the assumption that the dishonest committer can only
measure a limited number of qubits coherently. Our analysis differs
in that we allow any coherent measurement at the very end.
Furthermore, the security analysis of our
protocol is considerably simpler and more promising to be extended to
cover more general cases.

\section{Preliminaries}

\subsection{Definitions}
We start by introducing some tools, definitions and technical lemmas.
To define the security of 1-2 OT, we need to express what it means for a
dishonest quantum player not to gain any information.  Let $\rho_{XE}$ be a
state that is part classical, part quantum, i.e.~a cq-state
$\rho_{XE}=\sum_{x \in \setX} P_X(x) \outp{x}{x} \otimes \rho_E^x$.
Here, $X$ is a classical random variable distributed over the finite
set $\setX$ according to distribution $P_{X}$. 
In this Chapter, we will write the {\em non-uniformity}\index{non-uniformity}\index{distance!from uniform}\label{def:uniform}
of $X$ given
$\rho_E = \sum_x P_X(x) \rho_E^x$
as
$$
d(X|\rho_E) := \frac{1}{2}
\left|\left|\,\id/|\mX| \otimes \rho_E-\sum_{x}P_{X}(x) \outp{x}{x} \otimes \rho_{E}^x\,\right|\right|_{1}.
$$
Intuitively, if
$d(X|\rho_E) \leq \eps$ the distribution of $X$ is $\eps$-close to
uniform even given $\rho_E$, i.e., $\rho_E$ gives hardly any
information about $X$.  A simple property of the non-uniformity which
follows from its definition is that for
any cq-state of the form $\rho_{XED}=\rho_{XE} \otimes \rho_D$, we have\index{cq-state}
\begin{equation} \label{eq:indep}
d(X|\rho_{ED})=d(X|\rho_E) \, .
\end{equation}
We prove the security of a randomized version of OT. In such a
protocol, Alice does not choose her input strings herself, but
instead receives two strings $S_0$, $S_1 \in \01^\ell$ chosen
uniformly at random by the protocol. Randomized OT (ROT) can easily
be converted into OT: after the ROT protocol is completed, Alice uses
her strings $S_0,S_1$ obtained from ROT as one-time pads to encrypt
her original inputs $\hat{S_0}$ and $\hat{S_1}$, i.e.~she sends an
additional classical message consisting of $\hat{S_0} \oplus S_0$
and $\hat{S_1} \oplus S_1$ to Bob. Bob can retrieve the message of
his choice by computing $S_C \oplus (\hat{S}_C \oplus S_C) =
\hat{S}_C$. He stays completely ignorant about the other message
$\hat{S}_{1-C}$ since he is ignorant about $S_{1-C}$. The security
of a quantum protocol implementing ROT is defined in \cite{serge:bounded,serge:new} for a standalone setting. A more involved definition allowing for composability
can be found in~\cite{wehner07b}.
In the following, we use $\rho_B$ to denote the complete quantum
state of Bob's lab at the end of the protocol including any additional
classical information he may have received directly from Alice.
Similarly, we use 
$\rho_{CS'_0S'_1A}$ and $\rho_{S'_0S'_1A}$ to denote
the c-q states corresponding to the state of Alice's lab at the end
of the protocol including her classical information about Bob's
choice bit $C$ and outputs $S'_0$ and $S'_1$ as defined below.

\begin{definition} \label{def:ROT}\index{ROT}\index{randomized oblivious transfer}
An $\eps$-secure 1-2 $\mbox{ROT}^\ell$ is a protocol between Alice
and Bob, where Bob has input $C \in \01$, and Alice has no input.
For any distribution of $C$:
\begin{itemize}
\item (Correctness) If both parties are honest, Alice gets output $S_0,S_1 \in \01^\ell$ and
Bob learns $Y = S_C$ except with probability $\eps$.
\item (Receiver-security) If Bob is honest and obtains output $Y$, then for any cheating strategy of Alice resulting in her state $\rho_A$,
there exist random variables $S'_0$ and $S'_1$ such that $\Pr[Y=S'_C] \geq 1- \eps$ and
$C$ is $\eps$-independent of $S'_0$,$S'_1$ and $\rho_A$, i.e.,
$D(\rho_{CS'_0,S'_1A},\rho_C \otimes \rho_{S'_0,S'_1A})\leq \eps$.
\item (Sender-security) If Alice is honest, then for any cheating strategy of Bob resulting in his state $\rho_B$,
there exists a random variable $C' \in \01$ such that $d(S_{1-C'}|S_{C'}C'\rho_B) \leq \eps$.
\end{itemize}
\end{definition}

Note that cheating Bob may of course not choose a $C$ beforehand. 
Intuitively, our requirement for security states that
whatever Bob does, he will be ignorant about at least one of Alice's inputs.
This input is determined by his cheating strategy. Our requirement for
receiver security states that $C$ is independent of Alice's output,
and hence Alice learns nothing about $C$.

The protocol makes use of two-universal hash functions that
are used for privacy amplification similar as in QKD,
which we already encountered in Section~\ref{PAused}.
For the remainder of this Chapter, we first define

\begin{definition}
For a measurement $M$ with POVM elements $\{M_x\}_{x \in {\cal X}}$
let $p_{y|x}^M={\rm Tr}(M_y \rho_E^x)$ the probability of outputting
guess $y$ given $\rho_E^x$. Then 
$$
P_g(X|\rho_E) \assign \sup_M \sum_x P_X(x) p_{x|x}^M
$$ 
is the maximal average success probability of
guessing $x \in \setX$ given the reduced state $\rho_E$ of the
cq-state $\rho_{XE}$. 
\end{definition}
We will employ privacy amplification in the form of the 
following Lemma, which
is an immediate consequence of Lemma~\ref{guessing-lemma}
and Theorem~\ref{theorem:renato} (Theorem 5.5.1 in~\cite{renato:diss}):

\begin{lemma}\label{simplifiedPA}
Let $\setF$ be a class of two-universal hash functions from $\01^n$ to $\01^\ell$.
Let $F$ be a random variable that is uniformly and independently
distributed over $\setF$, and let $\rho_{XE}$ be a cq-state. Then,
$$
d(F(X)|F,\rho_E) \leq 2^{\frac{\ell}{2}-1} \sqrt{P_g(X|\rho_E)} \, .
$$
If we have an additional $k$ bits of classical information $D$ about $X$,
$$
d(F(X)|F,D,\rho_E) \leq 2^{\frac{\ell+k}{2}-1}\sqrt{P_g(X|\rho_E)}.
$$
\end{lemma}

Furthermore, we will need the following lemma which states that the
optimal strategy to guess $X=x \in \set{0,1}^n$ given individual
quantum information about the bits of $X$ is to measure each
register individually.
\begin{lemma}\label{lemma:individual}
Let $\rho_{XE}$ be a cq-state with uniformly distributed $X=x \in
\01^n$ and $\rho_E^x = \rho_{E_1}^{x_1} \otimes \ldots \otimes
\rho_{E_n}^{x_n}$. Then the maximum probability of guessing $x$
given state $\rho_E$ is $P_g(X|\rho_E)  = \Pi_{i=1}^n
P_g(X_i|\rho_{E_i})$, which can be achieved by measuring each
register separately.
\end{lemma}
\begin{proof}
  For simplicity, we will assume that each bit is encoded using the
  same states $\rho_0 = \rho_{E_i}^0$ and $\rho_1 = \rho_{E_i}^1$.  The argument
  for different encodings is analogous, but harder to read.  First of
  all, note that we can phrase the problem of finding the optimal
  probability of distinguishing two states as a semi-definite program
  (SDP)
\begin{sdp}{maximize}{$\frac{1}{2}\left(\Tr(M_0\rho_0) + \Tr(M_1\rho_1)\right)$}
&$M_0,M_1 \geq 0$\\
&$M_0 + M_1 = \id$
\end{sdp}
with the dual program
\begin{sdp}{minimize}{$\frac{1}{2}\Tr(Q)$}
&$Q \geq \rho_0$\\
&$Q \geq \rho_1$.
\end{sdp}
Let $p_*$ and $d_*$ denote the optimal values of the primal and dual
respectively. From the weak duality of SDPs, we have $p_* \leq d_*$.
Indeed, since $M_0,M_1 = \id/2$ are feasible solutions, we even have
strong duality: $p_* = d_*$ \cite{VB:sp}.

Of course, the problem of determining the entire string $x$ from
$\hat{\rho}_x \assign \rho_E^x$ can
also be phrased as a SDP:
\begin{sdp}{maximize}{$\frac{1}{2^n} \sum_{x \in \01^n} \Tr(M_x\hat{\rho}_x)$}
&$\forall x, M_x \geq 0$\\
&$\sum_{x \in \01^n} M_x = \id$
\end{sdp}
with the corresponding dual
\begin{sdp}{minimize}{$\frac{1}{2^n} \Tr(\hat{Q})$}
&$\forall x, \hat{Q} \geq \hat{\rho}_x$.
\end{sdp}
Let $\hat{p}_*$ and $\hat{d}_*$ denote the optimal values of this new primal and dual
respectively. Again, $\hat{p}_* = \hat{d}_*$.

Note that when trying to learn the entire string $x$, we are of course free
to measure each register individually and thus $(p_*)^n \leq \hat{p}_*$. We now
show that $\hat{d}_* \leq (d_*)^n$ by constructing a dual solution $\hat{Q}$ from the optimal
solution to the dual of the single-register case, $Q_*$:
Take $\hat{Q} = Q_*^{\otimes n}$. Since $Q_* \geq \rho_0$ and $Q_* \geq \rho_1$
it follows that $\forall x, Q_*^{\otimes n} \geq \hat{\rho}_x$. Thus $\hat{Q}$ is satisfies
the dual constraints. Clearly, $2^{-n} \Tr(\hat{Q}) = (2^{-1} \Tr(Q_*))^n$ and thus we
have $\hat{d}_* \leq (d_*)^n$ as promised. But from $(p_*)^n \leq \hat{p}_*$,
$\hat{p}_* = \hat{d}_*$, and $p_* = d_*$ we immediately have $\hat{p}_* = (p_*)^n$.
\end{proof}

The next tool we need is an uncertainty relation for noisy channels
and measurements. Let $\sigma_{0,+} = \outp{0}{0}$, $\sigma_{1,+} =
\outp{1}{1}$, $\sigma_{0,\times} = \outp{+}{+}$ and $\sigma_{1,\times} =
\outp{-}{-}$ denote the BB84-states corresponding to the encoding of a bit $z
\in \01$ into basis $b \in \{+,\times\}$ (computational resp.
Hadamard basis). Let $\sigma_+ = (\sigma_{0,+} + \sigma_{1,+})/2$
and $\sigma_\times = (\sigma_{0,\times} + \sigma_{1,\times})/2$.
Consider the state ${\cal S}(\sigma_{z,b})$ for
some super-operator ${\cal S}$.
Note that $P_g(X|{\cal S}(\sigma_{b}))$ (see Lemma~\ref{lemma:individual})
denotes the maximal average success probability for guessing a uniformly distributed $X$ when $b=+$ or
$b=\times$. An uncertainty relation for such success probabilities can
be stated as
\begin{equation}
P_g(X|{\cal S}(\sigma_+)) \cdot P_g(X|{\cal S}(\sigma_\times)) \leq \Delta({\cal S})^2,
\label{eq:uncert}
\end{equation}
where $\Delta$ is a function from the set of superoperators to the real numbers.
For example, when ${\cal S}$ is a quantum measurement ${\cal M}$
mapping the state $\sigma_{z,b}$ onto purely classical information
it can be argued (e.g.~by using a purification argument and
Corollary 4.15 in \cite{chris:diss}) that $\Delta({\cal M}) \equiv
\frac{1}{2}(1+2^{-1/2})$ which can be achieved by a measurement in
the Breidbart basis,\index{basis!Breidbart}\index{Breidbart basis}
where the Breidbart basis is given by $\{\ket{0}_B,\ket{1}_B\}$ with 
\begin{eqnarray*}
\ket{0}_B &=&
\cos(\pi/8)\ket{0} + \sin(\pi/8) \ket{1}\\
\ket{1}_B &=&
\sin(\pi/8) \ket{0} - \cos(\pi/8) \ket{1}
\end{eqnarray*}
It is clear that for a unitary superoperator $U$ we have
$\Delta(U)^2=1$ which can be achieved. It is not hard to show that
\begin{lemma}\label{lem:iso}
The only superoperators ${\cal S}\colon \hil_{in} \rightarrow
\hil_{out}$ with $\dim(\hil_{in}) = 2$ 
for which $P_g(X|{\cal S}(\sigma_+)) \cdot P_g(X|{\cal
  S}(\sigma_{\times}))=1$ are reversible operations.
\end{lemma}
\begin{proof}
Using Helstrom's formula \cite{helstrom:detection} we have that
$P_g(Z|\mS(\sigma_b))=\frac{1}{2}[1+\left|\left|{\cal S}(\sigma_{0,b})-{\cal S}(\sigma_{1,b})\right|\right|_{1}/2]$ and thus for $\Delta({\cal S})=1$
we need that for both $b \in \{\times, +\}$, $||{\cal S}(\sigma_{0,b})-{\cal S}(\sigma_{1,b})||_{1}/2=1$. This implies that
${\cal S}(\sigma_{0,b})$ and ${\cal S}(\sigma_{1,b})$ are states which have support on orthogonal
sub-spaces for {\em both} $b$. Let ${\cal S}(\sigma_{0,+})=\sum_k p_k \ket{\psi_k}\bra{\psi_k}$ and
${\cal S}(\sigma_{1,+})=\sum_k q_k \ket{\psi_k^{\perp}}\bra{\psi_k^{\perp}}$ where for all $k,l$ $\bra{\psi_k^{\perp}}\psi_l\rangle=0$.
Consider the purification of ${\cal S}(\sigma_{i,b})$ using an ancillary system
i.e. $\ket{\phi_{i,b}}=U_{{\cal S}} \ket{i}_b\ket{0}$. We can write
$\ket{\phi_{0,+}}=\sum_k \sqrt{p_k} \ket{\psi_k,k}$ and $\ket{\phi_{1,+}}=\sum_k \sqrt{q_k} \ket{\psi_k^{\perp},k}$.
Hence $U_{{\cal S}}\ket{0}_{\times}\ket{0}=\frac{1}{\sqrt{2}}(\ket{\phi_{0,+}}+\ket{\phi_{1,+}})$ and similar for
$U_{\cal S}\ket{1}_{\times} \ket{0}$. So we can write
\vspace{-1mm}
$$
\left|\left|{\cal S}(\sigma_{0,\times})-{\cal S}(\sigma_{1,\times})\right|\right|_{1} =
\left|\left|\sum_k \sqrt{p_k q_k} (\ket{\psi_k}\bra{\psi_k^{\perp}}+\ket{\psi_k^{\perp}}\bra{\psi_k})\right|\right|_{1} \leq 
2 \sum_k \sqrt{p_k q_k}.
$$
For this quantity to be equal to 2 we observe that it is necessary that $p_k=q_k$. Thus we set $p_k=q_k$.
We observe that if any of the states $\ket{\psi_k}$ (or $\psi_k^{\perp}$) are non-orthogonal, i.e. $|\bra{\psi_k} \psi_l \rangle|>0$, then
we have
$||\sum_k p_k (\ket{\psi_k}\bra{\psi_k^{\perp}}+\ket{\psi_k^{\perp}}\bra{\psi_k})||_{1} < 2$.

Let $S_k$ be the two-dimensional subspace spanned by the orthogonal vectors $\ket{\psi_k}$ and $\ket{\psi_k^{\perp}}$.
By the arguments above, the spaces $S_k$ are mutually orthogonal. We can reverse the super-operator ${\cal S}$ by first
projecting the output into one of the orthogonal subspaces $S_k$ and then applying a unitary operator $U_k$
that maps $\ket{\psi_k}$ and $\ket{\psi_k^{\perp}}$ onto the states $\ket{0}$ and $\ket{1}$.
\end{proof}

Finally, we need the following little technical lemma:
\begin{lemma} \label{lem:bestdistribution}
For any $\frac12 \leq p_i \leq 1$ with $\prod_{i=1}^n p_i \leq p^n$, we
have
\begin{equation} \label{eq:product}
\frac{1}{2^{n}} \prod_{i=1}^n (1+p_i) \leq p^{\log(4/3) n} \, .
\end{equation}
\end{lemma}
\begin{proof}
With $\lambda \assign \log(4/3)$, it is easy to verify that $ p_i^{-\lambda} +
  p_i^{1-\lambda} \leq 2$ for $1/2 \leq p_i \leq 1$ and therefore,
$$
\frac{1}{2^{n}}\prod_{i=1}^n (1+p_i) = \frac{1}{2^{n}} \prod_{i=1}^n p_i^\lambda \left( p_i^{-\lambda} +
  p_i^{1-\lambda} \right)
\leq \frac{1}{2^{n}} \cdot p^{\lambda n} \cdot 2^n.
$$
\end{proof}

\section{Protocol and analysis}

\subsection{Protocol}
We use $\in_R$ to denote the uniform choice of an element from a set.
We further use $x_{|\setT}$ to denote the string $x=x_1,\ldots,x_n$
restricted to the bits indexed by the set $\setT \subseteq
\{1,\ldots,n\}$.  For convenience, we take $\{+,\times\}$ instead of
$\01$ as domain of Bob's choice bit $C$ and denote by $\overline{C}$ the
bit different from $C$.

\begin{protocol}{1-2 $\mbox{ROT}^\ell(C,T)$\cite{serge:new}}{}
\item Alice picks $X \in_R \01^n$ and $\Theta \in_R \{+,\times\}^n$.
  Let $\setI_b = \{i\mid \Theta_i = b\}$ for $b \in \{+,\times\}$. At
  time $t=0$, she sends $\sigma_{X_1,\Theta_1} \otimes \ldots \otimes
  \sigma_{X_n,\Theta_n}$ to Bob.
\item Bob measures all qubits in the basis corresponding to his choice
  bit $C \in \{+,\times\}$.  This yields outcome $X' \in \01^n$.
\item Alice picks two hash functions $F_+,F_\times \in_R
  \setF$, where $\setF$ is a class of two-universal hash functions. At time $t=T$, she sends $\setI_+$,$\setI_\times$, $F_+$,$F_\times$ to Bob.
Alice outputs $S_+ = F_+(X_{|\setI_+})$ and $S_{\times}
  = F_\times(X_{|\setI_\times})$ \footnote{If $X_{|\setI_b}$ is less
    than $n$ bits long Alice pads the string $X_{|\setI_b}$ with 0's
    to get an $n$ bit-string in order to apply the hash function to
    $n$ bits.}.
\item Bob outputs $S_C = F_C(X'_{|\setI_C})$.
\end{protocol}

\subsection{Analysis}
We now show that this protocol is secure according to
Definition~\ref{def:ROT}.

(i) Correctness: It is clear that the protocol is correct.  
Bob can
determine the string $X_{|\setI_C}$ (except with negligible
probability $2^{-n}$ the set ${\cal I}_C$ is non-empty) and hence
obtains $S_C$.

(ii) Security against dishonest Alice: this holds in the same way as
shown in~\cite{serge:new}. As the protocol is non-interactive, Alice
never receives any information from Bob at all, and Alice's input
strings can be extracted by letting her interact with an unbounded
receiver.

(iii) Security against dishonest Bob: Our goal is to show that there
exists a $C' \in \{+,\times\}$ such that Bob is completely ignorant
about $S_{\ol{C'}}$. In our adversary model, Bob's collective storage cheating
strategy can be described by some superoperator
$$
\mS=\bigotimes_{i=1}^n \mS_i
$$ that is applied on the qubits between
the time they arrive at Bob's and the time $T$ that Alice sends the
classical information.
We define the choice bit $C'$ as a fixed function of Bob's cheating
strategy $\mS$. 
Formally, we set $C' \equiv +$ if
$\prod_{i=1}^n P_g(X_i|\mS_i(\sigma_+)) \geq \prod_{i=1}^n
P_g(X_i|\mS_i(\sigma_{\times}))$ and $C' \equiv \times$ otherwise.

Due to the uncertainty relation for each ${\cal S}_i$ (from
Eq.~(\ref{eq:uncert})) it then holds that 
$$
\prod_i P_g(X_i|{\cal
S}_i(\sigma_{\ol{C'}})) \leq \prod_i \Delta({\cal S}_i) \leq
(\Delta_{\rm max})^n
$$ 
where $\Delta_{\rm max} \assign \max_{i} \Delta({\cal
S}_i)$. This will be used in the proof below.

In the remainder of this section, we show that the non-uniformity
$$
\delta_{\rm sec} := d(S_{\ol{C'}}|S_{C'}C'\rho_B)
$$ 
is negligible
in $n$ for a collective attack. 
Here $\rho_B$ is the complete quantum
state of Bob's lab at the end of the protocol including the classical
information $\setI_+, \setI_{\times}, F_+, F_{\times}$ he got from
Alice and his quantum information $\bigotimes_{i=1}^n
\mS_i(\sigma_{X_i,\Theta_i})$. Expressing the non-uniformity in terms
of the trace-distance allows us to observe that $\delta_{\rm
  sec}=2^{-n} \sum_{\theta \in \{+,\times\}^n} d(S_{\ol{C'}} |
\Theta=\theta , S_{C'} C' \rho_B)$.  Now, for fixed $\Theta=\theta$,
it is clear from the construction that $S_{C'},C',F_{C'}$ and
$\bigotimes_{i \in \setI_{C'}} \mS_i(\sigma_{X_i,C'})$ are independent
of $S_{\ol{C'}}=F_{\ol{C'}}(X_{|\setI_{\ol{C'}}})$ and we can use
Eq.~(\ref{eq:indep}).  Hence, one can bound the non-uniformity as in
Lemma~\ref{simplifiedPA}, i.e.~by the square-root of the probability
of correctly guessing $X_{|_{\setI_{\ol{C'}}}}$ given the state
$\bigotimes_{i \in \setI_{\ol{C'}}} \mS_i (\sigma_{X_i,\ol{C'}})$.
Lemma~\ref{lemma:individual} tells us that to guess $X$, Bob can
measure each remaining qubit individually and hence we obtain
\begin{align*}
\delta_{\rm sec} &\leq 2^{\frac{\ell}{2}-1} \cdot 2^{-n} \!\!\! \sum_{\theta \in \{+,\times\}^n}
\sqrt{ \prod_{i \in \setI_{\ol{C'}}} P_g(X_i | \mS_i(\sigma_{\ol{C'}}) )}\\
&\leq 2^{\frac{\ell}{2}-1} \sqrt{ 2^{-n} \sum_{\theta \in \{+,\times\}^n}
\prod_{i \in \setI_{\ol{C'}}} P_g(X_i | \mS_i(\sigma_{\ol{C'}}) )}\\
&\leq 2^{\frac{\ell}{2}-1} \sqrt{ 2^{-n} \prod_{i=1}^n \big(1+ P_g(X_i |
  \mS_i(\sigma_{\ol{C'}}) ) \big)} \, ,
\end{align*}
where we used the concavity of the square-root function in the last
inequality.  Lemma~\ref{lem:bestdistribution} together with the bound
$\prod_i P_g(X_i|{\cal S}_i(\sigma_{\ol{C'}})) \leq (\Delta_{\rm max})^n$
lets us conclude that
$$
\delta_{\rm sec} \leq 2^{\frac{\ell}{2}-1} \cdot (\Delta_{\rm
  max})^{\frac{\log(4/3)}{2} n}.
$$
Lemma~\ref{lem:iso} shows that for essentially any noisy superoperator
$\Delta({\cal S}) < 1$.  This shows that for any collective attacks
there exists an $n$ which yields arbitrarily high security.

\section{Practical oblivious transfer}\index{oblivious transfer!practical}
\index{1-2 OT!practical}

In this section, we prove the security of a ROT protocol that is
robust against noise for the honest parties. Our protocol is thereby
a small modification of the protocol considered
in~\cite{chris:diss}. Note that for our analysis, we have to assume
a worst-case scenario where a dishonest receiver Bob has access to a
perfect noise-free quantum channel and only experiences noise during
storage. First, we consider erasure noise (in practice corresponding
to photon loss) during preparation, transmission and measurement of
the qubits by the honest parties. Let $1-p_{\rm erase}$ be the total
constant probability for an honest Bob to measure and detect a photon in the
$\{+,\times\}$ basis given that an honest Alice prepares a qubit (or weak
laser pulse) in her lab and sends it to him. The probability $p_{\rm
erase}$ is determined among others by the mean photon number in the
pulse, the loss on the channel and the quantum efficiency of the
detector. In our protocol we assume that the (honest) erasure rate
$p_{\rm erase}$ is {\em independent} of whether qubits were encoded
or measured in the $+$- or $\times$-basis. This assumption is
necessary to guarantee the correctness and the security against a
cheating \emph{Alice} only. Fortunately, this assumption is well matched with
physical capabilities.

Any other noise source during preparation, transmission and
measurement can be characterized as an effective classical noisy
channel resulting in the output bits $X'$ that Bob obtains at
Step~\ref{step:reception} of Protocol~\ref{prot:practical}.  For simplicity, we model
this compound noise source as a classical binary symmetric channel
acting independently on each bit of $X$. Typical noise sources for
polarization-encoded qubits are depolarization during transmission,
dark counts in Bob's detector and misaligned polarizing
beam-splitters. Let the effective bit-error probability of this binary
symmetric channel be $p_{\rm error} < 1/2$.

Before engaging in the actual protocol, Alice and Bob agree on the
system parameters $p_{\rm erase}$ and $p_{\rm error}$ similarly to
Step~1 of the protocol in~\cite{crepeau:practicalOT}. Furthermore,
they agree on a family $\set{C_n}$ of linear error correcting codes
of length $n$ capable of efficiently correcting $n \cdot p_{\rm
error}$ errors. For any string $x \in \set{0,1}^n$, error correction
is done by sending the syndrome information $syn(x)$ to Bob from
which he can correctly recover $x$ if he holds an output $x' \in
\set{0,1}^n$ obtained by flipping each bit of $x$ independently with
probability $p_{\rm error}$. It is known that for large enough $n$,
the code $C_n$ can be chosen such that its rate is arbitrarily close
to $1-h(p_{\rm error})$ and the syndrome length (the number of parity check bits) are
asymptotically bounded by $|syn(x)| < h(p_{\rm error})
n$~\cite{crepeau:efficientOT}, where $h(p_{\rm error})$ is the binary Shannon
entropy. We assume the players have synchronized clocks. In each
time slot, Alice sends one qubit (laser pulse) to Bob.

\begin{protocol}{Noise-Protected Photonic 1-2 $\mbox{ROT}^\ell(C,T)$}{} \label{prot:practical}
\item Alice picks $X \in_R \01^n$ and $\Theta \in_R \{+,\times\}^n$.
\item For $i=1,\ldots,n$: In time slot $t=i$, Alice sends
  $\sigma_{X_i,\Theta_i}$ as a phase- or polarization-encoded weak pulse
  of light to Bob.
\item \label{step:reception} In each time slot, Bob measures the
  incoming qubit in the basis corresponding to his choice bit $C \in
  \{+,\times\}$ and records whether he detects a photon or not.  He
  obtains some bit-string $X' \in \01^m$ with $m \leq n$.
\item Bob reports back to Alice in which time slots he received a
  qubit. Alice restricts herself to the set of $m \leq n$ bits that Bob
  did not report as missing. Let this set of qubits be $S_{\rm
    remain}$ with $|S_{\rm remain}|=m$.
\item \label{step:alicecheck} Let $\setI_b = \{i \in S_{\rm
    remain}\mid \Theta_i = b\}$ for $b \in \{+,\times\}$ and let
  $m_b=|\setI_b|$.  Alice aborts the protocol if either $m_+$ or
  $m_\times \leq (1-p_{\rm erase})n/2-O(\sqrt{n})$.
  If this
  is not the case, Alice picks two hash functions
  $F_+,F_\times \in_R \setF$, where $\setF$ is a set of two-universal hash functions. At time $t=n+T$, Alice sends
  $\setI_+$,$\setI_\times$, $F_+$,$F_\times$, and the syndromes $syn(X_{|\setI_+})$
  and $syn(X_{|\setI_\times})$ according to codes of appropriate length $m_b$ to Bob.  Alice outputs $S_+ =
  F_+(X_{|\setI_+})$ and $S_{\times} = F_\times(X_{|\setI_\times})$.
\item Bob uses $syn(X_{|\setI_C})$ to correct the errors on his output $X'_{|\setI_C}$. He obtains the corrected bit-string
$X_{\rm cor}$ and outputs $S'_C = F_C(X_{\rm cor})$.
\end{protocol}

Let us consider the security and correctness of this modified protocol.\\
(i) Correctness: By assumption, $p_{\rm erase}$ is
independent of the basis in which Alice sent the qubits.
Thus, $S_{\rm remain}$ is with high
probability a random subset of the transmitted qubits
of of size $m \approx (1-p_{\rm erase})n \pm
O(\sqrt{n})$ qubits independent of the value of bases $\Theta$. This
implies that in Step~\ref{step:alicecheck} the protocol is aborted with a probability
exponentially small in $m$, and hence in $n$. The codes are chosen
such that Bob can decode except with negligible probability. These
facts imply that if both parties are honest the protocol is correct
(i.e. $S_C=S'_C$) with exponentially small
probability of error. \\
(ii) Security against dishonest Alice: Even though in this scenario Bob {\em does}
communicate to Alice, the information stating which qubits were erased is by assumption independent of the basis in which he
measured and thus of his choice bit $C$. Hence Alice does not learn
anything about his choice bit $C$. Her input strings can be extracted as in Protocol~1. \\
(iii) Security against dishonest Bob: 
Our analysis is essentially
identical to our analysis for Protocol~1 where we address the
error-correcting properties as in~\cite{chris:diss}.
First of all, we note that Bob
can always make Alice abort the protocol by reporting back an
insufficient number of received qubits. 
If this is not the case,
then we define $C'$ as in the analysis of Protocol~1 and we need to
bound the non-uniformity $\delta_{\rm sec}$ as before. Let us for
simplicity assume that $m_b=m/2$ (this is true with high
probability, up to a factor of $O(\sqrt{n})$ which becomes negligible 
for large $n$) with $m \approx (1-p_{\rm erase})n$.
We perform the same analysis, where we restrict ourselves
to the set of remaining qubits.
We first follow through the same steps simplifying the non-uniformity
using that the total attack superoperator ${\cal S}$ is a product of
superoperators. Then we use the bound in Lemma~\ref{simplifiedPA}
for each $\theta \in \{+,\times\}^n$ where we now have to condition on the additional information
$syn(X_{|\setI_{\ol{C'}}})$ which is $m h(p_{\rm error})/2$ bits long.
Note that Bob does not gain any information when Alice
aborts the
protocol, since her decision to abort is a function of the bits
Bob reported as being erased and he can thus compute Alice's decision
himself.
Using the second part of Lemma~\ref{simplifiedPA} and following
identical steps in the remainder of the proof implies
\begin{equation}
\delta_{\rm sec} \leq  2^{\frac{\ell}{2}-1 + h(p_{\rm error}) \frac{m}{4}}
(\Delta_{\rm
  max})^{\frac{\log(4/3)}{2} m} \, .
  \label{eq:sec}
\end{equation}
From this expression it is clear that the security depends crucially
on the value of $\Delta_{\rm max}$ versus the binary entropy $h(p_{\rm error})$.
The trade-off in our bound is not extremely favorable for security
as we will see.

\section{Example: depolarizing noise}\index{depolarizing noise}
Let us now consider the security in an explicit example, where Bob's
storage is affected by depolarizing noise, and he is not able to encode
the incoming qubits into a higher-dimensional system such as an error
correcting code. 

\begin{figure}
\begin{center}
\includegraphics{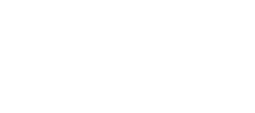}
\caption{{\small Bob performs a partial measurement $\mP_i$, followed by noise $\mN$, and outputs a guess bit
$x_g$ depending on his classical measurement outcome, the remaining quantum state, and the
additional basis information.}} \label{figure:depolModel}
\end{center}
\end{figure}

Again, we first address the simpler setting 
where the honest players experience
no noise themselves.
In order to explicitly bound $\Delta(\mS_i)$ we
should allow for intermediate strategies of Bob in which he partially
measures the incoming qubits leaving some quantum information
undergoing depolarizing noise. To model this noise we let $\mS_i =\mN
\circ \mP_i$, where $\mP_i$ is any noiseless quantum operation of
Bob's choosing from one qubit to one qubit that generates some
classical output. For example, $\mP_i$ could be a partial measurement
providing Bob with some classical information and a slightly disturbed
quantum state, or just a unitary operation. Let 
$$
\mN(\rho) := r \rho + (1-r)\frac{\id}{2}
$$ 
be the fixed depolarizing 'quantum storage'\index{depolarizing channel}\index{noise!depolarizing}
channel that Bob cannot influence (see Figure
\ref{figure:depolModel}).

To determine $\delta_{\rm sec}$, we have to find an uncertainty relation
similar to Eq.~(\ref{eq:uncert}) by optimizing over all possible
partial measurements $\mP_i$,
\begin{equation}\label{eq:target}
\Delta_{\max}^2 = 
\max_{\mS_i} \Delta(\mS_i)^2 = \max_{\mP_i} P_g(X|{\cal S}_i(\sigma_+)) \cdot
P_g(X|{\cal S}_i(\sigma_\times)).
\end{equation}
We solve this problem for depolarizing noise using
the symmetries inherent in our problem. In Section \ref{optimalProof} we prove the following.
\begin{theorem} \label{thm:depolarize}\label{thm:depolOptimal}
Let $\mN$ be the depolarizing channel and let $\max_{\mS_i} \Delta(\mS_i)$
be defined as above. Then
\begin{equation*}
\max_{\mS_i} \Delta(\mS_i) = \left\{
\begin{array}{ll}
\frac{1+r}{2} & \mbox{ for } r \geq \frac{1}{\sqrt{2}}\\
\frac{1}{2} + \frac{1}{2\sqrt{2}} & \mbox{ for } r <
\frac{1}{\sqrt{2}}
\end{array}\right.
\end{equation*}
\end{theorem}
Our result shows that for $r < 1/\sqrt{2}$ a direct measurement $\mM$ in the Breidbart basis is the best attack Bob can perform. For this measurement,
we have $\Delta(\mM)= 1/2 +
1/(2\sqrt{2})$.
If the depolarizing noise is low ($r \geq 1/\sqrt{2}$), then our result
states that the best strategy for Bob is to simply store the qubit as is.

\subsection{Optimal cheating strategy}\label{optimalProof}
We now prove Theorem~\ref{thm:depolOptimal} in a series of steps.
Recall, that to determine the security bound, we have to find an uncertainty relation similar to Eq.~(\ref{eq:uncert})
by optimizing over all possible partial measurements $\mP$ and final measurements $\mM$ as in
Eq.~\ref{eq:target}.
To improve readability, we will drop the index $i$ and use
$\mS$ in place of $\mS_i$ to denote the cheating operation acting on a single qubit.
For our analysis, it will be convenient to think of $\mP$ as a partial measurement of the incoming qubit.
Note that this corresponds to letting Bob perform an arbitrary CPTP map from the space of the incoming qubit to the space carrying the stored qubit. 
It will furthermore be convenient to consider the maximizing the sum instead:
$$
\Gamma(\mS) = \max_{\mM,\mP} P_g(X|{\cal S}(\sigma_+)) + P_g(X|{\cal S}(\sigma_\times)).
$$
This immediately gives us the bound $\Delta(\mS) \leq \Gamma(\mS)/2$.
In the following, we will use the shorthand 
\begin{eqnarray*}
p_+ &:=& P_g(X|\mS(\sigma_+))\\
p_\times &:=& P_g(X|\mS(\sigma_\times))
\end{eqnarray*}
for the probabilities that Bob correctly decodes the bit after Alice has announced the basis 
information.

Any measurement Bob may perform can be characterized by a set of measurement operators $\{F_k\}$ such that $\sum_k F_k^\dagger F_k = \id$. The probability that Bob succeeds in decoding the bit after the announcement of the basis is simply the average over the probability that he correctly decodes the bit, conditioned on the fact that he obtained outcome $k$. I.e., for $b \in \{+,\times\}$
\begin{eqnarray*}
p_b &=& 
\sum_k p_{k|b} \left(\frac{1}{2} + \frac{1}{4}||p_{0|kb} N(\tilde{\sigma}^k_{0,b}) - p_{1|kb} N(\tilde{\sigma}^k_{1,b})||_{1}\right)\\
\label{plainProb}
&=& \frac{1}{2} + \frac{1}{4} \sum_k p_{k|b} ||r (p_{0|kb} \tilde{\sigma}^k_{0,b} - p_{1|kb} \tilde{\sigma}^k_{1,b})
+ (1-r) (p_{0|kb} - p_{1|kb})\id/2||_{1},
\end{eqnarray*}
where 
$$
p_{k|b} = 
\Tr\left(F_k
\frac{\sigma_{0,b}+\sigma_{1,b}}{2} F_k^\dagger\right)=\frac12 \Tr(F_k
F_k^\dagger)
$$
is the probability of obtaining measurement outcome $k$
conditioned on the fact that the basis was $b$ (and we even see from
the above that it is actually independent of $b$),
$\tilde{\sigma}^k_{0,b} = F_k\sigma_{0,b}F_k^\dagger/p_{k|0b}$ is the
post-measurement state for outcome $k$, and $p_{0|kb}$ is the
probability that we are given this state. Definitions are analogous for
the bit $1$.

We now show that Bob's optimal strategy is to measure in the Breidbart basis for $r < 1/\sqrt{2}$, 
and to simply store the qubit for $r \geq 1/\sqrt{2}$. This then immediately allows us to evaluate $\Delta_{\max}$.
To prove our result, we proceed in three steps: First, we will simplify our problem considerably until
we are left with a single Hermitian measurement operator over which we need to maximize. Second, we show
that the optimal measurement operator is diagonal in the Breidbart basis. And finally, we show that
depending on the amount of noise, this measurement operator is either proportional to the identity,
or proportional to a rank one projector. Our individual claims are indeed very intuitive.

For any measurement $M = \{F_k\}$, let $B(M) = p_+^M + p_\times^M$ for the measurement $M$, where $p_+^M$ and $p_\times^M$
are the success probabilities similar to Eq.~(\ref{plainProb}), but restricted to using the measurement $M$.
First of all, note that we can easily combine two measurements. Intuitively,
the following statement says that if we choose one measurement with probability $\alpha$, and the other
with probability $\beta$ our average success probability will be the average of the success probabilities
obtained via the individual measurements:
\begin{claim}\label{convexity}
Let $M_1=\{F_k^1\}$ and $M_2=\{F_k^2\}$ be two measurements. Then
$B(\alpha M_1 + \beta M_2) = \alpha B(M_1) + \beta B(M_2)$, where  
where $\alpha M_1 + \beta M_2 = \{\sqrt{\alpha} F_k^1\} \cup \{\sqrt{\beta} F_k^2\}$ for $\alpha,\beta \geq 0$
and $\alpha + \beta=1$.
\end{claim}
\begin{proof}
  Let $F=\set{F_k}_{k=1}^f$ and $G=\set{G_k}_{k=1}^g$ be measurements,
  $0 \leq \alpha \leq 1$ and $M \assign \set{\sqrt{\alpha}
    F_k}_{k=1}^{f} \cup \set{\sqrt{1-\alpha} G_k}_{k=f+1}^{f+g}$ be
  the measurement $F$ with probability $\alpha$ and measurement $G$
  with probability $1-\alpha$. We denote by $p_{\cdot}^F, p_{\cdot}^G,
  p_{\cdot}^M$ the probabilities corresponding to measurements $F,G,M$
  respectively. Observe that for $1\leq k \leq f$, $p_{k|b}^M =
  \frac12 \Tr(\alpha F_k F_k^\dagger) = \alpha p_{k|b}^F$ and
  analogously for $f+1 \leq k \leq f+g$, we have $p_{k|b}^M =
  (1-\alpha)p_{k|b}^G$. We observe furthermore that for $1\leq k \leq
  f$ and $x \in \set{0,1}$, $\alpha$ cancels out by the normalization,
  $\tilde{\sigma}_{x,b}^{k,M} = \frac{\alpha F_k \sigma_{x,b} F_k^\dagger}{p_{k|xb}^M} = \frac{F_k
    \sigma_{x,b} F_k^\dagger}{p_{k|xb}^F} = \tilde{\sigma}_{x,b}^{k,F}$ and similarly for $f+1 \leq k
  \leq f+g$. Finally, we can convince ourselves that
  $p_{x|kb}^M=p_{x|kb}^F=p_{x|(k-f)b}^G$, as the probability to be
  given state $\tilde{\sigma}_{0,b}^k$ is the same when the measurement
  outcome and the basis is fixed.
Putting everything together, we obtain
\begin{align*}
  p_b^M &= \sum_{k=1}^{f+g} p_{k|b}^M \left(\frac{1}{2} +
    \frac{1}{4}||p_{0|kb}^M N(\tilde{\sigma}^{k,M}_{0,b}) - p_{1|kb}^M
    N(\tilde{\sigma}^{k,M}_{1,b})||_{1}\right) \\ 
 &= \sum_{k=1}^{f} \alpha p_{k|b}^F\left(\frac{1}{2} +
    \frac{1}{4}||p_{0|kb}^F N(\tilde{\sigma}^{k,F}_{0,b}) - p_{1|kb}^F
    N(\tilde{\sigma}^{k,F}_{1,b})||_{1}\right) \\ 
 &\quad +\sum_{k=f+1}^g (1-\alpha) p_{k|b}^G \left(\frac{1}{2}\right.
+
    \left.\frac{1}{4}||p_{0|kb}^G N(\tilde{\sigma}^{k,G}_{0,b}) - p_{1|kb}^G
    N(\tilde{\sigma}^{k,G}_{1,b})||_{1}\right) \\
 &= \alpha p_b^F + (1-\alpha) p_b^G \, .
\end{align*}
\end{proof}

We can now make a series of observations.
\begin{claim}\label{invarianceOfBunderG}
Let $M = \{F_k\}$ and $G = \{\id,X,Z,XZ\}$. Then for all $g \in G$ we have $B(M) = B(gMg^\dagger)$.
\end{claim}
\begin{proof}
This claim follows immediately from that fact that for the trace norm we have $||U A U^\dagger||_{1} = ||A||_{1}$
for all unitaries $U$, and by noting that for all $g \in G$, $g$ can at most exchange the roles
of $0$ and $1$. I.e., we perform a bit flip before the measurement which we can correct for afterwards
by applying classical post-processing: we have for all $g \in G$ that
\begin{eqnarray*}
&&p_{k|b}
\left|\left|p_{0|kb} N\left(
\frac{F_k g \sigma_{0,b} g^\dagger F_k^\dagger}{p_{k|0b}}\right)
- p_{1|kb} N\left(\frac{F_k g \sigma_{1,b} g^\dagger F_k^\dagger}{p_{k|1b}}\right)\right|\right|_{1}\\
&&= 
p_{k'|b}\left|\left|p_{0|kb} N\left(\frac{F_k \sigma_{0,b} F_k^\dagger}{p_{k|0b}}\right)
- p_{1|kb} N\left(\frac{F_k  \sigma_{1,b}  
F_k^\dagger}{p_{k|1b}}\right)\right|\right|_{1}.
\end{eqnarray*}
\end{proof}

It also follows that
\begin{corollary}\label{simplify}
For all $k$ we have for all $b \in \{+,\times\}$ and $g \in G$ that
\begin{eqnarray*}
&&\left|\left|p_{0|kb} N\left(\frac{F_k \sigma_{0,b} F_k^\dagger}{p_{k|0b}}\right)
- p_{1|kb} N\left(\frac{F_k \sigma_{1,b} F_k^\dagger}{p_{k|1b}}\right)\right|\right|_{1}\\
&& = 
\left|\left|p_{0|kb} N\left(\frac{F_{k} g\sigma_{0,b}g^\dagger F_{k}^\dagger}{p_{k|0b}}\right)
- p_{1|kb} N\left(\frac{F_{k} g \sigma_{1,b}g^\dagger  F_{k}^\dagger}{p_{k|1b}}\right)\right|\right|_{1}. 
\end{eqnarray*}
\end{corollary}
\begin{proof}
This follows from the proof of Claim~\ref{invarianceOfBunderG}.
\end{proof}

\begin{claim}\label{maximumInvariant}
Let $G = \{\id,X,Z,XZ\}$.
There exists a measurement operator $F$ such that the maximum of $B(M)$ over all measurements
$M$ is achieved by a measurement proportional to 
$\{gFg^\dagger \mid g \in G\}$.
\end{claim}
\begin{proof}
Let $M= \{F_k\}$ be a measurement. Let $K = |M|$ be the number of measurement operators. 
Clearly, $\hat{M} = \{\hat{F}_{g,k}\}$
with 
$$
\hat{F}_{g,k} = \frac{1}{4} g F_k g^\dagger,
$$
is also a quantum measurement since $\sum_{g,k} \hat{F}_{g,k}^\dagger\hat{F}_{g,k} = \id$.
It follows from Claims~\ref{convexity} and~\ref{invarianceOfBunderG} that $B(M) = B(\hat{M})$.
Define operators 
$$
N_{g,k} = \frac{1}{\sqrt{2\Tr(F_k^\dagger F_k)}}g F_k g^\dagger.
$$
Note that
$$
\sum_{g \in G} N_{g,k} = \frac{1}{\sqrt{2\Tr(F_k^\dagger F_k)}} \sum_{u,v \in \01}X^uZ^v 
F_k^\dagger F_k Z^v X^u = \id.
$$
(see for example Hayashi~\cite{hayashi:book}). 
Hence $M_k = \{N_{g,k}\}$ is a valid quantum measurement.
Now, note that $\hat{M}$ can be obtained from $M_1,\ldots,M_K$ by averaging. 
Hence, by Claim~\ref{convexity} we have
$$
B(M)= B(\hat{M}) \leq \max_k B(M_k).
$$
Let $M^*$ be the optimal measurement. 
Clearly, $m = B(M^*) \leq \max_k B(M^*_k) \leq m$ by the above and Corollary~\ref{simplify} 
from which our claim follows.
\end{proof}

Note that Claim~\ref{maximumInvariant} also gives us that we have at most 4 measurement operators.
Wlog, we will take the measurement outcomes to be labeled $1,2,3,4$.

Finally, we note that we can restrict ourselves to optimizing over positive-semidefinite (and hence Hermitian)
matrices only.
\begin{claim}
Let $F$ be a measurement operator, and let 
$$
g(F) := 1 + \sum_{b,k} p_{k|b}\left|\left|p_{0|b}N(\tilde{\sigma_{0,b}})- p_{1|b}N(\tilde{\sigma_{1,b}})\right|\right|_{1}
$$ with 
$\tilde{\sigma_{0,b}} = F \sigma_{0,b} F^\dagger/\Tr(F\sigma_{0,b}F^\dagger)$ and
$\tilde{\sigma_{1,b}} = F \sigma_{1,b} F^\dagger/\Tr(F\sigma_{1,b}F^\dagger)$.
Then there exists a Hermitian operator $\hat{F}$, such that $g(F) = g(\hat{F})$.
\end{claim}
\begin{proof}
Let $F^\dagger = \hat{F}U$ be the polar decomposition of $F^\dagger$,
where $\hat{F}$ is positive semidefinite and $U$ is unitary~\cite[Corollary 7.3.3]{horn&johnson:ma}.
Evidently, since the trace is cyclic, all probabilities remain the same. It follows immediately from
the definition of the trace norm that $||U A U^\dagger||_{1} = ||A||_{1}$ for all unitaries $U$, 
which completes our proof. 
\end{proof}

\noindent
To summarize, our optimization problem can now be simplified to
\begin{eqnarray*}
&&\max_M B(M) = \max_M p_+^M + p_\times^M \leq\\
&&\max_F 
1 + \sum_{b,k} p_{k|b} \left|\left|p_{0|b}N(\tilde{\sigma_{0,b}})- p_{1|b}N(\tilde{\sigma_{1,b}})\right|\right|_{1}\\
&&= 1 + 2\sum_{b} \left|\left|r(F(\sigma_{0,b} - \sigma_{1,b})F)
\qquad\qquad + (1-r)\Tr(F(\sigma_{0,b} - \sigma_{1,b})F)\frac{\id}{2}\right|\right|_{1}
\end{eqnarray*}
where the maximization is now taken over a single operator $F$, and we have used the fact
that we can write $p_{0|kb} = p_{k|0b}/(2p_{k|b})$ and we have 4 measurement operators.

\subsubsection{F is diagonal in the Breidbart basis}
Now that we have simplified our problem already considerably, we are ready to perform the actual optimization.
Since we are in dimension $d=2$ and $F$ is Hermitian, we may express $F$ as
$$
F = \alpha \outp{\phi}{\phi} + \beta \outp{\phi^\perp}{\phi^\perp},
$$
for some state $\ket{\phi}$ and real numbers $\alpha,\beta$. We first of all note that from
$\sum_{k} F_k F_k^\dagger = \id$, we obtain that
\begin{eqnarray*}
&&\Tr\left(\sum_k F_k F_k^\dagger\right) = \sum_k \Tr(F_k F_k) =\\
&& \sum_{g \in\{\id,X,Z,XZ\}} \Tr(gFgg^\dagger Fg^\dagger)
= 4 \Tr(FF) = \Tr(\id) = 2,
\end{eqnarray*}
and hence $\Tr(FF) = \alpha^2 + \beta^2 = 1/2$. Furthermore using that 
$\outp{\phi}{\phi} + \outp{\phi^\perp}{\phi^\perp} =\id$
we then have
\begin{equation}\label{Fdef}
F = \beta \id + (\alpha - \beta)\outp{\phi}{\phi},
\end{equation}
with $\beta = \sqrt{1-\alpha^2}$. Our first goal is now to show that $\ket{\phi}$ is a Breidbart vector (or the bit-flipped
version thereof). To this end, we first formalize our intuition that we may take $\ket{\phi}$ to lie in the XZ plane of 
the Bloch sphere only. Since we are only interested in the trace-distance term of $B(M)$, we restrict ourselves to considering
\begin{eqnarray*}
C(F) := \sum_{b}&&\left|\left|r(F(\sigma_{0,b} - \sigma_{1,b})F) + 
(1-r)\Tr(F(\sigma_{0,b} - \sigma_{1,b})F)\frac{\id}{2}\right|\right|_{1}.
\end{eqnarray*}

\begin{claim}\label{ignoreY}
Let $F$ be the operator that maximizes $C(F)$, and write $F$ as in Eq.(\ref{Fdef}). 
Then $\ket{\phi}$ lies in the XZ plane of the Bloch sphere. (i.e. $\Tr(FY)=0$).
\end{claim}
\begin{proof}
We first parametrize the state in terms of its Bloch vector:
$$
\outp{\phi}{\phi} = \frac{\id + x X + y Y + z Z}{2}.
$$
Since $\ket{\phi}$ is pure we can write $y = \sqrt{1 - x^2 - z^2}$. Hence, we can express
$F$ as
$$
F = \frac{1}{2}\left((\alpha + \beta) \id + (\alpha - \beta)(x X + y Y + z Z)\right).
$$
Noting that $\sigma_{0,+} - \sigma_{1,+} = Z$ and $\sigma_{0,\times} - \sigma_{1,\times} = X$
we can compute for the computational basis
\begin{eqnarray*}
P &\assign& r(FZF) + (1-r)\Tr(FZF)\frac{\id}{2}\\ &=&
\frac{1}{2}\left(\left(2\alpha^2 - \frac{1}{2}\right)z\id + 
r\left((\alpha-\beta)^2 xz X\right.\right.
+ \left.\left.(\alpha-\beta)^2 yz Y + 
\left((\alpha-\beta)^2 z^2 + 2 \alpha \beta\right)
Z\right)\right),
\end{eqnarray*}
and for the Hadamard basis:
\begin{eqnarray*}
T &\assign& r(FXF) + (1-r)\Tr(FXF)\frac{\id}{2}\\
&=&
\frac{1}{2}\left(\left(2\alpha^2-\frac{1}{2}\right)x \id + 
r\left(
\left((\alpha-\beta)^2 x^2 + 2 \alpha \beta\right)X\right)\right.\\
&+& \left.
(\alpha-\beta)^2 xy Y + (\alpha-\beta)^2 xz Z\right)
\end{eqnarray*}

Note that $||P||_{1} = \sum_j |\lambda_j(P)|$,
where $\lambda_j$ is the $j$-th eigenvalue of $P$.
A lengthy computation (using Mathematica), 
and plugging in $\beta = \sqrt{1/2 - \alpha^2}$
and $y = \sqrt{1-x^2-z^2}$ shows that we have
\begin{eqnarray*}
\lambda_1(P) &=& \frac{1}{4}\left(\left(4\alpha^2 - 1\right)z - 
r \sqrt{z^2 + 8 \alpha^2(2\alpha^2 - 1)(z^2-1)}\right)\\
\lambda_2(P) &=& \frac{1}{4}\left(\left(4\alpha^2 - 1\right)z +
r \sqrt{z^2 + 8 \alpha^2(2\alpha^2 - 1)(z^2-1)}\right)
\end{eqnarray*}
Similarly, we obtain for the Hadamard basis that
\begin{eqnarray*}
\lambda_1(T) &=& \frac{1}{4}\left(\left(4\alpha^2 - 1\right)x - 
r \sqrt{x^2 + 8 \alpha^2(2\alpha^2 - 1)(x^2-1)}\right)\\
\lambda_2(T) &=& \frac{1}{4}\left(\left(4\alpha^2 - 1\right)x +
r \sqrt{x^2 + 8 \alpha^2(2\alpha^2 - 1)(x^2-1)}\right)
\end{eqnarray*}
We define
\begin{eqnarray*}
f(\alpha,x) &\assign& \left(\alpha^2 - \frac{1}{4}\right)x\\
g(\alpha,x) &\assign& \frac{1}{4} \sqrt{x^2 + 8 \alpha^2(2\alpha^2-1)(x^2-1)}.\\
h(\alpha,x,r) &\assign& |f(\alpha,x) + r g(\alpha,x)| + |f(\alpha,x) - r g(\alpha,x)|
\end{eqnarray*}
Note that our optimization problem now takes the form
\begin{sdp}{maximize}{$h(\alpha,x,r) + h(\alpha,z,r)$}
&$x^2+z^2\leq 1$\\
&$0\leq x \leq 1$\\
&$0 \leq z \leq 1$,
\end{sdp}
where we can introduce the last two inequality constraints without loss of generality, since the remaining three
measurement operators will be given by $XFX$, $ZFZ$, and $XZFZX$.

To show that we can let $y=0$ for the optimal solution, we have to show that for all $\alpha$ and all $r$,
the function $h(\alpha,x,r)$ is increasing on the interval $0 \leq x \leq 1$ (and indeed Mathematica 
will convince you in an instant that this is the case). Our analysis is is further complicated
by the absolute values. We therefore first consider 
$$
h(\alpha,x,r)^2 = 2(f(\alpha,x)^2 + r^2 g(\alpha,x)^2 + |f(\alpha,x)^2 - r^2 g(\alpha,x)^2|,
$$
where we have used the fact that $f$ and $g$ are real valued functions. In principle, we can
now analyze 
$h_+(\alpha,x,r)^2 = 2(f(\alpha,x)^2 + r^2 g(\alpha,x)^2 + f(\alpha,x)^2 - r^2 g(\alpha,x)^2$
and 
$h_-(\alpha,x,r)^2 = 2(f(\alpha,x)^2 + r^2 g(\alpha,x)^2 - f(\alpha,x)^2 + r^2 g(\alpha,x)^2$
separately on their respective domains. By rewriting, we obtain
$$
h_+(\alpha,x,r)^2 = \frac{1}{4}r^2 (x^2 + 8 \alpha^2 (2 \alpha^2 - 1)(x^2-1)),
$$
and
$$
h_-(\alpha,x,r)^2 = 4\left(\alpha^2 - \frac{1}{4}\right)^2 x^2.
$$
Luckily, the first derivatives of
$h_+$ and $h_-$ turns out to be positive everywhere for our choice
of parameters $0 \leq \alpha \leq 1/\sqrt{2}$, and $0 \leq r,z \leq 1$.
Hence, by further inspection at the transitional points
we can conclude that $h$ is an increasing function of $x$.
But this means that to maximize our target expression, we must choose $x$ and
$z$ as large as possible. Hence, choosing $y=0$ is the best choice and our
claim follows.
\end{proof}

\noindent
We can now immediately extend this analysis to find
\begin{claim}
Let $F$ be the operator that maximizes $C(F)$, and write $F$ as in Eq.(\ref{Fdef}). 
Then 
$$
\ket{\phi} = g (\cos(\pi/8)\ket{0} + \sin(\pi/8)\ket{1}),
$$
for some $g \in \{\id,X,Z,XZ\}$.
\end{claim}
\begin{proof}
Extending our analysis from the previous proof, we can compute the second
derivative of both functions. It turns out that also the second
derivatives are positive, and hence $h$ is convex in $x$. 
By Claim~\ref{ignoreY}, we can rewrite our optimization problem as
\begin{sdp}{maximize}{$h(\alpha,x,r) + h(\alpha,z,r)$}
&$x^2+z^2 = 1$\\
&$0\leq x \leq 1$\\
&$0 \leq z \leq 1$
\end{sdp}
It now follows from the fact that $h$ is convex in $x$ and the constraint
$x^2 + z^2 = 1$ (by computing the Lagrangian of the above optimization problem),
that for the optimal solution we must have $x=z$, and our claim follows.
\end{proof}

\subsubsection{Optimality of the trivial strategies}
Now that we have shown that $F$ is in fact diagonal in the Breidbart basis 
(or the bit flipped version thereof) we have only a single parameter left in our optimization problem.
We must now optimize over all operators $F$ of the form
$$
F = \alpha \outp{\phi}{\phi} + \sqrt{1/2 - \alpha^2} \outp{\phi^\perp}{\phi^\perp},
$$
where we may take $\ket{\phi}$ to be $\ket{0}_B$ or $\ket{1}_B$. Our aim is now to show that
either $F$ is the identity, or $F = \outp{\phi}{\phi}$ depending on the value of $r$.
\begin{claim}\label{claim:maximizeAlpha}
Let $F$ be the operator that maximizes $C(F)$. Then 
$F = c \id$ (for some $c \in \Real$) for $r \geq 1/\sqrt{2}$, and
$F = \outp{\phi}{\phi}$ for $r < 1/\sqrt{2}$, where
$$
\ket{\phi} = g (\cos(\pi/8)\ket{0} + \sin(\pi/8)\ket{1}),
$$
for some $g \in \{\id,X,Z,XZ\}$.
\end{claim}
\begin{proof}
We can now plug in $x = z = 1/\sqrt{2}$ in the expressions for the eigenvalues in our previous proof.
Ignoring the constant positive factors which do not contribute to our argument, we can then write
\begin{eqnarray*}
\lambda_1(P) &=& 
\left(4\alpha^2 - 1\right) - 
r \sqrt{1 - 16 \alpha^4 + 8 \alpha^2},\\
\lambda_2(P) &=& 
\left(4\alpha^2 - 1\right) +
r \sqrt{1 - 16 \alpha^4 + 8 \alpha^2}.\\
\end{eqnarray*}
And similarly for the Hadamard basis. 
We again define functions
\begin{eqnarray*}
f(\alpha) &\assign& \left(4\alpha^2 - 1\right)\\
g(\alpha) &\assign& \sqrt{1 -16 \alpha^4 + 8 \alpha^2}\\
h(\alpha,r) &\assign& |f(\alpha,x) + r g(\alpha,x)| + |f(\alpha,x) - r g(\alpha,x)|
\end{eqnarray*}
Note that our optimization problem now takes the form
\begin{sdp}{maximize}{$2h(\alpha,r)$}
&$0\leq \alpha \leq \frac{1}{\sqrt{2}}$
\end{sdp}
Since we are maximizing, we might as well consider the square of our target function
and ignore the leading constant as it is irrelevant for our argument.
$$
h(\alpha,r)^2 = 2(f(\alpha)^2 + r^2 g(\alpha)^2 + |f(\alpha)^2 - r^2 g(\alpha)^2|,
$$
To deal with the absolute value, we now perform a case analysis similar to the one above.
Computing the zeros crossings of the function $f(\alpha)^2 - r^2 g(\alpha)^2$,
we analyze each interval separately. Computing the first and second derivatives on
the intervals we find that $h(\alpha,r)^2$ has exactly two peaks: The first at $\alpha=0$,
and the second at $\alpha=1/2$. We have that $h(0,r)^2 = 2$ for all $r$, and
$h(1/2,r)^2 = 4 r^2$. Hence, we immediately see that the maximum is located at
$\alpha=0$ for $r \leq 1/\sqrt{2}$, and at $\alpha=1/2$ for $r \geq 1/\sqrt{2}$.
\end{proof}

Theorem~\ref{thm:depolOptimal} now follows directly from Claim~\ref{claim:maximizeAlpha}:
Bob either measures in the Breidbart basis, or stores the qubit as is.
We believe that a similar analysis can be done for the dephasing channel, by first symmetrizing
the noise by applying a rotation over $\pi/4$ to our input states.

\subsection{Noise tradeoff}
We now consider the more practical setting, where the honest parties also experience noise.
Clearly, there is a strict tradeoff between the noise $p_{\rm error}$ on the channel experienced by the honest
parties, and the noise experienced by dishonest Bob. 
Our practical security bound is fairly weak. 
In the near-future we may anticipate that storage is better than direct measurement if good photonic memories
become available. However, we are free in our protocol to stretch the
waiting time $T$ between Bob's reception of the qubits and his reception of the classical basis information, say, to seconds, which means
that one has to consider the overall noise rate on a qubit that is stored for seconds. 

We again consider the case of depolarizing noise during storage.
For $r < 1/\sqrt{2}$ (when it is better for Bob to measure in the Breidbart basis), we obtain that
our protocol is secure as long as 
$$
h(p_{\rm error}) < 2 \log\left(\frac12 + \frac{1}{2 \sqrt{2}}\right) \log(3/4).
$$
Hence, we require
that $p_{\rm error} \lessapprox 0.029$.
This puts a strong restriction on the noise rate of the honest protocol. Yet, since
our protocols are particularly interesting at short distances (e.g. in the case of secure identification),
we can imagine very short free-space implementations such that depolarization noise during
transmission is negligible and the main depolarization noise source is due to Bob's honest measurements.

\begin{figure}
\begin{center}
\includegraphics[scale=0.5]{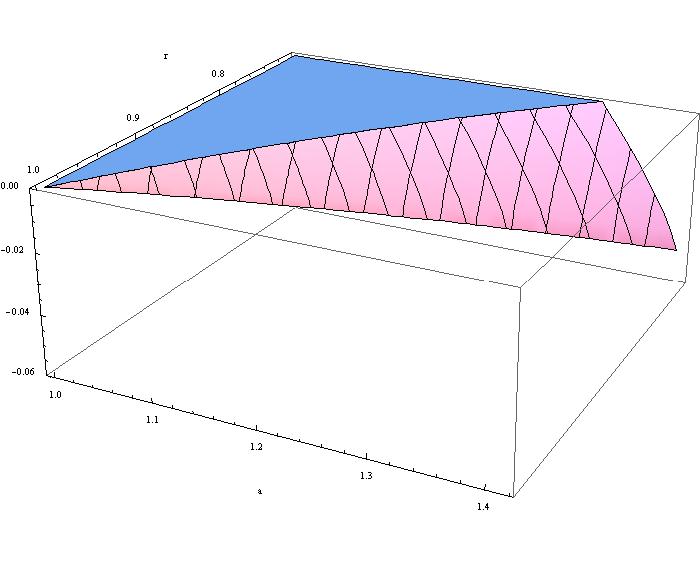}
\caption{
$h((1 - a r)/2)/4+\log(\frac{1+r}{2})\log(4/3)/2$, where we only show the region below 0, i.e.,
where security can be attained.}
\label{figure:tradeoff}
\end{center}
\end{figure}

For $r \geq 1/\sqrt{2}$ (when it's better for Bob to store the qubit as is) 
we also obtain a tradeoff involving $r$. As an example, suppose
that the qubits in the honest protocol are also subjected to depolarizing noise at rate $1-r_{d,{\rm honest}}$.
The effective classical error rate for a depolarizing channel
is then simply $p_{\rm error}=(1-r_{d,{\rm honest}})/2$. Thus we can consider when the function $h(p_{\rm error})/4+\log(\frac{1+r}{2})\log(4/3)/2$
goes below 0. If we assume that
$r_{d,{\rm honest}} = a r$, for some scaling factor $1 \leq a \leq 1/r$ 
(i.e., the honest party never has more noise than the dishonest party), 
we obtain a clear tradeoff between $a$ and $r$ depicted in Figure~\ref{figure:tradeoff}.

\section{Conclusion}

We have introduced the model of noisy-quantum storage.
In this model, 
we have determined security bounds for a perfect ROT
protocol given collective storage attacks by Bob. Furthermore, we
showed how to construct a practical ROT where we do allow the honest
parties to experience noise during transmissions and their operations
as well. We provided an explicit security tradeoff between the noise
affecting the honest parties, and the noise during storage for a dishonest
Bob.

Ideally, we would
like to show security against general coherent noisy
attacks. The problem with analyzing a coherent attack of Bob
described by some super-operator ${\cal S}$ affecting all his incoming
qubits is not merely a technical one: one first needs to determine a
realistic noise model in this setting.  It may be possible using variations of de
Finetti theorems as in the proof of QKD \cite{renato:diss} to prove
for a symmetrized version of our protocol that any coherent attack by
Bob is equivalent to a collective attack. Yet, the present
scenario differs in that it is not as straightforward to achieve 
a symmetrization of the protocol.
However, one can in fact analyze a
specific type of coherent noise, one that essentially corresponds to
an eavesdropping attack in QKD. Note that the 1-2 OT protocol can be
seen as two runs of QKD interleaved with each other. The strings
$f(x_{|\setI_+})$ and $f(x_{|\setI_\times})$ are then the two keys
generated. The noise must be such that it leaves Bob with exactly the
same information as the eavesdropper Eve in QKD. In this case, it
follows from the security of QKD that the dishonest Bob (learning
exactly the same information as the eavesdropper Eve) does not learn
anything about the two keys.

In terms of long-term security, {\em fault-tolerant} photonic
computation (e.g., with the KLM scheme \cite{KLM:lo}) might
allow a dishonest Bob to encode the incoming quantum information into
a fault-tolerant quantum memory. This implies that in storage, the
effective noise rate can be made arbitrarily small. However, the
encoding of a single unknown state is \emph{not} a fault-tolerant quantum
operation: already the encoding process introduces errors whose rates
cannot be made arbitrarily small with increasing effort. Hence, even
in the presence of a quantum computer, there is a residual storage
noise rate due to the unprotected encoding operations. The question of
security then becomes a question of a trade-off between this residual
noise rate versus the intrinsic noise rate. Finally, it
remains to address composability of the protocol within our model,
which has already been considered for the bounded-quantum-storage
model~\cite{wehner07b}.

\appendix

\chapter{Linear algebra and semidefinite programming}\label{semidefChapter}\index{semidefinite programming}\label{semidefIntro}
\label{appendix:semidef}
Semidefinite programming is a useful tool to solve optimization problems. Since we employed semidefinite
programming in Chapters~\ref{pistarChapter},~\ref{optimalStrategiesChapter}, and~\ref{storageErrorsChapter},
we briefly state the most important notions. We refer to~\cite{boyd:book} for an in-depth introduction.

\section{Linear algebra prerequisites}
Before turning to semidefinite programming in the next section, 
we first briefly recall
some elementary 
definitions from linear algebra. We thereby assume the reader is familiar with basic concepts, such as
matrix multiplication and addition.
Unless explicitly indicated, all vector spaces $V$ considered here are over the field of complex
numbers.
We use $V = \Complex^d$ to denote a $d$-dimensional complex vector space, and
$\Complex^{d \times d}$
to denote the space of complex $d \times d$ matrices.
A set of vectors $\ket{v_1},\ldots,\ket{v_d} \in V$
is \emph{linearly independent} if $\sum_{i=1}^d a_i \ket{v_i} = 0$ implies that
$a_1 = \ldots = a_d = 0$.
A \emph{basis}\index{basis} of a $d$-dimensional vector space $V$ is a set of linearly independent 
vectors $\ket{v_1},\ldots,\ket{v_d} \in V$, the \emph{basis vectors}\index{basis vectors}, 
such that any vector $\ket{u} \in V$ 
can be written as a linear combination of basis vectors. 
If there exists a vector $\ket{v} \in V$ with $\ket{v} \neq 0$ such that $A\ket{v} = \lambda \ket{v}$, 
we say that
$\ket{v}$ is an \emph{eigenvector}\index{eigenvector} of $A$ and the scalar $\lambda$ the 
corresponding \emph{eigenvalue}.\index{eigenvalue}

The \emph{inner product}\index{inner product} 
of two vectors $\ket{u}, \ket{v} \in V$ with 
$\ket{u} = (u_1,\ldots,u_d)$ and $\ket{v} = (v_1,\ldots,v_d)$ is given by 
$\inp{u}{v} = \sum_i u_i^* v_i$.
The 2-\emph{norm} of a vector is given by $||\ket{v}|| = \sqrt{\inp{v}{v}}$. 
Unless otherwise indicated, all norms of a vector are 2-norms in this text.
We also use $||\ket{v}||_V$ to denote emphasize that the norm is defined
on a vector space $V$.
Two vectors $\ket{u},\ket{v} \in V$ such that $\inp{u}{v} = 0$
are \emph{orthogonal}. If, in addition, $||\ket{u}|| = ||\ket{v}|| = 1$ 
then they are also called \emph{orthonormal}.

A \emph{Hilbert space}\index{Hilbert space} is defined as a vector space $V$ with an inner product, where
the vector space is complete. We refer to~\cite{conway:fa} for a formal definition
of the notion of completeness and merely note 
that informally a vector space is complete if for any sequence of vectors
in said space approaching a limit, the limit is also an element of the vector space. 
A \emph{bounded operator}\index{bounded operator} is an operator
$A: V \rightarrow V'$ such that there exists a $c \in \Real$
satisfying $||A\ket{v}||_{V'} \leq c ||\ket{v}||_{V}$ for all $v \in V$.
The smallest such $c$ is also called the \emph{operator norm}\index{operator
norm} of $A$.

The \emph{transpose}\index{transpose} of a matrix $A$ is written as $A^T$ and given by $A^T_{ij} = A_{ji}$,
where $A_{ij}$ denotes the entry of the matrix $A$ at column $i$ and row $j$.
Similarly, the conjugate transpose\index{conjugate transpose} $A^{\dagger}$ of $A$ is 
of the form $A^{\dagger}_{ij} = A_{ji}^*$.
We use $\id$ to denote the identity matrix defined as $\id = [\id_{ij}]$ with
$\id_{ij} = \delta_{ij}$.
A matrix $U$
is called \emph{unitary}\index{unitary} if
$UU^\dagger = U^\dagger U = \id$. 
Furthermore,
$M$ is called \emph{Hermitian}\index{Hermitian} if and only if $M = M^\dagger$.
Any Hermitian matrix can be decomposed in terms of its eigenvalues $\lambda_j$ and
eigenvectors $\ket{u_j}$ as $M = \sum_j \lambda_j \outp{u_j}{u_j}$, where
$\outp{u_j}{u_j}$ is a projector onto the vector $\ket{u_j}$.
We also call this the \emph{eigendecomposition} of $M$\index{eigendecomposition}.
The \emph{support} of $M$ is the space spanned by all its eigenvectors
with non-zero eigenvalue.

The \emph{tensor product}\index{tensor product} 
of an $m \times n$-matrix $A$ and an $m' \times n'$ matrix $B$ is given by
the $mm' \times nn'$-matrix 
$$
A \otimes B = \left(\begin{array}{ccc}A_{11}B & \ldots & A_{1n} B\\
A_{21}B & \ldots & A_{2n}B\\ &\ddots&\\
A_{n1}B & \ldots & A_{nn}B\\
\end{array}\right).
$$
The tensor product is also defined for two vector spaces $V$ and $V'$. In particular, 
if the basis of the $d$-dimensional vector space $V$ is given 
by $\{\ket{v_1},\ldots,\ket{v_d}\}$ and the basis of the $d'$-dimensional vector space $V'$ is given 
by $\{\ket{v'_1},\ldots,\ket{v'_{d'}}\}$, then $W = V \otimes V'$ denotes the $d \cdot d'$-dimensional vector
space $W$ with basis $\{\ket{v_i} \otimes \ket{v'_j}\mid i \in [d], j \in [d']\}$.

The \emph{direct sum} of an $m \times n$-matrix $A$ and an $m' \times n'$ matrix $B$ is given
by the $m+m' \times n + n'$ matrix 
$$
\left(\begin{array}{cc}A & 0\\
0 & B\end{array}\right).
$$
Two vector spaces $V$ and $V'$ defined as above can also be composed in an analogous fashion
yielding a $d+d'$ dimensional vector space $W = V \oplus V'$, where any $\ket{w} \in W$
can be written as $\ket{w} = \ket{v} \oplus \ket{v'}$ for some $\ket{v} \in V$ and $\ket{v'} \in V'$
with $\ket{v} \oplus \ket{v'} = 
(v_1,\ldots,v_d,v'_1,\ldots,v'_{d'})$ for $\ket{v} = (v_1,\ldots,v_d)$ and $\ket{v'} = (v'_1,\ldots,v'_{d'})$.

The \emph{trace}\index{trace} of a matrix $A$ is given by the sum of its diagonal entries 
$\Tr(A) = \sum_i A_{ii}$. 
Note that $\Tr(A + B) = \Tr(A) + \Tr(B)$, and $\Tr(AB) = \Tr(BA)$. If $A$
is an Hermitian matrix, then
$\Tr(A)$ is the sum of its eigenvalues.

Finally, the \emph{rank}\index{rank} of a matrix $A$ is denoted as $\rank(A)$ and
given by the maximal number of linearly independent columns (or rows) of $A$.

\section{Definitions}\label{psd}
We now turn to the definitions relevant for our discussion of semidefinite programming.
A Hermitian matrix $M$ is \emph{positive semidefinite}\index{positive semidefinite} if and only if
all of its eigenvalues are non-negative~\cite[Theorem 7.2.1]{horn&johnson:ma}.
Throughout this text, we use $M \geq 0$ to indicate that $M$ is positive semidefinite.
We know from~\cite[Theorem 7.2.11]{horn&johnson:ma}:
\begin{proposition}
For a Hermitian matrix $M \in \Complex^{d\times d}$ the following three
statements are equivalent:
\begin{enumerate}
\item $M \geq 0$,
\item $x^\dagger M x \geq 0$ for all vectors $x \in \Complex^d$,
\item $M = G^\dagger G$ for some matrix $G \in \Complex^{d\times d}$.
\end{enumerate}
\end{proposition}
$M$ is called \emph{positive definite}\index{positive definite} if and only if all of its
eigenvalues are positive: we have $x^\dagger M x > 0$ for all vectors
$x \in \Complex^d$. We use $M > 0$ to indicate that $M$ is positive definite.
We also encounter projectors, where a Hermitian matrix
$M$ is a \emph{projector}\index{projector} 
if and only if $M^2=M$. Note that this implies that $M \geq 0$.
We say that two projectors $M_1$ and $M_2$ are \emph{orthogonal projectors}
\index{orthogonal projector}\index{projector!orthogonal}
if and only if $M_1 M_2 = 0$.

Furthermore, we use $\setS^d$ to denote the set of all Hermitian matrices, 
$\setS^d = \{X \in \Complex^{d \times d}\mid X = X^\dagger\}$, and 
$\setS^d_+ = \{X \in \setS^d \mid X \geq 0\}$ for the set of all positive semidefinite
matrices. A set $\setT$ is a \emph{cone}\index{cone}, if for any $\alpha \geq 0$ and $T \in \setT$ 
we have $\alpha T \in \setT$.
A set $\setT$ is \emph{convex}, if for any $\alpha \in [0,1]$ and $T_1,T_2 \in \setT$
we have $\alpha T_1 + (1-\alpha)T_2 \in \setT$.
A set $\setT$ is called a \emph{convex cone}\index{convex cone}, if $\setT$ is convex and a cone: for any $\alpha_1,\alpha_2 \geq 0$
and $T_1,T_2 \in \setT$ we must have that $\alpha_1 T_1 + \alpha_2 T_2 \in \setT$.
Note that $\setS^d_+$ is a convex cone: Let $\alpha_1, \alpha_2 \geq 0$, and
$A, B \in \setS^d_+$. Then for any $x \in \Complex^d$ we have
$$
x^\dagger(\alpha_1 A + \alpha_2 B)x = \alpha_1 x^\dagger A x + \alpha_2 x^\dagger B x \geq 0.
$$
Hence, $\alpha_1 A + \alpha_2 B \in \setS^d_+$.
The following will be of use in Chapter~\ref{pistarChapter}.
\begin{proposition}\label{prop:positive}
Let $A,B \in S^d$.
Then $A \geq 0$ if and only if for all $B \geq 0$ 
$\Tr(AB) \geq 0$.
\end{proposition}
\begin{proof}
Suppose that $A \geq 0$.
Note that we can decompose $B = \sum_j \lambda_j \outp{u_j}{u_j}$ where for all $j$ $\lambda_j \geq 0$
since $B \geq 0$. Hence, $\Tr(AB) = \sum_j \lambda_j \Tr(A\outp{u_j}{u_j}) = 
\sum_j \lambda_j \bra{u_j}A\ket{u_j} \geq 0$, since $A \geq 0$.

To prove the converse, suppose on the contrary that for all $B \geq 0$ we have
$\Tr(AB) \geq 0$, but $A \not \geq 0$. If $A \not \geq 0$, then there exists
some vector $\ket{v}$ such that $\bra{v}A\ket{v} < 0$. Let $B = \outp{v}{v}$.
Clearly, $B \geq 0$ and $\Tr(AB) = \bra{v}A\ket{v} < 0$ which is a contradiction.
\end{proof}

\section{Semidefinite programming}

Semidefinite programming is a special case of convex optimization. 
Its goal is to solve the following semidefinite program (SDP)\index{semidefinite program (SDP)}\index{SDP}
\index{SDP!standard form}
in terms of the variable $M \in \setS^d$
\begin{sdp}{maximize}{$\Tr(CM)$}
&$\Tr(A_iM) = b_i, i = 1,\ldots,p$,
and $M \geq 0$
\end{sdp}
for given matrices $C,A_1,\ldots,A_p \in \setS^d$.
The above form is called the \emph{standard form} of an SDP, and any SDP can
be cast in this fashion (possibly at the expense of additional variables)~\cite{boyd:book}.
To gain some geometric intuition about this task, note that $M \geq 0$ means that
$M$ must lie in the cone $\setS^d_+$. The constraints $Tr(A_i M) = b_i$ determine a set of 
hyperplanes which further limit our possible solutions.
A matrix $M$ is called \emph{feasible},\index{SDP!feasible}
\index{feasible} 
if it satisfies all constraints.

An important aspect of semidefinite programming is duality.\index{SDP!duality}
\index{duality}
Intuitively, the idea behind Lagrangian duality is to extend the objective function
(here $\Tr(CM)$) with a weighted sum of the constraints in such a way, that we will be penalized
if the constraints are not fulfilled. The weights then correspond to the dual variables.
Optimizing over these weights then gives rise to the \emph{dual problem}\index{dual}.\index{SDP!dual}
The original problem is called the \emph{primal problem}.\index{primal}\index{SDP!primal}
For the above SDP in standard form, we can write
down the Lagrangian\index{SDP!Lagrangian}\index{Lagrangian} as
\begin{eqnarray*}
L(M,\lambda_1,\ldots,\lambda_p,K) &=& \Tr(CM) + \sum_{i=1}^p \lambda_i(b_i - \Tr(A_i M)) + \Tr(K M)\\
&=&\Tr((C - \sum_i \lambda_i A_i + K)M) + \sum_i \lambda_i b_i,
\end{eqnarray*}
where $K \geq 0$.
The dual function\index{dual function}\index{SDP!dual function} is then
\begin{eqnarray*}
g(\lambda_1,\ldots,\lambda_p,K) &=& \sup_M \left(\Tr((C - \sum_i \lambda_i A_i + K)M) + \sum_i \lambda_i b_i\right)\\
&=& \left\{\begin{array}{ll}
                       \sum_i \lambda_i b_i & \mbox{if } C - \sum_i \lambda_i A_i + K = 0\\[0.5mm]
                       \infty & \mbox{otherwise}
                      \end{array} \right.
\end{eqnarray*}
From $C - \sum_i \lambda_i A_i + K = 0$ and $K \geq 0$, we obtain that $K = - C + \sum_i \lambda_i A_i \geq 0$. 
This gives us the dual problem as
\begin{sdp}{minimize}{$\sum_i \lambda_i b_i$}
&$\sum_i \lambda_i A_i \geq C$,
\end{sdp}
where the optimization is now over the dual variables $\lambda_i$.

We generally use $d^{*}$ to denote the optimal value of the dual problem, and $p^{*}$ for the optimal value
of the primal problem. Weak duality says that $d^* \geq p^*$.\index{SDP!weak duality}\index{weak duality}\index{duality!weak}
Let's see why this is true in the above construction of the dual problem.
Let $M^{(*)}$ and $\{\lambda_i^{(*)}\}$ be the optimal solutions to the primal and dual problem respectively.
In particular, this means that $M^{(*)}$ and $\{\lambda_i^{(*)}\}$ must satisfy the constraints.
Then
\begin{eqnarray*}
d^*-p^* &=&
\sum_i \lambda_i^{(*)} b_i - \Tr(CM^{(*)})\\
&=& \sum_i \lambda_i^{(*)} \Tr(A_i M^{(*)}) - \Tr(C M^{(*)})\\
&=& \Tr\left(\left(-C + \sum_i \lambda_i^{(*)} A_i\right)M^{(*)}\right) \geq 0,
\end{eqnarray*}
by Proposition~\ref{prop:positive} since $M^{(*)} \geq 0$ and $\sum_i \lambda_i^{(*)} A_i \geq C$.
An important consequence of weak duality, is that if we have $d^* = p^*$ for a feasible dual and primal solution
respectively, we can conclude that both solutions are optimal. If solutions exist such that
$d^* = p^*$, we also speak of strong duality.\index{SDP!strong duality}\index{strong duality}\index{duality!strong} We know from Slater's conditions~\cite{boyd:book},\index{SDP!Slater's conditions}
that strong duality holds if there exists a feasible solution to the primal problem which
also satisfies $M > 0$.

\section{Applications}

In many quantum problems, we want to optimize over states, or measurement operators. Evidently, semidefinite programming
is very well suited to this case: When optimizing over a state $\rho$, we ask that $\rho \geq 0$ and $\Tr(\rho) = 1$.
When optimizing over measurement operators $M_1,\ldots,M_k$ belonging to one POVM, we ask that $M_j \geq 0$ for all $j \in [k]$
and $\sum_j M_j = \id$. Concrete examples can be found in Chapters~\ref{pistarChapter},~\ref{optimalStrategiesChapter}, and~\ref{storageErrorsChapter}.

\chapter{$C^*$-Algebra}\label{calgebraChapter}\index{$C^*$-Algebra}\label{cstar}\label{appendix:cstar}\label{chapter:cstar}

As $C^*$-algebras are not usually encountered in computer science, we briefly state the most important results
we will refer to for convenience. In particular, they help us understand the framework of post-measurement
information we encountered in Chapter~\ref{pistarChapter} as well as the structure of bipartite non-local games
in Chapter~\ref{entanglementIntroChapter}.

\section{Introduction}
Instead of starting out with the usual axioms of quantum states and their evolutions, any physical system
can be characterized by a $C^*$-algebra $\algA$ of observables. States of this system are now identified purely by means of
measurements of these observables. This starting point is rather beautiful in its abstraction: So far, nothing has 
been said how we can represent elements of this algebra. Yet, it turns out that all the usual axioms can
be derived from this abstract structure: we can represent observables as operators and states as vectors in a Hilbert space.
In fact, any such algebra $\algA$ is isomorphic to an algebra of bounded operators on a Hilbert space. 
So why should we bother adopting this abstract viewpoint? It turns out that $C^*$-algebras often make it
easier to understand the fundamental differences between the classical and the quantum setting. 
If the algebra $\algA$ is abelian, we have a classical system. Otherwise, our system is inherently quantum.
Commutativity leads to several nice structural properties of an algebra which have been exploited to
answer many central questions in quantum information: When can we clone physical states? What information
can be extracted without disturbing the system? That is, what part of a system is in fact classical and what
is truly quantum? 

Here, we will mere scratch the surface of this formalism. In particular, we will focus
on finite-dimensional $C^*$-algebras only, which is all we will need in Chapters~\ref{pistarChapter} 
and~\ref{entanglementIntroChapter}. For more information, consult any textbook on the topic~\cite{takesaki:cstar, robinson:cstar, arveson:cstar}.
We assume that the reader is familiar with the basic concepts such as 
a Hilbert space and refer to~\cite{conway:fa} for an introduction.
First, we need to introduce some essential definitions in Section~\ref{cstar:terminology}.
We then examine states and observables, and their familiar representation in a Hilbert space in Section~\ref{cstar:representation}.
In Section~\ref{cstar:commutation}, we then concentrate on commutation: We will sketch how from commutation relations
we in fact obtain a bipartite structure. It turns out that commutation relations also play an important role
in determining which operations leave states invariant.
Looking at the structure of the problem, it turns out that in fact many problems
ranging from cloning to post-measurement information and bipartite non-local games are quite closely related.

\section{Some terminology}\label{cstar:terminology}

\index{algebra!Banach}A Banach algebra $\algA$\index{Banach algebra} is a linear associative algebra\footnote{An associative algebra over the complex numbers is a vector space over the complex numbers with a multiplication that is associative.}
which is also a 
Banach space, with the property that for all $A$ and $B \in \algA$ we have 
$$\norm{AB} \leq \norm{A} \norm{B}.$$ The norm $\norm{A}$ of $A$
is thereby a real number satisfying the usual requirements that for all
$A \in \algA$ we have $\norm{A} \geq 0$ where $\norm{A} = 0$ if and only
if $A = 0$, $\norm{\alpha A} = \alpha \norm{A}$,
$\norm{A+B} \leq \norm{A} + \norm{B}$, and
$\norm{AB} \leq \norm{A} \norm{B}$.
$\algA$ is called a $*$-algebra\index{$*$-algebra}\index{algebra!$*$-algebra} if it has the additional 
property that it admits an involution $A \rightarrow A^\dagger \in \algA$ such that for 
all $A$ and $B \in \algA$ the following holds:
$(A^\dagger)^\dagger = A$, $(A+B)^\dagger = A^\dagger + B^\dagger$, $(\alpha A)^\dagger = \bar{\alpha} A^\dagger$, and $(A B)^\dagger = B^\dagger A^\dagger$.
A $C^*$-algebra is now an even more special case: in addition we also have that $\norm{A^\dagger A} = \norm{A}^2$ for all
$A \in \algA$. This also gives us $\norm{A^\dagger} = \norm{A}$. In the following we will simply 
use the term ``algebra'' to refer to a $C^*$-algebra.\index{$C^*$-Algebra}\index{algebra!$C^*$-algebra}
The trick is not to be intimidated. It is easier to have a more concrete picture in mind: For example, 
the algebra $\bop(\hil)$ of all bounded operators on a Hilbert space $\hil$
is a $C^*$-algebra,\index{Hilbert space!bounded operators}
when we take sums and products of operators in the usual way and
take our norm to be the 
operator norm $\norm{A} = \sup(\norm{Av}\mid v \in \hil, \norm{v} = 1)$, where $\norm{v}^2 = \inp{v}{v}$ for the
inner product $\inp{\cdot}{\cdot}$ of the Hilbert space.
This algebra is closed under all the usual operations such as addition, multiplication, and
multiplication by scalars\footnote{We will take the underlying field to be $\Complex$.} 
and the involution operation. This involution is now the adjoint operation $A \rightarrow A^\dagger$, which in physics is
usually denoted by $\dagger$ instead of $*$. In some physics papers, you will therefore also find the name $\dagger$-algebra
instead. As in the example of post-measurement information, we are also often interested in the $*$-algebra generated by\label{cstardef}
a given set of operators. Any operator $X$ in a Hilbert space $\hil$ 
determines a $C^*$-algebra $\algA$ which we will denote by $\algA = \langle X\rangle$. 
This is the smallest $C^*$-algebra which contains both
$X$ and the identity, i.e. $\langle X \rangle = \bigcap_{X,\id \in \algB}\algB$.
What's included in $\langle X \rangle$? Recall 
that $\algA$ is closed under the adjoint operation so we definitely have $X^\dagger$. In addition, our conditions above 
imply that we will see all possible polynomials in $X$ and $X^\dagger$. 
For example, $X+X^\dagger$ and $XX^\dagger$ are also elements of the algebra. We use $\langle X_1,\ldots,X_k \rangle$ to denote the 
$C^*$-algebra generated by operators $X_1,\ldots,X_k$,
and $\langle \mS \rangle$ to denote the algebra generated
by operators from the set $\mS$.

If an algebra $\algB$ satisfies
$\algB \subseteq \algA$, we call $\algB$ a \emph{subalgebra} of $\algA$. An algebra $\algA$ is \emph{unital}\index{algebra!unital}\index{unital!algebra} if
it contains the identity. We will always use $\id$ to denote the identity element. 
Since we restrict ourselves to the finite-dimensional case, we can assume that
any $C^*$-algebra is in fact unital~\cite{takesaki:cstar}. We will always take
$\algA$ to be unital here. 
An element $A \in \algA$ of a Banach algebra $\algA$ is called \emph{invertible}\index{invertible}\index{algebra!invertible} if there exists some
$A' \in \algA$ such that $A A' = A' A = \id$. 
Furthermore, for a $C^*$-algebra $\algA$, the \emph{spectrum} of\index{algebra!spectrum}\index{spectrum}
$A \in \algA$ is given by $\Sp_{\algA}(A) = \{\lambda \in \Complex\mid A - \lambda \id \mbox{ is not invertible}\}$. 
Note that for any $A \in \bop(\hil)$, this is just the spectrum
of the operator relative to $\bop(\hil)$ in the usual sense. 

A \emph{left ideal}\index{left ideal} in 
some algebra $\algA$ is a subalgebra $\algB \subseteq \algA$ such that for any elements $B \in \algB$ and $A \in \algA$ we have
that $AB \in \algB$. Similarly, $\algB$ is called a \emph{right ideal} if $BA \in \algB$. A \emph{two-sided ideal} or simply\index{right ideal}\index{two-sided ideal}\index{ideal}\index{algebra!ideal}
\emph{ideal} has both properties: $\algB$ is both a left and right ideal of $\algA$. 
An algebra $\algA$ is called \emph{simple}\index{algebra!simple}\index{simple} if its only ideals are $\{0\}$ and $\algA$ itself. 
An algebra $\algA$ is called \emph{semisimple},\index{algebra!semisimple}\index{semisimple}
if it can be written as the direct sum of simple algebras. To get a better feeling for what this actually means, it is
perhaps again helpful to think of a particular representation of the algebra in terms of bounded operators
on a Hilbert space. In terms of representations, being simple
means that the representation is irreducible. 
Being semisimple then means that the representation is completely reducible: i.e.
for the representation $\pi$ of $A$ we can express $\pi(A)$ as a sum of irreducible 
representations. We will examine this decomposition in more detail in Section~\ref{cstar:decompose}.

\section{Observables, states and representations}

\subsection{Observables and states}\index{observable}\index{observable!algebra}\index{state!algebra}\index{algebra!state}\index{algebra!observable}
A physical system is characterized by a set of measurable quantities, i.e.\ observables. As mentioned above, we will assume
that a physical system is in fact described by a $C^*$-algebra $\algA$ of observables. As we will see below, we can take the
observables to live in a Hilbert space $\hil$, and $\algA \subseteq \bop(\hil)$.
Where do the states come in? In the language of $C^*$-algebras, states are 
\emph{positive linear functionals} on $\algA$: A linear functional on an algebra is a function 
$f: \algA \rightarrow \Complex$
such that for all $A,B \in \algA$ we have $f(A+B) = f(A) + f(B)$ and $f(\alpha A) = \alpha f(A)$ where $\alpha \in \Complex$ is a 
scalar. A linear functional is called \emph{positive} if $f(A) \geq 0$ for any $A \in \algA$ whenever $A \geq 0$.\index{positive linear functional}\index{algebra!positive linear functional}
A \emph{state} on $\algA$ is a positive linear functional $f$ on $\algA$ with the additional property that it has norm 1, i.e., $f(\id) = 1$.
The set of states is a convex set of linear functionals and its extreme elements are called \emph{pure} states.\index{state!pure}\index{algebra!pure state}
The set of all states on an algebra $\algA$ 
is also called the \emph{state space}, often denoted by $\mE(\algA)$.\index{state space}\index{algebra!state space}
Any observable $A \in \algA$ in our algebra is uniquely characterized
by the expectation of all states when we measure $A$:
So the value of $f(A)$ for all states $f \in \mE(\algA)$ in our state space uniquely characterizes any element 
$A$ of our algebra. The converse is also true: the value of $f(A)$ for all $A \in \algA$ completely characterizes the state $f$.
To get a better feeling for this, it is again helpful to think of an algebra $\algA \subseteq \bop(\hil)$.
Given a vector $v$ living in the Hilbert space $\hil$, we can construct a linear functional on
$\algA$ by letting $f(A) = \inp{v}{Av}$. The same is true if we consider any abstract $\algA$ and its
representation $\pi$ on a Hilbert space, by letting $f(A) = \inp{v}{\pi(A)v}$ given $v \in \hil$.

\subsection{Representations}\label{cstar:representation}
We now examine how an abstract $C^*$-algebra can be represented by a set of operators on a Hilbert space, via
the famous construction by Gelfand, Naimark and Segal. 
An account of this construction can be found
in any standard textbook on $C^*$-algebra~\cite{takesaki:cstar, robinson:cstar, arveson:cstar}. 
For completeness, we here give a heavily annotated, largely self-contained, explanation of the GNS construction.
As it turns out, by the GNS construction,
any $C^*$-algebra is isomorphic to an algebra of bounded operators, a result which we will merely state here.
When trying to find a representation of a $C^*$-algebra $\algA$, our goal is to find a pair
$(\pi,\hil)$ where $\hil$ is a Hilbert space and $\pi: \algA \rightarrow \algB(\hil)$ is a $*$-homomorphism 
which maps any element
of our algebra to a bounded operator in the chosen Hilbert space.

\begin{theorem}[GNS]\index{GNS construction}\index{algebra!GNS construction}
Let $\algA$ be a unital $C^*$-algebra, and let $f$ be a positive linear functional on $\algA$. Then there exists a representation
$(\hil_f, \pi_f)$ of $\algA$ with a Hilbert space $\hil_f$,
a $*$-homomorphism\footnote{A homomorphism that
preserves the $*$.} $\pi_f: \algA \rightarrow \bop(\hil_f)$ 
and a vector $\Phi_f \in \hil_f$ such that for all $A \in \algA$
$$
f(A) = \inp{\Phi_f}{\pi_f(A) \Phi_f}.
$$
\end{theorem}
\begin{proof}
First, we construct the Hilbert space $\hil_f$. Since $\algA$ is a Banach space, we can turn it into a pre-Hilbert 
space\footnote{We take a \emph{pre-Hilbert space} to be a vector space with a 
positive semidefinite sesquilinear form, and a \emph{strict pre-Hilbert space} to be a vector
space with an inner product.}
by defining
the positive semidefinite sesquilinear form
$$
\inp{A}{B}_f = f(A^\dagger B),
$$
for all $A,B \in \algA$. Note that this form may be degenerate\footnote{Such a form is nondegenerate if 
and only if: $\inp{A}{B}_f = 0$ for all $B \in \algA$ implies that $A = 0$.}. In order to eliminate this degeneracy, consider
$$
\mI_f = \{A\mid A \in \algA\mbox{ and }f(A^\dagger A) = 0\}.
$$
Note that $\mI_f$ is a linear subspace
of $\algA$ since for all $I, J \in \mI_f$ we have 
$f((I+J)^\dagger(I+J)) = f(I^\dagger I) + f(J^\dagger I) + f(I^\dagger J) + f(J^\dagger J) \leq 2\sqrt{f(J^\dagger J) f(I^\dagger I)} = 0$,
where we used the Cauchy-Schwarz 
inequality\footnote{In this context
the CS-inequality gives us that for all $A,B \in \algA$ we have
$|f(A^\dagger B)|^2 \leq f(A^\dagger A) f(B^\dagger B)$}.
 
We now show that $\mI_f$ is a left ideal of $\algA$: Let $I \in \mI_f$ and $A,B \in \algA$. We then need to show that
$AI \in \mI_f$. Indeed, from $(AI)^\dagger(AI) \geq 0$ we have
$$
0 \leq f((AI)^\dagger(AI)) = f(I^\dagger A^\dagger AI) \leq \sqrt{f(I^\dagger I) f((A^\dagger AI)^\dagger(A^\dagger AI))} = 0,
$$
where the inequality follows from the Cauchy-Schwarz inequality. 

The Hilbert space $\hil_f$ is then constructed by completing the quotient space $\algA/\mI_f$. This works as follows:
Define the equivalence classes
$$
\Psi_A = \{A + I \mid I \in \mI_f\}.
$$
Note that these equivalence classes constitute a complex vector space on their own, where 
addition and scalar multiplication are defined via the following operations inherited from $\algA$.
We have $\Psi_{A+B} = \Psi_A + \Psi_B$ and $\Psi_{\alpha A} = \alpha \Psi_A$. We can then define
the inner product
$$
\inp{\Psi_A}{\Psi_B} = \inp{A}{B}_f = f(A^\dagger B).
$$
Note that $\Psi_A$ and $\Psi_B$ of course depend on $f$. 
One can verify that this a correct definition. Indeed, the inner product does not depend
on our choice of representative from each equivalence class: Let $I_1, I_2 \in \mI_f$, and let
$A, B \in \algA$. Then
$$
f((A+I_1)^\dagger(B+I_2)) = f(A^\dagger B) + f(A^\dagger I_2) + f(I_1^\dagger B) + f(I_1^\dagger I_2) = f(A^\dagger B),
$$
where the last equality follows again from the Cauchy-Schwarz inequality.
We can now obtain $\hil_f$ by forming the completion of this space.
It is well-known in functional analysis that any strict pre-Hilbert
space can be embedded as a dense subspace of a Hilbert space in such a way that the inner product is preserved.

Second, we must construct $\pi_f$. We first define the action of $\pi_f(A)$ on the vectors constructed above as
$$
\pi_f(A)\Psi_B = \Psi_{AB}.
$$
Note that this definition is again independent of our choice of representative from each equivalence class
since for all $A, B \in \algA$ we have
$$
\pi_f(A)\Psi_{B+I} = \Psi_{A(B+I)} = \Psi_{AB + AI} = \Psi_{AB} = \pi_f(A)\Psi_B,
$$
since $\mI_f$ is a left ideal of $\algA$
and we already saw that $AI \in \mI_f$. It remains to show that $\pi_f$ is a homomorphism and that $\pi_f(A)$ is indeed bounded.
To see that $\pi_f$ is a homomorphism, note that
$$
\pi_f(AB)\Psi_C = \Psi_{ABC} = \pi_f(A)\pi_f(B)\Psi_C
$$
and
$$
\pi_f(\lambda A + \gamma B)\Psi_C = \Psi_{\lambda A + \gamma B} = \lambda\Psi_A + \gamma \Psi_B = (\lambda \pi_f(A) + \gamma \pi_f(B)) \Psi_C,
$$
as desired. To see that $\pi_f(A)$ is bounded,
consider
\begin{eqnarray*}
\norm{\pi_f(A)\Psi_B}^2 &=& \inp{\Psi_{AB}}{\Psi_{AB}} = f((AB)^\dagger(AB))\\
&=& f(B^\dagger A^\dagger AB) \leq 
\norm{A}^2 f(B^\dagger B) \leq 
\norm{A}^2 \norm{\Psi_B}^2,
\end{eqnarray*}
where 
we used the fact that from 
$B^\dagger A^\dagger AB \leq \norm{A}^2 B^\dagger B$ we have 
$f(B^\dagger A^\dagger AB) \leq \norm{A}^2 f(B^\dagger B)$ (see for
example~\cite{takesaki:cstar}).

Finally, we need to construct the vector $\Phi_f$. Since $\algA$ is unital we can take $\Phi_f = \Psi_\id$. This gives us 
$\inp{\Phi_f}{\pi_f(A)\Phi_f} = \inp{\Psi_\id}{\pi_f(A)\Psi_\id} = \inp{\Psi_\id}{\Psi_A} = f(\id^\dagger A) = f(A)$.
Note that $\pi_f(A)\Psi_\id = \Psi_A$, i.e., $\Phi_f = \Psi_\id$ is cyclic for $(\hil_f,\pi_f)$.
\end{proof}

The resulting representation is irreducible if and only if $f$ is 
pure~\cite[Theorem 2.3.19]{robinson:cstar}. By considering a family of 
states $F$, and applying
the GNS construction to all $f \in F$ and taking the direct sum of representations it is then possible to show that:
\begin{theorem}(GN)\label{gelfandNaimarkTheorem}
Let $\algA$ be a unital $C^*$-algebra. Then $\algA$ is isomorphic to an algebra of bounded operators on a Hilbert space $\hil$.
\end{theorem}

\section{Commuting operators}\label{cstar:commutation}

$\algA$ is abelian if and only if the physical system corresponding to this algebra is classical. Thus to distinguish the quantum from the classical problems, commutation will be central to our discussion. In fact, it leads to very nice structural properties which 
we already exploited in Chapter~\ref{pistarChapter}.
First, however, we will need a bit more terminology.
The \emph{commutator} of two operators $A$ and $B$ is given by $[A,B] = AB - BA$. 
For quantum applications, two observables $A$ and $B$ 
are called \emph{compatible}\index{compatible}\index{observable!compatible} if they commute, i.e., $[A,B] = 0$. Conversely, $A$ and $B$ are called \emph{complementary}\index{complementary}\index{observable!complementary} if 
$[A,B] \neq 0$. 
The \emph{center} $\algZ_\algA$ of an algebra $\algA$ is the set of all elements in $\algA$
that commute with all elements of $\algA$, i.e. \index{algebra!center}\index{center}\label{center}
$$
\algZ_\algA = \{Z \mid Z \in \algA, \forall A \in \algA: [Z,A] = 0\}.
$$
It is easy to see that if $\algA$ only has a trivial center, i.e. $\algZ_\algA = \{c \id\mid c \in \Complex\}$,
$\algA$ is simple~\cite{takesaki:cstar}. 
If $\algA \subseteq \bop(\hil)$ for some Hilbert space $\hil$, then the \emph{commutant}\index{commutant}\index{algebra!commutant} of $\algA$ in $\bop(\hil)$\label{commutant}
is 
$$
\comm(\algA) = \{X\mid X \in \bop(\hil), \forall A \in \algA: [X,A] = 0\}.
$$
We have $\algZ_\algA = \algA \cap \comm(\algA)$.

\subsection{Decompositions}\label{cstar:decompose}
In any of our problems, the interesting case is when the algebra $\algA$ under consideration
is in fact simple: that is, ``fully quantum''. In all problems we will consider, it will turn out that we can always
break down the problem into smaller components by decomposing any 
$\algA$
into a sum of simple algebras.\footnote{Recall that we only consider the finite-dimensional case.}
Luckily, such a decomposition always exists in the finite-dimensional case:
\begin{lemma}\label{finiteMeansSemisimple}\index{semisimple!decomposition}
Let $\algA$ be a finite-dimensional $C^*$-algebra. Then there exists a decomposition 
$$
\algA = \bigoplus_j \algA_j,
$$
such that $\algA_j$ is simple.
\end{lemma}
\begin{proof} 
Let $\algZ_A$ be the center of $\algA$. Clearly, since $\algA$ is finite-dimensional, $\algZ_A$ is a finite-dimensional abelian $C^*$-algebra. Since $\algZ_A$ is finite, there exist
a finite set of positive linear functionals 
$\{f_1,\ldots,f_m\}$, such that $f_j(AB) = f_j(A)f_j(B)$ and $f_j(A) \in \Sp_{\algZ_A}(A)$ for
all $A,B \in \algZ_A$.\footnote{For a matrix algebra these are just the eigenvectors 
with equal eigenvalue} 
For all $1 \leq k \leq m$, choose $\Pi_k \in \algZ_A$
such that $f_j(\Pi_k) = \delta_{jk}$ for all $j$. Note that $\Pi_1,\ldots,\Pi_m$ are projectors and $\sum_j \Pi_j = \id$ since
for all $j$ we have
$f_j(\Pi_k \Pi_\ell) = f_j(\Pi_k) f_j(\Pi_\ell)$ since $\algZ_A$ is abelian.
Now we have
$$
\algA = \id \algA \id = \sum_{jk=1}^m \Pi_j \algA \Pi_k = \sum_{j=1}^m \Pi_j \algA \Pi_j,
$$
since for all $A \in \algA$ we have $\Pi_j A \Pi_k = \Pi_j \Pi_k A = 0$ since $\Pi_j,\Pi_j \in \algZ_A$.
Note that $\algA_j = \Pi_j \algA \Pi_j$ only has a trivial center: its only elements that commute with any element of $\algA_j$ are 
scalar multiples of $\Pi_j$. Hence, $\algA_j$ is simple.
\end{proof}

In fact, it is possible to show that~\cite{takesaki:cstar}:
\begin{corollary}
Let $\algA$ be a finite-dimensional $C^*$-algebra
Then there exists $\hil$ and a decomposition
$$
\hil = \bigoplus_j \hil_j,
$$
such that
$$
\algA \isomorph \bigoplus_j \bop(\hil_j),
$$
\end{corollary}
Note that this means that any element $A \in \algA$ can be written
as $A = \sum_j \Pi_j A \Pi_j$ where $\Pi_j$ is a projection onto $\hil_j$.

\subsection{Bipartite structure}\index{bipartite structure}

As we saw in Chapter~\ref{pistarChapter}, commutation relations induce a beautiful structure captured by the
Double Commutant theorem. We here sketch a proof of the parts of this theorem which is interesting for understanding
non-local games: Consider a bipartite system $\hil^1 \otimes \hil^2$, and operators 
$A = \hat{A} \otimes \id^{[2]}$ and $B = \id^{[1]} \otimes \hat{B}$ with $\hat{A} \in \bop(\hil^1)$ and $\hat{B} \in \bop(\hil^2)$.
Clearly, $[A,B] = 0$ since $A$ and $B$ act on two different subsystems. 
Curiously, however, we can essentially reverse the argument: A set of commutation
relations gives rise to a bipartite structure itself! 

\begin{lemma}\label{tensorProduct}
Let $\hil$ be a finite-dimensional 
Hilbert space, and let $\{\aoq \in \bop(\hil)\mid s \in S\}$ and $\{\boq \in \bop(\hil)\mid s \in T\}$.
Then the following two statements are equivalent:
\begin{enumerate}
\item For all $s \in S$, $t \in T$, $a \in A$ and $b \in B$ it holds that 
$[\aoq,\boq] = 0$.
\item There exist Hilbert spaces $\hil^A, \hil^B$ such that $\hil = \hil^A \otimes \hil^B$
and for all $s \in S$, $a \in A$ we have $\aoq \in \bop(\hil^A)$ and for all
$t \in T$, $b \in B$ we have $\boq \in \bop(\hil^B)$.
\end{enumerate}
\end{lemma}
This statement can easily be extended to more than two players. 
Here, we will only address the finite-dimensional case.

First of all, recall that by Lemma~\ref{simpleIsEnough}, we can 
greatly simplify our problem for non-local games and restrict 
ourselves to $C^*$-algebras that are simple.
As we saw earlier in Lemma~\ref{finiteMeansSemisimple}, it is well known 
that we can decompose any finite dimensional algebra into the
sum of simple algebras.
We furthermore need that for any simple algebra, the following holds:
\begin{lemma}\label{isomorphBop}\cite{takesaki:cstar}
Let $\hil$ be a Hilbert space, and let $\algA \subseteq \bop(\hil)$ be simple.
Then $\hil = \hil^{A} \otimes \hil^{B}$ and $\algA \isomorph \bop(\hil^A) \otimes \id^{B}$.
\end{lemma}

We are now ready to prove Lemma~\ref{tensorProduct}.
First, we examine the case where we are given a simple algebra $\algA \in \bop(\hil)$, for some Hilbert space
$\hil$. 
We will need the following version of Schur's lemma.
\begin{lemma}\label{schurLemma}
Let $\setZ$ be the center of $\bop(\hil)$. Then $\setZ = \{c \id|c \in \Complex\}$.
\end{lemma}
\begin{proof}
Let $C \in Z$ and let $d = \dim(\hil)$. Let $\setB = \{E_{ij}|i,j \in [d]\}$
be a basis for $\bop(\hil)$, where $E_{ij} = \outp{i}{j}$ is the matrix of all 0's
and a 1 at position $(i,j)$. Since $C \in \setZ$ and $E_{ij} \in \bop(\hil)$ we have
for all $i \in [d]$
$$
CE_{ii}  = E_{ii}C.
$$
Note that $C E_{ii}$ (or $E_{ii}C$) is the matrix of all 0's but
the $i$th column (or row) is determined by the elements of $C$. Hence
all off diagonal elements of $C$ must be 0. Now consider
$$
C (E_{ij} + E_{ji}) = (E_{ij} + E_{ji})C.
$$
Note that $C(E_{ij} + E_{ji})$ (or $(E_{ij} + E_{ji})C$) is the matrix
in which the $i$th and $j$th columns (rows) of $C$ have been swapped and the remaining 
elements are 0. Hence all diagonal elements of $C$ must be equal.
Thus there exists some $c \in \Complex$ such that $C = c \id$. 
\end{proof}

Using this Lemma, we can now show that
\begin{lemma}\label{simpleCommutant}
Let $C \in \bop(\hil^A \otimes \hil^B)$
such that for all $B \in \bop(\hil^B)$ we have
$$
[C,(\id^A \otimes B)]=0
$$
Then there exists an $A \in \bop(\hil^A)$ such that $C = A \otimes \id^B$.
\end{lemma}
\begin{proof}
Let $d_A = \dim(\hil^A)$ and $d_B = \dim(\hil^B)$. Note that we can write any $C$ as
$$
C = \left(\begin{array}{ccc} 
C_{11} &\ldots& C_{1d_A}\\
\vdots & &\vdots\\
C_{d_A1} &\ldots& C_{d_Ad_A}\end{array}\right),
$$
for $d_A \times d_A$ matrices $A_{ij}$.
We have $C(\id^A \otimes B) = (\id^A \otimes B)C$
if and only
if for all $i,j \in [d_A]$ $C_{ij}B = BC_{ij}$, i.e. $[C_{ij},B]=0$.
Since this must hold for all $B \in \bop(\hil^B)$, we have by Lemma~\ref{schurLemma}
that there exists some $a_{ij} \in \Complex$ such that $C_{ij} = a_{ij} \id^B$.
Hence $C = A \otimes \id^B$ with $A = [a_{ij}]$.
\end{proof}

For the case that the algebra generated by Alice and Bob's measurement operators is simple, 
Lemma~\ref{tensorProduct} now follows immediately:

\begin{proof}[Proof of Lemma~\ref{tensorProduct} if $\algA$ is simple]
Let $\algA = \langle \{\aoq\}\rangle \subseteq \bop(\hil)$ be the algebra generated by Alice's measurement operators.
If $\algA$ is simple, it follows from Lemma~\ref{isomorphBop} that $\algA \isomorph \bop(\hil^A) \otimes \id^B$
for $\hil = \hil^{A} \otimes \hil^{B}$. It then follows from Lemma~\ref{simpleCommutant} that
for all $t \in T$ and $b \in B$ we must have $\boq \in \bop(\hil^B)$. 
\end{proof}

Thus, we obtain a tensor product structure!
Recall that Lemma~\ref{simpleIsEnough} states that for
non-local games this is all we need.

In general, what happens if $\algA$ is not simple? 
We now sketch the argument in the case the $\algA$ is semisimple,
which by Lemma~\ref{finiteMeansSemisimple} we may always assume in the finite-dimensional case.
Fortunately, we can still 
assume that our commutation relations leave us with a bipartite structure. We 
can essentially infer this from van Neumann's
famous Double Commutant Theorem~\cite{takesaki:cstar,robinson:cstar}, partially stated here.
\begin{theorem}\label{doubleComm}
Let $\algA$ be a finite-dimensional $C^*$-algebra. Then there exists 
$\hil = \hil^A \otimes \hil^B$ and
a decomposition 
$$
\hil = \bigoplus_j \hil^A_j \otimes \hil^B_j
$$
such that
$$
\algA \isomorph \bigoplus_j \bop(\hil^A_j) \otimes \id^B_j
$$
and
\begin{equation}\label{commutantForm}
\comm(\algA) \isomorph \bigoplus_j \id^A_j \otimes \bop(\hil^B_j).
\end{equation}
\end{theorem}
\begin{proof}(Sketch)
We already now from Lemma~\ref{finiteMeansSemisimple} that $\algA$ can be decomposed into a sum of simple algebras.
Clearly, the RHS of Eq.~\ref{commutantForm} is an element of $\comm(\algA)$. To see that the LHS is contained
in the RHS, consider the projection $\Pi_j^A$ onto $\hil^A_j$. 
Note that $\Pi^A_j \in \algA$, and thus for any
$X \in \comm(\algA)$ we have $[X,\Pi_j^A] = 0$. Hence, we can write
$X = \sum_j (\Pi_j^A \otimes \id^B)X(\Pi_j^A \otimes \id^B)$, and thus we can restrict ourselves to considering
each factor individually. The result then follows immediately from Lemma~\ref{simpleCommutant}.
\end{proof}

If we have more than two players, the argument is essentially analogous, and we merely sketch it in the
relevant case when the algebra generated by the players's measurements is 
simple, since Lemma~\ref{simpleIsEnough} directly
extends to more than two players as well. Suppose we have $N$ players $\mP_1,\ldots,\mP_N$ and let $\hil$
denote their joint Hilbert space. Let $\algA$ be the algebra generated by all measurement
operators of players $\mP_1,\ldots.\mP_{N-1}$ respectively. Then it follows from Lemma~\ref{simpleCommutant}
and Lemma~\ref{isomorphBop} that $\hil = \hil^{1,\ldots,N-1} \otimes \hil^N$ where $\algA \isomorph \bop(\hil^{1,\ldots,N-1})$
and for all measurement operators $M$ of player $\mP_N$ we have that $M \in \bop(\hil^N)$. By applying
Lemma~\ref{simpleCommutant} recursively we obtain that there exists a way to partition the Hilbert
space into subsystems $\hil = \hil^1 \otimes \ldots \otimes \hil^N$ such that the measurement operators
of player $\mP_j$ act on $\hil^j$ alone.

In quantum mechanics, we will always obtain
such a tensor product structure from commutation relations, even if the Hilbert space is
infinite dimensional~\cite{summers:qftIndep}.
Here, we start out with a type-I    
algebra, the corresponding Hilbert space and operators can then be obtained by the famous GNS construction~\cite{takesaki:cstar},
an approach which is rather beautiful in its abstraction. In quantum statistical mechanics and
quantum field theory, we will also encounter factors of type-II and type-III. As it turns out, the above
argument does not generally hold in this case, however, there are a number of conditions that can lead to
a similar structure. Unfortunately, we cannot consider this case here and merely refer to the survey article by Summers~\cite{summers:qftIndep}.

\subsection{Invariant observables and states}\label{invariantStates}\index{state!invariant}

As we saw in Chapter~\ref{chapter:pistar}, expressing our problem in terms of commutation
relations enables us to exploit their structural consequences. Particularly interesting
is also the fact that we can characterize the set of states which are invariant by a quantum channel
by means of such relations, repeated here for convenience sake:
\begin{lemma}(HKL)~\cite{hkl:noiseCommutant}
Let $\Lambda: \hil \rightarrow \hil$ be a unital quantum channel with
$\Lambda(\rho) = \sum_m V_m \rho V_m^\dagger$, and let $\setS$
be a set of quantum states. Then
$$
\forall \rho \in \setS, \Lambda(\rho) = \rho \mbox{ if and only if } \forall m \forall \rho \in \mS, [V_m,\rho]=0.
$$
\end{lemma}

Let's see what this means for a specific unital channel $\Lambda(\rho) = \sum_m V_m \rho V_m^\dagger$ and
a particular ensemble given by states $\rho_1,\ldots,\rho_n \in \hil$. As in 
Chapter~\ref{chapter:pistar}, we now consider the $*$-algebra generated by $\rho_1,\ldots,\rho_n$.
Let $\algA$ denote the resulting algebra. By Theorem~\ref{doubleComm}, we know that we can
write
$$
\algA \isomorph \bigoplus_j \bop(\hil^1_j) \otimes \id^{[2]}_j
$$
and
$$
\comm(\algA) \isomorph \bigoplus_j \id^{[1]}_j \otimes \bop(\hil^{2}_j).
$$
Clearly, we have from the above that if $\Lambda$ leaves our ensemble of states untouched,
we must have $V_m \in \comm(\algA)$ for all $m$. Thus we know that $V_m$ must be of the form
$\id^{[1]}_j \otimes V^{[2]}_j$ on each factor. What does this mean operationally? Suppose
we can write $\hil = \bigoplus_j \hil_j$ such that 
$\rho_k = \sum_j \Pi_j \rho_k \Pi_j$ for each $\rho_k$, where $\Pi_j$ is a projector
onto $\hil_j$. That is, we can simultaneously block-diagonalize all $\rho_k$. 
Then we know that $V_m$ must be equal to the identity on each factor $\hil_j$, i.e.
$V_m$ must be of the form $\bigoplus_j c_j \Pi_j$ for some $c_j$ with $|c_j|=1$.
Another nice application of this viewpoint is an algebraic no-cloning theorem, as
put forward by Lindblad~\cite{lindblad:cloning}.

\section{Conclusion}

Even though the sheer number of new definitions may appear daunting, we saw that the language of $C^*$-algebras can help us get a
grip on some of the fundamental properties of quantum states quite easily. 
Of course, the language of $C^*$-algebras is not the most
convenient one for all problems. 
Yet, there are many 
cases for which the language of $C^*$-algebras is especially useful.
As we saw earlier, one of these cases is when we consider measurements performed by two parties on a bipartite system. 
Another class of problems deals with questions of the following forms:
Which operations leave a given set of states invariant? 
How much can we learn from a given state without disturbing it? What part of a state is ``truly'' quantum and which 
parts can we consider to be classical? How can we encode our states such that they are left untouched by a set of 
operations?

For example, another application is the compression of quantum states. Koashi and Imoto consider how a quantum state can be decomposed into a quantum, a classical and a redundant part to aid compression. In their paper, they provide an algorithm which in fact allows us to compute (with a lot of pain) the decomposition of an algebra and its commutant algebra~\cite{koashi&imoto:operations}.
It is probably not so surprising by now that other tasks involving invariance under operations are also closely related:
Choi and Kribs~\cite{choi:dfs} have phrased the principle of decoherence-free subspaces in terms of what they call
algebraic noise commutant formalism. In this text, we have exploited $C^*$-algebras to investigate the use of post-measurement information in Chapter~\ref{pistarChapter}.
As we saw in Chapter~\ref{boundingEntanglementChapter}, the question of how much post-measurement information is needed is in fact closely related to
how much entanglement we need to succeed in non-local games. Whereas these two problem may appear unrelated at first sight, their structural similarities show their close connection. Likewise, these similarities also enabled us in Chapter~\ref{entanglementIntroChapter} to investigate how much we can really gain by receiving additional post-measurement information. Finally, the close connection of $C^*$-algebras and Clifford algebras discussed in Appendix~\ref{cliffordChapter} was one of 
the factors that led us to discover the uncertainty relations of Chapter~\ref{uncertaintyChapter}. Hence, $C^*$-algebras sometimes help us to understand the similarities between problems, and aid our intuition.

\chapter{Clifford Algebra}\label{cliffordChapter}\index{Clifford algebra}
\label{cliffordAppendix}\label{cliffordApp}\label{appendix:clifford}

Similar to $C^*$-algebra, Clifford algebra plays little role in computer science even though
it has recently found numerous applications in the area of computer graphics. Here, we informally
summarize the most important facts we need in this text. Our aim is merely to provide the reader
with some intuition underlying our uncertainty relation in Chapter~\ref{chapter:uncertainty}, and
refer to~\cite{lounesto:book} for an in-depth introduction.

\section{Introduction}

Clifford algebra is closely related to $C^*$-algebra. Yet, it exhibits many beautiful 
geometrical aspects which remain inaccessible to us otherwise. In particular, we will see that
commutation and anti-commutation carries a geometric meaning within this algebra. 

For any integer $n$, the unital associative algebra generated by
$\Gamma_1,\ldots,\Gamma_{2n}$, subject to the anti-commutation
relations\index{anti-commuting observables}\index{Clifford algebra!generators}
$$
\Gamma_i\Gamma_j = - \Gamma_j \Gamma_i, \mbox{   } \Gamma_i^2 = \id
$$
is called \emph{Clifford algebra}.\index{Clifford algebra} It has a unique representation by
Hermitian matrices on $n$ qubits (up to unitary equivalence) which we fix henceforth.
This representation can be obtained via the famous Jordan-Wigner 
transformation~\cite{JordanWigner}:\index{Jordan Wigner transformation}
\begin{align*}
  \Gamma_{2j-1} &= \sigma_y^{\otimes(j-1)} \otimes \sigma_x \otimes \id^{\otimes(n-j)}, \\
  \Gamma_{2j}   &= \sigma_y^{\otimes(j-1)} \otimes \sigma_z \otimes \id^{\otimes(n-j)},
\end{align*}
for $j=1,\ldots,n$.
A Clifford algebra of $n$ generators is isomorphic to a $C^*$-algebra of matrices of size $2^{n/2} \times 2^{n/2}$ for $n$ even 
and to the direct sum of two $C^*$-algebras of matrices of size 
$2^{(n-1)/2} \times 2^{(n-1)/2}$ for $n$ odd~\cite{tsirel:separated}. 

\section{Geometrical interpretation}\index{Clifford algebra!geometrical properties}

The crucial advantage of the Clifford algebra is that we can view the operators
$\Gamma_1,\ldots,\Gamma_{2n}$ as $2n$ orthogonal vectors forming a basis
for a $2n$-dimensional real vector space $\Real^{2n}$. Each vector $a = (a_1,\ldots,a_{2n}) \in \Real^{2n}$ can
then be written as linear combination of basis elements as $a = \sum_j a_j \Gamma_j$. 
The \emph{Clifford product}\index{Clifford product}\index{Clifford algebra!product}
of two vectors $a$ and $b$ is given 
by 
$$
ab = a \cdot b + a \wedge b,
$$
where $a \cdot b = \sum_j a_j b_j \id$ is the inner product of two vectors
and $a \wedge b$ is the outer product, as given below. We will write scalars as scalar multiples
of the identity element whose matrix representation is simply the identity matrix.
If we represent $\Gamma_1,\ldots,\Gamma_{2n}$ using
the matrices from above, then the Clifford product is simply the matrix product of the resulting
matrices. Hence, we will now adopt this viewpoint with the representation in mind.
Note that the Clifford product satisfies $a^2 = |a|^2 \id = \sum_j a_j^2 \id$,
where $|a| = ||a||_2 = \sqrt{\sum_j a_j^2}$
is the $2$-norm of the vector $a$ which we refer to as the \emph{length} of a vector. 

\subsection{Inner and outer product}
\begin{figure}[h]
\begin{center}
\includegraphics[scale=0.6]{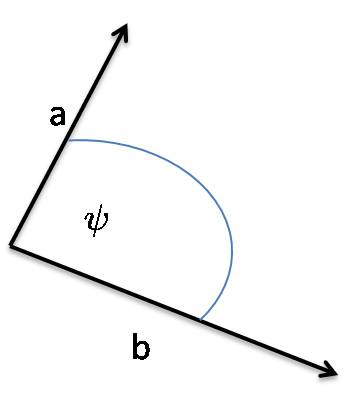}
\caption{Two vectors}
\label{figure:vectors}
\end{center}
\end{figure}

We can see immediately from the definition of the Clifford product
that the inner product of two vectors $a,b \in \Real^{2n}$ as depicted in Figure~\ref{figure:vectors}
is given
by $a\cdot b = |a| |b| \cos \psi \id$, and can be expressed as:
$$
a \cdot b = \frac{1}{2} \left\{a,b\right\} = \frac{1}{2}(ab + ba).
$$ 

Hence, anti-commutation takes a geometric
meaning within the algebra: two vectors anti-commute if and only if they are
orthogonal!

Similarly, we can write
$$
a \wedge b = \frac{1}{2} \left[a,b\right] = \frac{1}{2}(ab - ba).
$$
Geometrically, this means that two vectors are parallel if and only if they commute.

To gain some intuition, let's look at the simple example of $\Real^2$:
Here, we have $a = a_1 \Gamma_1 + a_2 \Gamma_2$ and $b = b_1 \Gamma_1 + b_2 \Gamma_2$.
The Clifford product of $a$ and $b$ is now given as
$$
ab = \sum_{jk} a_j b_k \Gamma_j \Gamma_k = (a_1b_1 + a_2b_2)\id + (a_1b_2 - b_1a_2)\Gamma_1\Gamma_2.
$$
The element $a \wedge b = (a_1b_2 - b_1a_2)\Gamma_1\Gamma_2$ represents the oriented plane segment
of the parallelogram determined by $a$ and $b$ in Figure~\ref{figure:outer1} below. The area of this parallelogram is exactly
$\left|a \wedge b\right| = |a_1b_2 - b_1a_2|$. Note that we have $a \wedge b = - b \wedge a$, as shown
in Figure~\ref{figure:outer2}.
Thus $a \wedge b$ not only gives us the area but also encodes a direction.\index{outer product}
\begin{figure}[h]
\begin{minipage}{0.45\textwidth}
\begin{center}
\includegraphics[scale=0.8]{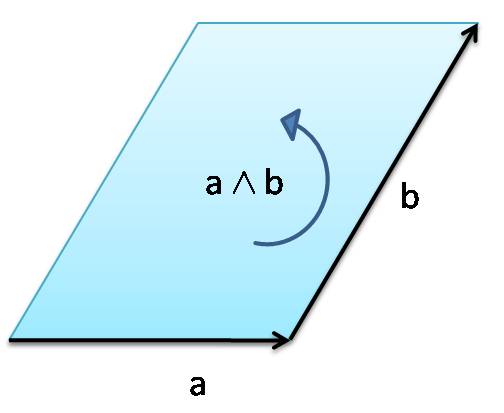}
\caption{$a \wedge b$}
\label{figure:outer1}
\end{center}
\end{minipage}
\begin{minipage}{0.45\textwidth}
\begin{center}
\includegraphics[scale=0.8]{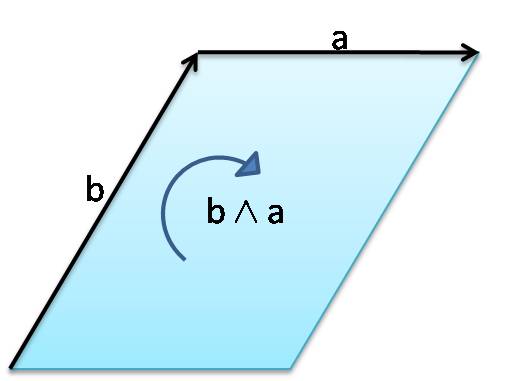}
\caption{$b \wedge a$}
\label{figure:outer2}
\end{center}
\end{minipage}
\end{figure}

In higher dimensions, the elements generated by $a \wedge b \wedge c$ etc similarly correspond
to oriented plane or volume segments. Note that we have $\Gamma_i \wedge \Gamma_j = \Gamma_i\Gamma_j$
for all basis vectors $\Gamma_i$ and $\Gamma_j$. We will refer to products of
$k$ elements of the form $\Gamma_{i_1}\ldots\Gamma_{i_k}$
as \emph{$k$-vectors}.\index{$k$-vector} 

\subsection{Reflections}

The power of the Clifford algebra mainly lies in the fact that we can express geometrical operations
involving any $k$-vector in an extremely easy fashion using the Clifford product. Here, we will only
be concerned with performing operations on $1$-vectors.

\begin{figure}[h]
\begin{center}
\includegraphics[scale=0.6]{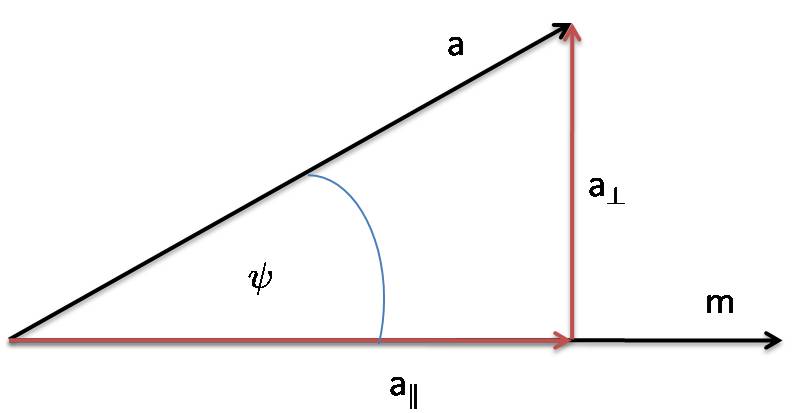}
\caption{Projections onto a vector}
\label{figure:projection}
\end{center}
\end{figure}

\begin{figure}[h]
\begin{center}
\includegraphics[scale=0.6]{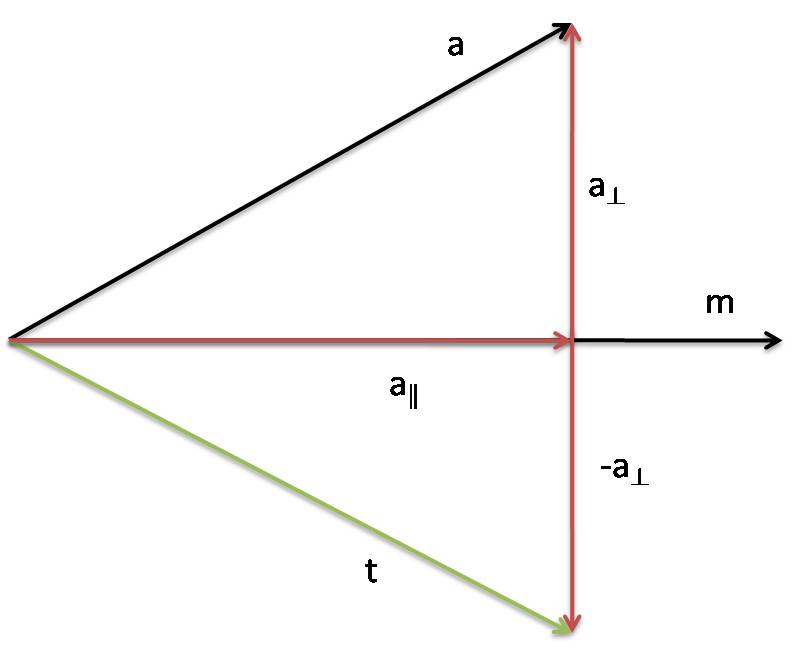}
\caption{Reflection of $a$ around $m$}
\label{figure:reflection1}
\end{center}
\end{figure}

\begin{figure}
\begin{center}
\includegraphics[scale=0.6]{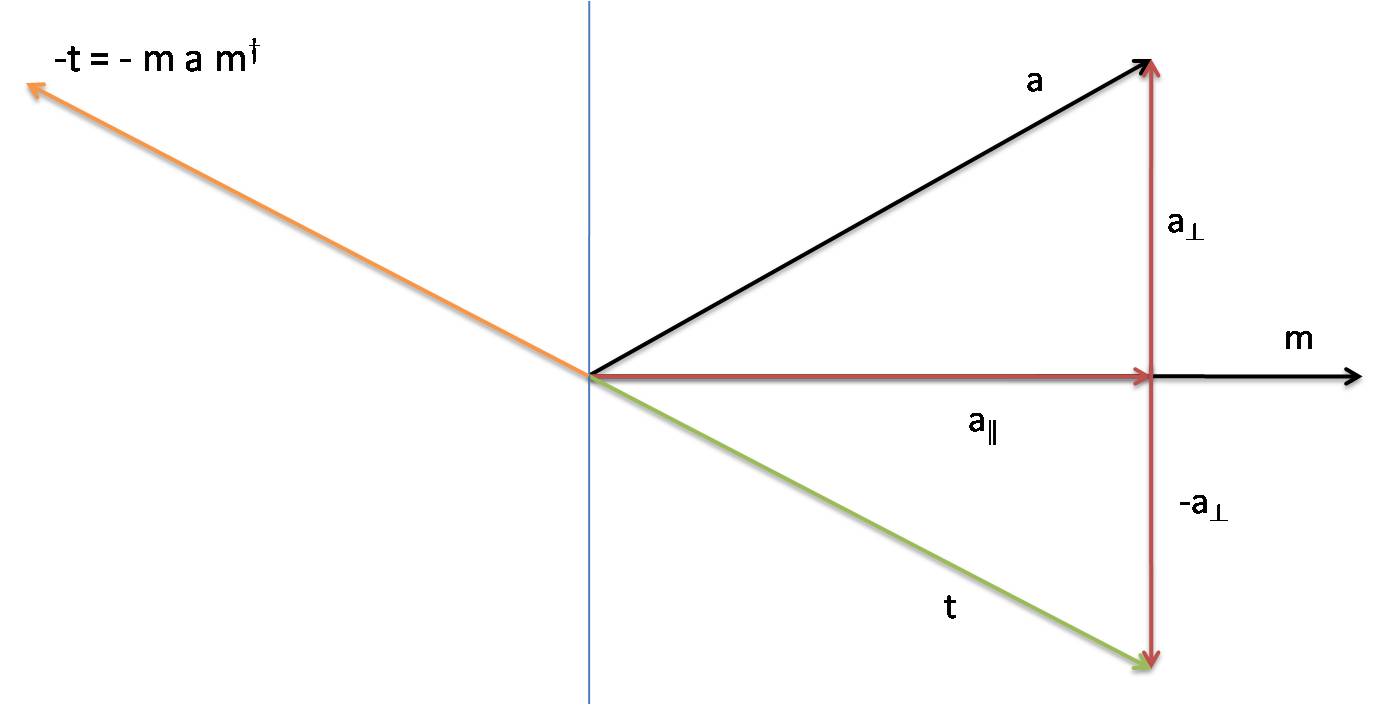}
\caption{Reflection of $a$ plane perpendicular to $m$}
\label{figure:reflection2}
\end{center}
\end{figure}

\begin{figure}
\begin{center}
\includegraphics[scale=0.6]{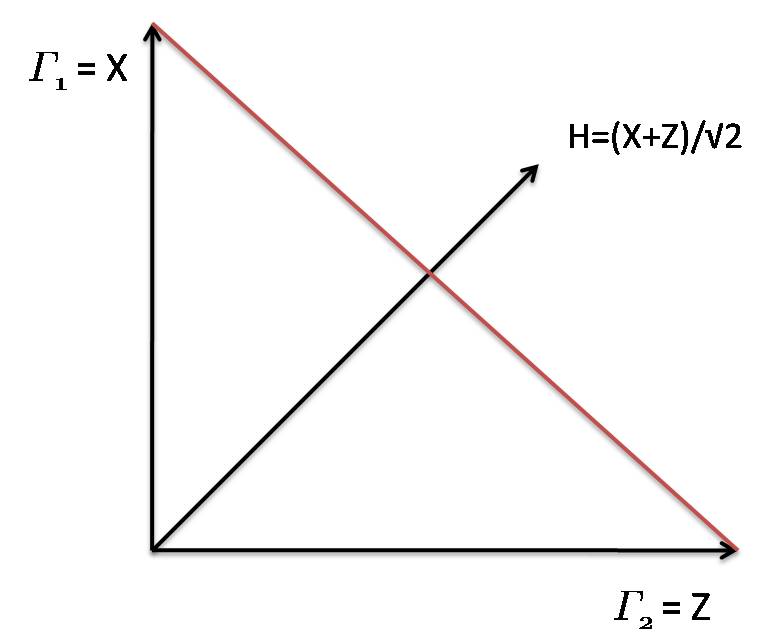}
\caption{Hadamard transform as reflection}
\label{figure:hadamard}
\end{center}
\end{figure}

Consider the projection of a vector $a$ onto a vector $m$ as depicted in Figure~\ref{figure:projection}.
Let $a_\parallel$ be the part of $a$ that is parallel to $m$, and $a_\perp$ the part of $a$ that lies
perpendicular to $m$. Clearly, we may write $a = a_\parallel + a_\perp$.
Using the definition of the Clifford product, we may write
$$
a_\parallel = |a| \cos \psi \frac{m}{|m|} = (a \cdot m)m^{\dagger},
$$
where we define $m^{\dagger}= m/|m|^2$ to be the inverse of $m$. 
Indeed, we have
$mm^\dagger = \id$. 
If $m$ is a unit vector,
then in terms of the matrix representation given above $m^\dagger$ is the adjoint
of the matrix $m$.
For the product of two vectors we define $(nm)^\dagger = m^\dagger n^\dagger$.
We can also write
$$
a_\perp = a - a_\parallel =  a - (a\cdot m)m^\dagger = (am - (a\cdot m))m^\dagger = (a \wedge m)m^\dagger.
$$
We can now easily determine the reflection of $a$ around the vector $m$, as depicted in Figure~\ref{figure:reflection1}:
$$
t = a_\parallel - a_\perp = (a \cdot m - a \wedge m)m^\dagger = (m \cdot a + m \wedge a)m^\dagger = mam^\dagger.
$$
Consider $n=1$. Then the $2$-dimensional real vector space is given by basis vectors $\Gamma_1 = X$ 
and $\Gamma_2 = Z$. Indeed, this is the familiar $XZ$-plane of the Bloch 
sphere depicted in Figure~\ref{blochSphere}. Consider the Hadamard
transform $H = (X + Z)/\sqrt{2}$. Figure~\ref{figure:hadamard} demonstrates that $H$ plays exactly
this role: it reflects $X$ around the vector $H$ to obtain $HXH = Z$.
Given $t$, we can also easily derive the vector obtained by reflecting $a$ around the plane
perpendicular to $m$ (in 0), as shown in Figure~\ref{figure:reflection2}.
$$
-t = - m a m^\dagger.
$$

\subsection{Rotations}

\begin{figure}[h]
\begin{center}
\includegraphics[scale=0.6]{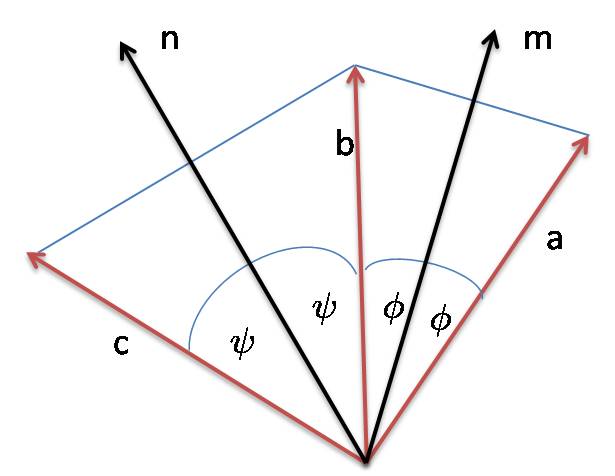}
\caption{Rotating in the plane $m \wedge n$.}
\label{figure:rotation}
\end{center}
\end{figure}

From reflections we may now obtain rotations as successive reflections. Suppose we are 
given vectors $m$ and $n$ as shown in Figure~\ref{figure:rotation}. To rotate the vector
$a$ by an angle that is twice the angle between $m$ and $n$, we now first reflect 
$a$ around $b$ to obtain $b = mam^\dagger$. We then reflect $b$ around $n$ to obtain
$$
c = nbn^\dagger = nmam^\dagger n^\dagger = RaR^\dagger,
$$
where we let $R = nm$. As desired, $R$ rotates $a$ by an angle of $2(\psi + \phi)$.

We can easily convince ourselves that $R$ does not affect any vector $d$ that is
orthogonal to both $n$ and $m$.
$$
R d R^\dagger = nm d m^\dagger n^\dagger = d nmm^\dagger n^\dagger = d,
$$
where we have used the fact that two vectors anti-commute if and only if they are orthogonal.
Note also that $RR^\dagger = \id$.
It can be shown that if $V$ is a $k$-vector, then $RVR^\dagger$ is also a $k$-vector
for any rotation $R$~\cite{doran:book}. Indeed, this is easy to see, for the
$k$-vector formed by orthogonal basis vectors:
\begin{eqnarray*}
R(\Gamma_{i_1} \wedge \ldots \wedge \Gamma_{i_k})R^\dagger &=& 
R(\Gamma_{i_1}\ldots\Gamma_{i_k})R^\dagger = 
R\Gamma_{i_1}R^\dagger\ldots R\Gamma_{i_k}R^\dagger\\
 &=& 
R\Gamma_{i_1}R^\dagger\wedge\ldots \wedge R\Gamma_{i_k}R^\dagger,
\end{eqnarray*}
where we have used the fact that rotations preserve the angles between vectors.
We will need this fact in our proof in Chapter~\ref{chapter:uncertainty}.

\begin{figure}[h!]
\includegraphics[scale=0.8]{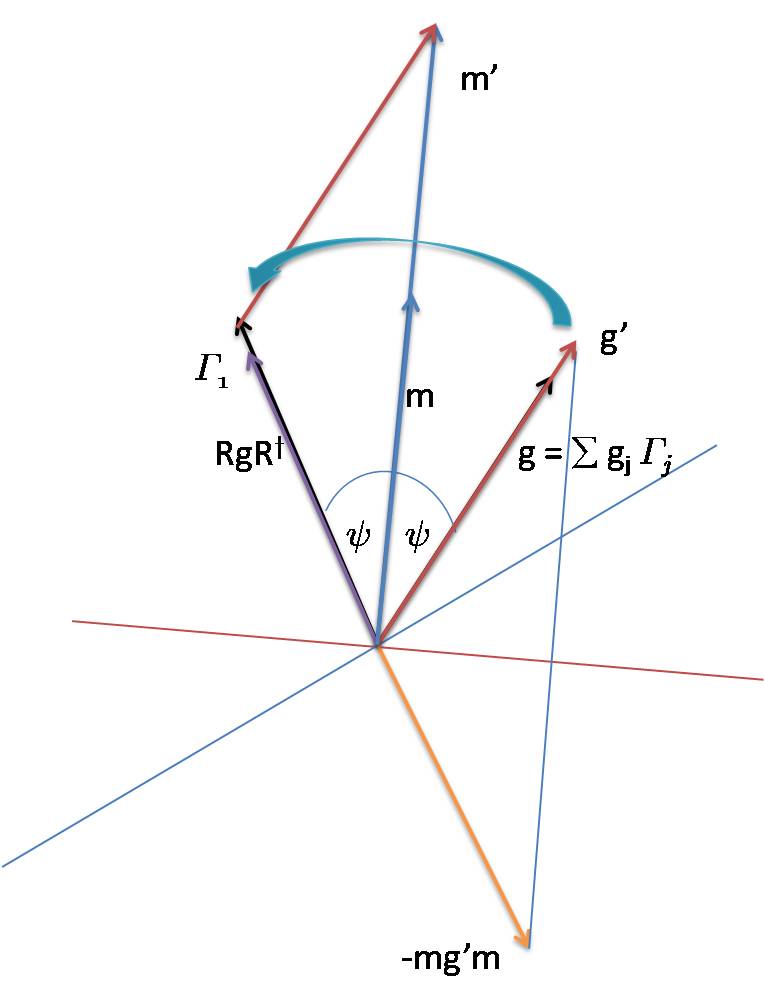}
\caption{Rotating $g$ to $|g| \Gamma_1$.}
\label{figure:tweakedRotation}
\end{figure}

Clifford algebra offers a very convenient way to express rotations around arbitrary
angles in the plane $m \wedge n$~\cite{lounesto:book}. In Chapter~\ref{chapter:uncertainty}, however,
we will only need to understand how we can find the rotation $R$ that takes us from a given
vector $g = \sum_j g_j \Gamma_j$ with length $|g|$ to the vector $|g|\Gamma_1$. Indeed,
our strategy works for finding the rotation of any vector $g$ to a target vector $t$ of the same length.
Consider Figure~\ref{figure:tweakedRotation}. 

For convenience, we first normalize $g$
to obtain the vector
$$
g' = \frac{g}{|g|}.
$$
We then compute the vector $m'$ lying exactly half-way between $g'$ and our target vector $\Gamma_1$
and normalize it to obtain
$$
m = \frac{g' + \Gamma_1}{|g'+\Gamma_1|} = \frac{g'+\Gamma_1}{\sqrt{2(1 + g_1/|g|)}}.
$$
We now first reflect $g'$ around the plane perpendicular to the vector $m$ to obtain $-m g' m$,
followed by a reflection around the plane perpendicular to the target vector $\Gamma_1$:
$$
-\Gamma_1 (-m g' m) \Gamma_1 = \Gamma_1 m g' m \Gamma_1 = R g' R^\dagger,
$$
with $R = \Gamma_1 m$, where we have used the fact that both $\Gamma_1$ and $m$ have unit length
and hence $m = m^\dagger$ and $\Gamma_1 = \Gamma_1^\dagger$.
Evidently,
\begin{eqnarray*}
R g' = \Gamma_1 m g' &=&
\frac{1}{|g'+\Gamma_1|} \Gamma_1 (g' + \Gamma_1) g'\\
&=& 
\frac{1}{|g'+\Gamma_1|} \Gamma_1 (g'^2 + \Gamma_1 g') = 
\frac{1}{|g'+\Gamma_1|} (\Gamma_1 + g') = m,
\end{eqnarray*}
and hence
$$
R g' R^\dagger = 
mm\Gamma_1 = \Gamma_1.
$$
We then also have that
$$
R g R^\dagger = |g| R g' R^\dagger = |g| \Gamma_1
$$
as desired. We will employ a similar rotation in Chapter~\ref{chapter:uncertainty}.

\section{Application}

Here, the primary benefit which we gain by considering a Clifford algebra, is that we can parametrize matrices
in terms of its generators, and products thereof. Suppose we are given some matrix
$$
\rho = \frac{1}{d}\left(\id + \sum_j b_j B_j\right),
$$
where $\id \cup \{B_j\}$ form a basis for the $d \times d$ complex matrices, such that
for all $j\neq j'$ we have $\Tr(B_jB_{j'}) = 0$,
$\Tr(B_j) = 0$, $B_j^2 = \id$, and $b_j \in \Real$. We saw in Chapter~\ref{chapter:uncertainty}
how to construct such a basis for $d=2^n$ based on mutually unbiased bases. In fact, this gives us
the well-known Pauli basis, given by the $2^{2n}$ elements of 
the form $B_j = B_j^1 \otimes \ldots \otimes B_j^{n}$ with $B_j^i \in \{\id,\sigma_x,\sigma_y,\sigma_z\}$.
When solving optimization problems within quantum information, we are often faced with the following
problem: When is $\rho$ a quantum state? That is, what are the necessary and sufficient conditions
for the coefficients $b_j$ such that $\rho \geq 0$?

For $d=2$, this is an easy problem: We can write $\rho = (\id + \sum_{j \in \{x,y,z\}} r_j \sigma_j)/2$
where $\vec{r} = (r_x,r_y,r_z)$ is the Bloch vector we encountered in Chapter~\ref{chapter:informationIntro}.
We have that $\rho \geq 0$ if and only if $-\id \leq \sum_j r_j \sigma_j \leq \id$, i.e. 
$$
\left(\sum_j r_j \sigma_j\right)^2 
= \frac{1}{2} \sum_{j,j'} r_j r_{j'} \{\sigma_j,\sigma_{j'}\} = \left(\sum_j r_j^2\right) \id \leq \id.
$$ 
Thus, we have $\rho\geq 0$ if and only if $\sum_j r_j^2 \leq 1$. Geometrically, this
means that any point on or inside the Bloch sphere corresponds to a valid quantum state as 
illustrated in Figure~\ref{blochSphere}.
Sadly, when we consider $d > 2$, our task becomes considerably more difficult.
Clearly, since $\Tr(\rho^2) \leq 1$ for any quantum state, we can always say
that
\begin{eqnarray*}
\Tr(\rho^2) &=& \frac{1}{d^2}\left(\Tr(\id) + 2 \sum_j b_j \Tr(B_j) + \sum_{jj'} b_jb_{j'} \Tr(B_jB_{j'})\right)\\
&=& \frac{1}{d^2} \left(d + \sum_j b_j^2 \Tr(\id)\right)\\ 
&=& \frac{1}{d}\left(1 + \sum_j b_j^2\right) \leq 1,
\end{eqnarray*}
giving us $\sum_j b_j^2 \leq d-1$. Unfortunately, this condition is too weak for almost all
practical applications. There exist many matrices which obey this condition, but nevertheless do not
correspond to valid quantum states. Luckily, we can say something much stronger using the Clifford algebra.

Let's consider the operators $\Gamma_1,\ldots,\Gamma_{2n}$ themselves. Evidently, 
each operator $\Gamma_i$ has exactly two eigenvalues $\pm 1$:
Let $\ket{\eta}$ be an eigenvector of $\Gamma_i$ with eigenvalue $\lambda$.
From $\Gamma_i^2 = \id$ we have that $\lambda^2 = 1$. Furthermore,
we have $\Gamma_i (\Gamma_j \ket{\eta}) = - \lambda \Gamma_j \ket{\eta}$.
Thus, if $\lambda$ is an eigenvalue of $\Gamma_i$ then so is $- \lambda$.
We can therefore express each $\Gamma_i$ as
$$
\Gamma_i = \Gamma_i^0 - \Gamma_i^1,
$$
where $\Gamma_i^0$ and $\Gamma_i^1$ are projectors onto the positive and
negative eigenspace of $\Gamma_i$ respectively. 
Furthermore, note that
we have for all $i,j$ with $i \neq j$
$$
\Tr(\Gamma_i \Gamma_j) = \frac{1}{2} \Tr(\Gamma_i \Gamma_j + \Gamma_j \Gamma_i) = 0,
$$
that is all such operators are orthogonal with respect to the Hilbert-Schmidt inner product.
We now use the fact that the collection of operators
\[\begin{split}
  \id           & \\
  \Gamma_j     & \phantom{===} (1\leq j\leq 2n) \\
  \Gamma_{jk}  &:= i\Gamma_j\Gamma_k \ (1\leq j < k \leq 2n) \\
  \Gamma_{jk\ell} &:= \Gamma_j\Gamma_k\Gamma_\ell \ (1\leq j < k < \ell \leq 2n) \\
    \vdots     &\\
  \Gamma_{12\ldots (2n)} &:= i\Gamma_1\Gamma_2 \cdots \Gamma_{2n} =: \Gamma_0
\end{split}\]
forms an orthogonal basis for the $d \times d$ matrices with $d = 2^n$~\cite{dietz:bloch}.
By counting, the above operators
form a complete operator basis with respect to the Hilbert-Schmidt
inner product. In fact, by working out the individual basis elements with respect to
the representation above, we see that this basis is in fact equal to the Pauli basis.
Notice that the products with an odd number of factors are
Hermitian, while the ones with an even number of factors are skew-Hermitian,
so in the definition of the above operators we introduce a factor of
$i$ to all with an even number of indices to make the whole set a
basis for the Hermitian operators.
Hence we can write every state $\rho \in \hil$ as
$$
  \rho = \frac{1}{d}\left( \id + \sum_j g_j\Gamma_j + \sum_{j<k} g_{jk}\Gamma_{jk}
                              + \ldots + g_0 \Gamma_0 \right),
$$
with real coefficients $g_j, g_{jk},\ldots$.

It is clear from the above that if we transform the generating set
of $\Gamma_j$ linearly,
\[
  \Gamma_k' = \sum_{j} T_{jk}\Gamma_j,
\]
then the set $\{ \Gamma_1',\ldots,\Gamma_{2n}'\}$ satisfies the anti-commutation
relations if and only if $(T_{jk})_{jk}$ is an orthogonal matrix: these are exactly the operations
which preserve the inner product. In that case there
exists a matching unitary $U(T) \in \bop(\hil)$ which transforms the operator 
basis as
$$
  \Gamma_j' = U(T) \Gamma_j U(T)^\dagger.
$$
As an importance consequence, it can be shown~\cite{dietz:bloch} that 
any operation $U(T)$ transforms the state $\rho$ as
$$
U(T)\rho U(T)^\dagger 
= \frac{1}{d}\left(\id + T(g)  + \sum_{j<k} g'_{jk}\Gamma_{jk} + \ldots + g'_0 \Gamma_0 \right),
$$
where we write $T(g)$ to indicate the transformation of the vector $g = \sum_j g_j \Gamma_j$ by $T$.
For example, for the rotation $R$ constructed earlier, we may immediately write
\begin{eqnarray*}
R \rho R^\dagger 
&=& \frac{1}{d}\left(\id + RgR^\dagger  + \sum_{j<k} g_{jk}R\Gamma_{jk}R^\dagger + \ldots + g_0 
R\Gamma_0R^\dagger \right),\\
&=& \frac{1}{d}\left(\id + |g|\Gamma_1+ \sum_{j<k} g'_{jk}\Gamma_{jk}+ \ldots + g'_0 
\Gamma_0 \right),
\end{eqnarray*}
Thus, we can think of the 1-vector components of $\rho$ as vectors in a generalized Bloch sphere.
In Chapter~\ref{chapter:uncertainty}, we will extend this approach to include the $\Gamma_0$
as an additional ``vector''.
There, we use these facts to prove a useful statement which
leads to our uncertainty relations:
\begin{lemma}[Lemma~\ref{metaUR}]
For any state $\rho$, we have 
$\sum_j g_j^2 \leq 1$.
\end{lemma}

With respect to our discussion above, this is indeed a generalization of what we observed for the
Bloch sphere in $d=2$. 
Note that we obtain a whole range of such statements
as we can find different sets of $2n$ anti-commuting matrices within the entire set of
$2^{2n}$ basis elements above.

\section{Conclusion}
Luckily, we made some progress to give a characterization of quantum states in terms of their basis coefficients
that was sufficient to prove our uncertainty relation from Chapter~\ref{chapter:uncertainty}. 
Parametrizing states using Clifford algebra elements provides us with additional structure to characterize 
quantum states that is not at all obvious when looking at them from a linear algebra point of view alone. 
We hope that parametrizing states in this fashion could enable us to make even stronger
statements in the future. 
It is also interesting to think about standard quantum gates as geometrical operations within
the Clifford algebra. Indeed, this is possible to a large extent, but lies outside the scope
of this text.

Clearly, the subspace spanned by the elements $\Gamma_1,\ldots,\Gamma_{2n}$ plays a special 
role. Note that when considering the state minimizing our uncertainty relation, only its 1-vector
coefficients played any role. The other coefficients do not contribute at all to the minimization
problem. It is interesting to observe that we have in fact already seen a similar effect 
in Chapter~\ref{chapter:entanglement}. Recall that we used Tsirelson's construction to turn 
vectors $a,b \in \Real^{2n}$ back into observables by 
letting $A = \sum_j a_j \Gamma_j$ and $B = \sum_j b_j \Gamma_j$.
The optimal strategy of Alice and Bob could then be implemented using the maximally entangled state of local
dimension $d = 2^n$
$$
\outp{\Psi}{\Psi} = \frac{1}{d}\left(\id + \sum_j g_j \Gamma_j \otimes \Gamma_j + \sum_j r_j R_j \otimes R_j\right),
$$
where $g_j =\pm 1$ and we used the $R_j$ simply as a remainder term. Clearly, the coefficients $r_j$ do 
not contribute to the term $\bra{\Psi}A \otimes B\ket{\Psi}$ at all, and only the coefficients $g_j$ matter. 
However, in dimension $d=2^{n}$ we have only $2n$ such terms. Curiously, the remaining terms are only needed
to ensure that $\rho \geq 0$. Numerical feasibility analysis using semidefinite programming for $d=4$ and $d=8$
reveals that we do indeed need to take the maximally entangled state, and cannot omit any of the remaining terms.

\bibliographystyle{alpha}
\bibliography{thesis}

\newcommand{\etalchar}[1]{$^{#1}$}
\begin{thebibliography}{BOGKW88}

\bibitem[ABRD04]{hein:coin}
A.~Ambainis, H.~Buhrman, H.~Roehrig, and Y.~Dodis.
\newblock Multiparty quantum coin flipping.
\newblock In {\em Proceedings of 19th IEEE Conference on Computational
  Complexity}, pages 250--259, 2004.

\bibitem[ADR82]{aspect:test2}
A.~Aspect, J.~Dalibard, and G.~Roger.
\newblock Experimental test of {B}ell's inequality using time-varying
  analyzers.
\newblock {\em Physical Review Letters}, 49(25):1804--1807, 1982.

\bibitem[AE01]{aharonov&englert:meanking}
Y.~Aharonov and B.~G. Englert.
\newblock The mean king's problem: Prime degrees of freedom.
\newblock {\em Physics Letters A}, 284:1--5, 2001.

\bibitem[AGR82]{aspect:test1}
A.~Aspect, P.~Grangier, and G.~Roger.
\newblock Experimental realization of {E}instein-{P}odolsky-{R}osen
  {G}edankenexperiment - a new violation of {B}ell inequalities.
\newblock {\em Physical Review Letters}, 49(2):91--94, 1982.

\bibitem[Amb01]{ambainis:coin}
A.~Ambainis.
\newblock A new protocol and lower bounds for quantum coin flipping.
\newblock In {\em Proceedings of 33rd ACM STOC}, pages 134--142, 2001.

\bibitem[AMTdW00]{ronald:pqc}
A.~Ambainis, M.~Mosca, A.~Tapp, and R.~de~Wolf.
\newblock Private quantum channels.
\newblock In {\em Proceedings of 41st IEEE FOCS}, pages 547--553, 2000.

\bibitem[ANTV99]{nayak:original}
A.~Ambainis, A.~Nayak, A.~{Ta-Shma}, and U.~Vazirani.
\newblock Quantum dense coding and a lower bound for 1-way quantum finite
  automata.
\newblock In {\em Proceedings of 31st ACM STOC}, pages 376--383, 1999.
\newblock quant-ph/9804043.

\bibitem[Arv76]{arveson:cstar}
W.~Arveson.
\newblock {\em An invitation to $C^*$-algebra}.
\newblock Springer, 1976.

\bibitem[AS83]{alpern:keyless}
B.~Alpern and F.~B. Schneider.
\newblock Key exchange using `keyless cryptography'.
\newblock {\em Information Processing Letters}, 16:79--1, 1983.

\bibitem[AS04]{smith:smallKey}
A.~Ambainis and A.~Smith.
\newblock Small pseudo-random families of matrices: Derandomizing approximate
  quantum encryption.
\newblock In {\em Proceedings of RANDOM 2004}, volume 3122 of {\em Lecture
  Notes in Computer Science}, pages 249--260. Springer, 2004.

\bibitem[Asp99]{aspect:overview}
A.~Aspect.
\newblock {B}ell's inequality test: more ideal than ever.
\newblock {\em Nature}, 398:189--190, 1999.

\bibitem[ATSVY00]{dorit:bitEscrow}
D.~Aharonov, A.~Ta-Shma, U.~Vazirani, and A.~Yao.
\newblock Quantum bit escrow.
\newblock In {\em Proceedings of the 32th ACM STOC}, pages 705--714, 2000.

\bibitem[Aza04]{azarchs:entropy}
A.~Azarchs.
\newblock Entropic uncertainty relations for incomplete sets of mutually
  unbiased observables.
\newblock quant-ph/0412083, 2004.

\bibitem[Bab85]{babai:ip}
L.~Babai.
\newblock Trading group theory for randomness.
\newblock In {\em Proceedings of 17th ACM STOC}, pages 421--429, 1985.

\bibitem[BB83]{bb84:isit}
C.~H. Bennett and G.~Brassard.
\newblock Quantum cryptography and its application to provably secure key
  expansion, public-key distribution and coin tossing.
\newblock In {\em Proceedings of IEEE ISIT 83}, page~91, 1983.

\bibitem[BB84]{bb84}
C.~H. Bennett and G.~Brassard.
\newblock Quantum cryptography: Public key distribution and coin tossing.
\newblock In {\em Proceedings of the IEEE International Conference on
  Computers, Systems and Signal Processing}, pages 175--179, 1984.

\bibitem[BB89]{bb84:experimental}
C.~H. Bennett and G.~Brassard.
\newblock The dawn of a new era for quantum cryptography: The experimental
  prototype is working.
\newblock {\em Sigact News}, 20(4):78--82, 1989.

\bibitem[BB06]{bia:uncertaintyRenyi}
I.~Bialynicki-Birula.
\newblock Formulation of the uncertainty relations in terms of the r{\'e}nyi
  entropies.
\newblock {\em Physical Review A}, 74:052101, 2006.

\bibitem[BBB{\etalchar{+}}92]{bb84:experimental2}
C.H. Bennett, F.~Bessette, G.~Brassard, L.~Salvail, and J.~Smolin.
\newblock Experimental quantum cryptography.
\newblock {\em Journal of Cryptology}, 5(1):3--28, 1992.

\bibitem[BBBW82]{bbOne}
C.H Bennett, G.~Brassard, S.~Breidbart, and S.~Wiesner.
\newblock Quantum cryptography, or unforgeable subway tokens.
\newblock In {\em Advances in Cryptology CRYPTO '82}, pages 267--275, 1982.

\bibitem[BBCS92a]{BBCS92}
C.~H. Bennett, G.~Brassard, C.~Cr{\'e}peau, and H.~Skubiszewska.
\newblock Practical quantum oblivious transfer.
\newblock In {\em Advances in Cryptology --- CRYPTO '91}, volume 576 of {\em
  Lecture Notes in Computer Science}, pages 351--366. Springer, 1992.

\bibitem[BBCS92b]{crepeau:practicalOT}
C.~H. Bennett, G.~Brassard, C.~Cr\'epeau, and M.-H. Skubiszewska.
\newblock Practical quantum oblivious transfer.
\newblock In {\em CRYPTO '91: Proceedings of the 11th Annual International
  Cryptology Conference on Advances in Cryptology}, pages 351--366.
  Springer-Verlag, 1992.

\bibitem[BBF{\etalchar{+}}]{anne:anon}
G.~Brassard, A.~Broadbent, J.~Fitzsimons, S.~Gambs, and A.~Tapp.
\newblock Anonymous quantum communication.
\newblock arXiv:0706.2356.

\bibitem[BBL{\etalchar{+}}06]{falk:nonlocal}
G.~Brassard, H.~Buhrman, N.~Linden, A.~Methot, A.~Tapp, and F.~Unger.
\newblock A limit on nonlocality in any world in which communication complexity
  is not trivial.
\newblock {\em Physical Review Letters}, 96:250401, 2006.

\bibitem[BBM75]{bm:uncertainty}
I.~Bialynicki-Birula and J.~Mycielski.
\newblock Uncertainty relations for information entropy in wave mechanics.
\newblock {\em Communications in Mathematical Physics}, 44:129--132, 1975.

\bibitem[BBRV02]{boykin:mub}
S.~Bandyopadhyay, P.O. Boykin, V.P. Roychowdhury, and F.~Vatan.
\newblock A new proof for the existence of mutually unbiased bases.
\newblock {\em Algorithmica}, 34(4):512--528, 2002.

\bibitem[BC90a]{brassard:bcAndCoin}
G.~Brassard and C.~Cr\'epeau.
\newblock Quantum bit commitment and coin tossing protocols.
\newblock In {\em Advances in Cryptology -- Proceedings of Crypto '90}, pages
  49--61, 1990.

\bibitem[BC90b]{braunstein:inequ}
S.L. Braunstein and C.M. Caves.
\newblock Wringing out better {B}ell inequalities.
\newblock {\em Annals of Physics}, 202:22--56, 1990.

\bibitem[BCJL93]{brassard:bc}
G.~Brassard, C.~Cr\'epeau, R.~Jozsa, and D.~Langlois.
\newblock A quantum bit commitment protocol provavly unbreakable by both
  parties.
\newblock In {\em Proceedings of 34th IEEE FOCS}, pages 362--371, 1993.

\bibitem[BCMS97]{brassard:brief}
G.~Brassard, C.~Cr\'epeau, D.~Mayers, and L.~Salvail.
\newblock A brief review on the impossibility of quantum bit commitment.
\newblock quant-ph/9712023, 1997.

\bibitem[BCU{\etalchar{+}}06]{wehner05b}
H.~Buhrman, M.~Christandl, F.~Unger, S.~Wehner, and A.~Winter.
\newblock Implications of superstrong nonlocality for cryptography.
\newblock {\em Proceedings of the Royal Society A}, 462(2071):1919--1932, 2006.

\bibitem[Bel65]{bell:epr}
J.~S. Bell.
\newblock On the {E}instein-{P}odolsky-{R}osen paradox.
\newblock {\em Physics}, 1:195--200, 1965.

\bibitem[Ben92]{b92}
C.~Bennett.
\newblock Quantum cryptography using any two nonorthogonal states.
\newblock {\em Physical Review Letters}, 68:3121--3124, 1992.

\bibitem[BFL91]{babai:ipNexp}
L.~Babai, L.~Fortnow, and C.~Lund.
\newblock Non-deterministic exponential time has two-prover interactive
  protocols.
\newblock {\em Computational Complexity}, 1(1):3--40, 1991.

\bibitem[BH05]{bergou:filtering2}
J.~Bergou and M.~Hillery.
\newblock Quantum-state filtering applied to the discrimination of boolean
  functions.
\newblock {\em Physical Review A}, 72:012302, 2005.

\bibitem[Bha97]{bathia:ma}
R.~Bhatia.
\newblock {\em Matrix Analysis}.
\newblock Springer, 1997.

\bibitem[BHH03]{bergou:filtering0}
J.~Bergou, U.~Herzog, and M.~Hillery.
\newblock Quantum state filtering and discrimination between sets of boolean
  functions.
\newblock {\em Physical Review Letters}, 90:257901, 2003.

\bibitem[BHH04]{bergou:survey}
J.~Bergou, U.~Herzog, and M.~Hillery.
\newblock Discrimination of quantum states.
\newblock In M.~Paris and J.~Rehacek, editors, {\em Quantum State Estimation},
  volume~3, pages 417--465. Springer, Berlin, 2004.

\bibitem[BHH05]{bergou:filtering}
J.~Bergou, U.~Herzog, and M.~Hillery.
\newblock Optimal unambiguous filtering of a quantum state: An instance in
  mixed state discrimination.
\newblock {\em Physical Review A}, 71:042314, 2005.

\bibitem[BK02]{barnumknill}
H.~Barnum and E.~Knill.
\newblock Reversing quantum dynamics with near-optimal quantum and classical
  fidelity.
\newblock {\em Journal of Mathematical Physics}, 43:2097, 2002.

\bibitem[Bla79]{blakley:secret}
G.~R. Blakley.
\newblock Safeguarding cryptography keys.
\newblock In {\em Proceedings of the National Computer Conference 48}, pages
  313--317, 1979.

\bibitem[Blu83]{blum:coin}
M.~Blum.
\newblock Coin flipping by telephone a protocol for solving impossible
  problems.
\newblock {\em SIGACT News}, 15(1):23--27, 1983.

\bibitem[BM05]{massar:tsirel}
H.~Buhrman and S.~Massar.
\newblock Causality and {C}irel'son bounds.
\newblock {\em Physical Review A}, 72:052103, 2005.

\bibitem[BOGKW88]{benor:ip}
M.~Ben-Or, S.~Goldwasser, J.~Kilian, and A.~Wigderson.
\newblock Multi prover interactive proofs: How to remove intractability.
\newblock In {\em Proceedings of 20th ACM STOC}, pages 113--131, 1988.

\bibitem[BOHL{\etalchar{+}}05]{benor:qkd}
M.~Ben-Or, M.~Horodecki, D.W. Leung, D.~Mayers, and J.~Oppenheim.
\newblock The universal composable security of quantum key distribution.
\newblock In {\em Proceedings of the 2nd Theory of Cryptography Conference},
  volume 3378 of {\em Lecture Notes in Computer Science}, pages 386--406.
  Springer, 2005.

\bibitem[BOM04]{benor:compose}
M.~Ben-Or and D.~Mayers.
\newblock General security definition and composability for quantum and
  classical protocols.
\newblock quant-ph/0409062, 2004.

\bibitem[Boy02]{boykin:thesis}
P.~Boykin.
\newblock {\em Information Security and Quantum Mechanics: Security of Quantum
  Protocols}.
\newblock PhD thesis, University of California, Los Angeles, 2002.

\bibitem[BP03]{qseals1}
H.~Bechmann-Pasquinucci.
\newblock Quantum seals.
\newblock {\em International Journal of Quantum Information}, 1(2):217--224,
  2003.

\bibitem[BPDM05]{qseals:nogo}
H.~Bechmann-Pasquinucci, G.M. D'Ariano, and C.~Macchiavello.
\newblock Impossibility of perfect quantum sealing of classical information.
\newblock {\em International Journal of Quantum Information}, 3:435--440, 2005.

\bibitem[BR02]{robinson:cstar}
O.~Bratteli and D.~Robinson.
\newblock {\em Operator Algebras and Quantum Statistical Mechanics I}.
\newblock Springer, 2002.

\bibitem[BR03]{boykin:pqc}
P.~O. Boykin and V.~Roychowdhury.
\newblock Optimal encryption of quantum bits.
\newblock {\em Physical Review A}, page 042317, 2003.
\newblock quant-ph/0003059.

\bibitem[Bra05]{brassard:history}
G.~Brassard.
\newblock Brief history of quantum cryptography: A personal perspective.
\newblock In {\em Proceedings of IEEE Informtion Theory Workshop on Theory and
  Practise in Information Theoretic Security}, pages 19--23, 2005.

\bibitem[BS05]{bouda:anon}
J.~Bouda and J.~Sprojcar.
\newblock Anonymous transmission of quantum information.
\newblock quant-ph/0512122, 2005.

\bibitem[BV04]{boyd:book}
S.~Boyd and L.~Vandenberghe.
\newblock {\em Convex Optimization}.
\newblock Cambridge University Press, 2004.

\bibitem[BW07]{wehner06c}
M.~Ballester and S.~Wehner.
\newblock Entropic uncertainty relations and locking: tight bounds for mutually
  unbiased bases.
\newblock {\em Physical Review A}, 75:022319, 2007.

\bibitem[Cac97]{cachin:diss}
C.~Cachin.
\newblock {\em Entropy Measures and Unconditional Security in Cryptography}.
\newblock PhD thesis, {ETH} Zurich, Switzerland, 1997.

\bibitem[CBH03]{qcVsbcnogo}
R.~Clifton, J.~Bub, and H.~Halvorson.
\newblock Characterizing quantum theory in terms of information-theoretic
  constraints.
\newblock {\em Foundations of Physics}, 33:1561--1591, 2003.

\bibitem[CCL90]{cai:ip}
J.~Cai, A.~Condon, and R.~Lipton.
\newblock On bounded round multi-prover interactive proof systems.
\newblock In {\em Proceedings of the Fifth Annual Conference on Structure in
  Complexity Theory}, pages 45--54, 1990.

\bibitem[CCM98]{cachin:boundedOT}
C.~Cachin, C.~Cr{\'e}peau, and J.~Marcil.
\newblock Oblivious transfer with a memory-bounded receiver.
\newblock In {\em Proceedings of 39th IEEE FOCS}, pages 493--502, 1998.

\bibitem[CG96]{canetti:coerce}
R.~Canetti and R.~Gennaro.
\newblock Incoercible multiparty computation (extended abstract).
\newblock In {\em Proceedings of 37th IEEE FOCS}, pages 504--513, 1996.

\bibitem[CGL99]{gottesmann:secret}
R.~Cleve, D.~Gottesman, and H-K. Lo.
\newblock How to share a quantum secret.
\newblock {\em Physical Review Letters}, 83(3):648--651, 1999.

\bibitem[CGL{\etalchar{+}}02]{gisin:nonmaxEnt}
D.~Collins, N.~Gisin, N.~Linden, S.~Massar, and S.~Popescu.
\newblock Bell inequalities for arbitrarily high dimensional systems.
\newblock {\em Physical Review Letters}, 88:040404, 2002.

\bibitem[CGS02]{smith:qsmp}
C.~Cr{\'e}peau, D.~Gottesman, and A.~Smith.
\newblock Secure multiparty quantum computation.
\newblock In {\em Proceedings of 34th ACM STOC}, 2002.

\bibitem[CGS05]{smith:qsmp2}
C.~Cr{\'e}peau, D.~Gottesman, and A.~Smith.
\newblock Approximate quantum error-correcting codes and secret sharing
  schemes.
\newblock In {\em Proceedings of Advances in Cryptology - EUROCRYPT '05},
  volume 3494 of {\em Lecture Notes in Computer Science}, pages 285--301.
  Springer, 2005.

\bibitem[Cha81]{chaum:mixnet}
D.~Chaum.
\newblock Untraceable electronic mail, return addresses, and digital
  pseudonyms.
\newblock {\em Communications of the ACM}, 24(2):84--88, 1981.

\bibitem[Cha88]{chaum:dc}
D.~Chaum.
\newblock The dining cryptographers problem: Unconditional sender and recipient
  untraceability.
\newblock {\em Journal of Cryptology}, 1:65--75, 1988.

\bibitem[Cha03]{qseals2}
H.F. Chau.
\newblock Sealing quantum meassage by quantum code.
\newblock quant-ph/0308146, 2003.

\bibitem[Chi05]{childs:pce}
A.~Childs.
\newblock Secure assisted quantum computation.
\newblock {\em Quantum Information and Computation}, 5:456, 2005.

\bibitem[Chr05]{matthias:thesis}
M.~Christandl.
\newblock {\em The structure of bipartite quantum states - Insights from group
  theory and cryptography}.
\newblock PhD thesis, University of Cambridge, 2005.
\newblock quant-ph/0604183.

\bibitem[CHSH69]{chsh:nonlocal}
J.~Clauser, M.~Horne, A.~Shimony, and R.~Holt.
\newblock Proposed experiment to test local hidden-variable theories.
\newblock {\em Physical Review Letters}, 23:880--884, 1969.

\bibitem[CHTW04a]{cleve:nonlocal}
R.~Cleve, P.~H{\o}yer, B.~Toner, and J.~Watrous.
\newblock Consequences and limits of nonlocal strategies.
\newblock In {\em Proceedings of 19th IEEE Conference on Computational
  Complexity}, pages 236--249, 2004.

\bibitem[CHTW04b]{cleve:nonlocalTalk}
R.~Cleve, P.~H{\o}yer, B.~Toner, and J.~Watrous.
\newblock Consequences and limits of nonlocal strategies.
\newblock Presentation at 19th IEEE Conference on Computational Complexity,
  2004.

\bibitem[CK88]{crepeau:weakened}
C.~Cr\'epeau and J.~Kilian.
\newblock Achieving oblivious transfer using weakened security assumptions.
\newblock In {\em Proceedings of 29th IEEE FOCS}, pages 42--52, 1988.

\bibitem[CK06]{choi:dfs}
M.~Choi and D.~Kribs.
\newblock A method to find quantum noiseless subsystems.
\newblock {\em Physical Review Letters}, 96:050501, 2006.

\bibitem[CL98]{lo:promise}
H.F. Chau and H-K. Lo.
\newblock Making an empty promise with a quantum computer.
\newblock {\em Fortsch. Phys.}, 46:507--520, 1998.
\newblock Republished in 'Quantum Computing, where do we want to go tomorrow?'
  edited by S. Braunstein,Wiley-VCH, Berlin, 1999.

\bibitem[Cla76]{ch:experiment}
J.F. Clauser.
\newblock Experimental investigation of a polarization correlation anomaly.
\newblock {\em Physical Review Letters}, 36(21):1223--1226, 1976.

\bibitem[CMW04]{CMW04}
Claude Cr{\'e}peau, Kirill Morozov, and Stefan Wolf.
\newblock Efficient unconditional oblivious transfer from almost any noisy
  channel.
\newblock In {\em International Conference on Security in Communication
  Networks (SCN)}, volume~4 of {\em Lecture Notes in Computer Science}, 2004.

\bibitem[Col07]{colbeck:coin}
R.~Colbeck.
\newblock An entanglement-based protocol for strong coin tossing with bias 1/4.
\newblock {\em Physics Letters A}, 362(5):309--392, 2007.

\bibitem[Con90]{conway:fa}
J.~B. Conway.
\newblock {\em A course in functional anlysis}.
\newblock Springer, 1990.

\bibitem[Cra99]{cramer:smp}
R.~Cramer.
\newblock Introduction to secure computation.
\newblock In {\em Lectures on Data Security - Modern Cryptography in Theory and
  Practise}, volume 1561 of {\em Lecture Notes in Computer Science}, pages
  16--62, 1999.

\bibitem[Cr{\'e}94]{crepeau:qot}
C.~Cr{\'e}peau.
\newblock Quantum oblivious transfer.
\newblock {\em Journal of Modern Optics}, 41(12):2455--2466, 1994.

\bibitem[Cr{\'e}97]{crepeau:efficientOT}
C.~Cr{\'e}peau.
\newblock Efficient cryptographic protocols based on noisy channels.
\newblock In {\em Advances in Cryptology -- Proceedings of EUROCRYPT '97},
  1997.

\bibitem[CSUU07]{falk:paralell}
R.~Cleve, W.~Slofstra, F.~Unger, and S.~Upadhyay.
\newblock Strong paralell repetition theorem for quantum xor proof systems.
\newblock In {\em Proceedings of the 22nd IEEE Conference on Computational
  Complexity}, pages 109--114, 2007.

\bibitem[CvdGT95]{crepeau:committedOT}
C.~Cr\'epeau, J.~van~de Graaf, and A.~Tapp.
\newblock Committed oblivious transfer and private multi-party computation.
\newblock In {\em CRYPTO '95: Proceedings of the 15th Annual International
  Cryptology Conference on Advances in Cryptology}, pages 110--123.
  Springer-Verlag, 1995.

\bibitem[CW79]{CarWeg79}
J.~L. Carter and M.~N. Wegman.
\newblock {Universal classes of hash functions}.
\newblock {\em Journal of Computer and System Sciences}, 18:143--154, 1979.

\bibitem[CW05a]{wehner04b}
M.~Christandl and S.~Wehner.
\newblock Quantum anonymous transmissions.
\newblock In {\em Proceedings of 11th ASIACRYPT}, volume 3788 of {\em LNCS},
  pages 217--235, 2005.

\bibitem[CW05b]{m:locking}
M.~Christandl and A.~Winter.
\newblock Uncertainy, monogamy and locking of quantum correlations.
\newblock {\em {IEEE} Transactions on Information Theory}, 51(9):3159--3165,
  2005.

\bibitem[Dav78]{davies:access}
E.~Davies.
\newblock Information and quantum measurement.
\newblock {\em {IEEE} Transactions on Information Theory}, 24(5):596--599,
  1978.

\bibitem[Deu83]{deutsch:uncertainty}
D.~Deutsch.
\newblock Uncertainty in quantum measurements.
\newblock {\em Phys. Rev. Lett.}, 50:631--633, 1983.

\bibitem[DFMS04]{serge:unfairNoisy}
I.~Damg{\aa}rd, S.~Fehr, K.~Morozov, and L.~Salvail.
\newblock Unfair noisy channels and oblivious transfer.
\newblock In {\em Proceedings of TCC 2004}, volume 2951 of {\em Lecture Notes
  in Computer Science}, pages 355--373. Springer, 2004.

\bibitem[DFR{\etalchar{+}}07]{serge:new}
I.~Damg{\aa}rd, S.~Fehr, R.~Renner, L.~Salvail, and C.~Schaffner.
\newblock A tight high-order entropic uncertainty relation with applications in
  the bounded quantum-storage model.
\newblock Proceedings of CRYPTO 2007, 2007.

\bibitem[DFSS05]{serge:bounded}
I.~Damg{\aa}rd, S.~Fehr, L.~Salvail, and C.~Schaffner.
\newblock Cryptography in the {B}ounded {Q}uantum-{S}torage {M}odel.
\newblock In {\em Proceedings of 46th IEEE FOCS}, pages 449--458, 2005.

\bibitem[DFSS07]{DFSS:secureid}
I.~Damg{\aa}rd, S.~Fehr, L.~Salvail, and C.~Schaffner.
\newblock Secure identification and {QKD} in the bounded-quantum-storage model.
\newblock In {\em Proceedings of CRYPTO 2007}, pages 342--359, 2007.

\bibitem[DFSS08]{DFSS08journal}
I.~B. Damg{\aa}rd, S.~Fehr, L.~Salvail, and C.~Schaffner.
\newblock Cryptography in the bounded-quantum-storage model.
\newblock {\em special issue of SIAM Journal of Computing}, 2008.
\newblock To appear.

\bibitem[DHL{\etalchar{+}}04]{terhal:locking}
D.~DiVincenzo, M.~Horodecki, D.~Leung, J.~Smolin, and B.~Terhal.
\newblock Locking classical correlation in quantum states.
\newblock {\em Physical Review Letters}, 92(067902), 2004.

\bibitem[DHT03]{barbara:hiding2}
D.~P. DiVincenzo, P.~Hayden, and B.~Terhal.
\newblock Hiding quantum data.
\newblock {\em Foundations of Physics}, 33(11):1629--1647, 2003.

\bibitem[Die06]{dietz:bloch}
K.~Dietz.
\newblock Generalized bloch spheres for m-qubit states.
\newblock {\em Journal of Physics A: Math. Gen.}, 36(6):1433--1447, 2006.

\bibitem[DKS99]{damgard:weakened}
I.~Damg{\aa}rd, J.~Kilian, and L.~Salvail.
\newblock On the (im)possibility of basing oblivious transfer and bit
  committment on weakened security assumptions.
\newblock In {\em Advances in Cryptology - EUROCRYPT '99}, volume 1592 of {\em
  Lecture Notes in Computer Science}, pages 56--73. Springer, 1999.

\bibitem[DKSW06]{kretch:bc}
G.~D'Ariano, D.~Kretschmann, D.~Schlingemann, and R.F. Werner.
\newblock Quantum bit commitment revisited: the possible and the impossible.
\newblock quant-ph/0605224, 2006.

\bibitem[DL03]{doran:book}
C.~Doran and A.~Lasenby.
\newblock {\em Geometric algebra for physicists}.
\newblock Cambridge University Press, 2003.

\bibitem[DLT02]{barbara:hiding1}
D.~P. DiVincenzo, D.~W. Leung, and B.~M. Terhal.
\newblock Quantum data hiding.
\newblock {\em IEEE Trans. Inf Theory}, 48(3):580--599, 2002.
\newblock arXiv e-print quant-ph/0103098.

\bibitem[DLTW08]{andrew:qmp}
A.~C. Doherty, Y-C. Liang, B.~Toner, and S.~Wehner.
\newblock The quantum moment problem.
\newblock Submitted., 2008.

\bibitem[DN06]{nayak:smallKey}
P.~A. Dickinson and A.~Nayak.
\newblock Approximate randomization of quantum states with fewer bits of key.
\newblock In {\em Quantum Computing Back Action, IIT Kanpur}, volume 864 of
  {\em AIP Conference Proceedings}, pages 18--36. Springer, 2006.

\bibitem[DPS02]{andrew:sep1}
A.~C. Doherty, P.~A. Parrilo, and F.~M. Spedalieri.
\newblock Distinguishing separable and entangled states.
\newblock {\em Physical Review Letters}, 88(18):187904, 2002.

\bibitem[DPS04]{andrew:sep2}
A.~C. Doherty, P.~A. Parrilo, and F.~M. Spedalieri.
\newblock A complete family of separability criteria.
\newblock {\em Physical Review A}, 69:022308, 2004.

\bibitem[DPS05]{andrew:sep3}
A.~C. Doherty, P.~A. Parrilo, and F.~M. Spedalieri.
\newblock Detecting multipartite entanglement.
\newblock {\em Physical Review A}, 71:032333, 2005.

\bibitem[EF01]{eldar:pgm}
Y.~Eldar and G.~Forney.
\newblock On quantum detection and the square-root measurement.
\newblock {\em {IEEE} Transactions on Information Theory}, 47:858--872, 2001.

\bibitem[EGL85]{even:firstOT}
S.~Even, O.~Goldreich, and A.~Lempel.
\newblock A randomized protocol for signing contracts.
\newblock {\em Communications of the ACM}, 28(6):637--647, 1985.

\bibitem[Eis01]{eisert:thesis}
J.~Eisert.
\newblock {\em Entanglement in quantum information theory}.
\newblock PhD thesis, University of Potsdam, 2001.
\newblock quant-ph/0610253.

\bibitem[Eke91]{ekert:91}
A.~Ekert.
\newblock Quantum cryptography based on {B}ell's theorem.
\newblock {\em Physical Review Letters}, 67:661--663, 1991.

\bibitem[Eld03]{eldar:sdp}
Y.~Eldar.
\newblock A semidefinite programming approach to optimal unambiguous
  discrimination of quantum states.
\newblock {\em {IEEE} Transactions on Information Theory}, 49:446--456, 2003.

\bibitem[EMV03]{eldar:sdpDetector}
Y.~Eldar, A.~Megretski, and G.~Verghese.
\newblock Designing optimal quantum detectors via semidefinite programming.
\newblock {\em {IEEE} Transactions on Information Theory}, 49:1017--1012, 2003.

\bibitem[EMV04]{eldar:symmetric}
Y.~Eldar, A.~Megretski, and G.~Verghese.
\newblock Optimal detection of symmetric mixed quantum states.
\newblock {\em {IEEE} Transactions on Information Theory}, 50:1198--1207, 2004.

\bibitem[EPR35]{epr:original}
A.~Einstein, B.~Podolsky, and N.~Rosen.
\newblock Can quantum-mechanical description of physical reality be considered
  complete?
\newblock {\em Physical Review}, 47:777--780, 1935.

\bibitem[EW02]{werner:hiding}
T.~Eggeling and R.F. Werner.
\newblock Hiding classical data in multi-partite quantum states.
\newblock {\em Physical Review Letters}, 89(9):097905, 2002.

\bibitem[Feh07]{serge:personal}
S.~Fehr.
\newblock Personal communication, 2007.

\bibitem[Fei91]{feige:twoProverOneRound}
U.~Feige.
\newblock On the success probability of two provers in one-round proof systems.
\newblock In {\em Proceedings of the Sixth Annual Conference on Structure in
  Complexity Theory}, pages 116--123, 1991.

\bibitem[Fei95]{feige:stateOfArt}
U.~Feige.
\newblock Error reduction by parallel repetition - the state of the art.
\newblock Technical Report CS95-32, Weizmann Institute, 1, 1995.

\bibitem[FL92]{feige:mip}
U.~Feige and L.~Lov\'asz.
\newblock Two-prover one-round proof systems: their power and their problems.
\newblock In {\em Proceedings of 24th ACM STOC}, pages 733--744, 1992.

\bibitem[FS04]{filipp:bell}
S.~Filipp and K.~Svozil.
\newblock Tracing the bounds on {B}ell-type inequalities.
\newblock In {\em Proceedings of Foundations of Probability and Physics-3},
  pages 87--94, 2004.

\bibitem[Fuc95]{fuchs:diss}
C.~A. Fuchs.
\newblock {\em Distinguishability and Accessible Information in Quantum
  Theory}.
\newblock PhD thesis, University of New Mexico, Albuquerque, 1995.
\newblock quant-ph/9601020.

\bibitem[GC01]{gottesman:signature}
D.~Gottesman and I.~Chuang.
\newblock Quantum signatures.
\newblock quant-ph/0105032, 2001.

\bibitem[Gis91]{gisin:pureStateViolates}
N.~Gisin.
\newblock Bell's inequality holds for all non-product states.
\newblock {\em Physics Letters A}, 154:201--202, 1991.

\bibitem[Gis99]{gisin:chsh}
N.~Gisin.
\newblock {B}ell inequality for arbitrary many settings of the analyzers.
\newblock {\em Physics Letters A}, 260:1--3, 1999.

\bibitem[GKK{\etalchar{+}}06]{ronald:qmem}
D.~Gavinsky, J.~Kempe, I.~Kerenidis, R.~Raz, and R.~de~Wolf.
\newblock Exponential separations for one-way quantum communication complexity.
\newblock In {\em Proceedings of 39th ACM STOC}, pages 516--525, 2006.

\bibitem[GMR89]{goldwasser:ip}
S.~Goldwasser, S.~Micali, and C.~Rackoff.
\newblock The knowledge complexity of interactive proof systems.
\newblock {\em SIAM Journal on Computing}, 1(18):186--208, 1989.

\bibitem[Gol01]{goldreich:book1}
O.~Goldreich.
\newblock {\em Foundations of Cryptography}, volume Basic Tools.
\newblock Cambridge University Press, 2001.

\bibitem[Got00]{gottesmann:secret2}
D.~Gottesman.
\newblock On the theory of quantum secret sharing.
\newblock {\em Physical Review A}, 61:042311, 2000.

\bibitem[Gra71]{circulant}
R.~M. Gray.
\newblock {\em Toeplitz and Circulant Matrices: A review}.
\newblock 1971.

\bibitem[Gra04]{grassl:mub}
M.~Grassl.
\newblock On {SIC-POVM}s and {MUB}s in dimension 6.
\newblock In {\em Proceedings ERATO Conference on Quantum Information Science},
  pages 60--61, 2004.

\bibitem[GRTZ02]{GRTZ:qkd_review}
N.~Gisin, G.~Ribordy, W.~Tittel, and H.~Zbinden.
\newblock Quantum cryptography.
\newblock {\em Reviews of Modern Physics}, 74:pp. 145--195, 2002.

\bibitem[Gur03]{gurvits:nphard}
L.~Gurvits.
\newblock Classical deterministic complexity of {E}dmund's problem and quantum
  entanglement.
\newblock In {\em Proceedings of 35th ACM STOC}, pages 10--19, 2003.

\bibitem[GW95]{goemans:maxcut}
M.X. Goemans and D.P. Williamson.
\newblock Improved approximation algorithms for maximum cut and satisfiability
  problems using semidefinite programming.
\newblock {\em J. Assoc. Comput. Mach.}, 42:1115--1145, 1995.

\bibitem[Hay06]{hayashi:book}
M.~Hayashi.
\newblock {\em Quantum Information - An introduction}.
\newblock Springer, 2006.

\bibitem[HBB99]{hillary:secret}
M.~Hillery, V.~Buzek, and A.~Bethiaume.
\newblock Quantum secret sharing.
\newblock {\em Physical Review A}, 3:1829--1834, 1999.

\bibitem[Hei27]{heisenberg:uncertainty}
W.~Heisenberg.
\newblock {\"U}ber den anschaulichen inhalt der quantentheoretischen kinematik
  und mechanik.
\newblock {\em Zeitschrift f{\"u}r Physik}, 43:172--198, 1927.

\bibitem[Hel67]{helstrom:detection}
C.~W. Helstrom.
\newblock Detection theory and quantum mechanics.
\newblock {\em Information and Control}, 10(1):254--291, 1967.

\bibitem[HHHO05]{karol:locking}
K.~Horodecki, M.~Horodecki, P.~Horodecki, and J.~Oppenheim.
\newblock Locking entanglement measures with a single qubit.
\newblock {\em Physical Review Letters}, 94:200501, 2005.

\bibitem[HJ85]{horn&johnson:ma}
R.~A. Horn and C.~R. Johnson.
\newblock {\em Matrix Analysis}.
\newblock Cambridge University Press, 1985.

\bibitem[HK04]{kent:cheatSensitive}
L.~Hardy and A.~Kent.
\newblock Cheat sensitive quantum bit commitment.
\newblock {\em Physical Review Letters}, 92(157901), 2004.

\bibitem[HKL03]{hkl:noiseCommutant}
J.A. Holbrook, D.~W. Kribs, and R.~Laflamme.
\newblock Noiseless subsystems and the structure of the commutant in quantum
  error correction.
\newblock {\em Quantum Information and Computation}, 2(5):381--419, 2003.

\bibitem[HLS05]{graeme:hiding}
P.~Hayden, D.~Leung, and G.~Smith.
\newblock Multiparty data hiding of quantum information.
\newblock {\em Physical Review A}, 71:062339, 2005.

\bibitem[HLSW04]{winter:randomizing}
P.~Hayden, D.~Leung, P.~Shor, and A.~Winter.
\newblock Randomizing quantum states: Constructions and applications.
\newblock {\em Communications in Mathematical Physics}, 250(2):371--391, 2004.

\bibitem[Hol73]{H73}
A.~S. Holevo.
\newblock Information theoretical aspects of quantum measurements.
\newblock {\em Probl. Inf. Transm.}, 9:110--118, 1973.

\bibitem[HR07]{HaiRei07}
I.~Haitner and O.~Reingold.
\newblock Statistically-hiding commitment from any one-way function.
\newblock In {\em Proceedings of 39th ACM STOC}, pages 1--10, 2007.

\bibitem[HRP{\etalchar{+}}06]{qkd:optical}
P.A. Hiskett, D.~Rosenberg, C.G. Peterson, R.J. Hughes, S.~Nam, A.E. Lita, A.J.
  Miller, and J.E. Nordholt.
\newblock Long-distance quantum key distribution in optical fibre.
\newblock {\em New Journal of Physics}, 8:193, 2006.

\bibitem[Hun03]{hunter:nouse}
K.~Hunter.
\newblock Measurement does not always aid state discrimination.
\newblock {\em Physical Review A}, 6:012306, 2003.

\bibitem[HW94]{hausladen:pgm}
P.~Hausladen and W.~Wootters.
\newblock A pretty good measurement for distinguishing quantum states.
\newblock {\em Journal of Modern Optics}, 41:2385--2390, 1994.

\bibitem[Ioa07]{lawrence:thesis}
L.~Ioannou.
\newblock {\em Computing finite-dimensional bipartite quantum separability}.
\newblock PhD thesis, University of Cambridge, 2007.
\newblock cs/0504110.

\bibitem[IT06]{lawrence:sep}
L.~M. Ioannou and B.~C. Travaglione.
\newblock Quantum separability and entanglement detection via
  entanglement-witness search and global optimization.
\newblock {\em Physical Review A}, 73:052314, 2006.

\bibitem[ITCE04]{lawrence:sep2}
L.~M. Ioannou, B.~C. Travaglione, D.~Cheung, and A.~Ekert.
\newblock Improved algorithm for quantum separability and entanglement
  detection.
\newblock {\em Physical Review A}, 70:060303, 2004.

\bibitem[Jai05]{jain:qbsc}
R.~Jain.
\newblock Stronger impossibility results for quantum string commitment.
\newblock quant-ph/0506001, 2005.

\bibitem[JW28]{JordanWigner}
P.~Jordan and E.~Wigner.
\newblock {\"U}ber das paulische {\"a}quivalenzverbot.
\newblock {\em Zeitschrift f{\"u}r Physik}, 47:631, 1928.

\bibitem[Kah96]{cryptoHistory}
D.~Kahn.
\newblock {\em The Codebreakers: The comprehensive history of secret
  communication from ancient times to the internet}.
\newblock Simon and Schuster, 1996.

\bibitem[Ken99]{kent:bc}
A.~Kent.
\newblock Secure classical bit commitment using fixed capacity communication
  channels.
\newblock {\em Journal of Cryptology}, 18(4):313--335, 1999.

\bibitem[Ken03]{kent:sc}
A.~Kent.
\newblock Quantum bit string commitment.
\newblock {\em Physical Review Letters}, 90(237901), 2003.

\bibitem[KI02]{koashi&imoto:operations}
M.~Koashi and N.~Imoto.
\newblock Operations that do not disturb partially known quantum states.
\newblock {\em Physical Review A}, 66:022318, 2002.

\bibitem[Kil88]{kilian:foundingOnOT}
J.~Kilian.
\newblock Founding cryptography on oblivious transfer.
\newblock In {\em Proceedings of 20th ACM STOC}, pages 20--31, 1988.

\bibitem[Kit02]{kitaev:coin}
A.~Kitaev.
\newblock Quantum coin flipping.
\newblock Talk at QIP 2002, 2002.

\bibitem[KKM{\etalchar{+}}07]{julia:mip}
J.~Kempe, H.~Kobayashi, K.~Matsumoto, B.~Toner, and T.~Vidick.
\newblock On the power of entangled provers: immunizing games against
  entanglement.
\newblock arXiv:0704.2903, 2007.

\bibitem[KLM01]{KLM:lo}
E.~Knill, R.~Laflamme, and G.~Milburn.
\newblock A scheme for efficient quantum computation with linear optics.
\newblock {\em Nature}, 409:46--52, 2001.

\bibitem[KM03]{kobayashi:mip}
H.~Kobayashi and K.~Matsumoto.
\newblock Quantum multi-prover interactive proof systems with limited prior
  entanglement.
\newblock {\em Journal of Computer and Systems Sciences}, 66(3):429--450, 2003.

\bibitem[KMP04]{kitaev:super}
A.~Kitaev, D.~Mayers, and J.~Preskill.
\newblock Superselection rules and quantum protocols.
\newblock {\em Physical Review A}, 69:052326, 2004.

\bibitem[KMR05]{KoMaRe05}
Robert Koenig, Ueli Maurer, and Renato Renner.
\newblock On the power of quantum memory.
\newblock {\em {IEEE} Transactions on Information Theory}, 51(7):2391--2401,
  2005.

\bibitem[KN04]{iordanis:coin}
I.~Kerenidis and A.~Nayak.
\newblock Weak coin flipping with small bias.
\newblock {\em Information Processing Letters}, 89(3):131--135, 2004.

\bibitem[KR03]{klappenecker:mubs}
A.~Klappenecker and M.~R{\"o}tteler.
\newblock Constructions of mutually unbiased bases.
\newblock In {\em Finite Fields and Applications: 7th international conference
  Fq7}, pages 137--144. Lecture Notes in Computer Science, 2003.

\bibitem[KR04]{galois:mub}
A.~Klappenecker and M.~R{\"o}tteler.
\newblock Constructions of mutually unbiased bases.
\newblock In {\em International Conference on Finite Fields and Applications
  (Fq7)}, volume 2948 of {\em Lecture Notes in Computer Science}, pages
  137--144. Springer, 2004.

\bibitem[KR05]{klappenecker&roetteler:meanking}
A.~Klappenecker and M.~R{\"o}tteler.
\newblock {\frakfamily New Tales of the Mean King}.
\newblock quant-ph/0502138, 2005.

\bibitem[Kra87]{kraus:entropy}
K.~Kraus.
\newblock Complementary observables and uncertainty relations.
\newblock {\em Physical Review D}, 35(10):3070--3075, 1987.

\bibitem[KRBM07]{rk:locking}
R.~Koenig, R.~Renner, A.~Bariska, and U.~Maurer.
\newblock Small accessible quantum information does not imply security.
\newblock {\em Physical Review Letters}, 98:140502, 2007.

\bibitem[KT87]{tsirel:obscure}
L.A. Khalfin and B.S. Tsirelson.
\newblock A quantitative criterion of the applicability of the classical
  description within the quantum theory.
\newblock In {\em Symposium on the Foundations of Modern Physics}, pages
  369--401, 1987.

\bibitem[KW00]{kitaev&watrous:qip}
A.~Kitaev and J.~Watrous.
\newblock Parallelization, amplification, and exponential time simulation of
  quantum interactive proof systems.
\newblock In {\em Proceedings of 32nd ACM STOC}, pages 608--617, 2000.

\bibitem[KW03]{kerenidis&wolf:qldc}
I.~Kerenidis and R.~{de} Wolf.
\newblock Exponential lower bound for 2-query locally decodable codes via a
  quantum argument.
\newblock In {\em Proceedings of 35th ACM STOC}, pages 106--115, 2003.

\bibitem[Lan87]{landau:bell}
L.J. Landau.
\newblock On the violation of bell's inequality in quantum theory.
\newblock {\em Physics Letters A}, 123(3):115--118, 1987.

\bibitem[Lan88]{landau:compat}
L.J. Landau.
\newblock Empirical two-point correlation functions.
\newblock {\em Foundations of Physics}, 18(4):449--460, 1988.

\bibitem[Lar90]{larsen:entropy}
U.~Larsen.
\newblock Superspace geometry: the exact uncertainty relationship between
  complementary aspects.
\newblock {\em J. Phys. A: Math. Gen.}, 23:1041--1061, 1990.

\bibitem[LBZ02]{lawrence:mub}
J.~Lawrence, C.~Brukner, and A.~Zeilinger.
\newblock Mutually unbiased binary observable sets on n qubits.
\newblock {\em Physical Review A}, 65:032320, 2002.

\bibitem[LC96]{lo&chau:bitcom2}
H-K. Lo and H.F. Chau.
\newblock Why quantum bit commitment and ideal quantum coin tossing are
  impossible.
\newblock In {\em Proceedings of PhysComp96}, 1996.
\newblock quant-ph/9605026.

\bibitem[LC97]{lo&chau:bitcom}
H-K. Lo and H.~F. Chau.
\newblock Is quantum bit commitment really possible?
\newblock {\em Physical Review Letters}, 78:3410, 1997.

\bibitem[LC98]{lo:coin}
H-K. Lo and H.F. Chau.
\newblock Why quantum bit commitment and ideal quantum coin tossing are
  impossible.
\newblock {\em Physica D}, 120:177--187, 1998.

\bibitem[LC99]{lochau:qkd}
H-K. Lo and H.~F. Chau.
\newblock Unconditional security of quantum key distribution over arbitrarily
  long distances.
\newblock {\em Science}, 283:2050--2056, 1999.

\bibitem[LD06]{andrew:paralell}
Y-C. Liang and A.C. Doherty.
\newblock Better bell inequality violation by collective measurements.
\newblock {\em Physical Review A}, 73:052116, 2006.

\bibitem[LD07]{andrew:states}
Y-C. Liang and A.C. Doherty.
\newblock Bounds on quantum correlations in bell inequality experiments.
\newblock {\em Physical Review A}, 75:042103, 2007.

\bibitem[Lin99]{lindblad:cloning}
G.~Lindblad.
\newblock A general no-cloning theorem.
\newblock {\em Letters in Mathematical Physics}, 47:189--196, 1999.

\bibitem[Lo97]{lo:insecurity}
H-K. Lo.
\newblock Insecurity of quantum secure computations.
\newblock {\em Physical Review A}, 56:1154, 1997.

\bibitem[Lou01]{lounesto:book}
P.~Lounesto.
\newblock {\em Clifford Algebras and Spinors}.
\newblock Cambridge University Press, 2001.

\bibitem[LS91]{lapidot:ip}
D.~Lapidot and A.~Shamir.
\newblock Fully parallelized multi prover protocols for {NEXP}-time.
\newblock In {\em Proceedings of 32nd FOCS}, pages 13--18, 1991.

\bibitem[Mas06]{masanes:blocks}
L.~Masanes.
\newblock Asymptotic violation of bell inequalities and distillability.
\newblock {\em Physical Review Letters}, 97:050503, 2006.

\bibitem[Mau92]{maurer:storage}
U.~Maurer.
\newblock Conditionally-perfect secrecy and a provably-secure randomized
  cipher.
\newblock {\em Journal of Cryptology}, 5(1):53--66, 1992.

\bibitem[May96a]{mayers:qkdproof}
D.~Mayers.
\newblock Quantum key distribution and string oblivious transfer in noisy
  channels.
\newblock In {\em Proceedings of Advances in Cryptology - CRYPTO '96}, pages
  343--357, 1996.

\bibitem[May96b]{mayers:trouble}
D.~Mayers.
\newblock The trouble with quantum bit commitment.
\newblock quant-ph/9603015, 1996.

\bibitem[May97]{mayers:bitcom}
D.~Mayers.
\newblock Unconditionally secure quantum bit commitment is impossible.
\newblock {\em Physical Review Letters}, 78:3414--3417, 1997.
\newblock quant-ph/9605044.

\bibitem[Moc04]{mochon:coin}
C.~Mochon.
\newblock Quantum weak coin-flipping with bias of 0.192.
\newblock In {\em Proceedings of 45th IEEE FOCS}, pages 2--11, 2004.

\bibitem[Moc05]{mochon:coinFamily}
C.~Mochon.
\newblock A large family of quantum weak coin-flipping protocols.
\newblock {\em Physical Review A}, 72:022341, 2005.

\bibitem[Moc07a]{mochon:pgm}
C.~Mochon.
\newblock A family of generalized `§pretty good' measurements and the
  minimal-error pure-state discrimination problems for which they are optimal.
\newblock {\em Physical Review A}, 75:042313, 2007.

\bibitem[Moc07b]{mochon:newCoin}
C.~Mochon.
\newblock Quantum weak coin flipping with arbitrarily small bias.
\newblock 2007.
\newblock arXiv:0711.4114.

\bibitem[MSC99]{mayers:coin}
D.~Mayers, L.~Salvail, and Y.~{Chiba-Kohno}.
\newblock Unconditionally secure quantum coin tossing.
\newblock quant-ph/9904078, 22 Apr 1999.

\bibitem[MU88]{maassen:entropy}
H.~Maassen and J.~Uffink.
\newblock Generalized entropic uncertainty relations.
\newblock {\em Physical Review Letters}, 60(1103), 1988.

\bibitem[MvOV97]{handbook:crypto}
A.~Menezes, P.~van Oorschot, and S.~Vanstone.
\newblock {\em Handbook of Applied Cryptography}.
\newblock CRC Press, 1997.

\bibitem[MW05]{mariott:qip}
C.~Marriott and J.~Watrous.
\newblock Quantum {A}rthur-{M}erlin games.
\newblock cs.CC/0506068, 2005.

\bibitem[Nao91]{Naor91}
M.~Naor.
\newblock Bit commitment using pseudorandomness.
\newblock {\em Journal of Cryptology}, 4(2):151--158, 1991.

\bibitem[Nay99]{nayak:rac}
A.~Nayak.
\newblock Optimal lower bounds for quantum automata and random access codes.
\newblock In {\em Proceedings of 40th IEEE FOCS}, pages 369--376, 1999.
\newblock quant-ph/9904093.

\bibitem[NC00]{nielsen&chuang:qc}
M.~A. Nielsen and I.~L. Chuang.
\newblock {\em Quantum Computation and Quantum Information}.
\newblock Cambridge University Press, 2000.

\bibitem[NPA07]{acin:bell}
M.~Navascues, S.~Pironio, and A.~Acin.
\newblock Bounding the set of quantum correlations.
\newblock {\em Physical Review Letters}, 98:010401, 2007.

\bibitem[Per93]{peres:book}
A.~Peres.
\newblock {\em Quantum Theory: Concepts and Methods}.
\newblock Kluwer Academic Publishers, 1993.

\bibitem[Per96]{peres:paralell}
A.~Peres.
\newblock Collective tests for quantum nonlocality.
\newblock {\em Physical Review A}, 54:2685, 1996.

\bibitem[PR94]{popescu:nonlocal}
S.~Popescu and D.~Rohrlich.
\newblock Quantum nonlocality as an axiom.
\newblock {\em Foundations of Physics}, 24(3):379--385, 1994.

\bibitem[PR96]{popescu:nonlocal2}
S.~Popescu and D.~Rohrlich.
\newblock Nonlocality as an axiom for quantum theory.
\newblock In {\em The dilemma of Einstein, Podolsky and Rosen, 60 years later:
  International symposium in honour of Nathan Rosen}, 1996.

\bibitem[PR97]{popescu:nonlocal3}
S.~Popescu and D.~Rohrlich.
\newblock Causality and nonlocality as axioms for quantum mechanics.
\newblock In {\em Proceedings of the Symposium of Causality and Locality in
  Modern Physics and Astronomy: Open Questions and Possible Solutions}, 1997.

\bibitem[Pre05]{preda:talk}
D.~Preda.
\newblock Non-local multi-prover interactive proofs.
\newblock CWI Seminar, 21 June, 2005.

\bibitem[Qua]{idQuantique}
ID~Quantique.
\newblock http://www.idquantique.com.

\bibitem[Rab81]{rabin:ot}
M.~Rabin.
\newblock How to exchange secrets by oblivious transfer.
\newblock Technical report, Aiken Computer Laboratory, Harvard University,
  1981.
\newblock Technical Report TR-81.

\bibitem[Raz05]{raz:quantumPCP}
R.~Raz.
\newblock Quantum information and the {PCP} theorem.
\newblock In {\em Proceedings of 46th FOCS}, pages 459--468, 2005.

\bibitem[Reg03]{oded:lattice}
O.~Regev.
\newblock New lattice-based cryptographic constructions.
\newblock {\em Journal of the ACM}, 51(6):899--942, 2003.

\bibitem[R{\'e}n60]{renyi:entropy}
A.~R{\'e}nyi.
\newblock On measures of information and entropy.
\newblock In {\em Proceedings of the 4th Berkeley Symposium on Mathematics,
  Statistics and Probability}, pages 547--561, 1960.

\bibitem[Ren05]{renato:diss}
R.~Renner.
\newblock {\em Security of Quantum Key Distribution}.
\newblock PhD thesis, ETH Zurich, 2005.
\newblock quant-ph/0512258.

\bibitem[Riv99]{trustedInit}
R.~L. Rivest.
\newblock Unconditionally secure commitment and oblivious transfer schemes
  using private channels and a trusted initializer.
\newblock http://people.csail.mit.edu/rivest/Rivest-commitment.pdf, 1999.

\bibitem[RK05]{renato:compose}
R.~Renner and R.~Koenig.
\newblock Universally composable privacy amplification against quantum
  adversaries.
\newblock In {\em Proceedings of TCC 2005}, volume 3378 of {\em Lecture Notes
  in Computer Science}, pages 407--425. Springer, 2005.

\bibitem[RKM{\etalchar{+}}01]{bellDetectors}
M.~A. Rowe, D.~Kielpinski, V.~Meyer, C.~A. Sackett, W.~M. Itano, C.~Monroe, and
  D.~J. Wineland.
\newblock Experimental violation of a bell's inequality with efficient
  detection.
\newblock {\em Nature}, 409:791--794, 2001.

\bibitem[Rob29]{robinson:uncertainty}
H.P. Robertson.
\newblock The uncertainty principle.
\newblock {\em Physical Review}, 34:163--164, 1929.

\bibitem[SA]{sedumi}
J.~Sturm and AdvOL.
\newblock {S}e{D}u{M}i.
\newblock http://sedumi.mcmaster.ca/.

\bibitem[SA99]{stajano99cocaine}
F.~Stajano and R.~J. Anderson.
\newblock The cocaine auction protocol: On the power of anonymous broadcast.
\newblock In {\em Information Hiding}, pages 434--447, 1999.

\bibitem[Sal98]{salvail:physical}
L.~Salvail.
\newblock Quantum bit commitment from a physical assumption.
\newblock In {\em Proceedings of CRYPTO'98}, volume 1462 of {\em Lecture Notes
  in Computer Science}, pages 338--353, 1998.

\bibitem[San93]{sanchez:entropy}
J.~Sanchez.
\newblock Entropic uncertainty and certainty relations for complementary
  observables.
\newblock {\em Physics Letters A}, 173:233--239, 1993.

\bibitem[Sch35a]{schroedinger:eprGerman}
E.~Schr{\"o}dinger.
\newblock Die gegenw{\"a}rtige {S}ituation der {Q}uantenmechanik.
\newblock {\em Naturwissenschaften}, 23:807,823,840, 1935.

\bibitem[Sch35b]{schroedinger:eprEnglish}
E.~Schr{\"o}dinger.
\newblock Discussion of probability relations between separated systems.
\newblock {\em Proceedings of the Cambridge Philosophical Society},
  31:555--563, 1935.

\bibitem[Sch07]{chris:diss}
C.~Schaffner.
\newblock {\em Cryptography in the Bounded-Quantum Storage Model}.
\newblock PhD thesis, University of Aarhus, 2007.

\bibitem[Sha48]{shannon:info}
C.~E. Shannon.
\newblock A mathematical theory of communication.
\newblock {\em Bell System Technical Journal}, 27:379--423, 623--656, 1948.

\bibitem[Sha79]{shamir:secret}
A.~Shamir.
\newblock How to share a secret.
\newblock {\em Communications of the ACM}, 22(2):612--613, 1979.

\bibitem[Sha92]{shamir:ipPspace}
A.~Shamir.
\newblock {IP} = {PSPACE}.
\newblock {\em Journal of the ACM}, 39(4):869--877, 1992.

\bibitem[She92]{shen:ipPspace}
A.~Shen.
\newblock {IP} = {PSPACE}: simplified proof.
\newblock {\em Journal of the ACM}, 39(4):878--880, 1992.

\bibitem[Sho97]{shor:factoring}
P.~W. Shor.
\newblock Polynomial-time algorithms for prime factorization and discrete
  logarithms on a quantum computer.
\newblock {\em SIAM Journal on Computing}, 26(5):1484--1509, 1997.
\newblock Earlier version in FOCS'94.

\bibitem[SIGA05]{gisin:cloning}
V.~Scarani, S.~Iblisdir, N.~Gisin, and A.~Ac{\'i}n.
\newblock Quantum cloning.
\newblock {\em Reviews in Modern Physics}, 77:1225, 2005.

\bibitem[SO06]{jsmo:locking}
J.~Smolin and J.~Oppenheim.
\newblock Information locking in black holes.
\newblock {\em Physical Review Letters}, 96:091302, 2006.

\bibitem[SP00]{sp:qkdproof}
P.W. Shor and J.~Preskill.
\newblock Simple proof of security of the bb84 quantum key distribution
  protocol.
\newblock {\em Physical Review Letters}, 85(2):441--444, 2000.

\bibitem[SR95]{sanchez:entropy2}
J.~Sanchez-Ruiz.
\newblock Improved bounds in the entropic uncertainty and certainty relations
  for complementary observables.
\newblock {\em Physics Letters A}, 201:125--131, 1995.

\bibitem[SR02a]{spekkens:tradeoffBc}
R.~Spekkens and T.~Rudolph.
\newblock Degrees of concealment and bindingness in quantum bit commitment
  protocols.
\newblock {\em Physical Review A}, 65(012310), 2002.

\bibitem[SR02b]{spekkens:coin}
R.~Spekkens and T.~Rudolph.
\newblock A protocol for cheat-sensitive weak coin flipping.
\newblock {\em Physical Review Letters}, 89:227901, 2002.

\bibitem[SS05]{qseals3}
S.~Singh and R.~Srikanth.
\newblock Quantum seals.
\newblock {\em Physica Scripta}, 71:433, 2005.

\bibitem[Sum90]{summers:qftIndep}
S.J. Summers.
\newblock On the independence of local algebras in quantum field theory.
\newblock {\em Reviews in Mathematical Physics}, 2(2):201--247, 1990.

\bibitem[Tak79]{takesaki:cstar}
M.~Takesaki.
\newblock {\em Theory of Operator Algebras I}.
\newblock Springer, 1979.

\bibitem[Tec]{magicq}
MagicQ Technologies.
\newblock http://www.magicqtech.com.

\bibitem[Ter99]{barbara:thesis}
B.~M. Terhal.
\newblock {\em Quantum Algorithms and Quantum Entanglement}.
\newblock PhD thesis, CWI and University of Amsterdam, 1999.

\bibitem[THLD02]{terhal:minfo}
B.~Terhal, M.~Horodecki, D.W. Leung, and D.P.DiVincenzo.
\newblock The entanglement of purification.
\newblock {\em J. Math. Phys.}, 43:4286--4298, 2002.

\bibitem[Tsi80]{tsirel:original}
B.~Tsirelson.
\newblock Quantum generalizations of {B}ell's inequality.
\newblock {\em Letters in Mathematical Physics}, 4:93--100, 1980.

\bibitem[Tsi87]{tsirel:separated}
B.~Tsirelson.
\newblock Quantum analogues of {B}ell inequalities: The case of two spatially
  separated domains.
\newblock {\em Journal of Soviet Mathematics}, 36:557--570, 1987.

\bibitem[Tsi93]{tsirel:hadron}
B.~Tsirelson.
\newblock Some results and problems on quantum {B}ell-type inequalities.
\newblock {\em Hadronic Journal Supplement}, 8(4):329--345, 1993.

\bibitem[TV06]{toner:blocks}
B.~Toner and F.~Verstraete.
\newblock Monogamy of bell correlations and tsirelson's bound.
\newblock quant-ph/0611001, 2006.

\bibitem[Uhl76]{Uhlmann:1976}
A.~Uhlmann.
\newblock The ``transition probability'' in the state space of a $*$-algebra.
\newblock {\em Rep. Math. Phys.}, 9(2):273--279, 1976.

\bibitem[Unr04]{unruh:compose}
D.~Unruh.
\newblock Simulatable security for quantum protocols.
\newblock quant-ph/0409125, 2004.

\bibitem[UTSM{\etalchar{+}}]{qkd:freespace}
R.~Ursin, F.~Tiefenbacher, T.~Schmitt-Manderbach, H.~Weier, T.~Scheidl,
  M.~Lindenthal, B.~Blauensteiner, T.~Jennewein, J.~Perdigues, P.~Trojek,
  B.~Oemer, M.~Fuerst, M.~Meyenburg, J.~Rarity, Z.~Sodnik, C.~Barbieri,
  H.~Weinfurter, and A.~Zeilinger.
\newblock Free-space distribution of entanglement and single photons over 144
  km.
\newblock quant-ph/0607182.

\bibitem[VB96]{VB:sp}
L.~Vandenberghe and S.~Boyd.
\newblock Semidefinite programming.
\newblock {\em SIAM review}, 38:49, 1996.

\bibitem[vD00]{wim:thesis}
W.~van Dam.
\newblock {\em Nonlocality \& Communication Complexity}.
\newblock PhD thesis, University of Oxford, Department of Physics, 2000.

\bibitem[vD05]{wim:nonlocal}
W.~van Dam.
\newblock Impossible consequences of superstrong nonlocality.
\newblock quant-ph/0501159, 2005.

\bibitem[vdG98]{vGraaf98}
J.~van~de Graaf.
\newblock {\em Towards a formal definition of security for quantum protocols}.
\newblock PhD thesis, épartment d'informatique et de r.o., Université de
  Montréal, 1998.
\newblock http://www.cs.mcgill.ca/~crepeau/PS/these-jeroen.ps.

\bibitem[Vya03]{vyalyi:qma}
M.~Vyalyi.
\newblock {QMA}={PP} implies that {PP} contains {PH}.
\newblock {\em Electronic Colloquium on Computational Complexity}, TR03-021,
  2003.

\bibitem[Wat99]{watrous:qip}
J.~Watrous.
\newblock {PSPACE} has constant-round quantum interactive proof systems.
\newblock In {\em Proceedings of 40th IEEE FOCS}, pages 112--119, 1999.
\newblock cs.CC/9901015.

\bibitem[WB05]{wocjan:mub}
P.~Wocjan and T.~Beth.
\newblock New construction of mutually unbiased bases in square dimensions.
\newblock {\em Quantum Information and Computation}, 5(2):93--101, 2005.

\bibitem[WdW05]{wehner04a}
S.~Wehner and R.~de~Wolf.
\newblock Improved lower bounds for locally decodable codes and private
  information retrieval.
\newblock In {\em Proceedings of the 32nd ICALP}, volume 3580 of {\em LNCS},
  pages 1424--1436, 2005.

\bibitem[Wer81]{werner:state}
R.F. Werner.
\newblock Quantum states with {E}instein-{P}odolsky-{R}osen correlations
  admitting a hidden-variable model.
\newblock {\em Physical Review A}, 40:4277--4281, 1981.

\bibitem[WF89]{wootters:mub}
W.K. Wootters and B.~Fields.
\newblock Optimal state-determination by mutually unbiased measurements.
\newblock {\em Ann. Phys.}, 191(368), 1989.

\bibitem[Wie83]{wiesner:conjugate}
S.~Wiesner.
\newblock Conjugate coding.
\newblock {\em Sigact News}, 15(1), 1983.

\bibitem[Wik]{ancientCrypto}
Wikibooks.
\newblock History of cryptography.
\newblock http://wikibooks.org/wiki/Cryptography:History\_of\_Cryptography.

\bibitem[Wul07]{Wullsc07}
J.~Wullschleger.
\newblock Oblivious-transfer amplification.
\newblock In {\em Advances in Cryptology --- {EUROCRYPT}~'07}, Lecture Notes in
  Computer Science, pages 555--572. Springer, 2007.

\bibitem[WW01a]{werner:bellBound}
R.F. Werner and M.M. Wolf.
\newblock All-multipartite bell-correlation inequalities for two dichotomic
  observables per site.
\newblock {\em Physical Review A}, 64:032112, 2001.

\bibitem[WW01b]{werner:overview}
R.F. Werner and M.M. Wolf.
\newblock Bell inequalities and entanglement.
\newblock {\em Quantum Information and Computation}, 1(3), 2001.

\bibitem[WW07]{wehner07b}
S.~Wehner and J.~Wullschleger.
\newblock Security in the bounded quantum storage model.
\newblock arxiv:0710.1185, 2007.

\bibitem[WY06]{wang:classification}
M.~Wang and F.~Yan.
\newblock Conclusive quantum state classification.
\newblock quant-ph/0605127, 2006.

\bibitem[Yao82]{yao:sfe}
A.~C. Yao.
\newblock Protocols for secure computations.
\newblock In {\em Proceedings of the 23rd Annual IEEE FOCS}, pages 160--164,
  1982.

\bibitem[Yao95]{yao:otFromBc}
A.~C.-C. Yao.
\newblock Security of quantum protocols against coherent measurements.
\newblock In {\em Proceedings of 20th ACM STOC}, pages 67--75, 1995.

\bibitem[YKL75]{yuen:maxState}
H.~P. Yuen, R.~S. Kennedy, and M.~Lax.
\newblock Optimum testing of multiple hypotheses in quantum detection theory.
\newblock {\em {IEEE} Transactions on Information Theory}, 21, 1975.

\bibitem[Zau99]{zauner:diss}
G.~Zauner.
\newblock {\em Quantendesigns - Grundz{\"u}ge einer nichtkommutativen
  Designtheorie}.
\newblock PhD thesis, Universit{\"a}t Wien, 1999.

\bibitem[ZLG00]{zhang:coin}
Y-S. Zhang, C-F. Li, and G-C. Guo.
\newblock Unconditionally secure quantum coin tossing via entanglement
  swapping.
\newblock quant-ph/0012139, 2000.

\end{thebibliography}

\printindex
\addcontentsline{toc}{chapter}{Figures}
\listoffigures
\symbols
\label{symbolsPage}
\begin{longtable}{|l|l|r|}
\hline
\hline
Symbol& & Page \\
\hline
\hline
$\log$ & binary logarithm & \\
$\ln$ & natural logarithm &\\
$a^*$ & complex conjugate of $a$ & \\
$|a|$ & absolute value of $a$ & \\
$|\mS|$ & number of elements of the set $\mS$ & \\
$\Natural$ & set of natural numbers $1,2,3,\ldots$&  \\
$\Real$ & set of real numbers&  \\
$\Complex$ & set of complex numbers& \\
$[n]$ & set of numbers $\{1,\ldots,n\}$&\\
$x_{|\setS}$ & string $x$ restricted to the indices in $\setS$&\\
$\delta_{ij}$ & $\delta_{ij} = 1$ if $i = j$ and $\delta_{ij} = 0$ otherwise&\\
\hline
$\id$ & identity matrix&\\
$\id_d$ & $d\times d$ identity matrix&\\
$A^{[1]}$ & matrix $A$ acting on subsystem 1&\\
$A^{-1}$ & inverse of the matrix $A$ &\\
$A^T$ & transpose of the matrix $A$& \\
$A^*$ & conjugate of the matrix $A$& \\
$A^\dagger$ & conjugate transpose of the matrix $A$& \\
$[a_{ij}]$ & matrix whose entry in the $i$-th row and $j$-th column is $a_{ij}$&\\
$\Tr(A)$ & trace of $A=[a_{ij}]$ given by $\sum_j a_{jj}$&\\
$\rank(A)$& rank of the matrix $A$&\\
$A > 0$&$A$ is positive definite&\pageref{psd}\\
$A \geq 0$&$A$ is positive semidefinite&\pageref{psd}\\
$||A||_1$ & trace norm of $A$, given by $\Tr\sqrt{A^\dagger A}$&\\
$\vec{a}$ & real vector $\vec{a} = (a_1,\ldots,a_d)$&\\
$\ket{\Psi}$ & complex vector $\ket{\Psi} = (\alpha_1,\ldots,\alpha_d)$&\\
$\ket{x_b}$ & string $x$ encoded in basis $b$&\\
$\inp{\Psi}{\Phi}$ & inner product of $\ket{\Psi}$ and $\ket{\Phi}$&\\
$x \cdot y$ & standard inner product of real vectors $x$ and $y$&\\
$\outp{\Psi}{\Phi}$ & outer product of $\ket{\Psi}$ and $\ket{\Phi}$&\\
$\outp{\Psi}{\Psi}$ & projector onto the vector $\ket{\Psi}$&\\
$||\ket{\Psi}||$ & 2-norm given by $\sqrt{\inp{\Psi}{\Psi}}$&\\
$\hil$ & a Hilbert space&\\
$\bop(\hil)$ & set of all bounded operators on $\hil$&\\
$\mS(\hil)$ & set of states on $\hil$&\\
$[A,B]$ & commutator $AB - BA$&\\
$\{A,B\}$ & anti-commutator $AB + BA$&\\
$\Comm(\algA)$ & commutant of the algebra $\algA$&\pageref{commutant}\\
$\setZ_\algA$ & center of the algebra $\algA$&\pageref{center}\\
$\langle A_1,\ldots,A_n\rangle$ & algebra generated by $A_1,\ldots,A_n$&\pageref{cstardef}\\
$\langle \mS \rangle$ & algebra generated by operators from the
set $\mS$ &\pageref{cstardef}\\
\hline
$D(\rho,\sigma)$& trace distance of $\rho$ and $\sigma$&\pageref{def:trdist}\\
$F(\rho,\sigma)$& fidelity of $\rho$ and $\sigma$&\pageref{def:fidelity}\\
$d(X|\rho)$ & distance from uniform of r.v.\ $X$ given state $\rho$&\pageref{def:uniform}\\
$h(p)$ & binary entropy&\pageref{def:binentropy}\\
$H(X,Y)$ & joint entropy of $X$ and $Y$&\pageref{def:jointentropy}\\
$H(X|Y)$ & conditional entropy of $X$ given $Y$&\pageref{def:condentropy}\\
$\mI(X,Y)$ & mutual information of $X$ and $Y$&\pageref{def:mutualinfo}\\
$\mI_c(\rho_{AB})$ & classical mutual information of $\rho_{AB}$&
\pageref{def:classicalmutualinfo}\\
$\mI_{acc}(\ens)$ & accessible information of an ensemble $\ens$&\pageref{def:iacc}\\
$S(\rho)$ & von Neumann entropy of the state $\rho$&\pageref{def:vonneumann}\\
$\chi(\rho)$ & Holevo quantity&\pageref{def:holevo}\\
$H_\infty(X)$ & min-entropy &\pageref{def:minentropy}\\
$H_2(X)$ & collision entropy&\pageref{def:collision}\\
$H_2(\rho_{AB}|\rho_B)$ & collision entropy of $\rho_{AB}$ given $\rho_B$&\pageref{def:qcollision}\\
\hline
\end{longtable}

\samenvatting

Quantum computing heeft een grote invloed op cryptografie gehad. Met de ontdekking van Shors quantum algorithme voor het factoriseren van grote getallen kunnen opeens bijna alle klassieke systemen gebroken worden zodra een quantum computer is gebouwd. Het is daarom belangrijk om andere manieren te verzinnen om veilige cryptografische protocollen te kunnen implementeren. 
Dit proefschrift draagt ertoe bij om zowel de fysieke beperkingen, als ook de mogelijkheden van cryptographie in een quantum omgeving beter te begrijpen.
Wij bekijken eerst twee aspecten die een cruciale rol spelen voor de veiligheid van quantum protocollen: onzekerheidsrelaties en quantum entanglement.
Hoe kunnen wij goede onzekerheidsrelaties voor een groot aantal meetinstellingen vinden? Wat is het effect van entanglement op klassieke protocollen? En, welke beperkingen
legt entanglement quantum protocollen op? Ten slotte, kunnen wij deze beperkingen omzeilen onder realitische aanames?

\section*{Informatie in quantum toestanden}

In dit deel houden wij ons bezig met het extraheren van informatie uit quantum toestanden.
Een van de meest fundamentele doelen is het onderscheiden van quantum toestanden. Gegeven een set
van mogelijke toestanden, wat is de toestand die wij op dit moment voor handen hebben? Wij bestuderen een variant
van dit probleem dat van belang is voor de veiligheid van protocollen in het bounded quantum storage model.
We ontvangen na de meting, of
meer algemeen nadat een quantum memory bound
toegepast wordt, 
nog extra informatie.
Wij introduceren een algemeen algebraisch
raamwerk, dat het mogelijk maakt om dit probleem voor elke set van toestanden
op te lossen en geven twee voorbeelden.

Verder onderzoeken wij entropische onzekerheidsrelaties, die
een andere manier vormen om
Heisenberg's onzekerheids principe te beschrijven.
Dit is meestal een beter manier om ``onzekerheid'' te beschrijven aangezien de ondergrens
niet afhangt van 
een bepaald toestand maar alleen van de metingen zelf.
Entropische onzekerheidsrelaties hebben recentelijk meer invloed gekregen
binnen het veld van
quantum cryptografie in het bounded
storage model, waar de veiligheid van protocollen uiteindelijk afhangt van zulke onzekerheidsrelaties.
Dus nieuwe onzekerheidsrelaties kunnen tot nieuwe protocollen leiden.

Onzekerheidrelaties zijn bekend voor twee of $d+1$ wederzijds ``unbiased measurements''.
Wij bewijzgn eerst nauwe entropische onzekerheidsrelaties voor metingen met een
groot aantal ``mutually unbiased bases'' (MUBs) in dimensionen $d=s^2$.
Wij laten ook zien dat MUBs geen goede keuze zijn voor ``locking'' van klassieke informatie in quantum toestanden;
ook als wij meer dan twee van zulke MUBs gebruiken neemt het locking effect niet toe.

Onze resultaten laten zien dat men heel voorzichtig dient te zijn om ``maximaal incompatibele'' metingen
als wederzijds ``unbiased'' te veronderstellen. 
Maar welke eigenschappen moeten een meting hebben om heel `incompatibel'
te zijn?
Gelukkig kunnen wij zulke eigenschappen vinden voor metingen met twee uitkomsten.
Voor anti-commuterende metingen die generatoren van een Clifford algebra 
vormen, bewijzen wij optimale onzekerheidsrelaties voor de Shannon entropie,
en bijna optimale relaties voor de collision entropie.
Onze resultaten kunnen worden toegepast op quantum cryptographie.

\section*{Entanglement}

In dit deel onderzoeken wij quantum entanglement. Allereerst, 
kijken wij naar Tsirelson inequalities. 
Wij laten zien hoe wij de optimale strategie
voor spelletjes met twee uitkomsten met behulp van semidefinite programming kunnen bepalen. 
Als voorbeeld laten wij een upper bound voor de gegeneraliseerde CHSH ongelijkheid zien. 

Verder laten wij zien hoe klassieke interactieve bewijssystemen met twee 
spelers (provers)
kunnen veranderen als de spelers
entanglement
kunnen delen. Dit is een voorbeeld van hoe de veiligheid van klassieke systemen kan veranderen, ook al 
is het alleen
mogelijk een beperkt soort quantum operaties uit te voeren: Het bewijssysteem wordt significant verzwakt ook al hebben
de spelers geen toegang tot een quantum computer. 

\section*{Applicaties voor de cryptografie}

In deel IV onderzoeken wij de consequenties van onzekerheidsrelaties 
en entanglement in quantum systemen
voor de cryptografie.
Traditioneel houdt de cryptografie zich vooral bezig met het veilig versturen van berichten.
Maar met de opkomst van het internet zijn nieuwe taken van belang geworden. 
Wij willen protocollen creeren voor het elektronisch stemmen, online veilingen,
ondertekenen van contracten en vele andere applicaties,
waarbij de deelnemers elkaar niet vertrouwen. 
De focus ligt daarbij op twee primitieven, met behulp waarvan wij al deze problemen kunnen oplossen:
bit commitment en oblivious transfer.
Klassieke protocollen voor deze primitieven zijn gebaseerd op computationele 
aanames die met behulp van een
quantum computer gebroken kunnen worden. Helaas is het bekend dat zelfs in de quantum wereld deze primitieven
niet helemaal zonder aannames geimplementeerd kunnen worden. Wat hopen wij dan 
wel te kunnen bereiken?

Als bit commitment onmogelijk is, kunnen wij misschien de taak een klein beetje aanpassen en dan 
nuttige protocollen
vinden? Hier bekijken wij commitments van een hele string van bits tegelijk, waar de tegenstander niet 
is beperkt.
Als bit commitment onmogelijk is, is perfecte string commitment ook niet mogelijk. Maar wij geven 
elke tegenstander
de mogelijkheid om een beetje vals te spelen.
Wij geven een raamwerk voor een familie van string commitment protocollen. Hoe wij informatie meten blijkt
een cruciale rol te spelen; voor een heel sterke maat van informatie laten wij zien dat zelfs deze imperfecte
string commitments niet mogelijk zijn. Maar voor een zwakkere manier om informatie te meten construeren wij
toch niet-triviale protocollen die klassiek niet mogelijk zijn.

Ten slotte 
laten wij zien dat bit commitment en oblivious transfer wel mogelijk worden, indien
wij de tegenstander realistische 
beperkingen opleggen. Wij introduceeren het noisy-storage model, dat nauw
is gerelateerd aan het bounded-storage model.
Wij laten zien dat het mogelijk is om oblivious transfer
te implementeren, zolang de tegenstander qubits niet zonder fouten kan 
opslaan.
Gegeven de status van
de experimentele mogelijkheden vandaag de dag, lijkt dit een realistische aanname, maar is afhankelijk van de implementatie moeilijk
te bepalen. Dezelfde problemen 
die het ook zo moeilijk maken om een quantum computer te bouwen komen
ons hier ten goede! 

\abstract

Quantum computing had a profound impact on cryptography. Shor's discovery of an efficient quantum algorithm for factoring large integers implies that nearly all existing classical systems based on computational assumptions can be broken, once a quantum computer is built. It is therefore imperative to find other means of implementing secure protocols. This thesis aims to contribute to the understanding of both the physical limitations, as well as the possibilities of cryptography in the quantum setting. To this end, we first investigate two notions that are crucial to the security of quantum protocols: uncertainty relations and entanglement. How can we find good uncertainty relations for a large number of measurement settings? How does the presence of entanglement affect classical protocols? And, what limitations does it impose on implementing quantum protocols? 
Finally, can we circumvent some of those limitations using realistic assumptions?

\section*{Information in Quantum States}

In this part, we start by investigating how to extract information from quantum states. 
One of the most basic tasks is the discrimination of quantum states. Given an ensemble
of known quantum states, which one do we hold in our hands? We study a variant of this problem which is of
central importance for the security of protocols in the bounded-quantum-storage model.
Here, we are given additional
information about the state after the measurement or, more generally, after a quantum
memory bound is applied. We prove general bounds on the success probability
which answer in the negative the question whether deterministic privacy 
amplification is possible in all known protocols in the bounded-quantum-storage model. 
To this end, we introduce a general algebraic framework which allows us
to solve this problem for any set of states and provide two explicit examples.

We then turn to examine entropic uncertainty relations, which are an alternative way to state Heisenberg's uncertainty principle.
They are frequently a more useful characterization, because the ``uncertainty'' is lower
bounded by a quantity that does not depend on the state to be measured.
Recently, entropic uncertainty relations have gained importance in the context of quantum cryptography
in the bounded-storage model, where proving the security of protocols ultimately
reduces to bounding such relations. Proving new entropic uncertainty relations could thus give rise to new protocols.
Such relations are known for two or $d+1$ mutually unbiased measurements.
We prove tight entropic uncertainty relations for measurements in a large number of
specific mutually unbiased bases (MUBs) in square dimensions.
In a similar way, we show that such MUBs are unsuitable for locking classical correlations in quantum states: Using 2 or
all of them does not increase the locking effect.

Our result shows that one needs to be careful about thinking of ``maximally incompatible'' measurements
as being necessarily mutually unbiased. But what properties do measurements
need to have in order to give strong uncertainty relations?
We find very strong
uncertainty relations from the generators of a Clifford algebra.
In particular, we prove that for $k$ such anti-commuting observables $X_1,\ldots,X_k$ we obtain optimal uncertainty relations for the Shannon entropy
and nearly optimal relations for the collision entropy.
Our results have immediate applications to quantum cryptography in the
bounded-storage model.

\section*{Entanglement}

In this part, we investigate the intriguing notion of quantum entanglement. 
We demonstrate how to find the optimal quantum strategies for correlation
inequalities where each measurement has exactly two outcomes using semidefinite programming. As an example, we prove a tight upper
bound for a well-known generalized CHSH inequality. 

Furthermore, we consider how a classical two-prover interactive proof system changes if the provers are allowed to share entanglement. 
In this setting, a polynomial time bounded verifier is allowed to ask questions to two
unbounded provers, who are trying to convince the verifier of the validity of a specific statement, even if the statement is false. The provers may thereby 
agree on any strategy ahead of time, but can no longer communicate once the
protocol starts.
Surprisingly, it turns out that, when the provers are allowed to share entanglement,
it is possible to simulate two such classical provers using a single quantum prover. This indicates that
entanglement among provers truly weakens the proof system, and 
provides an example of how classical systems can be affected, 
even if we only allow a very limited set of quantum operations.

\section*{Applications to Cryptography}

In this part, we consider the consequences of 
uncertainty relations and entanglement in quantum systems to cryptography. 
Traditional cryptography is concerned with the secure and reliable transmission of messages. 
With the advent of widespread electronic communication and the internet, however, new cryptographic tasks have 
become increasingly important. We would like to construct secure protocols for electronic voting, online auctions, 
contract signing and 
many other applications where the protocol participants themselves do not trust each other. 
main focus is on two primitives, which
form an important building block for constructing multi-party protocols: bit commitment and oblivious transfer.
Classical protocols for such problems are usually based on computational assumptions which do not stand up to a 
quantum computer. Unfortunately, it has been shown that even quantum computers do not help in this case and that 
perfect quantum bit commitment and oblivious transfer are impossible. In the face of such negative statements, 
what can we still hope to achieve?

Given that perfect bit commitment is impossible, perhaps we can alter the task slightly and obtain useful protocols?
Here, we considered commitments to an entire string of bits at once, when the attacker has unbounded resources at his disposal. Evidently, if perfect bit commitment is impossible, perfect string commitment is also impossible as well. However, we showed that we can obtain non-trivial quantum protocols, where the participants have a small ability to cheat.
To this end, we introduced a framework for the classification of string commitment protocols.
In particular, we proved that the measure of information is crucial to the security: For a very strong notion of security, we
showed that even slightly imperfect quantum string commitments are also impossible. Nevertheless, we showed
that for a weaker measure of information we can indeed obtain nontrivial protocols, which are impossible in a classical world. 

Luckily, it turns out that we can implement oblivious transfer if we
are willing to assume that storing qubits is noisy.
We introduce the model of noisy quantum storage, which
is similar to the model of bounded quantum storage. Here, however, we consider an explicit noise model inspired by present day technology. If the honest parties can perform perfect quantum operations, then we show that the protocol is secure for any amount of noise. In case the honest participants are only able to perform noisy operations themselves, we analyze a practical protocol that can be implemented using present-day hardware. We show how to derive explicit tradeoffs between the amount of storage noise, the amount of noise in the operations performed by the honest participants and the security of the protocol. 
Here, the very problem that makes it so hard to implement a quantum computer
can actually be turned to our advantage.


\cleardoublepage 

\pagestyle{empty}

\noindent
{\em Titles in the ILLC Dissertation Series:}

\newcommand{\illcpublication}[3]{\item[ILLC #1: ]{\bf #2}\\{\em #3}}

\begin{list}{}{ \settowidth{\leftmargin}{ILL}
		\setlength{\rightmargin}{0in}
		\setlength{\labelwidth}{\leftmargin}
		\setlength{\labelsep}{0in}
}

\illcpublication{DS-2001-01}{Maria Aloni}{Quantification under Conceptual Covers}
\illcpublication{DS-2001-02}{Alexander van den Bosch}{Rationality in Discovery - a study of Logic, Cognition, Computation and Neuropharmacology}
\illcpublication{DS-2001-03}{Erik de Haas}{Logics For OO Information Systems: a Semantic Study of Object Orientation from a Categorial Substructural Perspective}
\illcpublication{DS-2001-04}{Rosalie Iemhoff}{Provability Logic and Admissible Rules}
\illcpublication{DS-2001-05}{Eva Hoogland}{Definability and Interpolation: Model-theoretic investigations}
\illcpublication{DS-2001-06}{Ronald de Wolf}{Quantum Computing and Communication Complexity}
\illcpublication{DS-2001-07}{Katsumi Sasaki}{Logics and Provability}
\illcpublication{DS-2001-08}{Allard Tamminga}{Belief Dynamics. (Epistemo)logical Investigations}
\illcpublication{DS-2001-09}{Gwen Kerdiles}{Saying It with Pictures: a Logical Landscape of Conceptual Graphs}
\illcpublication{DS-2001-10}{Marc Pauly}{Logic for Social Software}
\illcpublication{DS-2002-01}{Nikos Massios}{Decision-Theoretic Robotic Surveillance}
\illcpublication{DS-2002-02}{Marco Aiello}{Spatial Reasoning: Theory and Practice}
\illcpublication{DS-2002-03}{Yuri Engelhardt}{The Language of Graphics}
\illcpublication{DS-2002-04}{Willem Klaas van Dam}{On Quantum Computation Theory}
\illcpublication{DS-2002-05}{Rosella Gennari}{Mapping Inferences: Constraint Propagation and Diamond Satisfaction}
\illcpublication{DS-2002-06}{Ivar Vermeulen}{A Logical Approach to Competition in Industries}
\illcpublication{DS-2003-01}{Barteld Kooi}{Knowledge, chance, and change}
\illcpublication{DS-2003-02}{Elisabeth Catherine Brouwer}{Imagining Metaphors: Cognitive Representation in Interpretation and Understanding}
\illcpublication{DS-2003-03}{Juan Heguiabehere}{Building Logic Toolboxes}
\illcpublication{DS-2003-04}{Christof Monz}{From Document Retrieval to Question Answering}
\illcpublication{DS-2004-01}{Hein Philipp R\"ohrig}{Quantum Query Complexity and Distributed Computing}
\illcpublication{DS-2004-02}{Sebastian Brand}{Rule-based Constraint Propagation: Theory and Applications}
\illcpublication{DS-2004-03}{Boudewijn de Bruin}{Explaining Games. On the Logic of Game Theoretic Explanations}
\illcpublication{DS-2005-01}{Balder David ten Cate}{Model theory for extended modal languages}
\illcpublication{DS-2005-02}{Willem-Jan van Hoeve}{Operations Research Techniques in Constraint Programming}
\illcpublication{DS-2005-03}{Rosja Mastop}{What can you do? Imperative mood in Semantic Theory}
\illcpublication{DS-2005-04}{Anna Pilatova}{A User's Guide to Proper names: Their Pragmatics and Semanics}
\illcpublication{DS-2005-05}{Sieuwert van Otterloo}{A Strategic Analysis of Multi-agent Protocols}
\illcpublication{DS-2006-01}{Troy Lee}{Kolmogorov complexity and formula size lower bounds}
\illcpublication{DS-2006-02}{Nick Bezhanishvili}{Lattices of intermediate and cylindric modal logics}
\illcpublication{DS-2006-03}{Clemens Kupke}{Finitary coalgebraic logics}
\illcpublication{DS-2006-04}{Robert \v{S}palek}{Quantum Algorithms, Lower Bounds, and Time-Space Tradeoffs}
\illcpublication{DS-2006-05}{Aline Honingh}{The Origin and Well-Formedness of Tonal Pitch Structures}
\illcpublication{DS-2006-06}{Merlijn Sevenster}{Branches of imperfect information: logic, games, and computation}
\illcpublication{DS-2006-07}{Marie Nilsenova}{Rises and Falls. Studies in the Semantics and Pragmatics of Intonation}
\illcpublication{DS-2006-08}{Darko Sarenac}{Products of Topological Modal Logics}
\illcpublication{DS-2007-01}{Rudi Cilibrasi}{Statistical Inference Through Data Compression}
\illcpublication{DS-2007-02}{Neta Spiro}{What contributes to the perception of musical phrases in western classical music?}
\illcpublication{DS-2007-03}{Darrin Hindsill}{It's a Process and an Event: Perspectives in Event Semantics}
\illcpublication{DS-2007-04}{Katrin Schulz}{Minimal Models in Semantics and Pragmatics: Free Choice, Exhaustivity, and Conditionals}
\illcpublication{DS-2007-05}{Yoav Seginer}{Learning Syntactic Structure}
\illcpublication{DS-2008-01}{Stephanie Wehner}{Cryptography in a Quantum World}
\end{list}

\end{document}